\documentclass[12pt, reqno]{amsart}
\usepackage{amsmath, amsthm, amscd, amsfonts, amssymb, graphicx, color}
\usepackage[bookmarksnumbered, colorlinks, plainpages]{hyperref}
\usepackage{pstricks}
\usepackage{pspicture,pst-all,multido}
\usepackage[utf8]{inputenc}

\theoremstyle{definition}

\theoremstyle{remark}

\numberwithin{equation}{section}

\newcommand{\e}{\operatorname{e}}

\renewcommand{\i}{\operatorname{i}}

\newcommand{\C}{\mathbb{C}}
\newcommand{\R}{\mathbb{R}}
\newcommand{\N}{\mathbb{N}}

\DeclareMathOperator{\im}{Im}

\newcommand{\norm}[1]{\parallel\!\!#1\!\!\parallel}

\newcounter{envcount}%

\newenvironment{Def}%
{\vspace{\bigskipamount}\refstepcounter{envcount}\textbf{(\theenvcount)\enspace Definition.}}%
  {\vspace{\bigskipamount}}

\newenvironment{POLWP}%
{\vspace{\bigskipamount}\refstepcounter{envcount}\textbf{(\theenvcount)\enspace POL for Weyl particles.}}%
  {\vspace{\bigskipamount}}

\newenvironment{Sum}%
{\vspace{\bigskipamount}\refstepcounter{envcount}\textbf{(\theenvcount)\enspace Summary.}}%
  {\vspace{\bigskipamount}}

\newenvironment{Tir}%
{\vspace{\bigskipamount}\refstepcounter{envcount}\textbf{(\theenvcount)\enspace Time Reversal.}}%
  {\vspace{\bigskipamount}}
  
\newenvironment{L-CS}%
{\vspace{\bigskipamount}\refstepcounter{envcount}\textbf{(\theenvcount)\enspace Large-\,$t_{\overline{e}}$\,-states.}}%
  {\vspace{\bigskipamount}}

\newenvironment{AC}%
{\vspace{\bigskipamount}\refstepcounter{envcount}\textbf{(\theenvcount)\enspace Asymptotic causality.}}%
  {\vspace{\bigskipamount}}

\newenvironment{Exa}%
{\vspace{\bigskipamount}\refstepcounter{envcount}\textbf{(\theenvcount)\enspace Example.}}%
  {\vspace{\bigskipamount}}

\newenvironment{Exam}%
{\vspace{\bigskipamount}\refstepcounter{envcount}\textbf{(\theenvcount)\enspace Example}}%
  {\vspace{\bigskipamount}}

\newenvironment{Con}%
{\vspace{\bigskipamount}\refstepcounter{envcount}\textbf{(\theenvcount)\enspace Construction}}%
  {\vspace{\bigskipamount}}

\newenvironment{DMBC}%
{\vspace{\bigskipamount}\refstepcounter{envcount}\textbf{(\theenvcount)\enspace Discussion. }}%
  {\vspace{\bigskipamount}}  

\newenvironment{The}%
{\vspace{\bigskipamount}\refstepcounter{envcount}\textbf{(\theenvcount)\enspace Theorem.}\itshape}%
  {\vspace{\bigskipamount}\upshape}
  
\newenvironment{Theo}%
{\vspace{\bigskipamount}\refstepcounter{envcount}\textbf{(\theenvcount)\enspace Theorem}\itshape}%
  {\vspace{\bigskipamount}\upshape}
  
\newenvironment{Pro}%
{\vspace{\bigskipamount}\refstepcounter{envcount}\textbf{(\theenvcount)\enspace Proposition.}\itshape}%
  {\vspace{\bigskipamount}\upshape}

\newenvironment{Cor}%
{\vspace{\bigskipamount}\refstepcounter{envcount}\textbf{(\theenvcount)\enspace Corollary.}\itshape}%
  {\vspace{\bigskipamount}\upshape}

\newenvironment{CFDNWL}%
{\vspace{\bigskipamount}\refstepcounter{envcount}\textbf{(\theenvcount)\enspace Frame-dependence of NWL.}\itshape}%
  {\vspace{\bigskipamount}\upshape}

 \newenvironment{Constr}%
{\vspace{\bigskipamount}\refstepcounter{envcount}\textbf{(\theenvcount)\enspace Construction of a late-change state.}\itshape}%
  {\vspace{\bigskipamount}\upshape}

\newenvironment{Lem}%
{\vspace{\bigskipamount}\refstepcounter{envcount}\textbf{(\theenvcount)\enspace Lemma.}\itshape}%
  {\vspace{\bigskipamount}\upshape}

\newenvironment{Examp}%
{\vspace{\bigskipamount}\refstepcounter{envcount}\textbf{(\theenvcount)\enspace Example.}\itshape}%
  {\vspace{\bigskipamount}\upshape}

\newenvironment{Exampl}%
{\vspace{\bigskipamount}\refstepcounter{envcount}\textbf{(\theenvcount)\enspace Example}\itshape}%
  {\vspace{\bigskipamount}\upshape}

  {\vspace{\bigskipamount}}

\newenvironment{Int}%
{\vspace{\bigskipamount}\refstepcounter{envcount}\textbf{(\theenvcount)\enspace Interpretation.}\itshape}%
  {\vspace{\bigskipamount}\upshape}

\newenvironment{CCS}%
{\vspace{\bigskipamount}\refstepcounter{envcount}\textbf{(\theenvcount)\enspace Canonical cross section.}}%
  {\vspace{\bigskipamount}}

\theoremstyle{definition}
\swapnumbers

\theoremstyle{remark}

\begin{document}
\setcounter{page}{1}


\vspace{3mm}

\title{Dirac and Weyl Fermions - the Only Causal   Systems}  

\author[D.P.L. Castrigiano]{Domenico P.L. Castrigiano$^*$}
\address{$^{*}$Technische Universit\"at M\"unchen, Fakult\"at f\"ur Mathematik, M\"unchen, Germany}
\email{\textcolor[rgb]{0.00,0.00,0.84}{castrig@ma.tum.de}}

\begin{abstract} 
Causal systems describe the localizability of  relativistic quantum systems complying  with the principles of special relativity and elementary causality. At their classification  we restrict ourselves to real mass and finite spinor systems. It follows that (up to certain  not yet discarded unitarily related systems) the only irreducible causal systems are the Dirac and the Weyl fermions.  
Their wave-equations are established as a mere consequence of causal localization. --- The  bounded localized Dirac and Weyl wave-functions are studied in detail.  One finds that, at the speed of light,  the carriers shrink in the past and  expand in the future. For every direction in space  there is a definite time at which the change from shrinking to expanding occurs. A late changing time characterizes those states, which shrink to a $\delta$-strip if  boosted in the opposite direction. Using a density result for these late-change states one shows that all Dirac and Weyl wave-functions are subjected to Lorentz contraction. The latter is discussed in some detail. --- We tackle the question whether a causal system induces  a representation of a causal logic and thus provides a localization in proper space-time regions rather than on spacelike hyperplanes. The causal logic generated  by the spacelike relation is shown to do not admit representations at all. But the logic generated by the non-timelike relation in general does, and the necessary condition is derived that there is a projection valued measure on every non-timelike not spacelike hyperplane being the high boost limit of the localization on the spacelike hyperplanes.  Dirac and Weyl systems are shown to satisfy this condition and thus to extend to all non-timelike hyperplanes, which implies more profound properties of the causal systems. The  bounded localized eigenstates of the projections to not spacelike flat strips are  late-change states.\\

{\it Keywords: Relativistic quantum systems, localization, causality, causal time evolution, Lorentz contraction, representation of causal logic}

\end{abstract}

\maketitle

\tableofcontents 

\section{Introduction}\label{I}



These investigations are concerned with causal systems $(W,E)$ (\ref{CS}), which describe the localizability of  relativistic quantum systems following Wightman \cite{W62}. What is going beyond the physical ideas developed by Newton, Wigner \cite{NW} and Wightman \cite{W62} is the inclusion  of special relativity and causality. So the representation $W$ of the  Poincar\'e group $\tilde{\mathcal{P}}$ describing the relativistic quantum system under consideration and  the projection valued measure (PM)  $E$ describing the localization of the system on every spacelike hyperplane of Minkowski space  satisfy Poincar\'e covariance (i) and the causality condition (ii), namely
\begin{itemize}
\item[(i)] $W(g)E(\Delta)W(g)^{-1}=E(g\cdot \Delta)$ 
\item[(ii)] $E(\Delta)\le E(\Delta_\tau)$ 
\end{itemize}
Poincar\'e covariance (i) refers to every Poincar\'e transformation $g$ and every spacelike flat region $\Delta$. As to the causality condition (ii), $\tau$ is any spacelike hyperplane and $\Delta_\tau$ denotes the region of influence of $\Delta$ in $\tau$ (\ref{RIHPR}).  Equivalent formulations of the causality condition, as local orthogonality $E(\Delta)E(\Gamma)=0$ for spacelike separated $\Delta$ and $\Gamma$, are given in (\ref{VEFCPOL}).

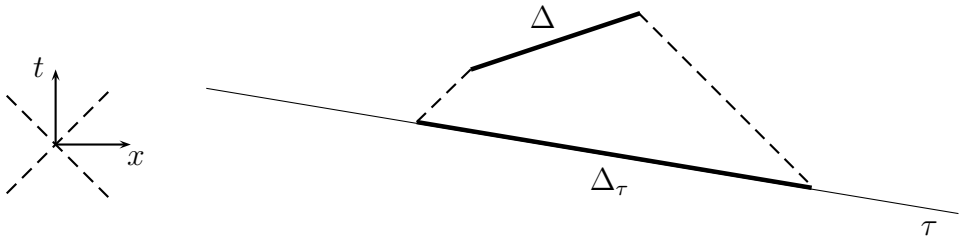
\begin{center}
\setlength{\unitlength}{1cm}
\begin{pspicture}(-2,0.5)(5,3.5)

\psline[linestyle=dashed](0.52,2.5)(-0.2,1.8)
\psline[linestyle=dashed](2.75,3.25)(5,1)
\put(-3,2.25){\line(6,-1){10}}

\psline{->}(-5,1.5)(-5,2.5)
\psline{->}(-5,1.5)(-4,1.5)
\psline[linestyle=dashed](-5.65,0.85)(-4.3,2.2)
\psline[linestyle=dashed](-5.65,2.15)(-4.3,0.8)

\linethickness{0,6mm}
\put(0.52,2.5){\line(3,1){2.25}}
\put(-0.2,1.8){\line(6,-1){5.25}}

\put(6.5,0.3){$\tau$}
\put(2.1,0.9){$\Delta_\tau$}
\put(1.3,3.05){$\Delta$}
\put(-5.3,2.4){$t$}
\put(-4.05,1.232){$x$}
\end{pspicture}
\end{center}
Hence the concept of a causal system is the unique one satisfying a number of minimal physical requirements. Moreover, the postulate (i) implies no restriction for relativistic systems  which are localized according to Wightman \cite{W62}, since as shown in \cite{CM82} and (\ref{PCPOL}) the localization automatically and unambiguously extends to all spacelike hyperplanes in a Poincar\'e covariant manner. What is really restrictive is the causality condition (ii). 
In particular there is no causal system $(W,E)$ if the energy $H$ is semi-bounded (\ref{AOLSBR}).  Since bounded below energy  is essential for  the  stability of the system one could infer from  this fact that localization in terms of position operators is incompatible with causality in relativistic quantum physics \cite{Y12}.\\
\hspace*{6mm}
 We hasten to expound  already at this point  that  this conclusion might  be not definitive. Indeed, in sec.\,\ref{NES} as in  \cite[sec.\,J]{CL15}, we argue that applying the localization operator $E(\Delta)$ of a causal system  to a particle (or antiparticle) state does not create a pure state but a mixed state of particle and antiparticle states, which of course have non-negative energy. We think this feature of causal systems is fascinating as it puts first the fundamental interplay in nature of matter and antimatter, as summed up   
 at the creation of matter from energy, where inevitably  matter and antimatter occur in a bookkeeping manner.\\ 
\hspace*{6mm}
Furthermore,
the localization properties of the resulting particle and antiparticle states are described by a POL (positive operator valued localization), which is the trace of the causal system on the particle and antiparticle subspace, respectively. It is causal since it still satisfies  Poincar\'e covariance (i) and the causality condition (ii). POL are studied in \cite[sec.\,G]{CL15} and in sec.\,\ref{SePOL},\,\ref{CTECL},\,\ref{SPLSS}. The results are exploited for the causal electron POL $T^e$ in \cite[sec.\,I]{CL15} and sec.\,\ref{CEL} proving among other things that $T^e$ is separated which implies that arbitrarily well localized   electrons in any neighborhood of every point exist. The causal  POL $T^p$ for the positron  differs from $T^e$ merely by a label change and shares all the properties of the latter.
The analogous properties are derived for the causal POL $T^{\chi\eta}$ of the four Weyl particles in (\ref{POLWP}).\\
\hspace*{6mm}
 The  existence of a causal POL $T$ describing the localization of a positive-energy system is not at all a trivial fact. As shown in \cite[Lemma 4]{CL15}, the localization operators $T(\Delta)\ne 0$ for bounded regions $\Delta$ are not local observables of a local quantum theory,  included the case of modular localization.  They might not be  observables at all,  although their expectation values give the localization probabilities. We do not pursue these questions here.\\
\hspace*{6mm}
Last and most important, the physical interpretation of relativistic quantum field theory necessitates a free particle theory equipped with a  causal and hence frame-independent localization  for the description of the incoming and outgoing particle states of a collision process.  One adopts  the Newton-Wigner localization, which is  frame-dependent (\ref{CFDNWL}) but is asymptotically causal (\ref{AC}) and, referring to \cite[sec.\,1]{S11}, is at least asymptotically  frame-independent;  nevertheless, alluding to an expression in \cite[sec.\,1]{S11} we think one should not content oneself with the glass only half full.

Proceeding in order, the first  sections of this article are concerned with the basic considerations on  localizability, mainly in order to recapitulate and elucidate   the well-known  difficulties  one meets  in relativistic quantum mechanics. Let us mention that the outstanding problems maintain a  vivid interest  in a theory of relativistic particle localization so that from the beginning of relativistic quantum theory up to now a vast literature is devoted to this subject.   For a selection of about fifty  articles from  1930 up to 2000 see   
 \cite{BFM05} 2005. For further references  see 
 \cite{EM17} 2017. Also more recently,  in 
 \cite{CL15} 2015 a thorough investigation of causal time evolution for massive relativistic quantum systems is presented, to which we will refer  recurrently. Some indicated parts of it are simply taken over for the sake of completeness.

We start in sec.\,\ref{NWPOWL}  with a brief presentation of the Wightman localization (WL) and the particular Newton-Wigner localization (NWL). In sec.\,\ref{POCOPO} Poincar\'e covariance of any WL   for a relativistic system (and the associated position operator) is established, i.e., the ingredient (i) for a causal system is ensured, and thus  a widespread objection against WL, in particular NWL, is removed. But despite of being Poincar\'e covariant, NWL does not localize frame-independently (\ref{CFDNWL}). We do not renounce to give a rigorous proof of this property (which certainly will exist in literature).  This is one of the inadequacies of  NWL.  Another shortcoming, still compatible with Poincar\'e covariance,  is the  instantaneous spreading to infinity of the wave-functions.  On this it is worked intensively. In essence,  {\it causal time evolution} and {\it localized state} are contrasted under the premise of semi-bounded energy. Particularly in the  work of Hegerfeldt this is done in a concise manner. This egged us on to use Occam's razor for a further elaboration (similar in the spirit to \cite{PW77}). The achievement  in sec.\,\ref{PELS} is an unbiased view at the notion of {\it localized state} being rid of  localization operators. \\
\hspace*{6mm}
Instantaneous spreading and frame-dependent localization both violate the causality condition (ii). The former attracts up to now much more attention, perhaps because  it is perceived to be   the quintessence of acausality. However, it is not. Indeed, systems with causal time evolution (SCT)  (\ref{SCTER}), i.e., systems satisfying (i) and the causality condition (ii)  for time translations, need not be causal.  The massive respectively massless SCT in (\ref{SCTLFD}) and  (\ref{FDWTL}) do not localize frame-independently  and hence are not causal. Moreover, not even causal time evolution and frame-independent localization  together imply causality as the example (\ref{ESTFINC}) shows.

So far this brief exposition already makes clear that  the alternative to pursue maintaining {\it causality\,} is either to  suppose a {\it semi-bounded Hamiltonian (stability)} or  to suppose {\it states localized in bounded regions}. However,  as already indicated, these assumptions seemingly opposed to each other are simultaneously realized and  intrinsically linked in causal systems.

One of our main results is the classification of the causal systems $(W,E)$.  We restrict ourselves to systems with non-negative mass squared operator  and finite spinor dimension. This is justified by the fact that representations of $\tilde{\mathcal{P}}$ not satisfying these spectral conditions have not found applications so far. (Only more recently it is suggested that the quantum fields, which are covariant with respect to the massless infinite spin representations, could be attributed to the galactic dark matter.) Within this frame the result obtained in sec.\,\ref{DCRS}, \ref{MCSR}, \ref{SMLCS} is that the Dirac system for every positive mass and the two massless right- and left-handed Weyl systems are the only irreducible causal systems.\\
\hspace*{6mm}
As to this result, actually we did not yet succeed to rule out certain  unlikely systems. See the conclusions at the end of sec.\,\ref{MCSR} and sec.\,\ref{SMLCS}.
At any rate the Dirac equation and the Weyl equations are established as a mere consequence of the principle of causality in relativistic quantum mechanics. In particular, the requirement of a causal localization  determines the right  handedness of particle and antiparticle constituting a Weyl system.
\\
\hspace*{6mm}
 The classification of the causal systems proceeds in three steps. The first one (a) consists in determining the causal time evolutions for WL. This is  the most involved part of the classification. Then (b) one establishes all possible extensions of the representations in (a) of the little kinematical group (i.e., the Euclidean transformations plus time translations)  to  representations of $\tilde{\mathcal{P}}$. Moreover one applies (\ref{PCPOL}) in order to define in a Poincar\'e covariant manner the localizations operators for all  flat spatial regions. Thus one obtains the SCT. They are more general than causal systems. In the last step (c) one has to select the causal systems among the SCT. In particular one discards all SCT which do not localize frame-independently. \\
\hspace*{6mm}
For massive systems the steps (a) and (b) are done  in \cite{CL15} yielding a complete and explicit description of the SCT reported in sec.\,\ref{CMSCT}.  The 
massless SCT are also completely classified in an analogous manner (see sec.\,\ref{MWLWCTE} for step (a) and sec.\,\ref{MLST} for step (b)), although the group theoretical part is more involved. Step (c) (see sec.\,\ref{FDMCS} and  sec.\,\ref{MLCS}) 
 is tedious and, as mentioned,  not yet fully accomplished.

There are some causal systems for which the mass squared operator has a spectrum containing $]-\infty,0]$. They are considered 
to be not physically relevant for its imaginary mass spectrum. Nevertheless we present them in sec.\,\ref{SRLBSGNTLR}, since they induce  representations  of the causal logic generated by the non-timelike relation. To our knowledge these are the only known examples of representations of a causal logic.   This topic, addressed below,  is treated in sec.\,\ref{HLCCR}, \ref{LOCCR}, and  \ref{LGNTLR}, \ref{SRLBSGNTLR}, \ref{ECLNTLHP}.

Much attention we pay to the fundamental causal systems, the Dirac system for every positive mass in sec.\,\ref{DS} up to sec.\,\ref{NES} and the two Weyl systems in sec.\,\ref{OWL}, \ref{LOCCRS}
obtaining important results. As the first act of course we verify  the causality condition (ii) for these systems (see (\ref{CSDF}),\,(\ref{CSWF})).\\
\hspace*{6mm}
   We study  in detail the free motion of the border of a bounded localized (i.e., localized in a bounded region) wave function. One finds that at any time the border moves at the speed of light. As expected, over the long term it moves radially to infinity. In the short term however the movement of the wave border is more complicated as parts of it may move in the opposite direction. \\
\hspace*{6mm}
To be more explicit, for every bounded localized wave function $\psi$ and
any direction in  position space   given by  a unit vector $e\in\R^3$, let
 $e(\psi)$ denote the largest real number such that the half-space $\{x\in\R^3: xe \le e(\psi)\}$ and the carrier $\{x\in\R^3:\psi(x)\ne 0\}$ of $\psi$ are  disjoint up to a null set. The plane $\{x\in\R^3: xe = e(\psi)\}$
 is perpendicular to $e$ and tangential to the carrier. \\
 
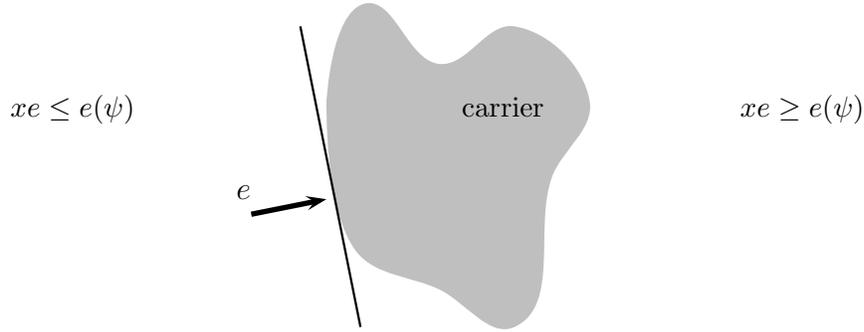
\begin{center}
\begin{pspicture}(3,3.5)
\pscurve*[linecolor=lightgray](0.5,3)(1,4.3)(2,3.5)(3,4)(4,3)(3.5,2)(3,0)(2,0.5)(0.9,1)(0.5,3)

\put(2.3,2.8){\small{carrier}}
\put(-3.7,2.8){\small{$xe\le e(\psi)$}}
\put(6,2.8){\small{$xe\ge e(\psi)$}}
\psline[linewidth=2pt]{->}(-0.5,1.5)(0.5,1.7)
\put(-0.7,1.7){$e$}
\linethickness{0.3mm}
\put(0.95,0){\line(-1,5){0.8}}
\end{pspicture}
\end{center} 
 Now the result (\ref{SETIIED}) for the evolution $\psi_t$ of $\psi$ is that there is a unique time $t_e=t_e(\psi)$ with
  \begin{equation*}
e(\psi_t)= e(\psi)+|t_e|-|t-t_e|
\end{equation*}

for all times $t\in\R$.

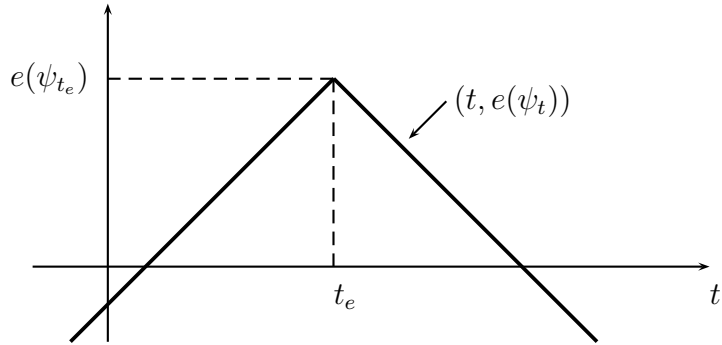
\begin{pspicture}(-6,-1)(3,3.5)
\psline{->}(-2,-1)(-2,3.5)
\psline{->}(-3,0)(6,0)
\psline{->}(2.5,2.2)(2,1.7)
\put(2.6,2.1){$(t,e(\psi_t))$}
\linethickness{0,5mm}
\put(1,2.5){\line(-1,-1){3.5}}
\put(1,2.5){\line(1,-1){3.5}}
\psline[linestyle=dashed](1,2.5)(1,0)
\psline[linestyle=dashed](1,2.5)(-2,2.5)
\put(1,-0.5){$t_e$}
\put(6,-0,5){$t$}
\put(-3.3,2.4){$e(\psi_{t_e})$}
\end{pspicture}

As long as $t<t_e$ one has 
$e(\psi_t)=e(\psi_{t_e})-t_e +t$. This means that, at the speed  of light,
   the carrier of $\psi_t$ retreats in the direction $e\in\R^3$. At time $t_e$ there is an abrupt change of motion. For $t>t_e$ the carrier advances in direction $-e$ with light velocity  as $e(\psi_t)=e(\psi_{t_e})+t_e -t$. 

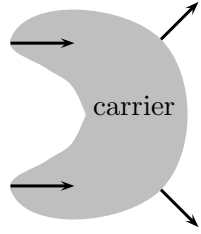
\begin{center}
\begin{pspicture}(0,-1,5)(0,1.6)
\pscurve*[linecolor=lightgray](0,0)(-0.2,0.4)(-0.5,0.6)(-1,1)(1,1)(1,-1)(-1,-1)(-0.5,-0.6)(-0.2,-0.4)(0,0)
\put(0.1,0){\small{carrier}}
\psline[linewidth=0.4mm]{->}(-1,-0.95)(-0.15,-0.95)

\psline[linewidth=0.4mm]{->}(-1,0.95)(-0.15,0.95)
\psline[linewidth=0.4mm]{->}(1,1)(1.5,1.5)
\psline[linewidth=0.4mm]{->}(1,-1)(1.5,-1.5)
\end{pspicture}
\end{center}

Only then the wave function expands in  direction $-e$ as expected. Hence, despite of this simple motion, 
 for a short period the picture is complicated since  the time of change $t_e$ depends in general on the direction $e$ (see (\ref{NISE})). The carrier of the wave function performs the change from shrinking to expanding not isotropicly. But not later than  the time corresponding to the diameter of the carrier a simultaneous  isotropic expansion  of the wave function at light speed  takes place (\ref{LTSIE}). \\
\hspace*{6mm}
The changes  of the carrier due to a boost are quite different. Let $\psi_{\rho e}$ denote the wave function $\psi$ boosted along the direction $e$ with rapidity $\rho$. Suppose that $\psi$ is bounded localized in the half-space $\{x\in\R^3: xe\ge 0\}$. Then by causality 
$e(\psi_{\rho e})\ge \operatorname{e}^{-|\rho|}e(\psi)$ and $-\overline{e}(\psi_{\rho e})\le -\operatorname{e}^{|\rho|}\overline{e}(\psi)$  for $\overline{e}:=-e$. This means that  the expansion of the carrier in direction $e$ is limited by the factor $\operatorname{e}^{|\rho|}$, whereas in direction $-e$ it is limited by the factor $\operatorname{e}^{-|\rho|}$ thus not overcoming the barrier $\{x\in\R^3:xe=0\}$. Actually one finds that $\psi_{\rho e}$ for  $\rho\in\R$ does not stay confined in any strip 
$\{x\in\R^3:0\le xe\le c\}$. If $|t_{\overline{e}}|<-\frac{1}{2}\overline{e}(\psi)$ then $\psi_{\rho e}$ is confined neither for $\rho\to \infty$ nor for $\rho\to - \infty$. See (\ref{XCTLRCL}).\\
\hspace*{6mm}
 Then there are the remarkable states assuming the maximum value $|t_{\overline{e}}|=-\frac{1}{2}\overline{e}(\psi)$, which we call late-change states or, more specifically,  large\,-\,$t_{\overline{e}}$\,-\,states (\ref{L-CS}). If $\psi$ is a large\,-\,$t_{\overline{e}}$\,-\,state with  $ t_{\overline{e}}=-\frac{1}{2}\overline{e}(\psi)$ then  by (\ref{XCTLRCL}) the carrier is shrinking  in direction $-e$ to a $\delta$-strip at the origin by the factor $\operatorname{e}^{-\rho}$ for $\rho\ge 0$, i.e., $- \overline{e}(\psi_{\rho e})= -\operatorname{e}^{-\rho}\overline{e}(\psi)$, or equivalently, still by (\ref{XCTLRCL}), the time evolution $\psi_{ t}$ is localized in $\{x\in\R^3:0\le xe\le t\}$ for some (and hence all)  $ t\ge -\frac{1}{2}\overline{e}(\psi)$. What is more, every   state localized in $\{x\in\R^3:xe\ge0\}$ can be approximated by   large\,-\,$t_{\overline{e}}$\,-\,states, see (\ref{DLTES}) and the comment following (\ref{ECCPWS}). The proof of this density result involves a  property of causal systems shown in sec.\,\ref{PPCS} and an asymptotic behavior of the evolution of the probability of localization. For the Dirac systems it is  the asymptotic causality (\ref{VPOBT}),\,(\ref{AC}), which is known for massive systems endowed with the NWL.  For the massless Weyl systems one has (\ref{GTERSWF})(c), which is harder to prove. 
 \\
\hspace*{6mm} 
 Eventually, as a consequence one obtains in (\ref{LCDWF}), (\ref{LCWWF})
 the Lorentz contraction  
  \begin{equation*}\label{ILC}
 \norm{E\big(\{x\in\R^3: -\delta\le xe \le \delta\}\big)\,\psi_{\rho e}}\, \to 1, \;\; |\rho|\to\infty
 \end{equation*}
for every wave-function $\psi$, as  in particular of the electron or positron or  of any of the four massless Weyl fermions, and any $\delta>0$. One recalls that in classical mechanics it is  discussed   whether the  length contraction  really exists or not. Therefore  we find it is worth attaching  to (\ref{LCDWF}) the rather obvious considerations concerning  (\ref{LCDSM}) about the objectiveness of the Lorentz contraction  in quantum theory. On the other hand, as a consequence of the Poincar\'e covariance of  localization, one ascertains the frame-dependence of the Lorentz contraction (\ref{FDLCDL}) and the  fact that also in quantum mechanics  Lorentz contraction does not exist for the comoving observer (\ref{NLCFCM}).

Causality is used to  define various  orthocomplemented lattices of subsets of Minkowski spacetime (see e.g. \cite{CJ13} 2013 and the literature cited therein),   known under the general name of causal logics. The link to quantum theory is thought to be provided attributing to the  sets of the lattice  projections on the Hilbert space of states. By physical reasons as well mathematical consequences this assignment in general cannot be required to be a lattice homomorphism into the quantum logic as argued subsequent to  (\ref{RLCCRRCL}). But it is expected to be a normalized monotone locally orthogonal $\sigma$-orthoadditive Poincar\'e covariant map, called a representation (rep)  for short.\\
\hspace*{6mm} 
Most adequate from a physical point of view seems to be the lattice $\mathcal{M}$ in sec.\,\ref{HLCCR} of causally complete regions generated by the spacelike relation.  However for structural deficiencies $\mathcal{M}$ does not allow for representations, see sec.\,\ref{LOCCR}. Mathematically more convenient is the lattice $\mathcal{M}^{'borel}$ in sec.\,\ref{LMMBOR} generated by the non-timelike relation   \cite{CJ77}. And, indeed there are representations of $\mathcal{M}^{'borel}$. Unfortunately the representations we know  so far (\ref{RLBSGNTLR}) refer to unphysical systems with imaginary mass spectrum. Moreover, every representation of 
$\mathcal{M}^{'borel}$ determines  a causal system (\ref{RLCCRRCL}). By (\ref{NRLCCRRCLP}) this is uniquely extendable as a high boost limit  to a PM on every non-timelike not spacelike hyperplane $\chi$,  obeying Poincar\'e covariance.

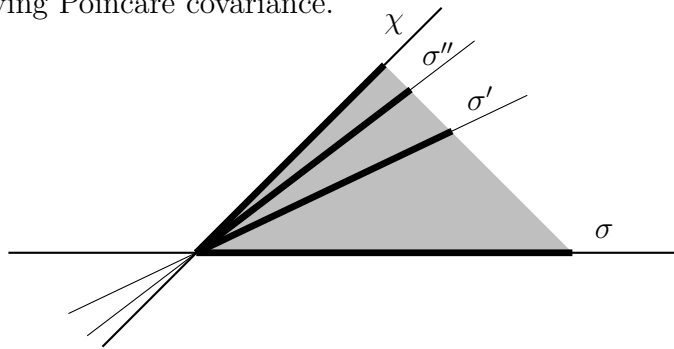
\begin{pspicture}(-8,-1.1)(1,2.9)
\psline(-4.5,0)(4.5,0)
\psline(-3.25,-1.25)(1.26,3.26)
\put(-2,0){\line(2.1,1){4.4}}
\put(-2,0){\line(2.1,1){-1.7}}
\put(-2,0){\line(2.1,1.6){3.7}}
\put(-2,0){\line(2.1,1.6){-1.45}}


\psline*[linecolor=lightgray]
(-2,0)(0.5,2.5)(3,0)(-2,0)
\linethickness{0,9mm}
\put(-2,0){\line(1,0){5}}
\put(-2,0){\line(2.1,1){3.4}}
\put(-2,0){\line(1,1){2.5}}
\put(-2,0){\line(2.1,1.6){2.85}}

\put(3.3,0,2){$\sigma$}
\put(0.5,3){$\chi$}
\put(1.,2.5){$\sigma''$}
\put(1.6,1.9){$\sigma'$}
\end{pspicture}

Hence this extendability is a necessary condition for a causal system to induce a representation of $\mathcal{M}^{'borel}$. It implies more profound properties of causal systems as e.g. the existence of  dense sets of  
late-change states.
And, indeed, the Dirac and Weyl systems allow for such an extension (\ref{CNGENTLNSLHP}). The somewhat involved proof is based on the results on group representations of the sec.\,\ref{A:SI} - \ref{A:ESRWS} in the appendix. In view of these insights,  $\mathcal{M}^{'borel}$ and its representations  apparently have some physical relevance.

{\it Acknowledgements:} The author would like  to  thank Ulrich Mutze and Andreas D. Leiseifer for many valuable discussions.\\

 \section{Newton-Wigner  Position Operator and Localization by  Wightman} \label{NWPOWL}
 Two main achievements  in particle localization  are the position operator $X^{\textsc{nw}}$   of Newton, Wigner \cite{NW}  in 1949  and Wightman's formulation  of  localization  \cite{W62} in 1962.  The latter furnishes also a rigorous mathematical frame of the former using group representation theory.  \\ 
\hspace*{6mm}  The central notion in  \cite{W62}, which we like to call  a    Wightman localization (WL),  is   a  projection valued measure (PM) $E$ on the Borel sets of the Euclidean space  $\R^3$ and a representation $U$ of the covering group $ISU(2)$ of the group of Euclidean motions, both acting on the Hilbert space of states $\mathcal{H}$, such that $(U,E)$  satisfies the covariance
 \begin{equation}\label{ECL}
U(g)E(\Delta)U(g)^{-1}=E(g\cdot \Delta)
\end{equation}
for all $g\in ISU(2)$ and Borel sets $\Delta\subset \R^3$. Note that here a representation of a topological group is always unitary continuous on a separable complex Hilbert space.\\
\hspace*{6mm}  With every WL  a Euclidean covariant position  operator $X=(X_1,X_2,X_3)$ is associated by spectral integration 
\begin{equation}\label{ACPO}
X_k := \int \pi_k \operatorname{d} E, \quad\pi_k(x):= x_k
\end{equation}

Clearly, $\pi_k(E)$ is the spectral measure of $X_k$ and  $E$ is the joint spectral measure of the three components of $X$. Hence $E$ is uniquely determined by  $X$.  \\
\hspace*{6mm} 
For every  irreducible  relativistic system, which is massive or   massless helicity zero,   a WL exist. 
Let this system be described by the representation $W$  of  the universal covering  group $\tilde{\mathcal{P}}=ISL(2,\C)$ of the Poincar\'e group. Then the former means that there is a PM $E$ which is covariant with respect to $U:=W|_{ISU(2)}$.
Any other covariant PM $E'$ is \textbf{unitarily related} to $E$, i.e.,  there is a unitary transformation $S$ commuting with $U$ such that $E'=SES^{-1}$.
A particular representative  is the \textbf{Newton-Wigner localization} (NWL)  $E^{\textsc{nw}}$, which   is singled out by  the requirement of time inversion invariance and by a mathematical smoothness condition. We prefer to distinguish NWL from all other unitarily related WL by the Newton-Wigner-Pryce formula (see \cite{MM97}), also called first Bakamjian-Thomas-Foldy formula   Eq.\,(\ref{BTFX}) below. The Newton-Wigner operator
  $X^{\textsc{nw}}$    is the position operator associated with  $E^{\textsc{nw}}$.  
\\ 
\hspace*{6mm} We add that 
   $E^{\textsc{nw}}$ and hence  $X^{\textsc{nw}}$ 
 exist  also for the corresponding antisystems with negative  energy operator  $H$, and, plainly, for  every system, which is  an orthogonal integral of  the former systems,   antisystems included. Such a system is massive  if the mass-squared operator 
 $$C:=H^2-P^2$$
  is positive ($C>0$). Here $P= ( P_1 , P_2 , P_3 )$ is the linear momentum and $P^2=P_1^2+P_2^2+P_3^2$.
\\
\hspace*{6mm} For massive  systems the remarkable Bakamjian-Thomas-Foldy formulae  Eq.\,(\ref{BTFX}), Eq.\,(\ref{BTFN}) hold on a common core. The first\\
    \begin{equation}\label{BTFX}
X^{\textsc{nw}} =\textrm{\SMALL{$\frac{1}{2}$}}(H^{-1}N+NH^{-1})-C^{-1/2}(C^{1/2}+|H|)^{-1} \,P\times \big(J+H^{-1} (P\times N)\big)
\end{equation}

may be regarded as the defining equation for the Newton-Wigner position operator  as it expresses  $X^{\textsc{nw}}$ in terms of the  ten generators of the Poincar\'e group $\tilde{\mathcal{P}}=ISL(2,\C)$, i.e.,  in addition to  the Hamiltonian $H$ and  the linear momentum $P= ( P_1 , P_2 , P_3 )$,  the angular momentum $J= ( J_1 , J_2 , J_3 )$, which generates spatial rotations, and the Lorentz booster $N=(N_1,N_2,N_3)$.  
The second\\
\begin{equation}\label{BTFN}
N=\textrm{\SMALL{$\frac{1}{2}$}}(HX^{\textsc{nw}} +X^{\textsc{nw}}H)+(C^{1/2}+|H|)^{-1} \operatorname{sgn}(H)\,P\times (J-X^{\textsc{nw}}\times P)
\end{equation}

shows that the generators of the \textbf{little kinematical group} $\R\otimes ISU(2)$, representing   time translations plus Euclidean motions,  together  with   $X^{\textsc{nw}}$   determine  the booster and hence the whole representation of  $\tilde{\mathcal{P}}$.  
Note that Eq.\,(\ref{BTFN}) generalizes the original formula to not semi-bounded $H$, as for instance in the case of  the Dirac representation. 
\\  
\hspace*{6mm} It is characteristic for the NWL that it commutes with the three Casimir operators, i.e.,  the sign of the energy, the mass-squared  operator, and the Pauli-Lubanski  scalar. 
\\

\section{Poincar\'e Covariance of Position}\label{POCOPO}   
The principle of special relativity is automatically satisfied for a quantum system  if the Hilbert space of states of the system is the carrier space of a  representation  $W$ of 
$\tilde{\mathcal{P}}$. 
This is  thoroughly  explained in Currie, Jordan, Sudarshan 1962 \cite{CJS62} and, specifically for localization in space regions, in Amrein 1969 \cite[sec. VIII]{A69}. There is also clearly pointed out that relativistic symmetry must be distinguished from  {\it manifest relativistic invariance}. This  rather formal concept, moreover,  has not found a generally accepted formulation, above all not for the localization problem (see e.g.  \cite{CJS62}, \cite{A69}, \cite{K71}, \cite{R81}, \cite{FKW02}).       $X^{\textsc{nw}}$   and  $E^{\textsc{nw}}$ are not manifest relativistic invariant, as argued e.g. in \cite[sec. III]{CM82} and in Monahan, McMillan 1997 \cite{MM97}, respectively. \\

In the following for the group elements $g=(\mathfrak{a},A)= (t,b,A)\in\tilde{\mathcal{P}}$  we often use  abbreviations as $\mathfrak{a}\equiv (\mathfrak{a},I_2)$,  $b\equiv (0,b,I_2)$, $t \equiv (t,0,I_2)$, $A\equiv (0,A)$. The elements in $ISU(2)$  are denoted by $(b,B)$. \\
\hspace*{6mm}
The group operation in $\tilde{\mathcal{P}}$ reads $gg'=(\mathfrak{a},A)(\mathfrak{a}',A'):= (\mathfrak{a}+\Lambda(A)\mathfrak{a}',AA')$, whence $g^{-1}=(-\Lambda(A^{-1})\mathfrak{a},A^{-1})$. Here $\Lambda:SL(2,\C)\to O(1,3)_0$ is  the universal covering homomorphism onto the proper orthochronous Lorentz group.\\


\subsection{Poincar\'e covariance of  localization}\label{SePCWL} By the foregoing remark it is clear that NWL does not violate the principle of special relativity. By means of $W$ the localization on the particular hyperplane $\R^3\equiv \{0\}\times \R^3$ is transported to any other spacelike hyperplane.
\\
\hspace*{6mm}
Independently  of this fact  we will reason  now following  \cite[sec.\,C]{CL15}   that  for a relativistic system,  rather than  a Euclidean covariant localization, physics requires a fully Poincar\'e covariant localization.  This  generalization  is more than a merely  manifest Poincar\'e covariant formulation of localization in the above sense. On its necessity  we are in accord with \cite[sec.\,7.3]{K71} (restricting the permissive view in  \cite[sec. I]{CM82}).  It is a  consequence of  the physical ideas on  Euclidean covariance in \cite{W62}  applied to general Poincar\'e transformations, thus giving  grounds for a unique extension  of a WL to  a  Poincar\'e covariant localization in regions, i.e., Borel subsets, of spacelike hyperplanes.
In Castrigiano, Mutze 1982 \cite[Theorem]{CM82}  existence and uniqueness of this extension is proven. The result is proven more generally for POL in (\ref{PCPOL}).\\
\hspace*{6mm}  First let us briefly report the physical idea behind Euclidean covariance of the localization  in  \cite{W62}. Imagine an apparatus  $\mathcal{A}$ suited for 
the measurement of the position of the system  in the \textbf{region} (i.e. Borel subset) $\Delta$   and thus realizing the observable $E(\Delta)$. If this apparatus is shifted by $b\in\R^3$ then relativistic symmetry means that it realizes the observable $W(b)E(\Delta)W(b)^{-1}$. On the other hand, the shifted apparatus obviously measures
 the position in the translated region $b+\Delta$ and thus realizes the observable $E(b+\Delta)$. Consequently translational covariance  
$W(b)E(\Delta)W(b)^{-1}=E(b+\Delta)$ must hold. Analogously one deduces  covariance  under all Euclidean transformations.\\
\hspace*{6mm} If now the apparatus  $\mathcal{A}$ is boosted by a Lorentz transformation $L=\Lambda(A)$ given by $A\in SL(2,\C)$ then, again by relativistic symmetry, the boosted apparatus realizes the observable $W(A)E(\Delta)W(A)^{-1}$.  Moreover, by special relativity, the boosted apparatus is still  suited for position measurement, namely in the \textbf{spatial region} $L\,\Delta$ of the spacelike hyperplane $L\,\R^3$   
  in Minkowski space $\R^4$. Here, for convenience, $\R^3$ is identified with $\{0\}\times \R^3$. 
 Let $E(L\,\Delta)$ denote the yes-no observable realized by the boosted apparatus. So 
$E(L\,\Delta)$ has to be equated with  $W(A)E(\Delta)W(A)^{-1}$. An analogous reasoning is valid for every transformation $g=(\mathfrak{a},A)\in \tilde{\mathcal{P}}$. \\
\hspace*{6mm} There arises the following question of consistency.  Let $\Delta'\subset \R^3$ be a region  and let $A'\in SL(2,\C)$ such that $L'\,\Delta'= L\,\Delta$. Further, let the apparatus $\mathcal{A}'$ realize the observable $E(\Delta')$. Then the apparatus $\mathcal{A}'$  boosted by $L'$ realizes the observable $W(A')E(\Delta')W(A')^{-1}$, which by special relativity equals the observable $E(L'\,\Delta')$ of position in $L'\,\Delta'$.  Hence physics   imposes on $E$ the condition $W(A)E(\Delta)W(A)^{-1}=W(A')E(\Delta')W(A')^{-1}$.\\
\hspace*{6mm} However, as shown in  \cite[Theorem]{CM82} and in (\ref{PCPOL}), in the final analysis it is due to Euclidean covariance that $E$ automatically   satisfies this condition. This allows  the claimed unique extension of the localization to  all  regions $\Gamma$ of  spacelike hyperplanes $\sigma$ such that the general \textbf{Poincar\'e covariance} 
\begin{equation}\label{PCWL}
W(g)E(\Gamma)W(g)^{-1}=E(g\cdot \Gamma)
\end{equation}
holds.  Here
\begin{equation}\label{PTUCH} 
g\cdot \mathfrak{x}:=\mathfrak{a}+\Lambda(A) \mathfrak{x}\quad  \text{ for } g=(\mathfrak{a},A)\in\tilde{\mathcal{P}}, \, \mathfrak{x}\in \R^4
\end{equation}

and  $g\cdot\Gamma:=\{g\cdot \mathfrak{x}: \mathfrak{x}\in\Gamma\}$.
  As a consequence of  the covariance Eq.\,(\ref{PCWL}),  one has a  WL  on every spacelike hyperplane $\sigma$ by  mapping $\Gamma\mapsto E(\Gamma)$. 
 In particular, $E^{\textsc{nw}}$ is Poincar\'e covariant.\\
\hspace*{6mm} 
The  above description of the WL $E$ refers to the frame  $\mathfrak{R}$ constituted by the origin and the standard basis $\big(\mathfrak{o},(\mathfrak{e}^0,\mathfrak{e}^1,\mathfrak{e}^2,\mathfrak{e}^3)\big)$ of $\R^4$. It remains to clarify  what the representation $E'$ of the localization looks like when referring to any other reference frame $\mathfrak{R}'\equiv g^{-1}\cdot\mathfrak{R}$ with $g\in \tilde{\mathcal{P}}$. (Find details on the change of  reference frame in sec.\,\ref{FDOLC}). The answer is 
\begin{equation}\label{PRWL} 
E'(\Gamma)=E(g\cdot\Gamma) 
\end{equation}
for every region $\Gamma$ of a spacelike hyperplane of Minkowski space. Indeed, according to relativistic symmetry the observable of localization in $\Gamma$, represented by $E(\Gamma)$ in $\mathfrak{R}$,   is represented by $E(\Gamma)'=W(g)E(\Gamma)W(g)^{-1} $ in $\mathfrak{R}'$. By the meaning of $E'$ one has $E'(\Gamma)=E(\Gamma)'$. Due to the covariance of $E$ the latter equals $E(g\cdot \Gamma)$, whence Eq.\,(\ref{PRWL}).\\
\hspace*{6mm}
One recalls that $E(\Gamma)$ referring to $\mathfrak{R}$ and $E'(\Gamma)$ referring to $\mathfrak{R}'$ both represent the observable of localization in $\Gamma$.


\subsection{Poincar\'e covariance of position operator} \label{PCPO}   
Poincar\'e covariance carries over from the WL to the  position operator associated with the WL. Indeed,
 let $\sigma$ be a spacelike hyperplane, choose an origin $\mathfrak{o}\in \sigma$ and a direction $\mathfrak{n}\in \sigma-\mathfrak{o}$. Then, by  \cite[(3.1),(3.2)]{CM82},
 the (four-component) position operator $X_{\sigma,\mathfrak{o}}$ for the three-dimensional Euclidean space $\sigma-\mathfrak{o}$   is given by the  integral of $\operatorname{id}_\sigma-\, \mathfrak{o}$ with respect to the WL   on $\sigma$, whence the position operator $X_{\sigma,\mathfrak{o},\mathfrak{n}}$ in direction $\mathfrak{n}$ follows to be $ - \mathfrak{n}\cdot\, X_{\sigma,\mathfrak{o}}$ with $\cdot$ the Minkowski product on Minkowski space.\footnote{ Representing Minkowski space by $\R^4$ the Minkowski product is given by $\mathfrak{a}\cdot \mathfrak{a}':=a_0a'_0-a_1a'_1-a_2a'_2-a_3a'_3$. Often we use the notation $\mathfrak{a}^{\cdot 2}:=\mathfrak{a}\cdot\mathfrak{a}$.}  Then 
  the Poincar\'e covariance
\begin{equation}
W(g)\, X_{\sigma,\mathfrak{o},\mathfrak{n}}\,W(g)^{-1}= X_{g\cdot\sigma,g\cdot\mathfrak{o},g\cdot \mathfrak{n}} 
\end{equation}
holds. This yields, in particular,  Poincar\'e covariance of the Newton-Wigner position operator $X^{\textsc{nw}}$. For this kind of covariance see Fleming 1965 \cite[(2.6)]{F65}. One more derivation of  the  Poincar\'e covariance of $X^{\textsc{nw}}$  can be found  in Farkacz, Kurucz, Weiner 2002  \cite{FKW02}. \\

\subsection{Frame-dependence of Newton-Wigner localization}\label{FDNWL}  Regarding  NWL  (introduced in sec.\,\ref{NWPOWL}) we are going to show 
 that   a  state $\varphi$, $||\varphi||=1$, localized in a bounded  region $\Delta\subset \R^3$,  i.e.,  $E^\textsc{nw}(\Delta)\varphi =\varphi$, when boosted or time translated    is no longer  localized in any bounded region. We give a rigorous proof.  Most probably any WL unitarily related to NWL has this property as well, but we do not pursue this question. \\
\hspace*{6mm}
Occasionally we will say that a state is \textbf{bounded localized} if it is localized in a bounded region.
 
 \begin{CCS}\label{PCCS} In the sequel we repeatedly will need the canonical cross section $Q(\mathfrak{k})$ for $ \mathfrak{k}\in\R^4$ with  $\mathfrak{k}\cdot\mathfrak{k}>0$. It is the unique positive $2\times 2$-matrix satisfying $Q(\mathfrak{k})\cdot (\eta m,0,0,0)=\mathfrak{k}$, where $m=\sqrt{\mathfrak{k} \cdot \mathfrak{k}}$, $\eta=\operatorname{sgn}(k_0)$.  
 This implies $Q(\mathfrak{\alpha k} )=Q(\mathfrak{k})$ for $\alpha\ne 0$ and $Q(B\cdot \mathfrak{k})=BQ(\mathfrak{k})B^{-1}$ for $B\in SU(2)$.
 Explicitly 
\begin{equation}\label{CCS}
Q(\mathfrak{k})=
 \sqrt{\frac{2m}{m+|k_0|}} \left(I_2+\frac{\eta}{m}\sum_{j=0}^3k_j\sigma_j\right)
\end{equation}
Here $\sigma_0=I_2$ and $\sigma_j$ are the Pauli matrices.\footnote{ $\sigma_1=\left( \begin{array}{cc} 0 & 1\\ 1 & 0 \end{array}\right)$, $\sigma_2=\left( \begin{array}{cc} 0 & -\operatorname{i}\\ \operatorname{i} & 0 \end{array}\right)$, $\sigma_3=\left( \begin{array}{cc} 1 & 0\\ 0 & -1 \end{array}\right)$}  Let $A\in SL(2,\C)$. Then $R( \mathfrak{k},A):=Q(\mathfrak{k})^{-1}AQ(A^{-1}\cdot \mathfrak{k})$ is called the \textbf{Wigner rotation}. It satisfies $R( \mathfrak{k},A)\in SU(2)$ and $R( \mathfrak{k},B)=B$ for $B\in SU(2)$. For $p\in\R^3$, $\epsilon(p):=\sqrt{|p|^2+m^2}$, $\eta=+,-$ and $\mathfrak{p}^\eta:=(\eta\epsilon(p),p)$, one has $Q(\mathfrak{p}^+)^{-1}=Q(\mathfrak{p}^-)$.
Further useful formulae are $Q(\mathfrak{k})^2=\frac{\eta}{m}\sum_{j=0}^3k_j\sigma_j$ and $AQ(\mathfrak{k})^2A^*=Q(A\cdot\mathfrak{k})^2$, $A^*Q(\mathfrak{k})^{-2}A=Q(A^{-1}\cdot\mathfrak{k})^2$.  Furthermore, $Q(\mathfrak{k})^{4\varsigma }=-I_2+\frac{2\epsilon(k)}{m}Q(\mathfrak{k})^{ 2\varsigma }$, $\varsigma=\pm1$.\qed
\end{CCS}

Let $[m,j,\eta]$ denote any irreducible representation of $\tilde{\mathcal{P}}$ for mass $m >0$,  spin $j\in\N_0/2$, and  sign of energy $\eta\in\{+,-\}$. In the massless case $[0,s,\eta]$ spin $j$ is replaced by helicity $s\in\mathbb{Z}/2$.

\begin{The}\label{LDNW} Let $W$ be an orthogonal sum of  irreducible  representations $[m,j,\eta]$ and $[0,0,\eta]$ with  $m>0$, $j\in\N_0/2$, $\eta\in\{+,-\}$  
provided with the NWL  $E^{\textsc{nw}}$. Let $\psi$ denote a state. Then  $\psi$ and $W(g)\psi$ are localized in a bounded region, if and only if
$g\in \tilde{\mathcal{P}}$  leaves the space $\{0\}\times\R^3$ invariant, i.e., $g\in ISU(2)$.
\end{The}\\
{\it Proof.}  First suppose $g\in ISU(2)$. Then due to the Euclidean covariance of the localization the claim is obvious. So we turn to  the converse implication.  Suppose $g\in\tilde{\mathcal{P}} \setminus ISU(2)$. Then there are $k,k'\in ISU(2)$
such that $k\,g\, k'$ either equals (a)  the boost $A_\rho$\footnote{ Frequently we use explicitly  $A_\rho:=\operatorname{e}^{\,\rho\,\sigma_3/2}$,  $\Lambda(A_\rho)=\left(\begin{array}{cccc}\cosh(\rho) & 0&0&\sinh(\rho)\\ 0&1&0&0\\0&0&1&0\\ \sinh(\rho)&0&0&\cosh(\rho)\end{array}\right)$ for $\rho \in \R$} in direction $(0,0,1)$ with rapidity $ \rho\ne 0$ or (b)  the time translation $t\ne0$. Hence it suffices to treat the cases (a) $g=A_\rho$, $\rho\ne 0$ and (b) $g=t\ne 0$. Moreover,  one may assume that  $W$ is irreducible.

(a) $g=A_\rho$, $\rho\ne 0$.  Consider first the massive case. 
Then $W$ acts  in  momentum space $L^2(\R^3,\C^{2j+1})$ by
\begin{equation}\label{RIMS}
\big(W^{mom}(\mathfrak{a},A)\varphi\big)(p)=\sqrt{\epsilon(q^\eta)/\epsilon(p)}\, \e^{\i \mathfrak{p}^{\eta}\cdot \,\mathfrak{a}} \,D^{(j)}\big(R(\mathfrak{p}^\eta,A)\big)\,\varphi(q^\eta)
\end{equation}
where $D^{(j)}$ is the irreducible $2j+1$  dimensional unitary representation of $SU(2)$,  $\mathfrak{p}^{\eta}=(\eta\epsilon(p),p)$, $\mathfrak{q}^\eta= (q_0^\eta,q^\eta):=A^{-1}\cdot \mathfrak{p}^\eta$. Note that $W^{mom}(\mathfrak{a},A)\varphi$ is continuous if $\varphi$ is so.\\
\hspace*{6mm} 
In position space representation, NWL is the canonical PM given by $E^\textsc{nw}(\Delta)\psi=1_\Delta\psi$.
 One switches to the momentum space representation by the Fourier transformation
$\varphi(p)=\big(\mathcal{F}\psi\big)(p)=(2\pi)^{-\frac{3}{2}}\int \operatorname{e}^{-ixp}\psi(x)\operatorname{d}^3x$.
\\
\hspace*{6mm} 
Assume now that  $\varphi$ and $W^{mom}(A_\rho)\varphi$ are bounded localized. For convenience only assume $\varphi(0,0,p_3)\ne 0$ for some  $p_3\in \R$.
By the Paley-Wiener Theorem each component of these functions is the restriction of an entire function on $\C^3$. 
 Note further that  $R(\mathfrak{p},A)\in SU(2)$ is diagonal and real for  $A=A_\rho$ and $p_1=p_2=0$, and equal to $I_2$ for $A=I_2$. One infers $R(\mathfrak{p}^\eta,A_\rho)=I_2$ if $p_1=p_2=0$ (see also Eq.\,(\ref{WRDB})).
Therefore it follows from Eq.\,(\ref{RIMS}) that some component of $\varphi$ yields an entire function $f\ne 0$ on $\C$ such that $z\mapsto 
\sqrt[4]{\frac{u(z)^2+m^2}{z^2+m^2}}f(u(z))$ with $u(z):=z\cosh(\rho)-\eta\sqrt{z^2+m^2}\sinh(\rho)$ has an entire extension. However,  this is impossible for the singularities at $z=\pm im\cosh(\rho)$.\\
\hspace*{6mm} We turn to the massless case $[0,0,\eta]$. Then, in momentum space  $L^2(\R^3,\C)$,
\begin{equation}\label{MLRIMS}
\big(W^{mom}(\mathfrak{a},A)\varphi\big)(p)=\sqrt{|q^\eta|/|p|}\, \operatorname{e}^{\i \mathfrak{p}^{\eta}\cdot \,\mathfrak{a}} \,\varphi(q^\eta)
\end{equation}
 with $\mathfrak{p}^{\eta}=(\eta |p|,p)$. Proceeding as in the massive case, one gets an entire function $f$ such that $f_\rho(z):= \sqrt{\frac{|u(z)|}{|z|}}f(u(z))$ with $u(z):=\alpha z+\beta|z|$ for $\alpha:=\cosh(\rho)$, $\beta:=-\eta\sinh(\rho)$  has an entire extension. Hence,  $f_\rho(z)=\sqrt{\alpha+\beta}f\big((\alpha+\beta)z\big)$ if $z>0$,  and $f_\rho(z)=\sqrt{\alpha-\beta}f\big((\alpha-\beta)z\big)$ if $z<0$, whence $\sqrt{\alpha+\beta}f\big((\alpha+\beta)z\big)=\sqrt{\alpha-\beta}f\big((\alpha-\beta)z\big)$ $\forall$ $z\in\C$. However, for $f\ne 0$, this is impossible.
 
 (b) $g=t\ne 0$. This case is  even easier. One has $\big(W^{mom}(t)\varphi\big)(p)=\e^{\i \eta \epsilon(p)t}\varphi(p)$ in the massive case and $ \epsilon(p):=|p|$ in the massless case. Proceeding as in (a) one gets an entire function $f$ such that $f_t(z):=\e^{\i \eta \epsilon(z)t} f(z)$ for $ \epsilon(z):=\sqrt{z^2+m^2},\,m\ge 0$  has an entire extension. Again, this is impossible for $f\ne 0$.
 \qed\\

The result in (\ref{LDNW}) does not single out a particular Lorentz frame. Indeed, let $\sigma$ and $\sigma'$ be any two spacelike hyperplanes. Then $\sigma'=g\cdot \sigma$ for some $g\in \tilde{\mathcal{P}}$. Further let $\varphi$ be a state, $h\in \tilde{\mathcal{P}}$, and $\Delta \subset \sigma$ a region such that $\varphi$ and $W(h)\varphi$ are localized in $\Delta$, i.e., $E^{\textsc{nw}}(\Delta)\varphi=\varphi$ and $E^{\textsc{nw}}(\Delta)W(h)\varphi=W(h)\varphi$. Then, due to Poincar\'e covariance Eq.\,(\ref{PCWL}), $\varphi':=W(g)\varphi$ and $W(h')\varphi'$ for $h':=ghg^{-1}$ are localized in $\Delta'\subset \sigma'$ for $\Delta':=g\cdot \Delta$. It remains to note that $\Delta'$ is bounded if  $\Delta$ is so and that $h'$ leaves $\sigma'$ invariant if $h$ leaves $\sigma$ invariant.\\
\hspace*{6mm} 
Moreover,  we like to recall that by relativistic symmetry  all Lorentz observers make the same ascertainments on the localization of a Newton-Wigner system. Consider two observers related to the inertial frames of reference $\mathfrak{R}$ and $\mathfrak{R}'=g^{-1}\cdot\mathfrak{R}$ and endowed with the 
NWL represented, according to (\ref{PRWL}), by $E^{\textsc{nw}}$ and $(E^{\textsc{nw}})'$, respectively. Now, e.g.,
 let $\Gamma$ be a bounded region of a spacelike hyperplane of Minkowski space. Referring to $\mathfrak{R}$ let $\psi$ represent a state localized in $\Gamma$, i.e., $E^{\textsc{nw}}(\Gamma)\psi=\psi$.  With respect to $\mathfrak{R}'$ the state is represented by $\psi'=W(g)\psi$. Then by (\ref{PRWL}) and the covariance of the localization $(E^{\textsc{nw}})'(\Gamma)\psi'=\psi'$, which as it should be confirms  that the system is localized in $\Gamma$.
 \\
\hspace*{6mm}
On the other hand (\ref{LDNW}) implies the frame-dependence of the NWL $E^{\textsc{nw}}$. Weidlich, Mitra 1963 \cite{WM63}
describe this dependence by  {\it a state strictly localized in one coordinate system $\sigma$ $\dots$ appears to be smeared out if observed from a system $\sigma'$ moving relative to $\sigma$}.  Frame-dependence of  NWL is still a current topic, see e.g. Farcacz et al. 2002  \cite{FKW02}
and Wagner et al. 2011 \cite{WSWSG11}.
 By virtue of the Poincar\'e covariance of the localization one can state the frame-dependence of NWL  without referring to a moving observer:
 
 \begin{CFDNWL}\label{CFDNWL} Let $\sigma$ and $\sigma'$ be two different spacelike hyperplanes in Minkowski space. Then, if $\psi$ is localized in a bounded region of $\sigma$, then $\psi$ is not localized in any bounded region of  $\sigma'$.
\end{CFDNWL}\\
{\it Proof.} One may identify $\sigma$ with $\{0\}\times \R^3\subset \R^4$. Let $\psi$ be bounded localized in $\sigma$. As  $\sigma\ne\sigma'$ there is a Poincar\'e transformation $g\not\in ISU(2)$ with $\sigma'=g^{-1}\cdot\sigma$. Now consider a bounded region  $\Gamma'$ of  $\sigma'$. Then $\Gamma:=g\cdot \Gamma'$ is bounded in $\sigma$. By (\ref{LDNW}), $W(g)\psi$ is not bounded localized,  whence $E^{\textsc{nw}}(\Gamma)W(g)\psi\ne W(g)\psi$, which means $E^{\textsc{nw}}(\Gamma')\psi\ne\psi$.\qed \\

Although the frame-dependence of  NWL is in accord with relativistic symmetry,  it is still a worrying feature. Indeed, let $\psi$ be localized in a bounded region. Then,  by (\ref{LDNW}), $E^\textsc{nw}(\{|x|\ge R\})W(A)\psi\ne0$ for all $R>0$ and every however small boost $g=A$, despite of  $E^\textsc{nw}(\{|x|\ge R_0\})\psi=0$ for some $R_0>0$. 
 So imagine  an electron localized in a bounded region. Exert temporarily on it  an  electric force. Then any moment later there is a non-vanishing probability to observe the electron arbitrarily far away. This is an acausal behavior. \\
 \hspace*{6mm} 
 Even more striking is the case  in (\ref{LDNW}) of a time translation $g=t$. It is the case of parallel hyperplanes $\sigma$ and $\sigma'$ in (\ref{CFDNWL}). The phenomenon is already mentioned 1955 in  \cite[sec.\,2]{WS} and studied 1963 in \cite[sec.\,3]{WM63}.  It is the \textbf{instantaneous spreading} to infinity of  wave-functions which initially vanish outside a bounded region.   
 In  Ruijsenaars 1981 \cite{R81}  one finds a conscientious discussion of quantum localization and the possibility to observe superluminal propagation with respect to Newton-Wigner position. (For more recent engagements in this topic see e.g.   \cite{WSWSG11} 2011,  \cite{EM17} 2017.)  The latter results in  an obvious  violation of causality and hence constitutes a serious objection against NWL. The following section \ref{PELS} studies in detail the phenomenon of instantaneous spreading as a consequence of  positivity of energy mainly referring to the work of  Hegerfeldt.\\
 \hspace*{6mm} 
Summarizing one states  that Newton-Wigner localization would be a satisfactory solution to  localizability of irreducible  relativistic systems, if there were not three  problems. First recall that
  massless irreducible  systems with non-zero helicity, like the photon  are not localizable  according to Newton-Wigner.  
 Then there is the  frame-dependence of NWL as discussed above.  Last not least there is the related acausal  instantaneous spreading to infinity. 
 \\
\hspace*{6mm}
From sec.\,\ref{SeCWL} on we will be concerned with  relativistic quantum systems, which implement Einstein causality. These systems are  endowed with a WL  for which there is no transmission of signals faster than light. In particular instantaneous spreading and  frame-dependence of localization are excluded on principle.

\section{Positive Energy and Localized States}\label{PELS}
 It is well-known by the work of Hegerfeldt (see   \cite{H01} for a  brief survey)   that the phenomenon  of superluminal propagation is closely related to positivity  of  energy,  
more precisely, to semi-boundedness of the Hamiltonian.  For  localized states  one has the  alternative, {\it either  instantaneous spreading or confinement}. The latter means that once the quantum system is localized in a  closed region   it stays there all the time. See   Hegerfeldt 2001 \cite[sec.\,4]{H01} and \cite[II.\,A]{HR80}.  Within the frame of  WL this alternative is shown in Schlieder 1971 \cite[Kap.\,3]{Sch71} (see also  Jadczyk 1977 in \cite[sec.\,1]{J77}, Castrigiano 1984 \cite[sec.\,I]{C84} )
using Borchers 1967 \cite[III.1 Theorem]{B67}.  \\

Indeed, semi-boundedness  of energy and little else already imply the above alternative.  We follow  Hegerfeldt   \cite[sec.\,4]{H01}
 reducing the assumptions.
 Let time translation be described by $V(t)= \operatorname{e}^{itH}$ with Hamiltonian $H$. Let $\Delta_0\subset \R^3$ and  $r_0>0$ be arbitrary but  \textbf{fixed}  throughout this section  \ref{PELS}.
 Set $\Delta_{r_0}:=\{y\in\R^3:\,\exists\, x\in\Delta_0  \text{ with } |y-x|\le r_0\}$. 
Then the  assumptions are as follows: 
\begin{itemize}
\item[($L1$)] {\it The set of all scalar multiples of the  states localized in $\Delta_{r_0}$   is a closed subspace.}

\item[($L2$)] {\it If the state $\varphi$ is localized in $\Delta_0$, then $V(t)\varphi$ is localized in $\Delta_{r_0}$ for all $t\in[0,\tau]$ for some $\tau>0$ depending on $\varphi$.} 
\end{itemize}

 The meaning of the  first  assumption is obvious. The second  one is a minimal requirement regarding non-instantaneous spreading. Note that the term \textbf{localized state} is not and need not be specified here. In the examples following  (\ref{CL}) the states localized  in a region $\Delta_0$ are the unit eigenvectors with eigenvalue $1$ of the localization operator assigned  to $\Delta_0$, which is  a positive  operator with spectrum contained in $[0,1] $, thus satisfying  $(L1)$. Within the frame of POL with causal time evolution \eqref{CCT}  the assumption
$(L2)$ is satisfied  for all Borel $\Delta_0$ and all $r_0>0$.

\begin{The} \label{CL} Let the Hamiltonian be semi-bounded and let the localized states of the quantum system satisfy the assumptions \emph{($L\,1,2$)}. Then, if the system is localized at some time $t_0$ in $\Delta_0$, it stays localized in $\Delta_{r_0}$ for all times $t\in\,]-\infty,+\infty\,[$.
\end{The}

{\it Proof.}  Without restriction assume $t_0=0$. Let $\varphi$ be a state localized in $\Delta_0$, and denote by $\mathcal{L}_{r_0}$ the subspace in ($L1$). By ($L2$), $V(t)\varphi\in \mathcal{L}_{r_0}$ for $t\in[0,\tau]$. Let $\chi\in \mathcal{L}_{r_0}^\perp$. Then $\langle \chi,V(t)\varphi\rangle =0$ holds for  $t\in[0,\tau]$ and hence for all $t$, see  \cite[II.\,A]{HR80} or \cite[Proof of Theorem]{H94}. The argument is as follows. Let without restriction $H\ge0$. Applying  the spectral theorem to $H$, one has $F(t):=\langle \chi,V(t)\varphi\rangle =\int\operatorname{e}^{it\lambda}\operatorname{d}m(\lambda)$ for a complex measure $m$  with $\operatorname{supp}(m)\subset [0,\infty[$. Therefore $F$ has a continuous extension onto $\{t\in\C:\Im (t)\ge0\}$, which is analytic on $\{t\in\C:\Im (t)>0\}$. Moreover, since $F(t)=0$ for $t\in I:=]0,\tau[$, by Schwarz reflection principle, 
there is a further analytic extension of $F$ onto $(\C\setminus \R) \cup I$. This is identical to zero by the principle of analytic continuation. By continuity, $F=0$. ---  It follows $V(t)\varphi\in  \mathcal{L}_{r_0}$ for all $t$.  
\hfill{$\Box$}\\

Essentially Hegerfeldt's   result in \cite[sec.\,4]{H01} is equivalent to  (\ref{CL}). Obviously (\ref{CL}) finds application  in case that  localizability is described  (i) by a WL, or  (ii) by a supra-additive, continuous from above, projection valued map  as used by Jauch, Piron 1967 \cite{JP67} for weak localization, or   (iii) by  a positive operator valued measure   (POM),  introduced by several authors (see e.g. Toigo 2005 \cite{T05} for a survey and sec.\,\ref{SePOL}). So, if the  Hamiltonian is  semi-bounded and  if $\Delta_0$ is closed and ($L\,1,2$) holds for  some $r_n$ in place of $r_0$ with $r_n\to 0$, then $\Delta_0=\bigcap_n \Delta_{r_n}$ and  in all three cases  states  localized in $\Delta_0$ obey the alternative, {\it either instantaneous spreading or confinement in $\Delta_0$}.  \\
\hspace*{6mm} Moreover,  in the  cases (i),\,(ii), confinement occurs in $\Delta_0$ (if and) only if {\it  the time translation operators commute with the localization operator for $\Delta_0$}.  Indeed,  by (\ref{CL}), $V(t)\big(E(\Delta_0)\mathcal{H}\big)\subset E(\Delta_{r_n})\mathcal{H}$ holds for all  $n$ and for all $t$.  Since $E$ is continuous from above,  $V(t)\big(E(\Delta_0)\mathcal{H}\big)\subset E(\Delta_0)\mathcal{H}$ follows by $n\to \infty$, and hence $V(t)\big(E(\Delta_0)\mathcal{H}\big) = E(\Delta_0)\mathcal{H}$ for all $t$.   This implies $V(t) E(\Delta_0) = E(\Delta_0)V(t) $ $\forall$ $t$. 
\\

According to (\ref{CL}), a quantum system with semi-bounded Hamiltonian without instantaneous spreading  and without confinement does not have  states, which are localized in bounded regions.
This result will now be refined exploiting translational symmetry. It applies to massive and massless relativistic 
 systems and also to systems with a more general energy-momentum spectrum. Partially we follow Thaller 1992 \cite[Theorem 1.6]{T92} minimizing the assumptions.\footnote{ and avoiding an error in \cite[Theorem 1.6]{T92}  regarding  the spectrum of $H=\lambda(P)$, as there the assumptions on $\lambda$ do not imply that $\lambda^{-1}([a,b])$ is compact}  Lemma (\ref{NLSCS}) is a version of   Hegerfeldt, Ruijsenaars 1980  \cite[II]{HR80}. Let $S$ be a representation of the group $\R^3$ of spatial translations. To ($L\,1,2$) we add the assumption
 
\begin{itemize}
\item[($L3$)] {\it If the state $\varphi$ is localized in $\Delta_{r_0}$, then $S(b)\varphi$ is localized in $\Delta_{2r_0}$ for all $|b|\le \beta$ for some $\beta>0$ depending on $\varphi$.} 
\end{itemize}

Let $E^\textsc{h}$ denote the spectral measure of $H$ and let  $E^\textsc{p}$ denote the joint spectral measure   on the Borel sets of $\R^3$ of the 
generators $P=(P_1,P_2,P_3)$
 of the spatial translations. 
 
 \begin{Def}\label{SCHP} The  \textbf{spectral condition} holds if there are Borel sets $A_n \subset \R$ with $\bigcup_{n\in\N}A_n=\R$ such that for every $n$ and  state $\varphi$ the vector valued measure $E^\textsc{p}(\cdot)E^\textsc{h}(A_n)\varphi$ on $\R^3$  has compact support. 
\end{Def}

Obviously the spectral condition  (\ref{SCHP}) holds if  there are Borel sets $A_n \subset \R$ with $\bigcup_{n\in\N}A_n=\R$ such that
$E^\textsc{p}(\R^3\setminus B_n)E^\textsc{h}(A_n)=0$  for some ball $B_n$, $n\in\N$. Hence the spectral condition   is satisfied, if $H=h(P)$  for some  measurable  semi-bounded function $h:\R^3\to \R$ with $|h(p)|\to \infty$  if $|p|\to \infty$. Take e.g. $A_n:=[-n,n]$. --- A  relativistic system with semi-bounded energy $H$ satisfies  (\ref{SCHP}). Indeed, by the representation theory of $\tilde{\mathcal{P}}$ it follows $C\ge 0$ so that $H^2\ge H^2-C=P^2$, whence $H\ge |P|$.

\begin{Lem}\label{NLSCS} Let the Hamiltonian be semi-bounded and let the  spectral condition hold. Let the localized states of the quantum system satisfy the assumptions \emph{($L\,1,2,3$)}, and $(L\,1)$ also for $2r_0$. Let $\varphi$ be a state localized in $\Delta_0$. Then  $S(b)\varphi$ is localized in $\Delta_{2r_0}$   for all $b\in \R^3$.
 \end{Lem}\\
{\it Proof.} Let $\varphi$ be a state localized in $\Delta_0$, and denote by $\mathcal{L}_{r_0}$ the subspace of states localized in $\Delta_{r_0}$ according to  ($L1$).  Let $\chi\in \mathcal{L}_{r_0}^\perp$.  Then $\int \operatorname{e}^{i\lambda t}\operatorname{d}\langle \chi,E^\textsc{h}(\lambda)\varphi\rangle =\langle \chi,V(t)\varphi\rangle =0$ holds  for all $t$ by (\ref{CL}). Fourier uniqueness theorem, valid also  for complex measures,\footnote{ Let $m$ be a complex measure on $\R$. By the Radon-Nikodym Theorem $m=u|m|$ holds with a measurable $u:\R\to \C$, $|u|=1$. Let $f\in L^1(\R)$. Then, by Fubini's theorem, 
 $\int \operatorname{e}^{-i\lambda t}\operatorname{d}m(\lambda)=0$ $\forall$ $t$ implies $\int f(t) \int \operatorname{e}^{-i\lambda t}\operatorname{d}m(\lambda)\operatorname{d}t =\int \hat{f}(\lambda)u(\lambda)\operatorname{d}|m|(\lambda)=0$. Hence $|m|=0$ since the set of Fourier transforms $\hat{f}$ is dense in $(C_0(\R), ||\cdot||_\infty)$ and $C_c(\R)\subset C_0(\R)$. } yields $\langle \chi,E^\textsc{h}(A)\varphi\rangle =0$ for all Borel sets $A\subset \R$. Hence $E^\textsc{h}(A)\varphi$ is localized in $\Delta_{r_0}$. So, by assumption ($L\,3$), $S(b)E^\textsc{h}(A)\varphi$ is localized in  $\Delta_{2r_0}$ for all $|b|\le \beta$. Let $\chi\in \mathcal{L}_{2r_0}^\perp$.  Then $\int \operatorname{e}^{-ib p}\operatorname{d}\langle \chi,E^\textsc{p}(p)E^\textsc{h}(A)\varphi\rangle =\langle \chi,S(b)E^\textsc{h}(A)\varphi\rangle =0$ holds  for all $|b|\le \beta$.
Obviously,  due to the spectral condition, for  $A=A_n$ the left hand side is entire in $b\in\C^3$, whence $\langle \chi,S(b)E^\textsc{h}(A_n)\varphi\rangle =0$ holds for all $b\in \R^3$ and every $n$. It follows $\langle \chi,S(b)\varphi\rangle =0$. Hence  $S(b)\varphi$ is localized in $\Delta_{2r_0}$   for all $b\in \R^3$. \hfill{$\Box$}\\

\begin{Cor}\label{AOLSBR} Suppose that  for every $b\in \R^3$ the state $S(b)\varphi$ in  \emph{(\ref{NLSCS})} is   localized in $\Delta_0+b$.  Suppose further that a state cannot be localized in each of two widely separated regions.  Then $\Delta_0$ is not bounded.
 \end{Cor}

Regarding  the result (\ref{AOLSBR}) see  also Hegerfeldt 1974 \cite{H74}. Obviously by (\ref{AOLSBR}) there is no causal system (see (\ref{CS})) in case of semi-bounded energy $H$. Under the premise of the relativistic spectral condition (i.e., the joint spectrum of $(H,P)$ lies in the closed forward cone) this  follows also from Perez, Wilde 1977 \cite[Theorem 1]{PW77} and it is shown in  \cite[Problems with Position operators, Theorem]{Y12} 2012 based on  \cite[Lemma]{PW77}.
\\

In the following theorem we refer to  POL   with causal time evolution introduced in sec.\,\ref{SePOL}. See in particular  Eq.\,(\ref{ECPOL}) and Eq.\,(\ref{CCT}). Actually, rotational covariance  is not needed for (\ref{NLS}). 
Call a Borel subset $\Delta\subset \R^3$  \textbf{essentially dense} if  $\overline{\Delta \setminus N}=\R^3$ for every  Lebesgue null set $N$. 

 \begin{The}\label{NLS} Let  localizability be described by a  POL  $T$ with causal time evolution.   
Let the spectral condition \emph{(\ref{SCHP})} hold.
 Suppose that there is a state localized in the region $\Delta_0$. Then $\Delta_0$ is essentially dense.
\end{The}\\
{\it Proof.} Let $\varphi$ be a state localized in $\Delta_0$, i.e., $T(\Delta_0)\varphi=\varphi$. As  $T$ is translational covariant it vanishes at Lebesgue null sets $N$ (see (\ref{PCPOL})(a)). So one checks that the assumptions in (\ref{NLSCS}) are satisfied for $\Delta_0\setminus N$ in place of $\Delta_0$ and for every  $r_0>0$. Then
from  (\ref{NLSCS}) we get immediately $T(A)S(b)\varphi =S(b)\varphi$ $\forall$ $b\in \R^3$  with $A:=\overline{\Delta_0\setminus N}$.     
For $\Gamma:=\R^3\setminus A$  this implies $T(\Gamma) S(b)\varphi =(I-T(A))S(b)\varphi =0$. By translation covariance $T(\Gamma -b)\varphi =0$ $\forall$ $b\in \R^3$ follows. If $\Gamma$ were not empty then, since $\Gamma$ is open,  $\sigma$-subadditivity of $T$ would  imply   the contradiction $\varphi=0$.  \hfill{$\Box$}

In view of (\ref{NLS}) see (\ref{SES}). The result (\ref{NLS}) does not answer the question if there is a  state localized in an open dense set $\Delta$, for which $\R^3\setminus \Delta$ has positive measure. 

So (\ref{NLS}) confirms the result in (\ref{LDNW})
that relativistic systems endowed with NWL show {\it instantaneous spreading} of localized states rather than confinement: if  a state $\varphi$ is bounded localized, then  the  state $V(t)\varphi$ evolved in time  $t\ne 0$ is not bounded localized. \\
\hspace*{6mm} 
This violation of causality  holds  also for
the weak localization of the two-component photon by  Amrein \cite{A69}, since by construction the  localized states are those of a NWL. Cf.  also \cite{S76}.  At last, localizability  of the two-component photon  described by an Euclidean system of covariance in Castrigiano   \cite[4.2.2]{C76}  and Kraus  \cite{K77} is not causal, too, since the localized states are the same as those of the weak localization in \cite{A69}.


\section{Three Contrasting Concepts}\label{SC} The discussion so far shows that  there are three contrasting concepts regarding relativistic quantum systems: {\it causality, positivity of energy (stability),   and localized state}. The least likely of the three to be discarded in the literature seems to be Einstein causality. It means that any signal, which could serve to synchronize clocks, cannot move faster than light.  It is hard to imagine a scenario where cause precede the effect, even if the detection probability of acausal events as predicted by the Newton-Wigner position operator is smaller than $10^{-10^{8}}$ according to Ruijsenaars \cite[sec.\,3]{R81}. Another reason why to uphold causality is the fact that in the Dirac theory it is  the  causal Dirac position operator which serves for the interaction of  the electron to the electromagnetic field through the minimal coupling rather then the Newton-Wigner one. Also in case of the massless Weyl fermions, for which the Newton-Wigner position operator does not even exist, the causal position operator plays the analogous role.\\
\hspace*{6mm} 
There is a third rather weighty argument in favor of causality.  We study relativistic quantum systems endowed with a WL  for which, by definition,  there is no transmission of signals faster than light, thus implementing
 Einstein causality.  Considering only physical systems, viz.  systems with non-negative mass-squared operator and finite spinor dimension, there are exactly three such irreducible systems, the Dirac system and the two Weyl systems.\footnote{Actually we did not succeed as yet to rule out  certain irreducible massive and massless SCT, see (\ref{DMCS}), (\ref{FDWTL}). } This proves  uniqueness and  fundamental  importance of the Dirac system for massive quantum systems. In view of the recently discovered Weyl fermions (see \cite{S et al15}, \cite{L et al15} 2015) this may equally hold true for the Weyl systems with regard to massless quantum systems.  One notes that the requirement of a causal localization also predicts the opposite handedness of particle and antiparticle constituting a Weyl system. 
 First, however, the above classification  shows the far-reaching impact of the principle of causality.\\

Maintaining causality  two possibilities are left for relativistic quantum systems: 
 \begin{itemize}
 \item[($C1$)] {\it causality plus semi-bounded Hamiltonian and  localization without localized states in bounded regions} 
  \item[($C2$)] {\it causality plus localized states  in bounded regions and  non-semi-bounded Hamiltonian} 
 \end{itemize}

Sec.\,\ref{SePOL}, \ref{CTECL}, \ref{SPLSS}  are  concerned with  ($C1$) in  detail. Referring to ($C2$),   in sec.\,\ref{SeCWL}  causal systems are introduced. Then in sec.\,\ref{DS} up to sec.\,\ref{NES} 
 and sec.\,\ref{OWL}, sec.\,\ref{LOCCRS}
 the  concepts developed to meet  ($C1$) and ($C2$) are illustrated by the most important Dirac system and the Weyl systems. A further development of the theory proceeds in sec.\,\ref{ECLNTLHP}.
It is an important issue of these investigations  that ($C1$) and ($C2$) are not opposed to each other but that they are simultaneously present and  intrinsically linked in causal systems.

 
 \section{Positive Operator Valued Localization }\label{SePOL}  
Regarding ($C1$),  in the literature there is the intensively studied concept  of   Euclidean systems of covariance, henceforth called  positive operator valued localization (POL).  See Neumann 1972 \cite{N72}, Scutaru 1977 \cite{S77},  Castrigiano, Henrichs 1980 \cite{CH80}, and \cite{T05} for several other authors. POL were  introduced  independently  by  Castrigiano 1976  \cite{C76},   and    Kraus 1977 \cite{K77} for the localization of massless particles with non-zero helicity and the two-component photon, respectively. This subject is also treated in  \cite[sections F-H]{CL15}, from which some parts of this section are  taken. Sec.\,\ref{CTECL} and \ref{SPLSS} are still concerned with POL. Particularly important is sec.\,\ref{GCPCPOL}, where causality for localizable relativistic quantum systems is introduced. This is the central notion, which will determine  the  investigations in the sequel. 
 
\hspace*{6mm} 
POL are characterized by the fact that the localization observables assigned to the space regions $\Delta$, which for Wightman localizations are projection operators  $E(\Delta)$ with spectrum in $\{0,1\}$, are permitted  to be positive operators $T(\Delta)$ with extended spectrum in $[0,1]$. Thus $T$ is a positive operator valued measure (POM), which is Euclidean covariant
\begin{equation}\label{ECPOL}
U(g)T(\Delta)U(g)^{-1}=T(g\cdot \Delta)
\end{equation}
Here $U$ is the representation of $ISU(2)$ describing the Euclidean symmetry of the system.

\subsection{Interpretation of POL}\label{IPOL}
 POL  is an unconventional concept and requires explanation. So \cite{N72}, \cite{N71},  \cite{K77} refer to Ludwig's reformulation of quantum theory (see  Ludwig 1970, 1983  \cite{L70}), where  a yes-no measurement, called an {\it effect},  is described by an operator $A$ with $0\le A\le I$. Being the system  in the state $\psi$, the probability  for the outcome yes is still given by $\langle \psi, A\psi\rangle$. Following \cite{N71},  \cite{K77} every single  localization observable $T(\Delta)$ is an effect with $\langle \psi, T(\Delta)\psi\rangle$ the probability for finding the particle in $\Delta$. Let us mention that in this theory an observable is described by a POM on the Borel sets of $\R$ and is called a {\it generalized observable}. The benefits for quantum physics of POM describing observables is fully  explained in Busch, Grabowski, Lahti 1997 \cite{BGL97}. There
these observables are called {\it unsharp observables}. According to Heinonen,  Lahti,  Ylinen 2004 \cite{HLY04}, the
POM obtained by smearing out PM represent so-called  
{\it fuzzy observables}  endowed with the interpretation of inaccurate measurements. Fuzzy observables were launched by Ali, Emch 1974 \cite{AE74}, who use the concept just for the construction of a POL for the two-component photon. \\
\hspace*{6mm} 
Within the frame of conventional quantum mechanics one possibility to interpret   POL  is the assumption that there are  described additional degrees of freedom for an elementary particle. But
about their nature the kinematic considerations give no information.  
One could help out intuition   imagining, as in  \cite{C76},  a property of the system like a substance which is spread over space so that 
$\langle \varphi, T(\Delta)\varphi\rangle\in [0,1]$ indicates the expected  fraction of the substance contained in $\Delta$, if    the system is in the state $\varphi$.  \\
\hspace*{6mm} However, in case that the POL is the trace of the WL  of a causal system, as the POL $T^e$ of the Dirac electron or $T^{\chi\eta}$ of the Weyl fermions, we propose 
in sec.\,\ref{NES}  an interpretation of the PO-localization operators   still within conventional quantum mechanics, which  attributes  an important role  to  the negative energy states. In this case, $\langle \varphi, T(\Delta)\varphi\rangle$ results to be the \textbf{probability of localization}  or \textbf{spatial probability} in $\Delta$ in the state $\varphi$. This interpretation will
reconcile the three contrasting concepts discussed in sec.\,\ref{SC}.\\ 
\hspace*{6mm}
Finally, as expressed in \cite{W62}, it might appear more natural  from a physical point of view  to require the existence of localizations operators $T(\Delta)$ only for boxes (cuboids)  and finite unions thereof and to weaken the $\sigma$-additivity of $T$ to finite additivity. But  it is shown in (\ref{EPOMPOVC}) that this generalization is only seeming as, essentially due to translational covariance, $T$ can anyway be extended uniquely to a POL. For the special case  of  WL  compare  \cite[Appendix I]{W62}.


\subsection{Poincar\'e covariance of POL}\label{SPCPOL} The discussion of Poincar\'e covariance for WL  in sec.\,\ref{SePCWL}  applies as well to POL.
So there is the necessity to extend a POL for a relativistic system to regions of spacelike hyperplanes in Minkowski space. Since this extension has to satisfy Poincar\'e covariance, it is obvious how to define this extension. But there is the  question  of  consistency of the definition as explained in sec.\,\ref{SePCWL}. Moreover, we will see that the formulation of causality in sec.\,\ref{CTECL} necessitates the further extension of POL to Lebesgue measurable sets.
\\
\hspace*{6mm} Let the physical system be described by the representation $W$ of $\tilde{\mathcal{P}}$ in the separable Hilbert space $\mathcal{H}$. Let $T$ be a POL  of the system, i.e.,  a POM on the Borel sets of the hyperplane $\varepsilon:=\R^3\equiv \{0\}\times \R^3$, which is Euclidean covariant, i.e.,
 $W(g)T(\Delta)W(g)^{-1}=T(g\cdot \Delta)$ for all $g\in ISU(2)$ (see Eq.\,(\ref{ECPOL})). \\
 \hspace*{6mm} 
  Let $\mathfrak{S}$ denote the set of all Lebesgue measurable sets of spacelike hyperplanes of Minkowski space. The elements of $\mathfrak{S}$  are called  \textbf{flat spatial regions}.
  
 \begin{The}\label{PCPOL} Let $T$ be a POL.
 \begin{itemize}
 \item[(a)] Let $\Delta\subset\R^3$ be a Borel set. Then $T(\Delta)=0$ if and only if $\Delta$ is a Lebesgue null set. Hence the completion of  $T$ is a POL defined for the Lebesgue measurable sets and vanishes just at the Lebesgue null sets.
 \end{itemize}
 Let  $W$ be a representation of $\tilde{\mathcal{P}}$ and let $T$ be Euclidean covariant with respect to 
 $W|_{ISU(2)}$.
\begin{itemize} 
 \item[(b)]  There is a unique positive-operator-valued extension of $T$ on $\mathfrak{S}$, still called $T$, such that Poincar\'e covariance 
 $W(g)T(\Gamma)W(g)^{-1}=T(g\cdot \Gamma)$ $\forall$
   $\Gamma\in \mathfrak{S}$, $g\in \tilde{\mathcal{P}}$ holds.
   \end{itemize}
  \end{The}
  {\it Proof.}  (a)  Choose a total set $\{\varphi_n:n\in\N\}$ of unit vectors  in $\mathcal{H}$. Then the probability  measure $\mu$ on the Borel sets of $\varepsilon$ defined by $\mu(\Delta):=\sum_n2^{-n}\langle \varphi_n,T(\Delta)\varphi_n\rangle$ has the same null sets as $T$, since generally $\langle \varphi,T(\Delta)\varphi\rangle=||\sqrt{T(\Delta)}\varphi||^2=0$ if and only if $T(\Delta)\varphi=0$.
So $\mu$ is quasi-invariant under translations $b\in\varepsilon$, since due to translational covariance of $T$ (see Eq.\,(\ref{ECPOL})) one has $\mu(b+\Delta)=\sum_n2^{-n}\langle U(b)^{-1}\varphi_n,T(\Delta)U(b)^{-1}\varphi_n\rangle=0$ if and only if $T(\Delta)=0$. Hence $\mu$ is equivalent to the Lebesgue measure (see e.g. \cite[\S \,9.1]{K76}).\\
\hspace*{6mm} 
   The well-known completion of an ordinary measure (see e.g. \cite[sec.\,1.1]{CR08}) is equally available for POM. Due to the foregoing result, the completion of $T$, still denoted by $T$, is defined on the $\sigma$-algebra of Lebesgue measurable subsets of $\R^3$, is zero just at Lebesgue null sets, and obviously preserves Euclidean covariance.\\
 \hspace*{6mm} 
 (b) Let $\Gamma \in \mathfrak{S}$. There are $\Delta\subset \varepsilon$ and $g\in \tilde{\mathcal{P}}$ with $\Gamma=g\cdot \Delta$. Note $\Delta\in\mathfrak{S}$. Set $T(\Gamma):=W(g)T(\Delta)W(g)^{-1}$. This definition of $T(\Gamma)$ obviously is forced from Poincar\'e covariance, and, if this definition is unambiguous, then
Poincar\'e covariance of $T$ holds. \\
\hspace*{6mm} 
So it remains to show that  $T(\Gamma)$ is well-defined. To this end let  $\Delta'\subset \varepsilon$ and $g'\in \tilde{\mathcal{P}}$ also satisfy  $\Gamma=g'\cdot \Delta'$. Introduce $h:=g'^{-1}g$. Then $h\cdot \Delta=\Delta'$. It suffices to show $T(\Delta')=W(h)T(\Delta)W(h)^{-1}$.
If $h \cdot \varepsilon=\varepsilon$ holds, then $h\in ISU(2)$. Hence $h$ yields an Euclidean transformation and the assertion follows from Euclidean covariance of $T$. Otherwise, the hyperplane $\sigma:=h\cdot \varepsilon$ is different from $\varepsilon$. Then $\Delta'\subset \varepsilon\cap\sigma$, which is a plane in $\varepsilon$ and hence a  null set with respect to the Lebesgue measure on $\varepsilon$.  Also $\Delta=h^{-1}\cdot \Delta'$ is contained in a plane and hence a Lebesgue null set. The proof is accomplished. \qed\\

In view of (\ref{PCPOL})  let us fix that henceforth a POL is  defined on the $\sigma$-algebra of  the Lebesgue measurable sets. Furthermore, a Lebesgue measurable set will be called simply \textbf{measurable set} or a \textbf{region}. 
\\

\begin{Def} \label{PCPOLB} 
Let $T$ be a positive operator valued map on $\mathfrak{S}$, which is a POM on every spacelike hyperplane. Let $W$ be a representation  of $\tilde{\mathcal{P}}$. 
Then $T$ is called a \textbf{Poincar\'e covariant} POL for $W$ if  the covariance  $W(g)T(\Gamma)W(g)^{-1}=T(g\cdot \Gamma)$ $\forall$
   $\Gamma\in \mathfrak{S}$, $g\in \tilde{\mathcal{P}}$ holds.
\end{Def}


\section{Causal Time Evolution and Causal Localization}\label{CTECL}

\subsection{POL with causal time evolution}\label{CPOL} For every region $\Delta\subset \R^3$ and  time $t\in\R$   the region of influence is
$$\Delta_t:=\{y\in\R^3:\,\exists \,x\in\Delta  \text{ with } |y-x|\le |t|\}$$
It is the set of points, which are  reached from $\Delta$  within time  $|t|$ with  velocity of light $c=1$. Every event $(t,x)\in\R^4$ determines its region of influence $\{x\}_t\subset \R^3$, which is  the closed ball with center $x$ and radius $|t|$. So  $\Delta_t=\bigcup_{x\in\Delta} \{x\}_t$.\\
\hspace*{6mm}  Causal time evolution requires that  after time $t$, respectively before time $t$,  the  probability of localization in $\Delta_t$ is not less than originally in $\Delta$. Hence $\langle \varphi, T(\Delta)\varphi\rangle \le \langle  V(t)\varphi, T(\Delta_t) V(t)\varphi\rangle$ holds for every state $\varphi$, where $V(t)$ describes the time translation by  time $t$. In other words, for causal time evolution one has
\begin{equation}\label{CCT}
V(t) T(\Delta) V(t)^{-1} \le T(\Delta_t)
\end{equation}

\begin{Def}\label{POLCTE}
One calls $(V,U,T)$ a  \textbf{POL with causal time evolution}, if $(V,U)$ is a representation of the little kinematical group, $(U,T)$ is a POL, and Eq.\,(\ref{CCT}) is satisfied.\footnote{ POL with causal time evolution  
 are called causal POL in \cite{CL15}, whereas causal POL in sec.\,\ref{GCPCPOL} are not considered in \cite{CL15}.}
\end{Def}

 In case of a Poincar\'e covariant POL   (\ref{PCPOLB}) let us verify explicitely that   Eq.\,(\ref{CCT}) does not single out a particular frame. Indeed, by covariance   one has a  POL with causal time evolution $(V_\sigma,U_\sigma,T_\sigma)$ on every spacelike hyperplane $\sigma$ 
 by $T_\sigma(\Gamma):= T(\Gamma)$ for all regions $\Gamma \subset \sigma$ with $U_\sigma(g) := W(g)$ for all $g\in g_0\, ISU(2) g_0^{-1}$ and
 $V_\sigma(g) := W(g)$ for all $g\in \{g_0 (t,0,I_2)g_0^{-1}: t\in \R\}$, where 
 $\sigma= g_0\cdot( \{0\}\times \R^3)$ for some 
 $g_0\in \tilde{\mathcal{P}}$. Using $g_0\cdot (g_0^{-1}\cdot \Gamma)_t=
  \Gamma_{g_0^{}(t,0,I_2)g_0^{-1}}$ one obtains the frame-independent formulation of Eq.\,(\ref{CCT}) 
 $$W(g)T(\Gamma)W(g)^{-1}\le T(\Gamma_g)$$
for $\Gamma \subset \sigma=g_0\cdot( \{0\}\times \R^3)$ and all time translations $g=g_0^{}(t,0,I_2)g_0^{-1}$  with respect to $\sigma$. So the concept of POL with causal time evolution obeys the principle of special relativity.\\

 \subsection{Causal localization}\label{GCPCPOL}  Although a Poincar\'e covariant POL with causal time evolution does not violate the principle of special relativity it is still a preliminary concept for the description of causality. Indeed, as shown in (\ref{CFDNWL}) the frame-dependence of NWL concerns not only time translations but as well boosts. Thus as explained after (\ref{CFDNWL})  also the latter constitute an acausal behavior of the Newton-Wigner system. Moreover, (\ref{ESTFINC})  is an example  of a relativistic system  
  (unitarily related to a Weyl system to be exact), for which time evolution is causal, which  localizes frame-independently, and which nevertheless  behaves acausal under boosts.
 \\
\hspace*{6mm}  
  Let $W$ be a representation  of $\tilde{\mathcal{P}}$ describing a relativistic system and let $T$ be a Poincar\'e covariant POL  (\ref{PCPOLB}) for $W$. Then Einstein causality requires more than the condition in Eq.\,(\ref{CCT}). Indeed, let $\Gamma$ be a measurable subset of a spacelike hyperplane and let $\sigma$ be any spacelike hyperplane. Then 
\begin{equation}\label{RIHPR}  
 \Gamma_\sigma :=\bigcup_{\mathfrak{z} \in \Gamma}\{\mathfrak{x}\in \sigma: (\mathfrak{x}-\mathfrak{z}) \cdot (\mathfrak{x}-\mathfrak{z}) \ge 0\}
\end{equation}  
   is the \textbf{region of influence} of $ \Gamma$ in $\sigma$.  It is the set of all points  $\mathfrak{x}$ in $\sigma$, which can be reached  from some point $\mathfrak{z}$ in $\Gamma$ by a signal not moving faster than light. Therefore, in particular, the  probability of localization of the system in $\Gamma_\sigma$ cannot be less than originally in $\Gamma$. Hence causality  requires
\begin{equation}\label{FICCT}
 T(\Gamma) \le T(\Gamma_\sigma)
\end{equation}

\begin{Def}\label{CPOLD}
A POL  $T$ for the system $W$  is \textbf{causal}  if it is Poincar\'e covariant and satisfies Eq.\,(\ref{FICCT}).
\end{Def}

In (\ref{FICCT}) the case of parallel  hyperplanes  yields the causal time evolution of the system with their temporal distance determining  the time translation. Indeed, due to the Poincar\'e covariance of the POL $T$ is suffices to ascertain this    for the standard hyperplane $\varepsilon$. Then for $\Delta\subset\varepsilon$,  $t\in\R$, and $\sigma:=\{-t\}\times \R^3$ check $\Delta_\sigma=(-t,0,I_2)\cdot \Delta_t$, whence (\ref{CCT}) by covariance of $T$.\\


  Let $\Delta$  be a measurable set of a spacelike hyperplane $\sigma$, let $g\in \tilde{\mathcal{P}}$. Then
 \begin{equation} 
 \Delta_g:=\bigcup_{\mathfrak{z} \in g\cdot \Delta}\{\mathfrak{x}\in\sigma: (\mathfrak{x}-\mathfrak{z})\cdot (\mathfrak{x}-\mathfrak{z})\ge 0\} 
 \end{equation}
is  the region of influence in $\sigma$ of $g\cdot \Delta$. One has  
 \begin{equation}\label{RIGD}
 \Delta_g=\bigcup_{\mathfrak{y}\in\Delta}\{\mathfrak{y}\}_g, \quad \{\mathfrak{y}\}_g=\{\mathfrak{x}\in\sigma: (\mathfrak{x}-g\cdot\mathfrak{y})^{\cdot 2}\ge 0\}
\end{equation}
where $\{\mathfrak{y}\}_g$ is the region of influence  in $\sigma$ of the single event $g\cdot\mathfrak{y}$. If $g=(t,0,I_2)\equiv t$ is a mere time translation then consistently $\Delta_g=\Delta_t$.
Note the relation
 \begin{equation}\label{RRI}
 \Gamma_\sigma=\big(g^{-1}\cdot \Gamma\big)_g=g^{-1}\cdot \Gamma_g \text{\quad\quad for \;}\Gamma\subset g\cdot \sigma
 \end{equation}
The second equality is a special case of the formula  $h\cdot\Gamma_g=(h\cdot \Gamma)_{hgh^{-1}}$ for all \,$\Gamma\in\mathfrak{S}$ and $h,g\in \tilde{\mathcal{P}}$.

We like to mention that in Jancewicz 1977 \cite{J77} a Poincar\'e covariant  probability density current, which is a 4-vector operator valued distribution, is related to what we call a causal POL and in particular causal WL. 

\begin{Exa}\label{ERIUB} 
{\it Let $y\in\R^3\equiv\{0\}\times\R^3$, $t\equiv (t,0,I_2)$ for $t\in\R$, and $A_\rho\equiv (0,A_\rho)$ for $A_\rho=\operatorname{e}^{\,\rho\,\sigma_3/2}$, $\rho\in\R$. Then the regions of influence in $\R^3$ of $t\cdot y$ and $A_\rho\cdot y$ are

\hspace*{6mm}
\emph{(a)} $\{y\}_t$ the ball with center $y$ and radius $|t|$.

\hspace*{6mm}
\emph{(b)} $\{y\}_{A_\rho}$ the ball with center $(y_1,y_2,\cosh(\rho)\,y_3)$ and radius $|\sinh(\rho)\,y_3|$.}

Let $e\in\R^3$, $|e|=1$, $0\le a<b$,  and  $A_{\rho e}\equiv (0,A_{\rho e})$ for $A_{\rho e}:=\operatorname{exp}(\frac{\rho}{2}\sum_{k=1}^3e_k\sigma_k)$ representing the boost in direction $e$ with rapidity $\rho$. We consider a strip.

\hspace*{6mm}
{\it \emph{(c)} $\{x\in\R^3: a\le xe \le  b\}_{A_{\rho e}}=\{x\in\R^3: a\operatorname{e}^{-|\rho|}\le xe \le b\operatorname{e}^{|\rho|}\}$.}

Finally we treat a cylinder. Let $c>0$ and let $c_\rho:=\cosh(\rho)$, $s_\rho:=\sinh(|\rho|)$, $t_\rho:=\tanh(|\rho|)$. Then

\begin{itemize}
\item[(d)] {\it $\{x\in\R^3: x_1^2+x_2^2\le c^2, \,a\le x_3 \le  b\}_{A_{\rho}}$ is the solid of revolution  obtained by rotating the  curve  around the $x_3$-axis, which joins the points $P_1:=(c,0,\operatorname{e}^{-|\rho|}a)$ and $P_2:=(c+t_\rho a,0,a/c_\rho)$ by the short segment of the circle around $(c,0,c_\rho a)$ with radius $s_\rho a$, $P_2$ and  $P_3:=(c+t_\rho b,0,b/c_\rho)$ by a straight line, and finally $P_3$ and $P_4:=(c,0,\operatorname{e}^{|\rho|} b)$ by the short segment of the circle around $(c,0,c_\rho b)$ with radius $s_\rho b$.}
\end{itemize}
\end{Exa}
{\it Proof.} (a) is obvious. As to (b), $A_\rho\cdot y=\big(\sinh(\rho)\,y_3,y_1,y_2, \cosh(\rho)\,y_3\big)$, whence the assertion by Eq.\,(\ref{RIGD}). Turn to (c). Note that $A_{\rho e}=B(e)\,A_\rho\, B(e)^{-1}$ due to Eq.\,(\ref{CRCS}) and that $B(e)\cdot\{x:a\le x_3\le b\}=\{x:a\le xe \le b\}$. Thus the assertion reduces to 
$\{ a\le x_3 \le  b\}_{A_\rho}=\{ a\operatorname{e}^{-|\rho|}\le x_3 \le b\operatorname{e}^{|\rho|}\}$, which holds by (b) and Eq.\,(\ref{RIGD}). Also (d) follows by (b) and Eq.\,(\ref{RIGD}) exploiting the rotational symmetry around the $x_3$-axis. \qed

\begin{Lem}\label{VEFCPOL} Let   $T$ be a Poincar\'e covariant POL for $W$. It is causal if and only if  one of the equivalent conditions \emph{(a)} -- \emph{(e)} holds.
\begin{itemize}
\item[(a)] $T(\Gamma) \le T(\Gamma_\sigma)$\quad for all \,$\Gamma\in\mathfrak{S}$ and all spacelike hyperplanes $\sigma$
\item[(b)] $T(g\cdot\Delta)\,\le \,T(\Delta_g)$ or, equivalently, $T(\Delta)\le T\big(g^{-1}\cdot \Delta_g\big)$ or, equivalently, $T(\Delta)\le T\big((g^{-1}\cdot\Delta)_g\big)$\quad 
for all \,$\Delta\in\mathfrak{S}$ and $g\in \tilde{\mathcal{P}}$
\item[(c)] $W(g)T(\Delta)W(g)^{-1}\le \,T(\Delta_g)$\quad for all \,$\Delta\in\mathfrak{S}$ and $g\in \tilde{\mathcal{P}}$

\item[(d)] $\langle \varphi, T(\Delta)\varphi \rangle \le \langle W(g)\varphi,T(\Delta_g)W(g)\varphi \rangle$\quad for all \,$\Delta\in\mathfrak{S}$, $g\in \tilde{\mathcal{P}}$, and states $\varphi$

\item[(e)]  $T(\Gamma)+T(\Gamma')\le I$\quad for all spacelike separated $\Gamma,\Gamma'\in\mathfrak{S}$
\end{itemize}
If $T$ is a WL  $E$, then 
\begin{itemize}
\item[(f)] $E(\Gamma)\, E(\Gamma')=0$\quad for all spacelike separated $\Gamma,\Gamma'\in\mathfrak{S}$
\end{itemize}
is a further equivalent condition.
\end{Lem}

{\it Proof.} (a) is the definition of  causal POL. --- The equivalence of (a) and (b) is obvious setting $g\cdot\Delta =\Gamma$ and $\Delta=g^{-1} \cdot\Gamma$, respectively, as $\Gamma_\sigma=\Delta_g$ holds. Recall Eq.\,(\ref{RRI}). --- (b) and (c) are equivalent by (\ref{PCPOLB}). ---  The equivalence of (c) and (d) is obvious. --- Now we show the equivalence of (a) and (e). Assume first (a). Let $\Gamma\subset \sigma$. Then $T(\Gamma')\le T(\Gamma'_\sigma)$. Since  $\Gamma$ and $\Gamma'$ are spacelike separated, $\Gamma \cap \Gamma'_\sigma=\emptyset$ and hence $T(\Gamma)+T(\Gamma'_\sigma)= T(\Gamma \cup \Gamma'_\sigma)\le I$. This yields (e). Conversely, for $\Gamma$ and $\sigma$ in (a),
  obviously $\Gamma$ and $\sigma \setminus \Gamma_\sigma$ are spacelike separated. Therefore, by (e), $I\ge T(\Gamma) + T(\sigma \setminus \Gamma_\sigma) = T(\Gamma) + I - T(\Gamma_\sigma)$, whence (a). --- Finally we turn to the equivalence of (e) for the WL $E$ and (f). By (e), $E(\Gamma) \le I - E(\Gamma')$, whence the projections $E(\Gamma)$ and $I - E(\Gamma')$ commute. Then $E(\Gamma)$ and $E(\Gamma')$ commute. So (f) follows. The converse implication is obvious.\qed\\

Let us comment on  (\ref{VEFCPOL}).  As to (a) note that Poincar\'e covariance of $T$  (\ref{PCPOLB}) implies \textbf{Poincar\'e covariance of causality} $W(h) T(\Gamma)W(h)^{-1}\le 
T\big((h\cdot \Gamma)_{h\cdot \sigma}\big)$ as $h\cdot \Gamma_\sigma= (h\cdot \Gamma)_{h\cdot \sigma}$ for all $h\in \tilde{\mathcal{P}}$. --- Regarding covariance for (b) use the analogous formula  $h\cdot\Delta_g=(h\cdot \Delta)_{hgh^{-1}}$. ---  (c) extends 
Eq.\,(\ref{CCT}) for causal time evolution. Indeed, recall the frame-independent formulation of causal time evolution given in sec.\,\ref{CPOL}. --- Property (f) is the  \textbf{local orthogonality},  well-known from local quantum physics. --- We add some obvious inequalities for a causal POL $T$ following from (\ref{VEFCPOL}), which may be useful: $T(g'g\cdot \Delta)$ is less or equal to each of the four positive operators $T(\Delta_{g'g})$,  $T((\Delta_g)_{g'})$, $T((g\cdot\Delta)_{g'})$, $T(g'\cdot\Delta_{g})$, where $\Delta_{g'g}\subset \sigma$, $(\Delta_g)_{g'}\subset \sigma$, $(g\cdot\Delta)_{g'}\subset g\cdot\sigma$, and $g'\cdot\Delta_{g}\subset g'\cdot \sigma$.
\\
\hspace*{6mm} 
According to (d) in (\ref{VEFCPOL}) the probability of localization of the system in the region of influence $\Delta_g$ in the state $W(g)\varphi$ is not less than that in the original region $\Delta$ and state $\varphi$.  Note that $\Delta_g$ is bounded if $\Delta$ is bounded. In case of a causal WL this implies the \textbf{frame-independence of localization}. This qualitative property means  that, if the system is localized  in a  bounded region with respect  to some  frame,   this holds true with respect to all  frames. Cf. the discussion after (\ref{LDNW}). So we keep

\begin{Cor} \label{FIDCWL}
A causal WL is frame-independent.
\end{Cor}

 As mentioned, the property (c) implies causal time evolution Eq.\,(\ref{CCT}). By (\ref{SCTLFD}) and (\ref{FDWTL}) there are  massive, respectively massless Poincar\'e covariant WL with causal time evolution, which   do not localize frame-independently. Consequently, by (\ref{FIDCWL}) they are not causal. Finally,  as mentioned there are  Poincar\'e covariant frame-independent WL with causal time evolution, which are not  causal, see (\ref{ESTFINC}). Hence the converse of (\ref{FIDCWL}) does not hold.

\subsection{Measurability of the region of influence}\label{MRI} For the causality condition Eq.\,(\ref{FICCT}) it is necessary that the region of influence $\Gamma_\sigma$ (\ref{RIHPR}) is measurable. 
Borel measurability has been abandoned in sec.\,\ref{SPCPOL}
just because  $\Gamma_\sigma$  might not be a Borel set although $\Gamma$ itself is Borel. Note that  $\Gamma_\sigma$ can be obtained from $\Gamma$ by harmless mappings as $k:\tau\times \sigma\to \tau\times \R$, $k(\mathfrak{y},\mathfrak{x}):=\big(\mathfrak{y}, (\mathfrak{x}-\mathfrak{y})^{\cdot 2}\big)$ and the projection $\pi_\sigma$. One has $\Gamma_\sigma=\pi_\sigma\big(k^{-1}(\Gamma\times [0,\infty[\,)\big)$. The problem is that the projection of a Borel set need not be Borel, but indeed, by the Measurable Projection Theorem, it is Lebesgue measurable. --- The measurability of  $\Gamma_\sigma$  follows also from the analysis  of  $\Gamma_\sigma$ given  in (\ref{LMRI}).
 Compare
Leiseifer 2014 \cite[chapter 2]{L14},  which treats a special case. 


\begin{Lem}\label{LMRI} Let $\sigma,\tau$ be two different spacelike hyperplanes. Let $\Gamma$ be any subset of $\tau$ and $\Gamma_\sigma =\bigcup_{\mathfrak{y} \in \Gamma}\{\mathfrak{x}\in \sigma: (\mathfrak{x}-\mathfrak{y})^{\cdot 2} \ge 0\}$ the region of influence of $\Gamma$ in $\sigma$. Define $M(\Gamma,\sigma):=\bigcup_{\mathfrak{y} \in \Gamma}\{\mathfrak{x}\in \sigma: (\mathfrak{x}-\mathfrak{y})^{\cdot 2} > 0\}$,  $N(\Gamma,\sigma):=\{\mathfrak{x}\in \sigma: \sup_{\mathfrak{y} \in \Gamma}(\mathfrak{x}-\mathfrak{y})^{\cdot 2} =0\}$. Then
\begin{itemize}
\item[(a)] $M(\Gamma,\sigma)\subset \Gamma_\sigma$
 and $M(\Gamma,\sigma)$ is open in $\sigma$ 
  \item[(b)] $N(\Gamma,\sigma)$ is closed in $\sigma$ and a  null set of $\sigma$
\item[(c)] $\Gamma_\sigma\subset M(\Gamma,\sigma)\cup N(\Gamma,\sigma)$  
\item[(d)] $\Gamma_\sigma$ is  measurable
\end{itemize}
\end{Lem}
{\it Proof.} (a) $M(\Gamma,\sigma)\subset \Gamma_\sigma$ is obvious. $M(\Gamma,\sigma)$ is open in $\sigma$, since $M(\Gamma,\sigma)=\sigma \cap \bigcup_{\mathfrak{y} \in \Gamma} (\mathfrak{y}+C)$ with $C:=\{\mathfrak{z}: \mathfrak{z}^{\cdot 2}>0\}$ the open  double cone.\\
\hspace*{6mm}
(c) Let $\mathfrak{x}\in \Gamma_\sigma\setminus M(\Gamma,\sigma)$. Then there is $\mathfrak{y} \in \Gamma$ with $(\mathfrak{x}-\mathfrak{y})^{\cdot 2} \ge 0$. Therefore $\sup_{\mathfrak{y}' \in \Gamma}(\mathfrak{x}-\mathfrak{y}')^{\cdot 2} \ge 0$. But $\sup_{\mathfrak{y}' \in \Gamma}(\mathfrak{x}-\mathfrak{y}')^{\cdot 2} > 0$ is excluded as $\mathfrak{x}\not\in M(\Gamma,\sigma)$. Hence $\mathfrak{x}\in N(\Gamma,\sigma)$.\\
\hspace*{6mm}
(d) is an obvious consequence of (a) -- (c).\\
\hspace*{6mm}
(b) We show that $N(\Gamma,\sigma)$ is closed. Let $\mathfrak{x}_n\in N(\Gamma,\sigma)$ with $\mathfrak{x}_n\to\mathfrak{x}$ for some $\mathfrak{x} \in\R^4$. Then, obviously,  $\mathfrak{x}\in\sigma$ and
$\sup_{\mathfrak{y}' \in \Gamma}(\mathfrak{x}-\mathfrak{y}')^{\cdot 2} \le0$.
For every $n$ there is $\mathfrak{y}_n\in\Gamma$ such that $-\frac{1}{n}\le (\mathfrak{x}_n-\mathfrak{y}_n)^{\cdot 2} \le 0$. We use now the representation $\tau=\{\mathfrak{y}: \mathfrak{y}\cdot \mathfrak{e}_\tau=\rho_\tau\}$  with  $\mathfrak{e}_\tau=(1,e_\tau)$, $|e_\tau|<1$ and $\rho_\tau\in\R$, see after (\ref{ESLHSF}). Hence $ \mathfrak{y}\in\tau \Leftrightarrow y_0=\rho_\tau+ ye_\tau$.
Then $-\frac{1}{n}\le (x_{n0}-\rho_\tau-y_ne_\tau)^2-(x_n-y_n)^2 \le 0$. As $|y_ne_\tau|\le|y_n||e_\tau|$ with $|e_\tau|<1$ this implies that $(y_n)_n$ is bounded. So there is a subsequence of $(\mathfrak{y}_n)_n$, which converges to some $\mathfrak{y}\in\overline{\Gamma}$. It satisfies $(\mathfrak{x}-\mathfrak{y})^{\cdot 2} = 0$. Hence $\sup_{\mathfrak{y}' \in \Gamma}(\mathfrak{x}-\mathfrak{y}')^{\cdot 2} =0$, whence $\mathfrak{x}\in N(\Gamma,\sigma)$.\\
\hspace*{6mm}
It remains to show that $N(\Gamma,\sigma)$ is a  null set of $\sigma$. Obviously $N(\Gamma,\sigma)=N(\overline{\Gamma},\sigma)$. Hence assume without restriction that $\Gamma$ is closed. Then  the foregoing considerations show that for every $\mathfrak{x}\in N(\Gamma,\sigma)$ there is $\mathfrak{y}\in \Gamma$ with $(\mathfrak{x}-\mathfrak{y})^{\cdot 2}=0$. Moreover, due to the covariance $g\cdot N(\Gamma,\sigma)=N(g\cdot \Gamma,g\cdot \sigma)$, without restriction  $\sigma=\{\mathfrak{x}:x_0=0\}$ and either (i) $\tau=\{\mathfrak{y}:y_0=s y_3\}$ for some $s\in ]0,1[$ or (ii) $\tau=\{\mathfrak{y}:y_0=s\}$ for some $s\ne 0$.\\
\hspace*{6mm}
 We turn to the  case (i).  Let $Q:=\{q\in\mathbb{Q}^3:q_3\ne 0\}$. For $q\in Q$ let $r_q:=\frac{s}{4+s}|q_3|$ and $K_q:=\{y\in\R^3: |y-q|\le r_q\}$.  Check that $\R^3\setminus \{y:y_3=0\}=\bigcup_{q\in Q} K_q$. Put 
  $\Gamma_q:=\Gamma\cap \{\mathfrak{y}\in\tau: y\in K_q\}$, and $\Gamma_0:=\{\mathfrak{y}\in\tau:y_3=0\}$.   We claim $N(\Gamma,\sigma)\subset\bigcup_{q\in Q\cup\{0\}} N(\Gamma_q,\sigma)$. Indeed, let $\mathfrak{x}\in N(\Gamma,\sigma)$. Choose $\mathfrak{y}\in \Gamma$ with $(\mathfrak{x}-\mathfrak{y})^{\cdot 2}=0$. There is 
$q\in Q\cup\{0\}$ with $\mathfrak{y}\in \Gamma_q$.  Plainly $\sup_{ \mathfrak{y}'\in\Gamma_q}(\mathfrak{x}-\mathfrak{y}')^{\cdot 2}\le 0$. Hence $\sup_{ \mathfrak{y}'\in\Gamma_q}(\mathfrak{x}-\mathfrak{y}')^{\cdot 2}= 0$, whence $\mathfrak{x}\in N(\Gamma_q,\sigma)$.\\
\hspace*{6mm}
Note that $N(\Gamma_0,\sigma)\subset \{\mathfrak{x}\in\sigma: x_3=0\}$ is a null set of $\sigma$. Hence for the following  fix $q\in Q$. Obviously it suffices to consider  $\Gamma\subset  \{\mathfrak{y}\in\tau: |y-q|\le r_q\}$ and
  to show that $N(\Gamma,\sigma)$ is a null set of $\sigma$.\\
\hspace*{6mm}
 For $\mathfrak{y}\in\Gamma$ one has $|q_3|-|y_3|\le |q_3-y_3|\le |y-q|\le r_q$, whence $|y_3|\ge |q_3|-r_q=\frac{4}{4+s}|q_3|$ and  hence $|y_0|=s|y_3|\ge 4r_q$. Further note that $N(\Gamma,\sigma)$ is bounded and hence compact. Indeed, every $\mathfrak{x}\in N(\Gamma,\sigma)$ satisfies $|s y_3|=|x-y|$ for some $\mathfrak{y}\in \Gamma$, whence $|x|<2(r_q+|q|)$.
\\
\hspace*{6mm}
For a further reductional step consider $U:=\{u\in\R^3:(u_1-q_1)^2+(u_2-q_2)^2<r_q^2\}$. Obviously $\{\mathfrak{x}\in\sigma: x\in R_q\,U\}$, where $R_q\equiv (q-Rq,R)\in ISO(3)$ is any rotation around $q$,  provides an open cover of $\sigma$. So finitely many of these sets  cover  $N(\Gamma,\sigma)$. Then there are also finitely many  $R_q$ such that the sets 
$\{\mathfrak{x}\in\sigma: x\in R_q\,H\}$ with $H:=\{u\in\R^3:(u_1-q_1)^2+(u_2-q_2)^2\le r_q^2, \,u_3-q_3\ge 0\}$ cover $N(\Gamma,\sigma)$. So it suffices to show that for every $R_q$ the set $S_0:=\{\mathfrak{x}\in\sigma: x\in R_q\,H\}\cap N(\Gamma,\sigma)$ is a null set of $\sigma$. Recall $\Gamma\subset  \{\mathfrak{y}\in\tau: |y-q|\le r_q\}$.\\
\hspace*{6mm}
So far the reductional steps. Now put $S_\alpha:=\mathfrak{a}+S_0$  with $\mathfrak{a}:=\alpha(0,R_qe)$, $e:=(0,0,1)$, $\alpha\ge 0$.   We show now that the sets $S_\alpha$ are mutually disjoint. Let $0\le\alpha\le\alpha'$ and let $\mathfrak{z}\in S_\alpha\cap S_{\alpha'}$. There are $\mathfrak{x},\mathfrak{x}'\in S_0$ such that $\mathfrak{z}=\mathfrak{a}+\mathfrak{x}=\mathfrak{a}'+\mathfrak{x}'$.
As $\mathfrak{x},\mathfrak{x}'\in N(\Gamma,\sigma)$, there are $\mathfrak{y},\mathfrak{y}'\in \Gamma$ with $(\mathfrak{x}-\mathfrak{y})^{\cdot 2}=0$, $(\mathfrak{x}'-\mathfrak{y}')^{\cdot 2}=0$. As $\mathfrak{x}'
\in N(\Gamma,\sigma)$ and $\mathfrak{y}\in \Gamma$ one has $(\mathfrak{x}'-\mathfrak{y})^{\cdot 2}\le0$. Use $\mathfrak{x}'=\mathfrak{x}+\mathfrak{a}-\mathfrak{a}'$. Then $0\ge (\mathfrak{x}-\mathfrak{y}+\mathfrak{a}-\mathfrak{a}')^{\cdot 2}=(\mathfrak{x}-\mathfrak{y})^{\cdot 2}+2(\mathfrak{x}-\mathfrak{y})\cdot (\mathfrak{a}-\mathfrak{a}')+(\mathfrak{a}-\mathfrak{a}')^{\cdot 2}=-2(\alpha-\alpha')(x-y)R_qe -(\alpha-\alpha')^2=(\alpha'-\alpha)\big( 2(x-y)R_qe+(x'-x)R_qe\big)=(\alpha'-\alpha)(x+x'-2y)R_qe$. Introduce $u:=R_q^{-1}x\in H$, similarly $u'$,  and $v:=R_q^{-1}y$. Note $|v-q|\le r_q$. Then we find $0\ge (\alpha'-\alpha)(u+u'-2v)e=(\alpha'-\alpha)(u_3+u'_3-2v_3)$. ---
We exploit now $(\mathfrak{x}-\mathfrak{y})^{\cdot 2}=0$.  Recall $4r_q\le |y_0|$. So $16r_q^2\le y_0^2=|x-y|^2=|(u-q)-(v-q)|^2\le 2(|u-q|^2+|v-q|^2)\le 2r_q^2+2(u_3-q_3)^2+2r_q^2$, whence $u_3-q_3>2r_q$. Analogously $u'_3-q_3>2r_q$ holds. Therefore, 
$u_3+u'_3-2v_3=u_3-q_3+u'_3-q_3-2(v_3-q_3)>2r_q+2r_q-2r_q=2r_q>0$. This implies $\alpha'=\alpha$.\\
\hspace*{6mm}
The closed sets $S_\alpha$, $0\le\alpha\le1$ lie in a compact set of $\sigma$, which has finite measure. By  the translational invariance of the Lebesgue measure each $S_\alpha$ has the same measure. Hence this measure is zero, thus accomplishing the proof for the  case (i).\\
\hspace*{6mm}
The proof for the case (ii) is simpler as $|y_0|=|s|>0$ holds independently of $y$. Therefore it suffices to consider the cover of $\R^3$ by $K_q:=\{y: |y-q|\le |s|/4\}$ with $q\in\mathbb{Q}^3$.\qed\\


\section{Point-Localized Sequences of States}\label{SPLSS}  
Provided the spectral condition (\ref{SCHP}) holds,   (\ref{NLS})  shows that for a  POL with causal time evolution  there cannot exist a  state  localized  in a non-essentially dense region $\Delta$, i.e., which is an eigenvector of $T(\Delta)$ with eigenvalue $1$. This means that  in any state  the  probability of localization in a closed $\Delta\ne \R^3$ is less than $1$.  So in \cite{C81}  a property is introduced, which is mathematically weaker than that of a localized state  in $\Delta$ but physically equivalent to it, i.e., $||T(\Delta)|| = 1$.  Indeed, as
\begin{equation*} \label{CID}
||T(\Delta)||= \sup \left\{\langle \varphi, T(\Delta)\varphi\rangle:\, ||\varphi||=1\right\}
\end{equation*}
 norm $1$  means that  the system can be localized within that region  $\Delta$ by a suitable preparation, not strictly but as accurately as desired.  \\
\hspace*{6mm} For obvious physical reasons one is interested in POM with $||T(B)||=1$ for every however small  open ball $B\ne \emptyset$.  In \cite[sec.\,G]{CL15} they are  called    \textbf{separated} POM. If the POM $T$ is separated, then obviously  for every point $b\in\R^3$
there is a sequence $(\varphi_n)$ of states satisfying 
\begin{equation}\label{SSL}
\big\langle \varphi_n,T(B )\, \varphi_n\big\rangle \to 1,  \quad n\to \infty
\end{equation}
for every  open ball $B$ around $b$. This means that by a suitable preparation the system can be localized around $b$ as good as desired, thus distinguishing $b$ from any other point. According to \cite[sec.\,G]{CL15},  any $(\varphi_n)$ satisfying  Eq.\,(\ref{SSL}) is called a \textbf{sequence of states localized  at} $b$.

\subsection{Properties of point-localized sequences}

  Euclidean covariance   Eq.\,(\ref{ECPOL}) and causal time evolution  Eq.\,(\ref{CCT})  of a POL $T$ imply the following transformation properties of point-localized sequences of states. 

\begin{The}\label{ECCPLSS} Let $(\varphi_n)$ be   localized at $b$. Then  $(U(g)\varphi_n)$ is localized at $g\cdot b$  for  $g\in ISU(2)$. If $V$ is a causal time evolution for $T$ then $(V(t)\varphi_n)$ for $t\in\R$ satisfies  $\langle V(t)\varphi_n,T(B) \,V(t)\varphi_n\rangle\to 1$ for every ball $B$ around $b$ with radius greater than $|t|$.
\end{The}

 {\it Proof.}  If $B$ is a ball around  $g\cdot b$, then $g^{-1}\cdot B$ is a ball around $b$ and, $\langle U(g)\varphi_n, T(B) U(g)\varphi_n\rangle=\langle \varphi_n, T(g^{-1}\cdot B)\varphi_n \rangle\to 1$.  --- Let  $V$ be causal time evolution and $B$ a ball around $b$ with radius $r>|t|$. Then $(\R^3\setminus B)_t=\R^3\setminus B'$ with $B'$ the ball around $b$ with radius $r-|t|$. Therefore $0\le 1-\langle V(t)\varphi_n,T(B) \,V(t)\varphi_n\rangle = \langle V(t)\varphi_n,T(\R^3\setminus B) \,V(t)\varphi_n\rangle \le \langle \varphi_n,T((\R^3\setminus B)_t) \varphi_n\rangle = \langle \varphi_n,T(\R^3\setminus B')\varphi_n\rangle = 1- \langle \varphi_n, T(B')\varphi_n\rangle \to 0$. 
 Hence $\langle V(t)\varphi_n,T(B) \,V(t)\varphi_n\rangle \to 1$. 
  \hfill{$\Box$}\\

 In (\ref{ECCPLSS})  one cannot expect  $\langle V(t)\varphi_n,T(B) \,V(t)\varphi_n\rangle\to 1$ 
 for a ball $B$ around $b$ with radius smaller than $|t|$, as $\{b\}_t=\{x\in\R^3:|x-b|\le|t|\}$ is the region of influence of the event $(t,b)$. If $T$ is even causal then due to causality (\ref{VEFCPOL})$(c)$ one has more generally
\\

\begin{The}\label{PCCPLSS} Let $T$ be a causal POL for $W$. Let $(\varphi_n)$ be    localized at $b$. Let $g\in\tilde{ \mathcal{P}}$   and put $(t,b'):=g\cdot(0,b)$. Then
$$ \langle W(g)\varphi_n, T(B)\,W(g)\varphi_n\rangle \to 1$$
for every ball $B$ around $b'$ with radius greater than $|t|$.  In particular, if $(\varphi_n)$ is localized at $0$, then  $(W(A)\varphi_n)$ is so for $A\in SL(2,\C)$.
\end{The} 
 
{\it Proof.}    Note that  $g=(\mathfrak{a},A)=(t,0,I_2)(0,b',I_2)(0,0,B')(0,0,A_\rho)(0,0,B)(0,-b,I_2)$ as $(t,b')= \mathfrak{a}+A\cdot (0,b)$ and $A=B'A_\rho B$ some $B,B' \in SU(2)$ and $A_\rho:=\operatorname{e}^{\frac{\rho}{2}\sigma_3}$, $\rho \in \R$.  Let $\varphi'_n:=W(B)W(-b)\varphi_n$. By (\ref{ECCPLSS}), $(\varphi'_n)$ is localized at $0$. Therefore it suffices to show that $(W(A_\rho)\varphi_n)$ is localized at $0$ if $(\varphi_n)$ is so. The result follows   applying (\ref{ECCPLSS}).\\
 \hspace*{6mm}  
 Let $B$ be a ball around $0$ with radius $r>0$. First we verify that $\R^3\setminus (\R^3\setminus B)_{A_\rho^{-1}}$ contains the ball $B'$ around $0$ with radius $\frac{r}{c+|s|}$, $c:=\cosh(\rho)$, $s:=\sinh(\rho)$. Indeed,
$\R^3\setminus (\R^3\setminus B)_{A^{-1}_\rho}=\{y\in\R^3: |y-z|   > |z_0| \text{ for all } (z_0,z)  \in A_\rho^{-1}\cdot (\R^3 \setminus B) \} $. Here $ A_\rho^{-1}\cdot(0,x)= (-x_3s ,x_1,x_2,x_3c)$.  Consider $y\in B'$. Then for all $x\in\R^3\setminus B$ one has $|y|<\frac{|x|}{c+|s|}\le \frac{|x|^2}{|z|+|z_0|}=|z|-|z_0|$, i.e., $|z_0|< |z|-|y|$, and hence $|z_0|< |z-y|$. --- Now, by (\ref{VEFCPOL})$(c)$, $0\le 1-\langle W(A_\rho)\varphi_n,  T(B)\,W(A_\rho)\varphi_n \rangle= \langle W(A_\rho)\varphi_n,  T(\R^3\setminus B)\,W(A_\rho)\varphi_n \rangle \le \langle \varphi_n, T\big((\R^3\setminus B)_{A_\rho^{-1}}\big)  \varphi_n\rangle \le \langle \varphi_n, T(\R^3\setminus B')  \varphi_n\rangle =1- \langle \varphi_n, T( B')  \varphi_n\rangle\to 0$. \hfill{$\Box$}\\

Finally we add

\begin{Lem}\label{LCPLS} A normalized linear combination of  two sequences of states localized at $b$ is  localized at $b$. More generally,   let $(\alpha_j)\in\ell^1$ and, for every $j\in\N$, let $(\varphi_{jn})$ be  a sequence of states
 localized at $b$. Suppose that $\varphi_n:= c_n\sum_j\alpha_j\varphi_{jn}$ with  $\sup_n|c_n|<\infty$ is normalized for every $n$. Then $(\varphi_n)$ is localized at $b$.
\end{Lem}

{\it Proof.} Let $B$ be a ball around $b$. One has to show that $||\sqrt{T(\R^3\setminus B)}\varphi_n||^2 =\langle \varphi_n, T(\R^3\setminus B)\varphi_n\rangle \to 0$.
Now, $||\sqrt{T(\R^3\setminus B)}\varphi_n||\le \sum_j |\alpha_j| \,|c_n|\, ||\sqrt{T(\R^3\setminus B)}\varphi_{jn}||\le  \sum_j |\alpha_j| \,c< \infty$ with $c:=\sup_n|c_n|$. By assumption $c > |c_n|\,||\sqrt{T(\R^3\setminus B)}\varphi_{jn}||\to 0$, $n\to \infty$ for every $j$. The result follows by dominated convergence. \hfill{$\Box$}\\

\subsection{Dilational covariance}\label{POMADC} Dilational covariance is a useful tool in order to show that a  POM is separated (cf. \cite{C81}, \cite{CL15}).

\begin{Def}\label{DCPOM}  A POM $T$ is said to admit dilational covariance if   there is a  representation
 $D$  of the group $\R_+$ of dilations satisfying $D_\lambda T(\Delta)D_\lambda^{-1}=T(\lambda \Delta)$ for all $\lambda>0$. 
\end{Def}\\
Plainly, since PM are   special  POM, (\ref{DCPOM}) concerns also  PM.  --- By \cite[Lemma 3]{C81}  a POM is separated if it admits dilational covariance. Such POM seem to be good candidates   for the norm-$1$-property, i.e. $T(\Delta)\ne 0\; \Rightarrow\; ||T(\Delta)||=1$ (cf. \cite{HLY04} for this property). The following criterion  (\ref{SPOL}) for separated POM  generalizes  \cite[Lemma 3]{C81}  and applies to cases like the  Dirac electron considered in sec.\,\ref{DS}, and the Weyl particles (\ref{POLWP}).

\begin{Def}\label{TPM}
A POM $T$ on $\mathcal{H}$ is said to be the \textbf{trace} of  the PM $E$  on a Hilbert space $\mathcal{K}$, if $\mathcal{H}$ is a subspace of $\mathcal{K}$ and $T(\Delta)\varphi =PE(\Delta)\varphi$ $\forall$ $\varphi\in\mathcal{H}$, where $P$ is the projection in $\mathcal{K}$ on  $\mathcal{H}$.
 \end{Def}\\
 One  easily checks that the trace of a PM  is a POM. It is well known that  actually every POM is the trace of some PM (see  \cite[Appendix]{RN90}).

\begin{Theo} \cite[Theorem 7]{CL15}\textbf{.}\label{SPOL} Let the POM $T$ be the trace of a PM  $E$ by the projection $P$. Suppose that $E$  admits dilation covariance.
 Suppose further that $Q:=\text{s-}\lim_{\lambda\to\infty} D_\lambda P D_\lambda^{-1}$ exists with $Q\ne 0$. 
 Let $\phi\in \mathcal{K}$ with $Q\phi\ne 0$.
 Then  $||PD_{\lambda}^{-1}\phi||  \to ||Q\phi||\ne 0$
 and  $\varphi_\lambda:=\frac{1}{||PD_{\lambda}^{-1}\phi ||} PD_{\lambda}^{-1}\phi $  satisfies 
 $$ ||T(B)\varphi_\lambda||\to 1, \quad \lambda\to \infty$$
for every ball $B$ around $0$. In particular, $ ||T(B) ||=1$ and  $(\varphi_n)$, up to finitely many $n$, is a sequence of states localized at $0$. 
 \end{Theo}
 
{\it Proof.} First, $||PD_{\lambda}^{-1}\phi|| =  ||D_\lambda PD_{\lambda}^{-1}\phi|| \to ||Q\phi||\ne 0$, whence $PD_{\lambda}^{-1}\phi \ne 0$ for all large $\lambda>0$. --- Now, put $P_\lambda:=D_\lambda PD_\lambda^{-1}$. Keep in mind that $Q^2=Q$. Then $||T(B)PD_{\lambda}^{-1}\phi|| =|| P_\lambda E(\lambda B)P_\lambda\phi|| $, and  $|| P_\lambda E(\lambda B)P_\lambda\phi -Q\phi|| \le 
|| P_\lambda E(\lambda B) \big(P_\lambda\phi-Q\phi\big) || + ||P_\lambda \big(E(\lambda B)-I\big)Q\phi|| +||\big(P_\lambda-Q\big)Q\phi|| \le  || P_\lambda\phi-Q\phi || + || \big(E(\lambda B)-I\big)Q\phi|| +||\big(P_\lambda-Q\big)Q\phi|| \to 0$ 
for $\lambda\to \infty$, since $E(\lambda B)\to I$ strongly. This proves $||T(B)PD_{\lambda}^{-1}\phi||\to ||Q\phi||$.\hfill{$\Box$}\\

For the proof of  (\ref{SPOL}) the existence of the limit $Q$ is crucial. ---  In  (\ref{SPOL}), in case of a POL,  $T$ is separated because of translational covariance.  Moreover,  by  translating $(\varphi_n)$ in (\ref{SPOL}) one gets a sequence of states localized at any  given point. If  $E$ in  (\ref{SPOL}) is Euclidean covariant, i.e., if $E$  is a WL, then $E$ automatically  admits dilation covariance \cite[sec.\,2]{C81}. This will be exploited for  the Dirac system and the Weyl systems, see sec.\,\ref{CEL} and (\ref{POLWP}).\\

\subsection{Decay of spatial probability}\label{SDLD}  According to  Hegerfeldt 1985 \cite{H85}  the spatial probability in any state $\varphi$ satisfies
\begin{equation}\label{LSD} 
\langle \varphi, T(\{x\in \R^3:\,|x|>r\})\varphi\rangle \notin \mathcal{O}(\operatorname{e}^{-Kr}), \quad r\to \infty
\end{equation}
 for $K >2\frac{mc}{\hbar}$, if the  causal time evolution for the   POL  $T$ is  relativistic  with mass $m\ge 0$  and semi-bounded energy. So causal time evolution not only forbids localized states in non-essentially dense regions  but  requires also a limited exponential decay of the spatial probability. The limit is determined by   the Compton wavelength  $\lambda_C=\frac{\hbar}{mc}$. This is an interesting behavior of free relativistic systems.  But certainly it is not a paradox as sometimes 
it is referred to in the literature, neither does it mean that ``arbitrarily good localization'' \cite[sec. 5]{H01} is impossible. We consider in sec.\,\ref{CEL} the  POL $T^e$ for the Dirac electron and  in (\ref{POLWP}) the POL $T^{\chi\eta}$ for the Weyl particles, which are causal {\it and} separated.\\

There are myriads of inequivalent POL for massive as well massless  systems. The POL studied for the two-component photon in  \cite{C76}, \cite{K77} are not causal and  have 
 states localized in bounded regions.  POL for massless irreducible systems with non-zero helicity never have  states localized in bounded regions. This is shown in  \cite[(2.5)]{C76} adapting the proof  in Galindo 1968 \cite[sec. 3]{G68}.  For the latter systems,  
a complete description of  POL admitting  dilation covariance  can be given, see   Castrigiano 1981 \cite[Theorem]{C81}.  \\
\hspace*{6mm} Up to here there are  two related problems concerning POL for irreducible relativistic systems, viz., their great  variety  and their  unknown nature. However, as we will see, imposing causality   will lead to uniqueness and furnish
an interpretation of the POL.\\


\section{Causal Systems}\label{SeCWL} We turn to ($C2$) from sec.\,\ref{SC}. Within this frame    there  is no a-priori-objection to returning to PM  describing localizability. In the following definition of a causal system we resume the considerations in sec.\,\ref{SePCWL},\,\ref{SPCPOL} and  (\ref{GCPCPOL}) on Poincar\'e covariance and causality of localization.  

\begin{Def}\label{CS} Let $W$ be a representation  of $\tilde{\mathcal{P}}$ describing a relativistic quantum system. Let $E$ be a causal WL for $W$ according to (\ref{CPOLD}). Then $(W,E)$  is called a \textbf{causal system}.\footnote{ Let us note that  Poincar\'e covariant WL with causal time evolution (sec.\,\ref{CPOL})  are called  causal   systems in \cite{CL15}, whereas causal systems according to (\ref{CS}) are not considered in \cite{CL15}. The present notations  leave the name causal system to the more  fundamental concept.}
\end{Def}\\

Recall that $E$ in (\ref{CS}) is a map on the set $\mathfrak{S}$ of all measurable subsets of spacelike hyperplanes, which is a PM on every spacelike hyperplane and which is Poincar\'e covariant, viz. $W(g)E(\Gamma)W(g)^{-1}=E(g\cdot \Gamma)$ for all $\Gamma\in\mathfrak{S}$,  $g\in\tilde{\mathcal{P}}$, and which is causal, viz. 
$$E(\Gamma) \le E(\Gamma_\sigma)$$
 for all \,$\Gamma\in\mathfrak{S}$ and spacelike hyperplanes $\sigma$. Recall that $\Gamma_\sigma= \{\mathfrak{x}\in \sigma:  (\mathfrak{x}-\mathfrak{z})^{\cdot2} \ge 0 \textrm{ for some } \mathfrak{z} \in \Gamma \}$ is the region of influence in $\sigma$ of $ \Gamma$, see Eq.\,(\ref{RIHPR}).  Recall also that by (\ref{VEFCPOL})(f) $E$ is causal if and only if it is locally orthogonal, i.e.,
$$  E(\Gamma)\, E(\Gamma')=0$$
for all spacelike separated $\Gamma,\Gamma'\in\mathfrak{S}$.
\\

The trace  (\ref{TPM})  of a causal WL  on a group-invariant subspace  clearly  yields a causal POL. This is worth to be stated explicitly.

\begin{Lem}\label{TCWL} 
Let  $\mathcal{K}$ be a Hilbert space and $\mathcal{H}$  a Hilbert subspace of  $\mathcal{K}$. Let $(W,E)$ be a causal system  in  $\mathcal{K}$.
Suppose that $\mathcal{H}$ is invariant under $W$. Then the trace $T$ of $E$ on  $\mathcal{H}$ is a causal POL for   the subrepresentation $W'$of $W$ on  $\mathcal{H}$.
\end{Lem}\\
 As mentioned after (\ref{TPM}), every POM is the trace of a PM. Similarly, every POL is the trace of a WL (see   e.g.  \cite{N72},  \cite{C76},  \cite{S77},  \cite{CH80}). However, it is most likely that the causal POL  of the massive scalar boson \cite{C24} are not the trace  of a causal WL.


\section{A Property  of Causal Systems } \label{PPCS}

Let $(W,E)$ be a causal system (\ref{CS}). 
By causality $E(\Delta_\tau)-E(\Delta)\ge 0$ holds for all $\Delta\in\mathfrak{S}$  and   spacelike hyperplanes $\tau$. Recall that $\Delta_\tau :=\{\mathfrak{x}\in \tau: (\mathfrak{x}-\mathfrak{z}) \cdot (\mathfrak{x}-\mathfrak{z}) \ge 0 \textrm{ for some } \mathfrak{z} \in \Delta  \}$ is the region of influence in $\tau$ of $ \Delta$. We study the particular case where $\Delta$ is an open spacelike  half-hyperplane $\sigma^>$. We write $\sigma^>_\tau:=(\sigma^>)_\tau$. \\
 \hspace*{6mm}
The section heading refers to the results (\ref{CCIDES}) -- (\ref{DSLSLHPTLC}) concerning $E(\sigma^>_\tau)-E(\sigma^>)$. 
 The proposition (\ref{CCIDES}) implies the no-go result in
 (\ref{CNELD})(c) about the representability of the lattice of causally complete regions, and  (\ref{DSLSLHPTLC})(f) is needed for the proof of the Lorentz contraction of the Dirac wave-functions in (\ref{LCDWF}) and of the Weyl wave-functions (\ref{LCWWF}).\\

Consider   a   spacelike hyperplane $\sigma$ and let $\pi\subset \sigma$ be a plane. Let $\sigma^>$ denote one of the two open half-hyperplanes constituting $\sigma\setminus \pi$.\footnote{ To be explicit write $\sigma=\{\mathfrak{x}: \mathfrak{x}\cdot \mathfrak{e}_\sigma=\rho_\sigma\}$ for  unique $\mathfrak{e}_\sigma=(1,e_\sigma)$ with $|e_\sigma|<1$ and $\rho_\sigma\in\R$ (cf. after (\ref{ESLHSF})). Moreover, there are unique $e\in\R^3$ with $|e|=1$  and $\rho\in\R$ such that $\pi=\{\mathfrak{x}\in\sigma: xe=\rho\}$ and $\sigma^> =\{\mathfrak{x}\in\sigma: xe>\rho\}$. } Let $\tau$ be a further spacelike hyperplane.\\

 \begin{Lem} \label{EHHP} If $\tau$    contains $\pi$ and if $\tau^>$ is the half-hyperplane in $\tau\setminus \pi$, which is at the same side of $\pi$ as $\sigma^>$,\footnote{ $M\subset\R^4$ is said to be at the \textbf{same side of $\pi$} as $\sigma^>$ if $M\subset\Pi^{\sigma, >}:=\{\mathfrak{x}:\frac{e_\sigma e}{1-e^2_\sigma}(\mathfrak{x}\cdot\mathfrak{e}_\sigma-\rho_\sigma)-xe>\rho\}$, which means that the spatial component 
 of every $\mathfrak{y}\in M$ lies in $\sigma^>$. Indeed, note that $\hat{\mathfrak{e}}_\sigma:= (1-e^2_\sigma)^{-1/2}\mathfrak{e}_\sigma$ is the direction of time referring to $\sigma$, as $\hat{\mathfrak{e}}_\sigma^{\cdot 2}=1$ and $\hat{\mathfrak{e}}_\sigma\cdot (\mathfrak{x}-\mathfrak{x}')=0$ for all $\mathfrak{x},\mathfrak{x}'\in\sigma$, and that $\mathfrak{y}-s\hat{\mathfrak{e}}_\sigma$ lies in $\sigma$ exactly for $s=(1-e^2_\sigma)^{-1/2}(\mathfrak{y}\cdot\mathfrak{e}_\sigma-\rho_\sigma)$. --- Note that $\{\mathfrak{y}\}$ is at the same side of $\pi$ as $\sigma^>$ if the timelike line $l=\mathfrak{y}+\R\mathfrak{e}_\sigma$ intersects $\sigma^>$.} then
$\tau^>_\sigma =\sigma^>, \quad \sigma^>_\tau = \tau^>$  and  $E(\tau^>) = E(\sigma^>)$. In particular, $E(\sigma^>_\tau)-E(\sigma^>)=0$.
\end{Lem}

{\it Proof.} Because of Poincar\'e symmetry it satisfies to check $\tau^>_\sigma =\sigma^>$, $ \sigma^>_\tau = \tau^> $ for  the half-hyperplanes of $\sigma:=\{\mathfrak{x}: x_0=0\}$ and $\tau :=\{\mathfrak{x}: x_0-\alpha x_3=0\}$, $\alpha \in ]0,1[$ determined by $x_3> 0$.\footnote{ Explicitly one has $\sigma=g\cdot \{\mathfrak{x}: x_0=0\}$   for $g\in\tilde{\mathcal{P}}$ if and only if $g=\big((\rho_\sigma,0),Q(\mathfrak{e}_\sigma)\big)(b,B)$ with arbitrary $(b,B)\in ISU(2)$ and $Q$ the canonical cross section. 
Put $\mathfrak{d}=(d_0,d):=Q(\mathfrak{e}_\sigma)^{-1}\cdot (0,e)$. Note $\mathfrak{d}^{\cdot 2}=-1$, whence $|d|^2=1+d_0^2>0$. Then, in addition to $\sigma=g\cdot \{\mathfrak{x}: x_0=0\}$, one has also $\pi=g\cdot \{\mathfrak{x}: x_0=x_3=0\}$ if and only if $bd=\rho$, $B^{-1}\cdot d=\pm|d|(0,0,1)$. If $(b,B)$ is specified in this way then one checks $\Pi^{\sigma,>}=g\cdot \{\mathfrak{x}:  x_3>0\}$ using $d_0=(1-e^2_\sigma)^{-1/2}e_\sigma e$, and one finds $\tau=g\cdot \{\mathfrak{x}: x_0-\alpha x_3=0\}$ for some  $|\alpha|<1$. If necessary then replace $B$ by $B\operatorname{i}\sigma_2$ to ensure $\alpha\ge 0$, as $\operatorname{i}\sigma_2$ acts on $\R^4$ by $\operatorname{diag}(1,-1,1,-1)$.} 
So  $E(\tau^>) = E(\sigma^>)$ holds by causality (\ref{VEFCPOL})(a) applied to $\sigma^>$ and $\tau^>$.\qed\\
   
Analogous equalities hold true for the closed  half-hyperplanes $\sigma^\ge$, $\tau^\ge$. Since the closed and respective open  half-hyperplanes differ by a Lebesgue null set only, one has also $E(\sigma^>) = E(\sigma^\ge) $,  $E(\tau^>) = E(\tau^\ge)$. --- Note that $\tau = g\cdot \sigma$ holds for some  $g\in \tilde{\mathcal{P}}$ satisfying $g\cdot \pi=\pi$. Then $W(g)$ commutes with $E(\sigma^>)$ by Eq.\,(\ref{PCWL}) as $E(\sigma^>)=E(g\cdot\sigma^>)$. \\

  \hspace*{6mm} 
  Now we turn to the more interesting case that  $\sigma$ and $\tau$ do not intersect.\footnote{ Explicitly, $\sigma=\{\mathfrak{x}: \mathfrak{x}\cdot \mathfrak{e}=\rho_\sigma\}$, $\tau=\{\mathfrak{x}: \mathfrak{x}\cdot \mathfrak{e}=\rho_\tau\}$ with $ \mathfrak{e}=(1,e)$, $|e|<1$. Then $\sigma=g\cdot \{\mathfrak{x}: x_0=0\}$ and $\tau=g\cdot \{\mathfrak{x}: x_0=t\}$  with $t:=(1-e^2_\sigma)^{-1/2}(\rho_\tau-\rho_\sigma)$ for $g=\big((\rho_\sigma,0),Q(\mathfrak{e})\big)(b,B)$ with arbitrary $(b,B)\in ISU(2)$.}
 
 \begin{Lem}\label{IFCC} If $\tau$ does not intersect $\sigma$, and if $g\in \tilde{\mathcal{P}}$ is any  rotation satisfying $g\cdot \pi=\pi$ and $g\cdot \sigma^>=\sigma \setminus \sigma^\ge$,
  then 
\begin{equation*}
 \big( E(\sigma^>_\tau)-E(\sigma^>)\big)+ \big(E(g\cdot\sigma^>_\tau)-E(g\cdot\sigma^>)\big)=E( \sigma^>_\tau\cap (g\cdot \sigma^>_\tau))
\end{equation*}
and the eigenspaces of the non-trivial projection  $E( \sigma^>_\tau \cap(g\cdot \sigma^>_\tau))$ are infinite dimensional.
\end{Lem}

{\it Proof.} Note $E(g\cdot \sigma^>) = E(\sigma \setminus \sigma^\ge) = I-E(\sigma^\ge)= I-E(\sigma^>)$,  $I-E(\sigma^>_\tau) =E(\tau\setminus \sigma^>_\tau) $, and $(g\cdot \sigma^>_\tau) \supset(\tau\setminus \sigma^>_\tau)$. As to the latter,  it suffices to check that $ \sigma^>_\tau =\{\mathfrak{x}:x_0=\alpha, x_3>-|\alpha|\}$ 
and
$g\cdot \sigma^>_\tau=\{\mathfrak{x}:x_0=\alpha, x_3<|\alpha|\}$
 for $\sigma^>=\{\mathfrak{x}:x_0=0, x_3>0\}$, $\tau=\{\mathfrak{x}:x_0=\alpha\}$, $\alpha\in\R\setminus \{0\}$, and $g=\operatorname{i}\sigma_2$, which acts on $\R^4$ by $\operatorname{diag}(1,-1,1,-1)$.
This shows in addition that neither $(g\cdot \sigma^>_\tau) \setminus (\tau\setminus \sigma^>_\tau) = (g\cdot \sigma^>_\tau)\cap \sigma^>_\tau$ nor its complement is a 
 Lebesgue null set. Hence the eigenspaces of $E( \sigma^>_\tau \cap (g\cdot \sigma^>_\tau))$ are infinite dimensional.
 \qed\\
 
 Using obvious notations the formula in (\ref{IFCC}) reads  more easily as
 \begin{equation}\label{LHPTLC}
 \big( E(\sigma^>_\tau)-E(\sigma^>)\big)+ \big(E(\sigma^<_\tau)-E(\sigma^<)\big)=E(\sigma^>_\tau \cap \sigma^<_\tau)
\end{equation}
due to the relations $g\cdot \sigma^> = \sigma^< $ and $g\cdot\sigma^>_\tau=\sigma^<_\tau$. Moreover, remember
 \begin{equation}\label{CLHPTLC}
   E(\sigma^<_\tau)-E(\sigma^<)= W(g)\big(E(\sigma^>_\tau)-E(\sigma^>)\big) W(g)^{-1}
\end{equation}
with $g$ from (\ref{IFCC}).

 \begin{Pro}\label{CCIDES} Let $\sigma$, $\tau$ be two non-intersecting spacelike hyperplanes. Then the eigenspaces of the 
 non-trivial projection  $E(\sigma^>_\tau)-E(\sigma^>)$ are infinite dimensional. The same holds true for the projection $E(\sigma^<_\tau)-E(\sigma^<)$. Moreover, $E(\sigma^>_\tau)-E(\sigma^>)$ and  $E(\sigma^<_\tau)-E(\sigma^<)$ are orthogonal.
\end{Pro}

{\it Proof.} Since $E(\sigma^>_\tau)\ge E(\sigma^>)$  by causality, $P^<:=E(\sigma^>_\tau)-E(\sigma^>)\ge 0$ is a projection.  Analogously define $P^>$. Let $g\in \tilde{\mathcal{P}}$ be as in (\ref{IFCC}). Due to Eq.\,(\ref{PCWL}), $E(g\cdot\sigma^>_\tau)-E(g\cdot\sigma^>)=W(g) P^< W(g)^{-1}$ is a projection, too. By \eqref{IFCC}  their  sum is  a projection with infinite dimensional eigenspaces.  Therefore $P^<$ and  $W(g) P^< W(g)^{-1}$ are orthogonal and none of the four eigenspaces is finite dimensional. It remains to note that $P^>=W(g) P^< W(g)^{-1}$. \qed\\

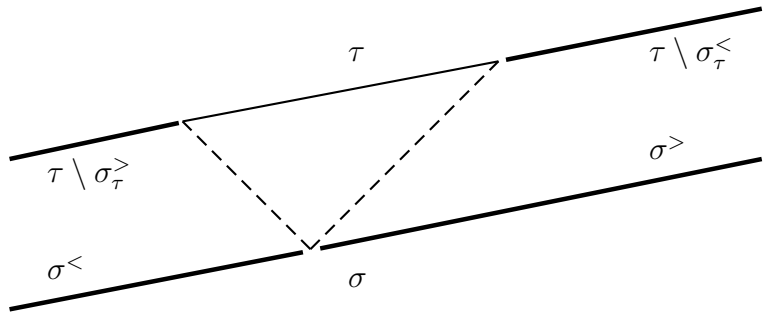
\begin{pspicture}(-5,0)(3.5,4.2)




\psline[linestyle=dashed](2,1.2)(4.5,3.7)
\psline[linestyle=dashed](2,1.2)(0.3,2.9)
\psline(0.3,2.9)(4.5,3.7)

\psline[linewidth=0.6mm](2.13,1.21)(8,2.4)
\psline[linewidth=0.6mm](-2,0.4)(1.9,1.16)

\psline[linewidth=0.6mm](-2,2.4)(0.25,2.88)
\psline[linewidth=0.6mm](4.6,3.72)(8,4.4)

\put(-1.5,0.8){$\sigma^<$}
\put(-1.5,2.1){$\tau\setminus \sigma^>_\tau$}

\put(6.5,2.4){$\sigma^>$}
\put(6.5,3.7){$\tau\setminus \sigma^<_\tau$}

\put(2.5,3.7){$\tau$}
\put(2.5,0.7){$\sigma$}

\end{pspicture}

Recall $E(\sigma^>_\tau)\ge E(\sigma^>)$. Hence
 \begin{equation}\label{ALHPTLC}
  E(\sigma^>_\tau)-E(\sigma^>)=E(\sigma^>_\tau)E(\sigma^<)
\end{equation}
as well the analogous formula, where  $ < $  and  $>$ are interchanged. \\
 \hspace*{6mm}
We are going to characterize the ranges of  
 these projections. Let $\mathfrak{e}\in\R^4$ determine the time coordinate for the spacelike hyperplane $\sigma$. It satisfies $\mathfrak{e}\cdot \mathfrak{e}=1$, $\mathfrak{e}\cdot (\mathfrak{x}-\mathfrak{x}')=0$ for all $\mathfrak{x},\mathfrak{x}'\in\sigma$, and $e_0>0$. $\mathfrak{e}$   is uniquely determined. Then  $\tau =\alpha \mathfrak{e} +\sigma$ for some unique $\alpha\in\R$. Let $\tau^>$ be determined  with respect to the plane $\alpha \mathfrak{e} +\pi$ in $\tau$ such that $\tau^>=\alpha \mathfrak{e} +\sigma^>$. Then $\sigma^>_\tau =\alpha \mathfrak{e}+ \tau^>_\sigma$. By Poincar\'e  symmetry it suffices  to check all the foregoing relations for 
$\sigma=\mathfrak{a}+\{\mathfrak{x}:x_0=0\}$,  $\sigma^>=\mathfrak{a}+\{\mathfrak{x}:x_0=0, x_3>0\}$, $\mathfrak{e}=(1,0,0,0)$, and $\tau=\mathfrak{a}+\{\mathfrak{x}:x_0=\alpha\}$, $\mathfrak{a}\in\R^4$. As a consequence one has 
\begin{equation}\label{BLHPTLC}
  E(\sigma^>_\tau)= W(\alpha \mathfrak{e}) E(\tau^>_\sigma)W(\alpha \mathfrak{e})^{-1}
 \end{equation}
 as well the formula for $>$ replaced with $<$.

\begin{Pro}\label{ACCIDES}  Let $\sigma$, $\tau$ be two non-intersecting spacelike hyperplanes. Then a state $\varphi$ is in the range of $E(\sigma^>_\tau)-E(\sigma^>)$ if and only if $\varphi$ is localized in $\sigma^<$ and $W(-\alpha \mathfrak{e})\varphi$ is localized in $\tau^>_\sigma$. The analogous statement with $>$ replaced with $<$ holds true, too.
\end{Pro}

{\it Proof.} By Eqs.\,(\ref{ALHPTLC}),\,(\ref{BLHPTLC}) $\varphi$ is in the range of $E(\sigma^>_\tau)-E(\sigma^>)$ if and only if $\varphi$ is in the range of $E(\sigma^<)$ and in the range of $W(\alpha \mathfrak{e}) E(\tau^>_\sigma)W(\alpha \mathfrak{e})^{-1}$. The proof is easily accomplished.\qed\\

  \hspace*{6mm} 
Recall  $\tau =\alpha \mathfrak{e} +\sigma$ with $\alpha\in\R$. By causality $E(\sigma^>_\tau)\le E(\sigma^>_{\tau'})$
if $0\le \alpha\le\alpha'$ or $\alpha'\le\alpha\le 0$.
Hence  the strong limit $P^<_\sigma:=\lim_{\alpha\to \infty}\big(E(\sigma^>_\tau)-E(\sigma^>)\big)=
\lim_{\alpha\to \infty}E(\sigma^>_\tau)E(\sigma^<)$ exists. Analogously define $P^>_\sigma$. For the strong limits for $\alpha\to -\infty$ see (\ref{DSLSLHPTLC})(g).

\begin{Pro}\label{DSLSLHPTLC}  Let $\sigma$ be  a spacelike hyperplane. Let $\tau :=\alpha \mathfrak{e} +\sigma$ with $\alpha\in\R$ and $\mathfrak{e}$ the four-vector of time with respect to $\sigma$. Recall $P^<_\sigma=\lim_{\alpha\to \infty}E(\sigma^>_\tau)E(\sigma^<)$ and  $P^>_\sigma=\lim_{\alpha\to \infty}E(\sigma^<_\tau)E(\sigma^>)$. Then
\begin{itemize}
\item[(a)] 
 the eigenspaces of $P^<_\sigma$ and $P^>_\sigma$ are infinite dimensional
 \item[(b)] $P^<_\sigma \le E(\sigma^<)$, $P^>_\sigma \le E(\sigma^>)$
\item[(c)] $P^<_\sigma P^>_\sigma=0$
\item[(d)] $P^>_\sigma=W(g)P^<_\sigma W(g)^{-1}$ with $g$ from \emph{(\ref{IFCC})}.

\item[(e)]
$P^<_\sigma +P^>_\sigma =\lim_{\alpha\to \infty} E(\sigma^>_\tau \cap \sigma^<_\tau)=\lim_{\alpha\to \infty} W(\alpha \mathfrak{e}) E(\tau^>_\sigma\cap \tau^<_\sigma)W(\alpha \mathfrak{e})^{-1}$
\item[(f)] $P^<_\sigma  = E(\sigma^<)$  $\Leftrightarrow$ $P^>_\sigma  =E(\sigma^>)$  $\Leftrightarrow$ $P^<_\sigma +P^>_\sigma =I$  $\Leftrightarrow$\\ $\not\exists$ state $\varphi$ satisfying $E(\tau^>_\sigma\cap \tau^<_\sigma)W(\alpha \mathfrak{e})^{-1}\varphi=0$ $\forall$ $\alpha>0$ $\Leftrightarrow$\\
$\not\exists$ state $\varphi$ satisfying $E(\tau^>_\sigma)W(\alpha \mathfrak{e})^{-1}\varphi=0$ $\forall$ $\alpha>0$ $\Leftrightarrow$\\
 $\not\exists$ state $\varphi$ satisfying $E( \tau^<_\sigma)W(\alpha \mathfrak{e})^{-1}\varphi=0$ $\forall$ $\alpha>0$
 \item[(g)] Analogously, the strong limits for $\alpha\to -\infty$ exist and they  satisfy  \emph{(a)\,--\,(f)} for $\alpha<0$ and $\alpha\to -\infty$.
\end{itemize} 
\end{Pro}

{\it Proof.} Keep the  definition of $P^\lessgtr_\sigma$ in mind. (g) is obvious. Now (a) follows from  (\ref{CCIDES}), Eq.\,(\ref{ALHPTLC}). (b) is obvious. (c) holds by (b). One has (d)  due to  Eq.\,(\ref{CLHPTLC}). (e) follows from  Eqs.\,(\ref{LHPTLC}),\,(\ref{ALHPTLC}),\,(\ref{BLHPTLC}).\\
\hspace*{6mm}
 We turn to (f). The first two  $\Leftrightarrow$ hold by (b),\,(d). As $W(g)E(\tau^>_\sigma)W(\alpha \mathfrak{e})^{-1}W(g)^{-1}\varphi=W(g)E(\tau^>_\sigma)W(g)^{-1}W(\alpha \mathfrak{e})^{-1}\varphi=E( \tau^<_\sigma)W(\alpha \mathfrak{e})^{-1}\varphi$, also the last  $\Leftrightarrow$ holds true. ---  As to  the forth $\Leftrightarrow$\,, the direction $\Rightarrow$  is obvious. For the proof of  the reverse direction assume the contrary and define $\varphi^<:=E(\sigma^<)\varphi$, $\varphi^>:=E(\sigma^>)\varphi$. Then $\varphi= \varphi^<+\varphi^>$, whence $ \varphi^<\ne 0$ or $ \varphi^>\ne 0$. Let $Q:=W(\alpha \mathfrak{e})^{-1}E(\sigma^<)W(\alpha \mathfrak{e})$. By causality $I-E(\tau^>_\sigma)\le Q\le E(\tau^<_\sigma)$. Therefore these three projections commute and $E( \tau^>_\sigma)W(\alpha \mathfrak{e})^{-1}\varphi^<=E( \tau^>_\sigma) QW(\alpha \mathfrak{e})^{-1}\varphi=QE( \tau^>_\sigma) W(\alpha \mathfrak{e})^{-1}\varphi=E(\tau^<_\sigma)QE( \tau^>_\sigma) W(\alpha \mathfrak{e})^{-1}\varphi=Q E(\tau^<_\sigma\cap \tau^>_\sigma) W(\alpha \mathfrak{e})^{-1}\varphi=0.$ Similarly $E( \tau^<_\sigma)W(\alpha \mathfrak{e})^{-1}\varphi^>=0$ holds. Hence the assertion follows, using the last $\Leftrightarrow$ in case that $\varphi^<= 0$. \\
\hspace*{6mm} 
  It remains to prove the third $\Leftrightarrow$ of (f). The implication $\Rightarrow$ is obvious by (e). As to  $\Leftarrow$, suppose $P^<_\sigma +P^>_\sigma<I$. Let $V:=\{\varphi: \big(I-P(\alpha)\big)\varphi_{-\alpha}\to 0, \alpha\to \infty\}$, where $P(\alpha):=E(\tau^>_\sigma\cap \tau^<_\sigma)$, $\varphi_{-\alpha}:=W(\alpha \mathfrak{e})^{-1}\varphi$. Check $V=\operatorname{ran}(P^<_\sigma +P^>_\sigma)$. Then $V^\perp\ne \{0\}$ by (e). Let $\varphi^*\in V^\perp\setminus\{0\} $. Let $\beta>0$ and $\chi\in\operatorname{ran}P(\beta)$. We claim $\chi_{\beta}\in V$. Indeed, by causality $\chi_{\beta-\alpha} \in \operatorname{ran}P(\beta+|\alpha-\beta|)$ for $\alpha\ge 0$, whence $\chi_{\beta-\alpha} \in \operatorname{ran}P(\alpha)$  for $\alpha\ge \beta$ and hence $\chi_{\beta}\in V$. So one infers $0=\langle \varphi^*, \chi_{\beta}\rangle=\langle \varphi^*_{-\beta}, \chi \rangle$ for all $\chi\in\operatorname{ran}P(\beta)$, $\beta>0$. This means $\varphi^*_{-\beta}\in\operatorname{ran}\big(I-P(\beta)\big)$ and hence $P(\beta)\varphi^*_{-\beta}=0$ for all $\beta>0$.  
\qed\\

To comment on  (\ref{DSLSLHPTLC}) consider $\sigma=\{\mathfrak{x}:x_0=0\}$,  $\sigma^>=\{\mathfrak{x}:x_0=0, x_3>0\}$, $\mathfrak{e}=(1,0,0,0)$,  $\tau=\{\mathfrak{x}:x_0=\alpha\}$,  
and $\tau^<_\sigma=\{x_0=0, x_3< |\alpha|\}$.
 The question   raised by  the last item of (f) is whether there exists a state $\varphi$ such that  the  states $W(-t)\varphi$ evolved in time  are localized in  $\{x_0=0, x_3>t\}$ for $t >0$. In case of the Dirac system, (\ref{CDSLSLHPTLC}) shows
that such a state  does not exist, thus proving $I= P^<_\sigma +P^>_\sigma$ by (\ref{DSLSLHPTLC})(f). In case of the Weyl systems, $I= P^<_\sigma +P^>_\sigma$ holds as well. However its proof  in (\ref{ECCPWS}) (see  the comments to (\ref{GTERSWF})  and (\ref{ECCPWS}))  requires a different method.  There is an interesting alternative proof of $I= P^<_\sigma +P^>_\sigma$  for Dirac and Weyl systems. It arises from  group representation theoretical considerations leading to farther-reaching results,  see (\ref{CNGENTLNSLHP}).
\\
 

 \section{The Lattice of Causally Complete Regions}\label{HLCCR}

 The lattice $\mathcal{M}$ of  causally complete regions  of Minkowski space,  generated and orthocomplemented by the spacelike relation
$$ \mathfrak{x} \perp    \mathfrak{y} \quad \Leftrightarrow \quad(\mathfrak{x} - \mathfrak{y})^{\cdot 2}<0$$
is well-known  from local quantum physics. In view of Poincar\'e covariance and local orthogonality (equivalent to causality by (\ref{VEFCPOL}))   one may ask whether a causal system  $(W,E)$ gives rise to a \textbf{representation} \textbf{(rep)} $(W,F)$ of  $\mathcal{M}$  by orthogonal projections on the space of states. The properties to impose on $F$ are Poincar\'e covariance with respect to $W$,  normalization $F(\emptyset)=0$, $F(\R^4)=I$, monotony, local orthogonality, and $\sigma$-orthoadditivity. 
 It is important to point out  that $F$ is not required to be a lattice homomorphism of $\mathcal{M}$ into the lattice of the projections on the Hilbert space of states.\\
\hspace*{6mm} 
  One is tempted to think such a rep $(W,F)$ to be possible just because of the fact that the energy  need not  be semi-bounded. But actually is does not exist, see (\ref{NRLCCR}) and (\ref{CNELD})(c). \\
\hspace*{6mm}  
 In local quantum theory to every finite open  contractible region $M$ a von Neumann ring $\mathcal{R}(M)$ is assigned. The causality principle relates the orthocomplement $M^\perp$   to the commutant $\mathcal{R}(M)'$. So in Haag 1992  \cite[III.4.2.1]{R.H92} it is postulated tentatively  that for the vacuum sector  the assignment  $\mathcal{R}$  actually extends to  a homomorphism from $\mathcal{M}$ into the lattice of von Neumann rings on a Hilbert space.  Obviously,  the tentative to construct a rep $F$ describing localization within the frame of local quantum theory may be inspired by these considerations. But a localization operator $E(\Delta)$ in general cannot be regarded to be an element of $\mathcal{R}(\hat{\Delta})$ by \cite[Lemma 4]{CL15} just because of the semi-boundedness of  energy.\\
   \hspace*{6mm}  
     Before showing the mentioned no-go results we like to study  $\mathcal{M}$  in some detail. Many examples at the level of exercises serve to become more familiar with 
$\mathcal{M}$ and can be skipped. Only (\ref{EMTVMH}) is  needed in the sequel.  In particular it shows that there are maximal spacelike sets, which are not  bases. In (\ref{MCCS}) it will be proven that  causally complete sets are automatically measurable, i.e., Lebesgue measurable subsets of $\R^4$.  Regarding the definition of $\mathcal{M}$   we follow  \cite[III.4.1 The Lattice of Causally Complete Regions]{R.H92}. \\

 \subsection{Causal completion} Let $M$ be  any subset of Minkowski space and  let $M^\perp$ denote the causal complement of $M$ consisting of all points, which lie spacelike to all points of $M$, i.e.,
 \begin{equation}\label{DCC}
 M^\perp:=\{ \mathfrak{x}: (\mathfrak{x}-\mathfrak{y})^{\cdot 2}<0\;\forall \,\mathfrak{y}\in M \}
 \end{equation}
 Clearly,   $M \cap M^\perp =\emptyset$, and $M^\perp\subset L^\perp$ if $L\subset M$. The set  $\hat{M} :=(M^\perp)^\perp$ is called the causal completion of $M$.  If $M=\hat{M}$ then $M$ is said to be causally complete.  
 
    \hspace*{6mm}  
 Clearly, $\emptyset$ and $\R^4$ are causally complete. Any causal complement $M^\perp$ is causally complete and, equivalently,   $M^\perp  = \hat{M}^\perp$ holds.  (Indeed, $M\subset \hat{M}$ implies $M^\perp \supset \hat{M}^\perp= M^{\perp\perp\perp}=(M^\perp)^\land\supset M^\perp$.) Therefore $(\hat{M})^\land=\hat{M}$. For any family $(M_\iota)_\iota$ of sets  one has 
 $\big(\bigcup_\iota M_\iota\big)^\perp =\bigcap_\iota M_\iota^\perp$, as both sides describe the set of points lying spacelike to all $M_\iota$. This implies $\big(\bigcup M_\iota^\perp \big)^\perp=\bigcap_\iota\hat{M}_\iota$, which proves that the intersection of any family of causally complete sets is causally complete.  Moreover,  $\big(\bigcup_\iota M_\iota\big)^\land =\big(\bigcap_\iota M_\iota^\perp\big)^\perp$  and $\big(\bigcup_\iota M_\iota^\perp\big)^\land =\big(\bigcap_\iota \hat{M}_\iota\big)^\perp$ follow. Finally one has $\big(\bigcup_\iota M_\iota\big)^\land =\big(\bigcup_\iota \hat{M}_\iota\big)^\land$. (Indeed, the inclusion $\subset$ holds by monotony of causal completion. Conversely, still by this monotony, $\big(\bigcup_\iota M_\iota\big)^\land \supset \hat{M}_\kappa$ for every $\kappa$. Hence $\big(\bigcup_\iota M_\iota\big)^\land  \supset\bigcup_\kappa  \hat{M}_\kappa $, where $\supset$ still holds for the causal completion of the right hand side.)\\
 \hspace*{6mm}   
   Note, however,  that in general $(M_1\cap M_2)^\perp\ne (M_1^\perp \cup M_2^\perp)^\land$. (Indeed, let $M_1:=\{\mathfrak{x}:x_0-x_3=0\}, M_2:=\{\mathfrak{x}:x_0+x_3=0\}$. Then $M_1^\perp=M_2^\perp=\emptyset$ and  $M_1\cap M_2=\{x_0=x_3=0\}$, whence $(M_1\cap M_2)^\perp=\{|x_0|<|x_3|\}$ and $ (M_1^\perp \cup M_2^\perp)^\land=\emptyset$.)\\
 \hspace*{6mm}   
    Moreover, in general, $M_1\cup M_2$ is not causally complete, even if $M_1$, $M_2$ are causally complete spacelike separated. (Indeed, let $M_1:=\{\mathfrak{x}:|x_0|\le -x_3\}$, $M_2:=\{\mathfrak{x}:|x_0|<x_3\}$. Then,  by (\ref{FHS})(b), $M_1=h(\ge 0)\cap k(\le 0)$, $M_2=h(< 0)\cap k(>0)$, whence $M_1=M_2^\perp$ are causally complete spacelike separated. As $M_1\cup M_2$ contains the spacelike hyperplane $\{\mathfrak{x}:x_0=0\}$, $(M_1\cup M_2)^\land=\R^4$ follows.)

 \begin{Lem}\label{AOPS}  The one-point sets $\{\mathfrak{x}\}$ are causally complete.
   \end{Lem}
 
{\it Proof.} It suffices to consider $\mathfrak{x}=0$. Since $\{0\}^\perp=\{\mathfrak{y}: |y_0|<|y|\}$,  one has $\mathfrak{x}\in\{0\}^\land$ if  $|x_0-y_0|<|x-y|$ for all $|y_0|<|y|$.  Then $x=0$ as otherwise $\mathfrak{y}:=(0,x)$ yields the contradiction $|x_0|<0$. Now $x_0=0$ follows similarly.\qed

\begin{Exampl}\label{CCTPS} for the causal completion of a two-point set.  Let $s\in\R$. Then $$\{0,(s,0,0,0)\}^\land=\{\mathfrak{x}: |x_0-\textnormal{\tiny{$\frac{s}{2}$}}|+|x|\le |\textnormal{\tiny{$\frac{s}{2}$}}|\}=\{\mathfrak{x}: x_0=\textnormal{\tiny{$\frac{s}{2}$}}, |x|\le |\textnormal{\tiny{$\frac{s}{2}$}}|\}^\land$$
\end{Exampl}\\
{\it Proof.}  $\{0,(s,0,0,0)\}^\land =(\{0\}^\perp \cap \{(s,0,0,0)\}^\perp)^\perp=(\{|y_0|<|y|\}\cap \{|y_0-s|<|y|\})^\perp=\{|y_0-\frac{s}{2}| +|\frac{s}{2}|<|y|\}^\perp=\{\mathfrak{x}:|x_0-y_0|<|x-y|  \textrm{ for all } \mathfrak{y} \textrm{ with } |y_0-\frac{s}{2}| +|\frac{s}{2}|<|y|\}=\{\mathfrak{x}:|x_0-\frac{s}{2}|+|y_0-\frac{s}{2}|<\big||y|-|x|\big| \textrm{ for all } \mathfrak{y} \textrm{ with } |y_0-\frac{s}{2}| +|\frac{s}{2}|<|y|\}$, whence $|x|\le|\frac{s}{2}|$ and hence the first equality $=$.  As to the second $=$, the inclusion $\supset$ holds, since $\{\mathfrak{x}: x_0=\frac{s}{2}, |x|\le |\frac{s}{2}|\}$ is contained in the left hand side. For the inclusion $\subset$ it suffices to show that $\{0,(s,0,0,0)\}$ is contained in the right hand side. Anticipating (\ref{MTVMH}), consider $\mathfrak{z}=(1,z)$ with $|z|\le 1$ and note that $\mathfrak{x}:=\frac{s}{2}\mathfrak{z}$ satisfies $x_0=\frac{s}{2}, |x|\le |\frac{s}{2}|$. Hence the assertion holds true for $0$. Similarly the assertion is verified for $(s,0,0,0)$.\qed\\

 \begin{Exampl}\label{FHS} for some causally  complete sets. Let $\gamma\in\R$. Consider the   half-spaces
  $h(\le\gamma):=\{\mathfrak{x}: x_0-x_3\le\gamma \}$,  $k(\le\gamma):=\{\mathfrak{x}: x_0+x_3\le\gamma \}$ and, generally,  $h(\sim \gamma)$, $k(\sim \gamma)$ for $\sim\,\in\{<,\,\le,\,>,\,\ge\}$.
  
  \hspace*{6mm}   
    Obviously, their causal complements $h^\perp$, $k^\perp$ are empty and, hence, their causal completions $\hat{h}$,  $\hat{k}$ equal $\R^4$. For the causal complement of the intersection of two half-spaces of the type $h$ and $k$ one finds 
\begin{itemize} 
\item[(a)]    $\big(h(\sim\gamma)\cap k(\backsim\delta)\big)^\perp=\emptyset$
\end{itemize}
if $\sim$ and $\backsim$ are both in $\{<,\,\le\}$ or both in $\{>,\,\ge\}$, and
\begin{itemize}    
\item[(b)]    $\big(h(\sim\gamma)\cap k(\backsim\delta)\big)^\perp= h(\not\sim\gamma)\cap  k(\not\backsim\delta)$
 \end{itemize}
 otherwise.  It follows that the intersections $h\cap k$  in \emph{(b)} are causally complete. Note $h( \not\sim\gamma)=\R^4\setminus h(\sim\gamma)$, $k(\not\backsim\,\delta)=\R^4\setminus k(\backsim\delta)$.
   \end{Exampl}

  {\it Proof.} As to  (a), for every $\mathfrak{y}\in\R^4$ there is $s\in\R$ such that  $(y_0+s)-y_3<\gamma$ and $(y_0+s)+y_3<\delta$. Hence $\mathfrak{x}:=(y_0+s,y)\in h(<\gamma)\cap k(<\delta)$ and $(\mathfrak{x}-\mathfrak{y})^{\cdot 2}=(s,0)^{\cdot 2}=s^2\ge 0$. Therefore $\mathfrak{y}\not\in\big(h(<\gamma)\cap k(<\delta)\big)^\perp$. This proves the cases 
  $\sim,\backsim\, \in\{<,\le\}$. The cases   $\sim,\backsim\, \in\{>, \ge\}$   follow analogously.  --- We turn to (b). Consider the case $(\sim,\backsim)=(\le,>)$. Let  
$\mathfrak{y}\in  \big(h(\le\gamma)\cap k(>\delta)\big)^\perp$. This means  that for every $\mathfrak{x}$ satisfying $x_0-x_3\le\gamma$, 
$x_0+x_3>\delta$ one has $|y_0-x_0|<|y_3-x_3|$, i.e., either \big[\;$y_0-y_3<x_0-x_3$ \,and\,  $y_0+y_3>x_0+x_3$\;\big]\; or\: \big[\;$y_0-y_3>x_0-x_3$ \,and\,  $y_0+y_3<x_0+x_3$\;\big].  Obviously this holds true if and only if  $y_0-y_3>\gamma$ and $ y_0+y_3\le \delta$. --- The other cases follow analogously.\qed\\

 \subsection{Set of determinacy}\label{SOD} For any subset $M$ of Minkowski space  its \textbf{set of  determinacy} 
 \begin{equation}
 \tilde{M}:=\{\mathfrak{x}:\forall\,\mathfrak{z}\ne 0,\, \mathfrak{z}^{\cdot2}\ge 0\;\exists\, s\in\R \textrm{ with } \mathfrak{x}+s\mathfrak{z}\in M\}
 \end{equation}
   consists of  all points $\mathfrak{x}$ such that every timelike or lightlike line\footnote{ throughout line means straight line } $\mathfrak{x}$  meets $M$. \\
   
  \hspace*{6mm}  
 Obviously,  $M\subset \tilde{M}$. Further, $\tilde{\tilde{M}}= \tilde{M}$ holds. (Indeed, let $\mathfrak{x}  \in \tilde{\tilde{M}}$. For $\mathfrak{z}\ne 0,\, \mathfrak{z}^{\cdot2}\ge 0$ there is $s\in \R$ with $\mathfrak{x}+s\mathfrak{z}\in \tilde{M}$. Hence, there is $s'\in\R$ with $(\mathfrak{x}+s\mathfrak{z})+s'\mathfrak{z}\in M$, whence $\mathfrak{x}\in \tilde{M}$.) Clearly $\tilde{M}_1\subset \tilde{M}_2$ if $M_1\subset M_2$.

 \begin{Lem}\label{MTVMH}  $\tilde{M}\subset \hat{M}$.
 \end{Lem}\\
 {\it Proof.} Fix $\mathfrak{x}  \in \tilde{M}$. Note first that $\mathfrak{x} \not\in M^\perp$, as $\mathfrak{x} +
\mathfrak{w} \in M$ for some  not spacelike $\mathfrak{w}$. Now let  $\mathfrak{y}  \in M^\perp$. Then   $\mathfrak{z}:=\mathfrak{x}  -\mathfrak{y}\ne 0$. Obviously, it suffices to show 
 $\mathfrak{z}^{\cdot 2}<0$.  Assume the contrary. Then there is $s\in\R$ with $\mathfrak{x}- s \mathfrak{z} \in M$. This implies  the contradiction $0>(\mathfrak{x}- s \mathfrak{z} - \mathfrak{y})^{\cdot 2}= (1-s)^2\mathfrak{z}^{\cdot2}\ge0$.\qed\\

 Thus, by  (\ref{MTVMH}), one has  $M\subset \tilde{M}=\tilde{\tilde{M}} \subset \hat{M}$, whence $M=\tilde{M}$ if $M=\hat{M}$. But $M=\tilde{M}$ does not imply that $M$ is causally complete.  A simple example is $M:=\R^4\setminus \R\mathfrak{e}$ for $\mathfrak{e}=(1,0,0,0)$ with $M=\tilde{M}$ and $\hat{M}=\R^4$. (Indeed, $M=\tilde{M}$ is obvious.  Next note $M^\perp\subset \R\mathfrak{e}$, since generally $M^\perp\cap M=\emptyset$, and $s  \mathfrak{e} \not\perp (2+s,0,0,1)$. Hence $M^\perp=\emptyset$, whence $\hat{M}=\R^4$.) Another example is $M$ in (\ref{FEMTVMH}). 
 For the computations we anticipate (\ref{FEDOD}).  Furthermore, $M:=\tilde{P}$ for $P$ from (\ref{EMTVMH}) is a  particularly interesting example as $P$ is maximal spacelike but not a base.

 \begin{Exampl} \label{FEMTVMH} for a non causally complete set of determinacy. 
Let $M:=K\cup L$ for  
$K:=\{\mathfrak{x}: -1\le x_0-x_3\le 1, -3\le x_0+x_3\le -1\}$,  $L:=\{ \mathfrak{x}: -1\le x_0-x_3\le 1, 1\le x_0+x_3\le 3\}$. $K,L$ are causally complete by \emph{(\ref{FHS})(b)}. 
One has $M=\tilde{M}$ and $\hat{M}=\{\mathfrak{x}: -1\le x_0-x_3\le 1, -3\le x_0+x_3\le 3\}$,
whence 
$\{\mathfrak{x}: -1<x_0-x_3<1, -1<x_0+x_3<1\}
\subset \hat{M}\setminus\tilde{M}$.
\end{Exampl}\\
 {\it Proof.} First compute $\hat{M}=M^{\perp\perp}$. As $M=K\cup L=\big(h(\le 1)\cap k(\ge -3)\cap
 h(\ge -1)\cap k(\le -1)\big) \cup \big(   h(\le 1) \cap k(\ge 1) \cap h(\ge -1)\cap k(\le 3) \big)$,  by   (\ref{FEDOD})(ii) one has $M^\perp= \big(\big(h(> 1)\cap k(< -3)\big) \cup \big(h(< -1)\cap k(> -1)\big)\big) 
 \cap \big(\big(   h(> 1) \cap k(< 1)\big) \cup \big(h(< -1)\cap k(> 3)\big) \big)
 =\big(h(> 1)\cap k(< -3)\cap   h(> 1) \cap k(< 1)\big)
  \cup
\big(h(> 1)\cap k(< -3)\cap    h(< -1)\cap k(> 3) \big)
\cup
 \big(h(< -1)\cap k(> -1) \cap    h(> 1) \cap k(< 1)\big)
 \cup
\big(h(< -1)\cap k(> -1) \cap  h(< -1)\cap k(> 3) \big)
 =
 \big(h(> 1)\cap k(< -3)\big) \cup \big(h(< -1)\cap k(> 3) \big)$, and further, $M^{\perp\perp}=h(\le 1)\cap k(\ge -3)\cap h(\ge -1)\cap k(\le 3)$ as asserted. \\
 \hspace*{6mm}  
  We turn to $\tilde{M}$. Recall $M\subset \tilde{M}\subset \hat{M}$. One easily checks $M\cap \big(k(>-1)\cap k(<1)\big)=\emptyset$ and   $\hat{M}\setminus M=
  h(\le 1)\cap h(\ge -1)\cap k(>-1)\cap k(<1)$. Now, $\mathfrak{x}\in  k(>-1)\cap k(<1)$ means $-1<x_0+x_3<1$. So the lightlike line $\mathfrak{x}+\R(1,0,0,-1)$ stays in  $k(>-1)\cap k(<1)$ and does not meet $M$. This proves $\tilde{M}=M$. Obviously  $\{\mathfrak{x}: -1<x_0-x_3<1, -1<x_0+x_3<1\}
\subset \hat{M}\setminus\tilde{M}$. \qed \\

 \subsection{The lattice $\mathcal{M}$} In oder to be clear in the sequel we recall the  definition (\ref{DLOCMPMOM}), see e.g. \cite[sec.\,11]{CD09}.
 
   \begin{Def} \label{DLOCMPMOM} Let $(L,\le) $ be a partially ordered set. Then $L$ is a \textbf{lattice},  resp. a $\sigma$-complete lattice, resp. a complete lattice if every finite subset, resp. every countable subset, resp. every subset $S\subset L$ has a least upper bound $\bigvee S$ and a greatest lower bound $\bigwedge S$. A lattice $L$ is \textbf{modular}, if for $a$, $b$, $c$ in $L$ with $a\le b$ one has $(a\lor c)\land b=a\lor (c\land b)$. A lattice $L$ is bounded from below, if there is a minimal element $0$ in $L$, and it is \textbf{bounded}, if there is also a maximal element $I$ in $L$. \\
 \hspace*{6mm}    
  Let $L$ have a minimal element $0$.  Then $P\in L$ is called an \textbf{atom}, if $P\ne 0$ and $a\le P$ implies $a=P$ for all $a\in L\setminus\{0\}$. $L$ is \textbf{atomic}, if for every $a\in L\setminus\{0\}$ there is an atom $P$ with $P\le a$. $L$ satisfies  the \textbf{covering property},  if
 for every $a,b\in L$ and atom $P$ with $a\le b\le a\lor P$ one has $b\in\{a,a\lor P\}$.\\
   \hspace*{6mm}    
  A lattice  $L$ is \textbf{orthocomplemented} if  it is bounded and if there is an order-reversing involution $\perp :L\to L$ such that $a \land a^\perp =0$, $a\lor a^\perp =I$ for  $a\in L$. An orthocomplemented lattice $L$ is \textbf{orthomodular}  if for $a$, $b$ in $L$ with $a\le b$   one 
  has $a=(a\lor b^\perp)\land b$ (or equivalently $b=a\lor (a^\perp \land b) )$.\qed
   \end{Def}\\
    Note that for an orthocomplemented lattice modularity implies orthomodularity. Indeed, choose simply $c:=b^\perp$ in the modularity relation. The reverse implication in general does not hold. For instance, the  lattice of the orthogonal projections on an infinite dimensional Hilbert space, occasionally called \textbf{quantum proposition system} or \textbf{quantum logic},  is not modular (see  \cite[p. 444]{P64}) but obviously orthomodular.\\ 
 \hspace*{6mm}  
Orthomodularity, also called \textbf{weak modularity},  is equivalent to the \textbf{axiom (P)} introduced and studied in Piron 1964 \cite[sec.\,5]{P64}. According to \cite{P64} orthomodularity and  atomicity  with the covering property constitute the fundamental properties of a quantum proposition system. For a thorough exposition of this topic see also Jauch 1968 \cite{J68}. \\ 
 \hspace*{6mm} 
In  \cite[sec.\,5.8]{J68} a further property, which   we call q-modularity, is asserted to be equivalent to orthomodularity. An orthocomplemented lattice $L$ is \textbf{q-modular} if,  for $a$, $b$, $c$ in $L$ with $a\le b$ and $c\le b^\perp$, the equality $c\lor a=c\lor b$ implies $a=b$. We  verify the equivalence. 
First assume  orthomodularity.   
As $a\le b\le c^\perp$ it yields $(c\lor a)\land c^\perp=a$ and $(c\lor b)\land c^\perp=b$. --- Now assume q-modularity. Set $d:=(a\lor b^\perp)\land b$. Then obviously $a\le d\le b$. Hence $c:=b^\perp$ satisfies $c\le d^\perp$ and $c\lor a \le c\lor d\le c\lor (a\lor b^\perp)=c\lor a$, whence $ c\lor a= c\lor d$, and, by q-modularity, the assertion $a=d$.\\

\hspace*{6mm} 
For later use in sec.\,\ref{ECLNTLHP} we note

\begin{Lem}\label{ODCDS} Let $L$ be a $\sigma$-complete 
orthomodular lattice. Let $a_n\in L$, $n\in\N$ with $a_1\ge a_2\ge a_3\ge \dots$. Then one has the orthogonal decomposition
$$a_1\land(\bigwedge_na_n)^\perp=\bigvee_n(a_n\land a_{n+1}^\perp)$$
 of the left hand side.
\end{Lem}\\
{\it Proof.} First use the general identity $(\bigwedge_na_n)^\perp=(\bigvee_na_n^\perp)$. Then, by  \cite[Th\'eor\`em VIII]{P64} due to orthomodularity, $a_1\land(\bigvee_na_n^\perp)=\bigvee_n(a_1\land a_n^\perp)$. Obviously the proof  is accomplished showing $a_1\land a_n^\perp=\bigvee_{i=1}^{n-1} (a_i\land a_{i+1}^\perp)$ for every $n\ge 2$. Hence, using induction, it suffices to verify $a_1\land a_{n+1}^\perp=(a_1\land a_{n}^\perp)\lor(a_n\land a_{n+1}^\perp)$. To this end  note $(a_1\land a_n^\perp)\le (a_1\land a_{n+1}^\perp)$ and $a_n\le a_1$. Therefore, applying twice orthomodularity, one has $(a_1\land a_{n+1}^\perp)=(a_1\land a_n^\perp)\lor \big(a_1\land a_{n+1}^\perp\land (a_1^\perp\lor a_n)\big)=(a_1\land a_n^\perp)\lor (a_{n+1}^\perp\land a_n)$ as claimed.\qed\\

\hspace*{6mm} 
According to the above definition,  the considerations so far   show that   the set $\mathcal{M}$ of all causally complete sets endowed with  set inclusion $\subset$ and causal complementation $\perp$ is a complete orthocomplemented lattice. The greatest lower bound $\bigwedge \mathcal{K}$ for any subset $\mathcal{K} \subset \mathcal{M}$ is given by the intersection $\bigcap \mathcal{K}$ of all its elements. Its smallest upper bound $\bigvee \mathcal{K}$ is $\big(\bigcup \mathcal{K}\big)^\land=\big(\bigcap\{M^\perp: M \in\mathcal{K}\}\big)^\perp$.

\hspace*{6mm}
   $\mathcal{M}$ is \textbf{atomic}. Indeed, $\emptyset$ is the minimal element, and due to (\ref{AOPS}), the atoms of  $\mathcal{M}$ are the one-point sets. Obviously, every $M\in  \mathcal{M}\setminus \{\emptyset\}$  contains an atom.

 \hspace*{6mm} 
  $\mathcal{M}$ is atomic \textbf{without  the covering property}. For an example let $L$ be the lightlike line $\R(1,0,0,1)$ and $L'$ the segment $\{s(1,0,0,1):s\in [0,1]\}$. Then $L\cap \{0\}^\perp=\emptyset$, since $\{0\}^\perp=\{\mathfrak{y}: |y_0|<|y|\}$. By (\ref{LLLCC}), $L$ and $L'$ are causally complete. Therefore
 $L\cap \{0\}^\perp\subsetneq L'\subsetneq L$ implies $M\subsetneq M'\subsetneq M\lor \{0\}$ for $M:=L^\perp$, and $M':=L'^\perp$, contradicting atomicity. --- Another example is given by $\{0\}\subsetneq \{0\}\lor \{(1,0,0,0)\}\subsetneq  \{0\}\lor \{(2,0,0,0)\}$. This holds true by (\ref{CCTPS}).

   \hspace*{6mm}  
 $\mathcal{M}$ is \textbf{ not orthomodular}.   For an example let $L:=h(< 0)\cap k(> 2)$ and $M:=h(< 0)\cap k(> 0)$. Then clearly $L\subset M$. By (\ref{FHS})(b),  $L,M \in\mathcal{M}$  with $L^\perp=h(\ge 0)\cap k(\le 2)$, $M^\perp=h(\ge 0)\cap k(\le 0)$, and $ L^\perp\cap M = h(\ge 0)\cap k(\le 2)\cap h(< 0)\cap k(> 0)=\emptyset$. Hence $(L\cup M^\perp)^\land =(L^\perp\cap M)^\perp=\R^4$. Therefore $(L\lor M^\perp)\land M=(L\cup M^\perp)^\land \cap M=M\ne L$, 
 and orthomodularity does not hold.\footnote{ The  presumed proof of orthomodularity  of $\mathcal{M}$ in  \cite[Proposition 4.1.3]{R.H92} is not correct, since there  one infers  $\mathfrak{x}\in (L\cup M^\perp)^\sim\cap M$ from  $\mathfrak{x}\in (L\cup M^\perp)^\land \cap M$. Indeed, this does not hold true in general.  In the above example one has $(L\cup M^\perp)^\land\cap M=M$ and $(L\cup M^\perp)^\sim\cap M=L$ with $L\subsetneq M$ by (\ref{EMTVMH}) as $L=\hat{\Gamma}, M=\Delta^\perp$.} \\
\hspace*{6mm} 
 The non-orthomodularity of  $\mathcal{M}$  is already stated in \cite[sec.\,3]{C02}. See also the general analysis in \cite{C02} following Theorem 1.
 \\

Note that a causally complete set need not be Borel. (For an example take $\Gamma:= \{0\}\times (A\setminus \mathbb{Q}^3)$ with $A\subset \R^3$ not Borel. So $\Gamma$ is not Borel, too. Let $\mathfrak{x}\in \tilde{\Gamma}$.  Then one easily checks   $ \{y\in\R^3: |x-y|\le |x_0|\} \subset (A\setminus \mathbb{Q}^3)$. This implies $x_0=0$, whence $x=y$ and hence $\mathfrak{x}\in\Gamma$. This proves $\tilde{\Gamma}=\Gamma$. As (\ref{ETHSSHP})(v) shows, $\tilde{\Gamma}=\hat{\Gamma}$ holds since $\Gamma$ is a subset of a spacelike hyperplane. Therefore $\Gamma$ is causally complete.) However  by (\ref{MCCS}),  causally complete sets are measurable. For $M\subset\R^4$ let $$ M^\lor:=\bigcup_{\mathfrak{y}\in M}\{\mathfrak{x}:(\mathfrak{x}-\mathfrak{y})^{\cdot 2}\ge 0\}$$
be the \textbf{maximal set of influence} of $M$. Obviously $ M^\lor=\R^4\setminus M^\perp$.

\begin{Lem}\label{MCCS} Let $M\subset \R^4$. 
Define $U(M):=\bigcup_{\mathfrak{y} \in M}\{\mathfrak{x}: (\mathfrak{x}-\mathfrak{y})^{\cdot 2} > 0\}$.  Then
\begin{itemize}
\item[(a)] $U(M)\subset M^\lor$ and  $U(M)$ is open  
  \item[(b)]  $M^\lor\setminus U(M)$ is contained in an $F_\sigma$-null-set
  \item[(c)] $M^\lor$, $M^\perp$, $\hat{M}$  are  measurable
\end{itemize}
\end{Lem}
{\it Proof.} (a) is obvious. (c) $M^\lor$ is measurable as an immediate consequence of (a), (b). The remainder is obvious.\\
\hspace*{6mm} 
 So we turn to (b). Put  $N(M):=\{\mathfrak{x}: \sup_{\mathfrak{y} \in M}(\mathfrak{x}-\mathfrak{y})^{\cdot 2} =0\}$.  Obviously (i) $N(M)=N(\overline{M})$. --- Note  (ii) $M^\lor\setminus U(M)\subset N(M)$. Indeed,
 let $\mathfrak{x}\in M^\lor\setminus U(M)$. Then there is $\mathfrak{y}\in M$ with $(\mathfrak{x}-\mathfrak{y})^{\cdot 2}\ge 0$ and $(\mathfrak{x}-\mathfrak{y}')^{\cdot 2}\le 0$ for all $\mathfrak{y}'\in M$. Hence $\sup_{\mathfrak{y}'\in M}(\mathfrak{x}-\mathfrak{y}')^{\cdot 2}= 0$, whence $\mathfrak{x}\in N(M)$.\\
\hspace*{6mm}
Suppose now that $M$ is compact. Then, (iii) for every $\mathfrak{x}\in N(M)$ there is $\mathfrak{y}\in M$ with $(\mathfrak{x}-\mathfrak{y})^{\cdot 2}=0$. Indeed, there are $\mathfrak{y}_n\in M$ satisfying $-\frac{1}{n}\le (\mathfrak{x}-\mathfrak{y}_n)^{\cdot 2}\le 0$. Without restriction $\mathfrak{y}_n\to \mathfrak{y}\in M$, whence the claim. --- Next claim that (iv) $N(M)$ is closed.
Indeed,  let $\mathfrak{x}_n\in N(M)$ with $\mathfrak{x}_n\to \mathfrak{x}_0$. There are $\mathfrak{y}_n\in M$ satisfying $-\frac{1}{n}\le (\mathfrak{x}_n-\mathfrak{y}_n)^{\cdot 2}\le 0$. Without restriction $\mathfrak{y}_n\to \mathfrak{y}_0\in M$. Hence $(\mathfrak{x}_0-\mathfrak{y}_0)^{\cdot 2}= 0$. Furthermore, for every $\mathfrak{y}\in M$, $(\mathfrak{x}_n-\mathfrak{y})^{\cdot 2}\le 0$, whence $(\mathfrak{x}_0-\mathfrak{y})^{\cdot 2}\le 0$. Therefore
$ \mathfrak{x}_0\in N(M)$ as claimed. --- Now claim that (v) for every $a\in\R^3$ the section $N(M)_a=\{\xi\in\R: (\xi,a)\in N(M)\}$ contains at most two points. Indeed, assume that there are $\alpha, \beta,\gamma \in \R$ with $\alpha<\beta<\gamma$ such that $(\alpha,a),(\beta,a),(\gamma,a)\in N(M)$. By (iii) there is $\mathfrak{y}\in M$ with $\big((\beta,a)-\mathfrak{y}\big)^{\cdot 2}=0$. Furthermore, $\big((\alpha,a)-\mathfrak{y}\big)^{\cdot 2}\le0$ and $\big((\gamma,a)-\mathfrak{y}\big)^{\cdot 2}\le0$ hold. Hence, one finds $(\beta-y_0)^2=|a-y|^2$,
$(\alpha-y_0)^2\le|a-y|^2$, and $(\gamma-y_0)^2\le|a-y|^2$, whence $(\alpha-y_0)^2\le(\beta-y_0)^2$, $(\gamma-y_0)^2\le(\beta-y_0)^2$, and further the contradiction $\gamma +\beta\le 2y_0\le \beta+\alpha$, proving the claim. --- Finally claim (vi) $\lambda^4\big(N(M)\big)=0$.\footnote{ $\lambda^d$ denotes the Lebesgue measure on $\R^d$} Indeed, this is a ready consequence of (iv), (v) due to Tonelli's theorem by which $\lambda^4\big(N(M)\big)=\int_{\R^3}\lambda^1\big(N(M)_x\big)\operatorname{d}\lambda^3(x)=\int 0\operatorname{d}\lambda^3=0$.\\
\hspace*{6mm}
We are ready to proof (b) for general $M\subset\R^4$. Let $M_n\subset M$ be bounded with $M=\cup_nM_n$. Then $M^\lor\setminus U(M)=\big(\cup_nM^\lor_n\big)\setminus \big(\cup_n U(M_n)\big)\subset \cup_n\big(M^\lor_n\setminus U(M_n)\big)\subset \cup_n N(M_n)$ by (ii), and further $=\cup_n N(\overline{M}_n)$ by (i), whence the assertion by (iv), (vi).\qed\\

  Summarizing the considerations so far  one has
 
  \begin{The} \label{LMCCSMS}  
  The causally complete subsets of Minkowski space are measurable and form a complete orthocomplemented atomic lattice $\mathcal{M}$  without  the covering property and which is not  orthomodular.
 \end{The}

$\mathcal{M}$ is called the \textbf{lattice of causally complete regions}. The lattice $\mathcal{M}$ and versions of it occasionally are addressed as  \textbf{relativistic causal logic}. We restrict ourselves to refer to Cegla, Jancewicz 2013 \cite{CJ13} and the literature cited  therein  for a partial overview on the subject. More relevance  than $\mathcal{M}$ regarding causal systems will have the lattice $\mathcal{M}'$ generated by the non-timelike relation, see sec.\,\ref{LGNTLR},\,\ref{SRLBSGNTLR},\,\ref{ECLNTLHP}.
\\

\subsection{Spacelike sets}\label{SLSCI} A subset $A$ of Minkowski space is said to be \textbf{spacelike} or  causally independent or a $\perp$-set if any two different points $\mathfrak{x}$, $\mathfrak{y}$ of $A$ are spacelike separated, i.e., $(\mathfrak{x}-\mathfrak{y})^{\cdot 2} <0$. Clearly, if $A$ is spacelike and $B\subset A$, then  $A\setminus B\subset B^\perp$.\\
  \hspace*{6mm} 
 A spacelike set $A$ is \textbf{maximal} if $A^\perp=\emptyset$. E.g., spacelike hyperplanes are maximal. Every spacelike set is contained in a maximal one. More generally, if $A$ is spacelike contained in some set $M$, then there is a spacelike set $A'$ satisfying $A\subset A'\subset M$, which is maximal in $M$, i.e.,  there is no spacelike set $B\subset M$ with $A' \subsetneq B$. This holds true by an obvious application of Zorn's lemma.

 \begin{Def}
 Let  $\mathcal{M}_1$ denote the set of all those subsets of Minkowski space which are the causal completion of a spacelike set.  
\end{Def} 
 
 Of course, $\mathcal{M}_1$ is interesting  with regard to a possible extension of the localization  $E$ of a causal system to   causally complete regions. One observes first that $\mathcal{M}_1 \subsetneq\mathcal{M}$.
 
 \begin{Examp}\label{LLLCC} There are causally complete sets $M$, which are not the causal completion of a spacelike set and, a fortiori, are not determined by a spacelike set, i.e.,  $M\ne \hat{\Delta}$ and, hence, $M\ne \tilde{\Delta}$ for every spacelike $\Delta$. Such sets are for instance \emph{(i)} the lightlike lines and \emph{(ii)} segments of them, and  \emph{(iii)} $M:=\{\mathfrak{x}: x_0=x_3\in[0,1]\}$.
\end{Examp}
 
{\it Proof.} As to (i), let without restriction $M:=\R(1,0,0,1)$. First we show $M=\hat{M}$. Note, $\mathfrak{y}\in M^\perp$ if and only if  $y_1^2+y_2^2>(y_0-y_3)(y_0+y_3-2s) $ for all $s\in\R$. Hence $M^\perp=\{\mathfrak{y}:y_0=y_3, (y_1,y_2)\ne (0,0)\}$. Therefore $\mathfrak{x}\in M^{\perp\perp}$ if and only if $(x_1-r)^2+(x_2-t)^2>(x_0-x_3)(x_0+x_3-2s) $ for all $r,t,s\in\R$ with $(r,t)\ne (0,0)$. This implies $x_1=x_2=0$ and further $x_0=x_3$. It follows $M=\hat{M}$. --- As $\big((s,0,0,s)-(r,0,0,r)\big)^{\cdot 2}=0$, every spacelike $\Delta\subset M$ contains at most one point, whence $\hat{\Delta}=\Delta\ne M$.\\
\hspace*{6mm} 
As to (ii), let $M:= \{(s,0,0,s): s\in[0,1]\}$. We show $M=\hat{M}$. Note $M^\perp=\{\mathfrak{y}: (y_0-s)^2<y_1^2+y_2^2+(y_3-s)^2,\;s\in[0,1]\}$. Now consider $\mathfrak{x}\in \R^4\setminus M$. This means either $x_0<x_3$, or $x_0>x_3$, or $x_0=x_3=r<0$, or $x_0=x_3=r>1$. If $x_0<x_3$, then  $\mathfrak{x}\not\in M^{\perp\perp}$, since $\mathfrak{y}:=(1,x_1,x_2,x_3-x_0+1)\in M^\perp$, but $(\mathfrak{x}-\mathfrak{y})^{\cdot 2}=(x_0-1,0,0,x_0-1)^{\cdot 2}=0$. 
If $x_0>x_3$, then  $\mathfrak{x}\not\in M^{\perp\perp}$, since $\mathfrak{y}:=(0,x_1,x_2,x_3-x_0)\in M^\perp$, but $(\mathfrak{x}-\mathfrak{y})^{\cdot 2}=(x_0,0,0,x_0)^{\cdot 2}=0$. 
If $x_0=x_3=r<0$, then  $\mathfrak{x}\not\in M^{\perp\perp}$, since $\mathfrak{y}:=(0,x_1,x_2,r)\in M^\perp$, but $(\mathfrak{x}-\mathfrak{y})^{\cdot 2}=(r,0,0,0)^{\cdot 2}>0$. Finally,  if $x_0=x_3=r>1$, then  $\mathfrak{x}\not\in M^{\perp\perp}$, since $\mathfrak{y}:=(1,x_1,x_2,r)\in M^\perp$, but $(\mathfrak{x}-\mathfrak{y})^{\cdot 2}=(r-1,0,0,0)^{\cdot 2}>0$. Thus  $M=\hat{M}$. --- Finish the proof as in (i).  

\hspace*{6mm} 
As to (iii), let $M_n:= \big(h(\ge  -\frac{1}{n})\cap k(\le 2)\big)\cap\big(h(\le 0)\cap k(\ge 0)\big)$. Then $M_n=\hat{M}_n$ by (\ref{FEDOD})(ii), whence $M=\hat{M}$, since $M=\bigcap_nM_n$. Now let $A\subset M$ be spacelike.  We show $\hat{A}\subsetneq M$. If $\mathfrak{a},\mathfrak{a'}\in A$, $\mathfrak{a}\ne\mathfrak{a'}$, then $(a_1,a_2)\ne (a'_1,a'_2)$. Hence there is $f:\R^2\to [0,1]$ such that $A\subset\{(f(r,s),r,s,f(r,s)):r,s\in\R\}$.  (Equality holds if $A$ is maximal.) ---  If $f=0$, then $A^\perp\supset \{(0,r,s,0): r,s\in\R\}^\perp=\{\mathfrak{y}: |y_0|<|y_3|\}= \big(h(<0)\cap k(>0)\big)\cup \big(h(>0)\cap k(<0)\big)$, whence   $\hat{A}\subset h(\ge 0)\cap k(\le 0)\cap h(\le 0) \cap k(\ge 0)=\{\mathfrak{x}: x_0=x_3=0\}\subsetneq M$ by (\ref{FEDOD})(ii). --- If $f\ne 0$, there is $f(r_0,s_0)\ne 0$. Then $\mathfrak{y}:=(t,r_0,s_0,-t)$ for $t>0$ lies in $A^\perp$. Indeed, $(f(r,s)-t)^2<(r-r_0)^2+(s-s_0)^2+(f(r,s)+t)^2 $ holds, since $|f(r,s)-t|\le f(r,s)+t$ and  $|f(r,s)-t| = f(r,s)+t$ only if $f(r,s)=0$, in which case $(r-r_0)^2+(s-s_0)^2>0$. But $\mathfrak{y}\not\in M^\perp$, as $\mathfrak{x}:=(0,r_0,s_0,0)\in M$ and $(\mathfrak{y}-\mathfrak{x})^{\cdot 2}= t^2-t^2=0$. Hence $M^\perp\subsetneq A^\perp$.
\qed\\

What is more, it turns out that 

\begin{Lem}
$\mathcal{M}_1$ is not a sublattice of $\mathcal{M}$ and it is not closed under causal complementation. 
\end{Lem}\\
{\it Proof.} See (\ref{LPNL}) and  (\ref{LPNIUOC}).\qed

\begin{Examp}\label{LPNL}  There are $M,L\in \mathcal{M}_1$ with $M\cap L\not\in \mathcal{M}_1$, for instance $M:=h(\le 0)\cap k(\ge 0)$ and $L:= h(\ge 0)\cap k(\le 2)$.
\end{Examp} \\
{\it Proof.}  $M,L\in \mathcal{M}_1$ by (\ref{FEDOD})(i). But $M\cap L=\{x_0-x_3=0, 0\le x_0+x_3\le 2\}=\{\mathfrak{x}: x_0=x_3\in [0,1]\}\not\in \mathcal{M}_1$ by (\ref{LLLCC})(iii).\qed
 
\begin{Examp}\label{LPNIUOC}  There is $M\in  \mathcal{M}_1$ with $M^\perp \not\in \mathcal{M}_1$, for instance $M:=\{\mathfrak{x}: x_0=x_3, (x_1,x_2)\ne (0,0)\}$.
\end{Examp}  

{\it Proof.} As shown in (\ref{LLLCC})(i) one has $M=\{(s,0,0,s):s\in\R\}^\perp$ and $M^\perp=\{(s,0,0,s):s\in\R\}\not\in \mathcal{M}_1$. It remains to show $M\in  \mathcal{M}_1$.\\
 \hspace*{6mm}
Clearly, $A:=\{(f(r,s),r,s,f(r,s)): (r,s)\ne (0,0)\}$ with $f(r,s):=\textrm{sgn}(r)(r^2+s^2)^{-1}$ is a spacelike subset of $M$. It suffices to  show $A^\perp \subset M^\perp$. Let $\mathfrak{y}\in A^\perp$. This means $(y_0-y_3)\big(y_0+y_3-2f(r,s)\big)<(y_1-r)^2+(y_2-s)^2$ for all $(r,s)\ne(0,0)$. Note $(y_1-r)^2+(y_2-s)^2\le C:=(|y_1|+1)^2+(|y_2|+1)^2$ for all $|r|\le 1$, $|s|\le 1$. --- Assume first $y_0>y_3$. Then for $(r,s):=(-\frac{1}{n},\frac{1}{n})$, $n\in\N$ one gets the contradiction  $(y_0-y_3)(y_0+y_3+n^2)\le C$ for all $n$. --- Now assume $y_0<y_3$. Then for $(r,s):=(\frac{1}{n},\frac{1}{n})$, $n\in\N$ one gets the contradiction  $(y_0-y_3)(y_0+y_3-n^2)\le C$ for all $n$. --- So $y_0=y_3$ remains. Therefore $0<(y_1-r)^2+(y_2-s)^2$ for all $(r,s)\ne(0,0)$. This implies $y_1=y_2=0$. Hence $\mathfrak{y}\in M^\perp$.\qed\\

  \subsection{Bases} 
A familiar idea of a causal evolution  is that the \textbf{Cauchy data} given on a spacelike set $\Delta$ determine by means of a hyperbolic PDE  the events in $\tilde{\Delta}$. Inspired by this we call  a spacelike set $\Sigma$  a (causal)  \textbf{base} if $\tilde{\Sigma}=\R^4$. Every base $\Sigma$ is maximal spacelike, since $\R^4= \tilde{\Sigma}\subset  \hat{\Sigma}=\Sigma^{\perp\perp}$, whence $\Sigma^\perp= (\R^4)^\perp=\emptyset$. Plainly, every spacelike hyperplane is a base.

\begin{Lem}\label{ESLHSF}   Let $\Sigma\subset \R^4$.  \emph{(i)} If $\Sigma$ is a base, then  there is $\xi:\R^3\to\R$ satisfying $|\xi(x)-\xi(y)| < |x-y|$ for all $x\ne y$ such that $\Sigma=\{(\xi(x),x):x\in\R^3\}$. \emph{(ii)} 
Conversely,  $\Sigma$ is a base if there is $\xi:\R^3\to\R$ satisfying $|\xi(x)-\xi(y)| < |x-y|$ for all $x\ne y$ and $\limsup_{|x|\to\infty}|\xi(x)| / |x|<1$ such that $\Sigma=\{(\xi(x),x):x\in\R^3\}$.
 \end{Lem}\\
 {\it Proof.} As to (i), for every $a\in\R^3$ there is $\xi(a)\in\R$ satisfying $(0,a)+\xi(a)(1,0)=(\xi(a),a)\in\Sigma$. Let  $(\xi(x),x)\ne \mathfrak{y}$ for $x\in\R^3$, $\mathfrak{y}\in\Sigma$. Since $\Sigma$ is spacelike, $|\xi(x)-y_0| < |x-y|$ follows. This proves also $\Sigma=\{(\xi(x),x):x\in\R^3\}$. \\
 \hspace*{6mm}
  As to (ii), clearly $\Sigma$ is spacelike. Assume  $\mathfrak{a}\not\in \tilde{\Sigma}$. Let $a_0>\xi(a)$. (The case $a_0<\xi(a)$ is analogous.) Then there is $\mathfrak{z}=(1,e)$ with $0<|e|\le 1$ such that $(\mathfrak{a}+\R\mathfrak{z}) \cap \Sigma=\emptyset$. It follows $a_0+s>\xi(a+se)$ for all $s$, since otherwise there were $s$ with $a_0+s\le \xi(a+se)$ and by Bolzano's theorem there were $s^*$ satisfying $a_0+s^*= \xi(a+s^*e)$, i.e., $ (\mathfrak{a}+s^*\mathfrak{z}) \in\Sigma$. Hence for $s<-a_0$ one has $ |\xi(a+se)|  /|a+se|>|a_0+s|/|a+se|\to 1/|e|\ge 1$ for $s\to -\infty$. This contradicts the assumption $\limsup_{|x|\to\infty}|\xi(x)| / |x|<1$.\qed\\

Clearly, in (\ref{ESLHSF})(i), $\xi$ is Lipschitz continuous and $\Sigma$ is a closed set.  By (\ref{ESLHSF})(i)$,  \sigma$ is a spacelike hyperplane if and only if it is the graph of $\xi(x)=\rho +xe$ for some $\rho\in\R$ and $e\in\R^3$ with $|e|<1$. Equivalently, $\sigma=\{\mathfrak{x}: \mathfrak{x}\cdot \mathfrak{e}=\rho\}$ for uniquely determined $\rho\in\R$ and timelike
$\mathfrak{e}=(1,e)$.\\
\hspace*{6mm} 
There are  maximal spacelike sets which are not a  base. Two examples follow.

\begin{Examp}\label{MFEMSLS} The mass shell $M:=\{(\xi(x),x): x\in\R^3\}$ for $\xi(x):=\sqrt{|x|^2+m^2}$ with mass $m>0$ is a maximal spacelike set but not a base. One has $\tilde{M}=\{\mathfrak{y}: y_0> |y|\}$. Note that $|\xi(x)-\xi(y)|<|x-y|$ for all $x\ne y$, cf. \emph{(\ref{ESLHSF})}.
\end{Examp}\\
 {\it Proof.} If $x\ne y$ then $|\xi(x)-\xi(y)|=\frac{|\,|x|^2-|y|^2|}{\xi(x)+\xi(y)}<\frac{|\,|x|^2-|y|^2|}{|x|+|y|}=\big||x|-|y|\big|\le|x-y|$. Hence, in particular, $M$ is spacelike.  --- $M$ is maximal, since for every $\mathfrak{y}\not\in M$,  $|y_0-\xi(x)|<|y-x|$ is not satisfied by $x:=y$.\\  
\hspace*{6mm} 
  We turn to $\tilde{M}$. Let $\mathfrak{y}\in\R^4$. If $y_0\le |y|$, then $y_0+s\le |y|+s <  \xi(y+se)=\sqrt{(|y|+s)^2+m^2}$ for all $s$ choosing $e=\frac{1}{|y|}y$ if $y\ne 0$ and any unit vector otherwise. This proves $\mathfrak{y}\not\in \tilde{M}$. Next consider the case  $y_0\ge\xi(y)$. Let $\mathfrak{z}=(1,e)$ with $|e|\le 1$. Then  the function $f(s):= y_0+s-\xi(y+se)$ has a zero for some $s\in\R$ by  the intermediate value theorem, since $f(y_0)=-\xi(y+y_0e)<0$ and $f(0)\ge0$ by the assumption. Hence 
$\mathfrak{y}\in \tilde{M}$. Finally let $|y|<y_0<\xi(y)$. In this case $f$ has a zero, since $f(0)<0$ by the assumption and $f(\bar{s})\ge 0$ for $\bar{s}:=\frac{1}{2}\big(\xi(y)^2-y_0^2\big)/(y_0-|y|)$. As to the latter, note first that $\bar{s}>0$ by the assumption and, hence, $f(\bar{s})=y_0+\bar{s}-\xi(y+\bar{s}e)\ge
y_0+\bar{s}-\sqrt{(|y|+\bar{s})^2+m^2}=0$. Hence $\mathfrak{y}\in \tilde{M}$.\qed\\

 \hspace*{6mm} 
Particularly important is

\begin{Examp} \label{EMTVMH} 
Let $P:=\Delta\cup\Gamma$ with $\Delta:=\{\mathfrak{x}: x_0=0, x_3\le 0\}$, $\Gamma:= \{\mathfrak{x}: x_0=1, x_3>1\}$.\\ 
\hspace*{6mm} 
Then $P$ is maximal spacelike, whence $\hat{\Delta}\perp \hat{\Gamma}$, i.e., $\hat{\Gamma}\subset  \Delta^\perp$, and
  $\Delta^\perp\cap\Gamma^\perp=\emptyset$. But $P$ is not a base. More precisely,
   $\tilde{P}=(\hat{\Delta}\cup\hat{\Gamma})^\sim=\hat{\Delta}\cup\hat{\Gamma}\cup \{\mathfrak{x}:2-x_3<x_0=x_3\}=\{\mathfrak{x}: |x_0|\le -x_3\}\cup \{ \mathfrak{x}: 2-x_3<x_0\le x_3\}\ne \hat{P}= \R^4$. 
  In particular,  $\tilde{P}\cap \Delta^\perp=\hat{\Gamma}$. 
\end{Examp}

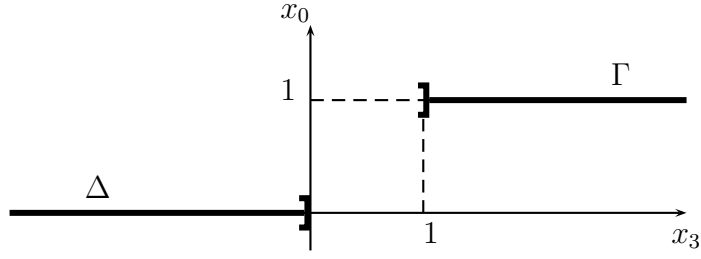
\begin{pspicture}(-8,-1)(2,2.5)
\psline{->}(-5,0)(3,0)
\put(2.8,-0.4){$x_3$}
\psline{->}(-2,-0.5)(-2,2.5)
\put(-2.4,2.6){$x_0$}
\psline[linewidth=0.8mm]{-]}(-6,0)(-2,0)
\psline[linewidth=0.8mm]{]-}(-0.5,1.5)(3,1.5)
\put(-2.4,1.5){$1$}
\psline[linestyle=dashed](-2,1.5)(-0.4,1.5)
\put(-0.5,-0.4){$1$}
\psline[linestyle=dashed](-0.5,0)(-0.5,1.25)

\put(-5,0.2){$\Delta$}
\put(2,1.7){$\Gamma$}
\end{pspicture}

{\it Proof.} $P$ is spacelike, since $\Delta$ and $\Gamma$ are spacelike and since $(\mathfrak{y}-\mathfrak{x})^{\cdot 2}=1-|y-x|^2\le 1-|y_3|^2<0$ for $\mathfrak{x}\in \Delta$, $\mathfrak{y} \in \Gamma$. Then check $\Delta^\perp=h(< 0)\cap k(>0)$, $\Gamma^\perp=h(\ge 0)\cap k(\le 2)$, whence $P^\perp=  \Delta^\perp\cap \Gamma^\perp=\emptyset$ is maximal spacelike. 
 --- We turn to $\tilde{P}$. First we show $A\subset \tilde{P}$ for $A:=\hat{\Delta}\cup\hat{\Gamma}\cup \{\mathfrak{x}:2-x_3<x_0=x_3\}$. Indeed, as $\Delta\subset P$ it follows $\hat{\Delta}=\tilde{\Delta}\subset  \tilde{P}$ by (\ref{ETHSSHP}). Similarly, $\hat{\Gamma}\subset  \tilde{P}$. Finally let $ 2-x_3<x_0=x_3$. For $\mathfrak{z}=(1,0,0,1)$ one has $\mathfrak{x}-x_0\mathfrak{z}=(0,x_1,x_2,0)\in P$. For  any other $\mathfrak{z}=(1,e)$ check 
$\mathfrak{x}-(x_0-1)\mathfrak{z}=(1,x-(x_0-1)e)$ with $(x-(x_0-1)e)_3=x_0-(x_0-1)e_3=(x_0-1)(1-e_3)+1>1$ as $x_0>1$ and $e_3<1$. Therefore $\mathfrak{x}-(x_0-1)\mathfrak{z} \in P$. Hence $\mathfrak{x}\in  \tilde{P}$. --- Next we show $B\cap\tilde{P}=\emptyset$ for $B:=\{\mathfrak{x}: 0<x_0+x_3\le 2\}\cup\{\mathfrak{x}: |x_3-1|<|x_0-1|\}$. Obviously $B\cap P=\emptyset$. If $0<y_0+y_3\le 2$ then the whole lightlike line $\mathfrak{y}+\R(1,0,0,-1)$ through  $\mathfrak{y}$ lies in $\{0<x_0+x_3\le 2\}$. If $|y_3-1|<|y_0-1|$, then the whole timelike line $\mathfrak{y}+\R(y_0-1,0,0,y_3-1)$ through  $\mathfrak{y}$ lies in $\{|x_3-1|<|x_0-1|\}\cup \{x_0=x_3=1\}$. 
Note $  \{x_0=x_3=1\}\cap P=\emptyset$. Thus  $B\cap\tilde{P}=\emptyset$. --- It remains to show $ A\cup B=\R^4$. 
Since $\hat{\Delta}=h(\ge0)\cap k(\le 0)$, $\hat{\Gamma}= h(<0)\cap k(> 2)$ by (\ref{FHS})(b),
one has $A=\big(h(\ge0)\cap k(\le 0)\big)\cup \big(h(\le0)\cap k(> 2)\big)$ and $B=\big(k(>0)\cap k(\le 2)\big)\cup \big(h(>0)\cap k(>2)\big) \cup \big(h(<0)\cap k(< 2)\big)$.  So $A\cup B=k(>2)\cup \big(h(\ge0)\cap k(\le 0)\big)\cup \big(h(<0)\cap k(< 2)\big)\cup \big(k(>0)\cap k(\le 2)\big)=k(>0)\cup \big(h(\ge0)\cap k(\le 0)\big)\cup \big(h(<0)\cap k(<2)\big)\supset k(>0)\cup \big(h(\ge0)\cap k(\le 0)\big)\cup \big(h(<0)\cap k(\le 0)\big)=k(<0)\cup k(\le 0)=\R^4$. 
 \qed \\

\begin{Lem}\label{ETHSSHP} Let $A$ be a subset of a base $\Sigma$. Then \emph{(i)} $\Sigma \setminus A=A^\perp\cap \Sigma$, \emph{(ii)} $(\Sigma \setminus A)^\perp=\hat{A}$, \emph{(iii)} $(\Sigma \setminus A)^\land=A^\perp$, \emph{(iv)} $A=\hat{A}\cap \Sigma$, \emph{(v)} $\hat{A}=\tilde{A}$, and finally \emph{(vi)}  $A=\hat{A}_\Sigma =\bigcup_{\mathfrak{y} \in \hat{A}}\{\mathfrak{x}\in\Sigma: (\mathfrak{x}-\mathfrak{y})^{\cdot 2}\ge 0\}$  the set of influence in $\Sigma$ of $\hat{A}$.
\end{Lem}

{\it Proof.}  Recall first $A\subset \Sigma$, $A\cap A^\perp =\emptyset$, and $\Sigma\setminus A\subset A^\perp$ as $\Sigma$ is spacelike. Hence (i). --- Then obviously $\hat{A}\subset (\Sigma \setminus A)^\perp$. We show now $(\Sigma \setminus A)^\perp\subset \tilde{A}$. Then (ii) follows by (\ref{MTVMH}). So let $\mathfrak{x}\in (\Sigma \setminus A)^\perp$. This means $(\mathfrak{x}-\mathfrak{y})^{\cdot 2}< 0$ for all $\mathfrak{y}\in \Sigma \setminus A$. Let $\mathfrak{z}\ne 0$, $\mathfrak{z}^{\cdot 2}\ge 0$. As $\tilde{\Sigma}=\R^4$ there is $s\in\R$ with $\mathfrak{x}-s\mathfrak{z} \in\Sigma$. Assume $\mathfrak{x}-s\mathfrak{z} \in\Sigma\setminus A$. Then the contradiction $0>(\mathfrak{x}-\mathfrak{x}+s \mathfrak{z})^{\cdot 2}=s^2\mathfrak{z}^{\cdot 2}\ge 0$ follows. Therefore $\mathfrak{x}-s\mathfrak{z} \in A$ showing $\mathfrak{x}\in\tilde{A}$. --- By orthocomplementation  (ii) implies (iii). --- Further, (iv) equals (i) for $\Sigma\setminus A$ in place of $A$ because of (ii). --- (v) is already shown in the proof of (ii). --- Finally, as to the less obvious inclusion $\supset$ in (vi), consider  $\mathfrak{x}\in\Sigma\setminus A$. This implies $\mathfrak{x}\in A^\perp$ and hence, by definition, $\mathfrak{x}$ lies spacelike to $\hat{A}$.
 \qed\\

The examples (\ref{MFEMSLS}), (\ref{EMTVMH}) also show that $\Sigma$ in (\ref{ETHSSHP})(v) in general cannot be replaced with a maximal spacelike set.\\

Contrary to $\mathcal{M}_1$, the set $\mathcal{M}_2$ of all causally complete sets of the kind $\hat{A}$ for $A$ a subset of a base is closed under causal complementation. This follows from (\ref{ETHSSHP})(iii). Moreover, $\emptyset,\,\R^4\in \mathcal{M}_2$. So $\mathcal{M}_2$ is an \textbf{orthoposet}, which is not orthomodular (see the definition \cite[sec.\,11, Def.\,129]{CD09}). Also $\mathcal{M}_2$ is  not a sublattice of $\mathcal{M}$ still by (\ref{LPNL}). \\

\subsection{Diamonds}   Recall that spacelike hyperplanes are special bases. A  set $M$ of the kind  $M=\tilde{\Delta }$ for  $\Delta$ in some spacelike hyperplane  $ \sigma$ is called  a \textbf{diamond}.  $\Delta$ is called a \textbf{flat base} of $M$. 
   If $M$ is a diamond then, by (\ref{ETHSSHP})(v) and (iii), $M=\hat{M}=\tilde{M}$, whence $M$ is causally complete, and $M^\perp =( \sigma\setminus \Delta)^\land=( \sigma\setminus \Delta)^\sim$ is a diamond, too.

\begin{Exampl}\label{FEDOD} for some diamonds and their flat bases.

  \hspace*{6mm}  
 \emph{(i)} Let $\alpha,\delta\in\R$. Then for $(\sim,\backsim)\in\{(\ge,\le),(>,<),(\le,\ge),(<,>)\}$
\begin{itemize}
\item[(a)]
$h(\sim\alpha)\cap k(\backsim\delta)$  
\end{itemize}
 is a diamond with causal complement
 \begin{itemize}
\item[(b)]
$h(\not\sim\alpha)\cap k(\not\backsim\delta)$  
\end{itemize} 
and  
  \begin{itemize}
\item[(c)]
$\{\mathfrak{x}: x_0=\frac{\delta+\alpha}{2}, \,x_3\backsim \frac{\delta-\alpha}{2}\}$
\end{itemize} 
a flat base of it.

   \hspace*{6mm}  
\emph{(ii)}   Let $\alpha,\beta,\gamma,\delta\in\R$ with $\alpha< \beta$ and $\gamma<\delta$. Then for $(\sim,\backsim)\in\{(\ge,\le),(>,<)\} $ and $(\sim',\backsim')\in\{(\le,\ge),(<,>)\}$
\begin{itemize}
\item[(d)]
 $h(\sim\alpha)\cap k(\backsim\delta) \cap h(\sim'\beta)\cap k(\backsim'\gamma)$
   \end{itemize} 
  is a  diamond with causal complement
\begin{itemize}
\item[(e)]
$\big(h(\not\sim\alpha)\cap k(\not\backsim\delta)\big) \cup \big(h(\not\sim'\beta)\cap k(\not\backsim'\gamma)\big)$
\end{itemize}  
and   with unique flat base
\begin{itemize}
\item[(f)]
$\{\mathfrak{x}: (x_0,x_3)=s (\frac{\delta+\alpha}{2},\frac{\delta-\alpha}{2})+(1-s)(\frac{\gamma+\beta}{2},\frac{\gamma-\beta}{2}),  \,0\sim's \backsim1\}$
\end{itemize}
\end{Exampl}

{\it Proof.} As to the proof of (i), we treat the case $(\sim,\backsim)=(\ge,\le)$. The other cases follow analogously. Let $\mathfrak{y}\in\Delta^\perp$ for $\Delta:=\{\mathfrak{x}: x_0=\frac{\delta+\alpha}{2}, \,x_3\le \frac{\delta-\alpha}{2}\}$. This means for  $x_0=\frac{\delta+\alpha}{2}$ and all  $x_3\le \frac{\delta-\alpha}{2}$ that $|y_0-x_0|<|y_3-x_3|$  or, equivalently, 
\big[\;$y_0-y_3<x_0-x_3$ \,and\,  $y_0+y_3>x_0+x_3$\;\big]\; or\: \big[\;$y_0-y_3>x_0-x_3$ \,and\,  $y_0+y_3<x_0+x_3$\;\big].  Obviously this holds if and only if  $y_0-y_3<\alpha$ and $ y_0+y_3> \delta$. So $\Delta^\perp= h(<\alpha)\cap k(>\delta)$, cf. (b). Then, by (\ref{FHS})(b), $\hat{\Delta}=\Delta^{\perp\perp}=h(\ge\alpha)\cap k(\le\delta)$. One infers (i). --- For the proof of (ii) we proceed analogously treating the case $(\sim,\backsim)=(\ge,\le)$, $(\sim',\backsim')=(<,>)$ explicitly.  Let $\mathfrak{y}\in\Delta^\perp$ for $\Delta:=\{\mathfrak{x}: (x_0,x_3)=s (\frac{\delta+\alpha}{2},\frac{\delta-\alpha}{2})+(1-s)(\frac{\gamma+\beta}{2},\frac{\gamma-\beta}{2}),  \,0<s \le1\}$. This means that for all $(x_0,x_3)=s (\frac{\delta+\alpha}{2},\frac{\delta-\alpha}{2})+(1-s)(\frac{\gamma+\beta}{2},\frac{\gamma-\beta}{2}),  \,0<s \le1$ one has  $|y_0-x_0|<|y_3-x_3|$  or, equivalently, 
\big[\;$y_0-y_3<x_0-x_3$ \,and\,  $y_0+y_3>x_0+x_3$\;\big]\; or\: \big[\;$y_0-y_3>x_0-x_3$ \,and\,  $y_0+y_3<x_0+x_3$\;\big]. Since $x_0-x_3=s \alpha+ (1-s)\beta$, $x_0+x_3=s \delta+ (1-s)\gamma$,  this holds if and only if \big[\,$y_0-y_3<\alpha$ and $y_0+y_3> \delta$\,\big]
or \big[\,$y_0-y_3\ge \beta$ and $y_0+y_3\le \gamma$\,\big]. So $\Delta^\perp= \big(h(<\alpha)\cap k(>\delta)\big)\cup \big(h(\ge\beta)\cap k(\le\gamma)\big)$, cf. (e). Therefore, by (\ref{FHS})(b), $\hat{\Delta}=\Delta^{\perp\perp}=h(\ge\alpha)\cap k(\le\delta)\cap h(<\beta)\cap k(>\gamma)$. Hence (ii) follows. Note that uniqueness of the flat base follows from (\ref{UPBD}). \qed

\begin{Lem}\label{UPBD} Let $M$  be a diamond. Then there is at most one flat basis $\Delta$ of $M$ in a given spacelike  hyperplane $\sigma$.    
 If $M$  has two different flat bases, then either  $M=\R^4$ (and the flat bases of $M$ are the spacelike hyperplanes $\sigma$) or  there are a spacelike $2$-dimensional plane $\pi$ and a subset $\Lambda \subset \pi$ such that the flat bases of $M$ are the intersections $M\cap \sigma$ with the spacelike hyperplanes $\sigma$ containing $\pi$. The intersections $M\cap \sigma$ are equal to 
 $\Lambda \cup\sigma^c$ or $\Lambda\cup (\sigma\setminus \pi)$ with  $\sigma^c$ one of the two components of $\sigma\setminus \pi$.
 \end{Lem}
 
 {\it Proof.} The first assertion holds since $M\cap\sigma=\hat{\Delta}\cap \sigma =\Delta$. --- Now, let $\Delta\subset \sigma$, $\Gamma\subset\tau$ be different flat bases of $M$ in spacelike hyperplanes $\sigma$, $\tau$. First note that $\Delta \cap \tau= \Gamma \cap \sigma$, since otherwise, without restriction, one had $\mathfrak{x} \in \Delta \cap \tau \in \tau\setminus \Gamma$, whence $\mathfrak{x}\in \Gamma^\perp$, which contradicts $\mathfrak{x}\in\hat{\Gamma}$. Next, $\Delta\setminus \tau\ne\emptyset$, since otherwise $\Delta, \Gamma\subset \tau$, whence the contradiction $\Delta=\Gamma$. In particular, $\tau\ne \sigma$. Now,  without restriction let $\sigma=\{0\}\times \R^3 \equiv \R^3$. Write $\tau=\{\mathfrak{x}:\mathfrak{x}\cdot \eta=\rho\}$ for some $\eta=(1,e)$ with $e\in\R^3$, $|e| < 1$, $\rho\in\R$, and $(\rho,e)\ne 0$.  For  $x\in\sigma\setminus \tau$  obviously $\xi:= xe+\rho\ne 0$. --- We claim that for every $x\in\Delta\setminus \tau$  the whole closed ball 
    $B(x,|\xi|)$ around $x$ with  radius  $|\xi|$ is contained in $\Delta$. Indeed, as $(0,x)\in \Delta\subset M=\hat{\Gamma}=\tilde{\Gamma}$, there is $s\in\R$ such that $(0,x)+
   s(\xi,0)\in\Gamma\subset\tau$. It follows $s=1$ and $(\xi,x)\in \Gamma\subset \hat{\Delta}=\tilde{\Delta}$. Hence for every $|z|\le 1$ there is $s\in\R$ such that $(\xi,x)+s(1,z)$ is in $\Delta$. This implies the claim. --- If $e=0$, then $\rho\ne 0$, $\sigma\cap \tau=\emptyset$, so that  $B(x,|\rho|)\subset \Delta$ for all $x\in\Delta$, whence $\Delta=\sigma$ and $M=\R^4$. It remains the case $e\ne 0$. Here $\sigma\setminus\tau=\sigma^>\cup \sigma^<$, where $\sigma^\gtrless:=\{x\in\sigma: \xi\gtrless 0\}$ are the components of $\sigma\setminus\tau$, which are open (in $\sigma$). 
   Accordingly, $\Delta\setminus\tau=\Delta^>\cup\Delta^<$ with $\Delta^\gtrless:=\Delta\cap\sigma^\gtrless$. --- We claim $\Delta^>=\sigma^>$ if $\Delta^>\ne\emptyset$. Indeed, let $x\in\Delta^>$. Then $B(x,\xi)\subset \Delta$. For every $x'\in B(x,\xi)$ one has $\xi'\ne 0$. Therefore, by continuity, $\xi'>0$, whence $B(x,\xi)\subset \Delta^>$. So $\Delta^>$ is open. It remains to show that  $\Delta^>$ is also closed in $\sigma^>$. Let $\bar{x}$ be in the closure of $\Delta^>$. As $\bar{x}\in \sigma^>$ one has $\bar{\xi}>0$. There is $x\in\Delta^>$ with $|x-\bar{x}|<\frac{1}{2}\bar{\xi}$. Then $\xi=\bar{\xi}+(x-\bar{x})e>\frac{1}{2}\bar{\xi}$. So $\bar{x}\in B(x,\frac{1}{2}\bar{\xi})\subset \Delta^>$. --- Analogously $\Delta^<=\sigma^<$ if $\Delta^<\ne\emptyset$. Therefore, $\Delta\setminus \tau$ equals  $\sigma^>$ or $\sigma^<$ or $\sigma^> \cup\sigma^<=\sigma\setminus\tau$. --- Finally the result follows for $\Lambda:=\Delta\cap\tau$ and $\pi:=\sigma\cap\tau$.\qed\\

In particular, the above result shows that a diamond has a \textbf{unique flat base}  if it does not contain a spacelike half-hyperplane or, a fortiori,  if it is bounded. \\
   
    \hspace*{6mm} 
 In general, the intersection of two diamonds is not a diamond. (For instance, let $\Delta :=\{x_0=0, -4<x_3<4\} \subset \sigma :=\{0\}\times \R^3$ and $\Gamma := \Gamma_1 \cup \Gamma_2 \subset  \tau:= \{2\}\times \R^3$ with $\Gamma_1:=  \{x_0=2, -2<x_3<0\}$, $\Gamma_2:=    \{x_0=2, 0<x_3<4\}$. Then the intersection $\hat{\Delta} \cap \hat{\Gamma}= \hat{\Lambda}\cup \hat{\Gamma}_1$ with $\Lambda=\{(2-s,0,0,3s): 0\le s\le 1\} $ is not a diamond, as $\Lambda$ and  $\Gamma_1$ do not lie in a common spacelike hyperplane.)\\

 \section{Localization Operators for Causally Complete Regions}\label{LOCCR}

 We turn  to the question about  a rep $(W,F)$    of  the lattice $\mathcal{M}$ of causally complete regions posed at the beginning of sec.\,\ref{HLCCR}.
 There is a technical reason  concerning measurability (see (\ref{NRLCCR})) and  also a structural reason  concerning maximal spacelike sets (see (\ref{CNELD})(c)) why such a rep does not exist.\\
 
 \begin{The}\label{NRLCCR} 
 There is no  rep $(W,F)$ of $\mathcal{M}$. 
\end{The} \\
{\it Proof.} Assume the existence of a rep $(W,F)$. 
For every subset $X$ of $\R^3$ put $Q(X):=F\big((\{0\}\times X)^\land\big)$. (Recall that causally complete sets are measurable (\ref{LMCCSMS}).) Clearly $Q(\R^3)=I$, $Q(\emptyset)=0$. Let $X_n\subset \R^3$ be disjoint. Then $(\{0\}\times X_n)^\land$ are spacelike separated and $\big(\cup_n(\{0\}\times X_n)\big)^\land =\lor_n(\{0\}\times X_n)^\land$. Hence $Q(\cup_n X_n)=F\big(\big(\{0\}\times (\cup_nX_n)\big)^\land\big)=F\big(\big(\cup_n(\{0\}\times X_n)\big)^\land\big)=F\big(\lor_n(\{0\}\times X_n)^\land\big)=\sum_nF\big((\{0\}\times X_n)^\land\big)=\sum_nQ(X_n)$. Furthermore, for every translation $b\in\R^3$ one has $W(b)Q(X)W(b)^{-1}=W(b)F\big((\{0\}\times X)^\land\big)W(b)^{-1}=F\big((0,b)+(\{0\}\times X)^\land\big)=F\big((\{0\}\times (b+X))^\land\big)=Q(b+X)$. \\
\hspace*{6mm} 
Hence $Q$ is a translation covariant PM on  the power set of $\R^3$. We are going to argue that such a measure does not exist. Choose any section $K\subset \R^3$ for $\R^3/\mathbb{Q}^3$, i.e., $\{b+K:b\in \mathbb{Q}^3\}$ is a countable disjoint cover of $\R^3$. Therefore $I=Q(\R^3)=\sum_{b\in \mathbb{Q}^3}Q(b+K)$. Since $Q(b+K)=W(b)Q(K)W(b)^{-1}$ it follows $Q(K)\ne 0$. Now for every eigenvector  $\varphi=Q(K)\varphi$ one has $W(b)\varphi=Q(b+K)W(b)\varphi$ and hence $\langle \varphi,W(b)\varphi\rangle=0$ for $b\in\mathbb{Q}^3$ since $Q(K)Q(b+K)=0$. Then by continuity $0=\langle \varphi,W(b)\varphi\rangle\to \norm{\varphi}^2$ for $b\in\mathbb{Q}^3$, $b\to 0$, whence the contradiction $\varphi=0$.\qed

Note that by the proof of (\ref{NRLCCR})  there are even no reps of $\mathcal{M}$ which are merely translation covariant rather than Poincar\'e covariant.\\

\hspace*{6mm}
The proof of (\ref{NRLCCR}) makes clear that the lattice $\mathcal{M}$ is too large for a covariant  rep in that all causally complete sets are measurable. However there is a more relevant obstruction for such a rep which already becomes apparent if we restrict our focus to the special set of diamonds with measurable flat bases.
Consider first  a causal system $(W,E)$. One observes

 \begin{Lem} \label{BOD} Let $\sigma$, $\tau$ be spacelike hyperplanes.
 Let $M=\hat{\Delta}$ be a diamond, where $\Delta$ is a measurable  subset of $\sigma$. Let $\Gamma \subset M\cap \tau$ be a measurable subset of $\tau$. Then $E(\Gamma)\le E(\Delta)$.
 If $\hat{\Gamma}=M$ then $E(\Delta)=E(\Gamma)$.
  \end{Lem}
  
 {\it Proof.} As $\Gamma \subset \hat{\Delta}$, obviously $\Gamma_\sigma \subset \hat{\Delta}_\sigma$. Since the latter equals $\Delta$ by (\ref{ETHSSHP})(vi), one has $\Gamma_\sigma \subset \Delta$. Hence the result  follows by 
 causality  (\ref{VEFCPOL})(a).\qed\\

Hence $E$ assigns to all measurable flat bases of a diamond the same operator. Therefore  $E$ induces  by $$F_\textsc{e}(M):=E(\Delta) \textrm{ for } \Delta\in\mathfrak{S} \textrm{ with } \hat{\Delta}=M$$
  a  map $F_\textsc{e}$ on the set of diamonds with measurable   flat base. Obviously it has the properties assumed for $F$ in (\ref{CNELD}).

\begin{The}\label{CNELD} Let $W$ be a representation of $\tilde{\mathcal{P}}$ and let $F$ be an orthogonal  projection-valued map on the set of  diamonds with measurable flat bases. Let $F$ be   Poincar\'e covariant by $W$ and suppose that $F$ is normalized, local orthogonal,  and $\sigma$-orthoadditive  $\sum_n F(M_n) = F(\lor_nM_n)$ if the diamonds $M_n$ are mutually spacelike separated with bases in a common spacelike hyperplane. Then \emph{(a), (b), (c)} hold.

\begin{itemize}
\item[(a)] $F$ is monotone and  $\sigma$-supraorthoadditive, i.e.,  $\sum_n F(M_n) \le F(M)$ for mutually spacelike separated $M_n\subset M$. 
\item[(b)] $(W,E)$ with  $E(\Delta):=F(\hat{\Delta})$ for $\Delta\in\mathfrak{S}$ is a causal system.
\item[(c)]  $F$ fails to be orthoadditive since there are  spacelike separated $M_1$ and $M_2$ satisfying  $M_1 \lor M_2=\R^4$ and $F(M_1)+F(M_2)<I$.
\end{itemize}
\end{The}

{\it Proof.} (b)  Let $\sigma$ be a spacelike hyperplane. Let $\Delta_n\subset\sigma$, $n\in\N$ be measurable  and  disjoint. Put $M_n:=\hat{\Delta}_n$. Then  $M_n^{\perp}=\Delta_n^{\perp}=(\sigma\setminus\Delta_n)^{\land}\supset \hat{\Delta}_m=M_m$ for $n\ne m$ by (\ref{ETHSSHP})(iii). Hence the diamonds $M_n$ are mutually spacelike separated, whence $\sum_n E(\Delta_n)=\sum_n F(M_n)=F(M)$ for $M:=\lor_nM_n$ by $\sigma$-orthoadditvity of $F$. For $\Delta:=\bigcup_n\Delta_n$ obviously one has $\hat{\Delta}=(\bigcup_n M_n)^{\land}=M$, whence $E(\Delta)=F(\hat{\Delta})=F(M)$. This proves $\sigma$-additivity for $E$. --- Moreover, $E(\sigma)=F(\R^4)=I$ and $E(\emptyset)=F(\emptyset)=0$. ---
Poincar\'e covariance carries over from $F$ to $E$ since $(g\cdot \Delta)^{\land} =g\cdot \hat{\Delta}$. --- Let $\Delta,\Gamma\in\mathfrak{S}$ be spacelike separated. So are  the completions $\hat{\Delta}$ and $\hat{\Gamma}$, whence $E(\Delta)\, E(\Gamma)=F(\hat{\Delta})\,F(\hat{\Gamma})=0$ by local orthogonality of $F$. Hence $E$ is causal by (\ref{VEFCPOL})(f). --- Thus $(W,E)$ is a causal system.\\
\hspace*{6mm}
(a) Let $M_1\subset M_2$. Let $\Delta_i\in\mathfrak{S}$ be a base of $M_i$. Then, by  (\ref{BOD}), $F(M_1)=E(\Delta_1)\le E(\Delta_2)=F(M_2)$. ---
$\sigma$-supraorthoadditivity of $F$  is obvious as $F(M_n)$ are mutually orthogonal projections $\le F(M)$.\\
\hspace*{6mm}
(c) Consider 
 $\Delta:=\{\mathfrak{x}: x_0=0, x_3\le 0\}$ and $\Gamma:=\{\mathfrak{x}: x_0=1, x_3> 1\}$.  By (\ref{EMTVMH}),
  $M_1:=\hat{\Delta}$ and $M_2:=\hat{\Gamma}$ are spacelike separated and satisfy $M_1\lor M_2=\R^4$. For   $\tau:=\{0\}\times\R^3$,  $\sigma:= \{1\}\times\R^3$, and $\sigma^>:=\Gamma$, one has  $F(M_1)+F(M_2)=E(\Delta) +E(\Gamma)=I-\big(E(\tau\setminus \Delta)-E(\Gamma)\big) =I-\big(E(\sigma^>_\tau)-E(\sigma^>)\big)<I$, 
since  $E(\sigma^>_\tau)-E(\sigma^>)>0$ by (\ref{CCIDES}).\qed\\

 $F$ in (\ref{CNELD}) is by no means a restricted lattice homomorphism. For instance, $\{0\}$, $\{2,0,0,0)\}$, and $\{0\} \lor \{2,0,0,0)\}$ are diamonds by (\ref{AOPS}), (\ref{CCTPS}). Clearly $F(\{0\})=F(\{2,0,0,0)\})=0$ and hence $F(\{0\}) \lor F(\{2,0,0,0)\})=0$, but by (\ref{CCTPS}) $F(\{0\} \lor \{2,0,0,0)\})=E(\{\mathfrak{x}:x_0=1,|x|\le 1\})>0$.\\

One feels that one should not content oneself with the no-go-result on  orthoadditivity (\ref{CNELD})(c) for $F_\textsc{e}$. What is really behind it is revealed by a thorough analysis of the eigenstates of the projections $E(\sigma^>_\tau)-E(\sigma^>)$ considered in (\ref{CCIDES}).  In (\ref{XCTLRCL}), (\ref{LBSLSET}) and Eq.\,(\ref{WLBSLSET})  for the Dirac and the Weyl systems an intrinsic  relation between time evolution and boosts is established. This relation in turn is inextricably linked with  the lattice $\mathcal{M}^{'borel}$ generated by the non-timelike relation (in place of the more restrictive spacelike relation) and  possible reps of $\mathcal{M}^{'borel}$. Sec.\,\ref{LGNTLR},\,\ref{SRLBSGNTLR},\,\ref{ECLNTLHP} are concerned with this topic.\\


\section{The Dirac System}\label{DS} 

For every mass $m>0$ the Dirac system is shown to be a causal system as defined in (\ref{CS}). It will turn 
out to be the only irreducible massive causal system with finite spinor dimension.\footnote{ Actually we did not succeed as yet to rule out  certain irreducible massive SCT, see (\ref{DMCS}).}   As being of fundamental importance we are going to study it in detail.\\
 \hspace*{6mm}
In momentum space $L^2(\R^3,\C^4)$, which is the image of position space under Fourier transformation $\mathcal{F}$,  the representation  $W^{\textsc{d}}$   of $\tilde{\mathcal{P}}$ describing the Dirac system with mass $m>0$
is given by
\begin{equation}\label{RDS}
\big(W^{\textsc{d}\,mom}(t,b,A)\varphi\big)(p) =\operatorname{e}^{-\operatorname{i}bp}
\sum_{\eta =+,-} \operatorname{e}^{\operatorname{i}t\eta\epsilon(p)}  \pi^\eta(p)s(A)^{*-1}    \varphi(q^\eta)
\end{equation}\\
(see e.g. \cite[Theorem 5, case $n=1$]{CL15})
with the Dirac Hamiltonian $H$   
\begin{equation}\label{DHMR}
\big(H^{mom}\varphi\big)(p)=  h(p)\varphi(p),\quad \; h(p):=\sum_{k=1}^3\alpha_k p_k + \beta m
\end{equation} 
being  the  matrix multiplication operator by $h$.  For the energy and the shell representation of the Dirac system see  sec. \ref{A:ESRDS}.\\
\hspace*{6mm} 
 As to the notations, $\beta,\alpha_k$ denote the Dirac matrices, which  in the Weyl representation read 
 $$\beta =\left( \begin{array}{cc} 0 & I_2\\ I_2 & 0 \end{array}\right),\quad \alpha_k =\left( \begin{array}{cc}  \sigma_k & 0\\ 0 & -\sigma_k \end{array}\right) $$ 
with $\sigma_k$   the Pauli matrices. So $h(p)^*=h(p)$,  $h(p)^2=\epsilon(p)^2I_4$ 
with $\epsilon(p)=\sqrt{|p|^2+m^2}$. As to $q^\eta$, recall $\mathfrak{p}^{\eta}=(\eta\epsilon(p),p)$ and $\mathfrak{q}^\eta= (q_0^\eta,q^\eta)=A^{-1}\cdot \mathfrak{p}^\eta$ from Eq.\,(\ref{RIMS}).
Furthermore, $\pi^\eta(p):=\frac{1}{2}\big(I_2+\frac{\eta}{\epsilon(p)}h(p)\big)$ for $\eta\in\{+,-\}$. The multiplication operator 
by $\pi^\eta$ is the orthogonal projection onto the subspace of positive energy 
($\eta=+$) and negative energy ($\eta=-$), respectively.
Finally, in the Weyl representation one has $s(A):=\operatorname{diag}(A,A^{* -1})$.\footnote{ Let  $S\in U(4)$ and let  $\text{\small{{\it S}}}$ be the unitary operator on $L^2(\R^3,\C^4)$ with $(\text{\small{{\it S}}}\psi )(x):=S\,\psi(x)$. Then
 $\text{\small{{\it S}}}\;W^{\textsc{d}\,mom}(g)\,\text{\small{{\it S}}}^{-1}$ is given by the right hand side of Eq.\,(\ref{RDS}) with respect to  
 $\beta':= S \,\beta\,S^{-1}$, $\alpha'_k:=S\,\alpha_k\,S^{-1}$, and  $s'(A):=S\,s(A)\,S^{-1}$. --- Moreover, there is a  formula  valid for any permissible choice (cf. \cite[Lemma 2]{CL15}) of the $\beta$, $\alpha$'s, cf. \cite[(1-44)]{SW64} $$s(A)=\big(a_01_4+\sum_{k=1}^3a_k\gamma_5 \alpha_k\big)\frac{1}{2}(1_4+\gamma_5)+\big(\bar{a}_01_4+\sum_{k=1}^3\bar{a}_k\gamma_5 \alpha_k\big)\frac{1}{2}(1_4-\gamma_5)$$ where $\gamma_5:=-i\alpha_1\alpha_2\alpha_3$ and $A=a_01_2+\sum_{k=1}^3a_k\sigma_k$ with $a_0,a_k\in\C$, $a^2_0-a_1^2-a_2^2-a_3^2=1$.}\\
 
 In (\ref{DRPS}) we will need the formula
 \begin{equation}\label{VSDPI}
 \pi^\eta(p) s(A^{* -1})=\big(\epsilon(q^\eta)/\epsilon(p)\big)s(A)\pi^\eta(q^\eta)
\end{equation}
which is mainly a consequence of the fundamental relation $A(\sum_{\mu=0}^3\mathfrak{p}_\mu\sigma_\mu)A^*=\sum_{\mu=0}^3(A\cdot\mathfrak{p})_\mu\sigma_\mu$ with $\sigma_0:=I_2$ for $A\in\C^{2,2}$, $\mathfrak{p}\in\R^4$.

\hspace*{6mm}
The Dirac representation $W^{\textsc{d}}$ is reducible. It is the sum of the  representations of the electron on the subspace of positive energy, and its antisystem, the positron, see appendix \ref{A:ESRDS}. Hence   
$$W^{\textsc{d}}=[m,\textnormal{\tiny{$\frac{1}{2}$}},+]\oplus [m,\textnormal{\tiny{$\frac{1}{2}$}},-]$$ with $m>0$. Any unit vector in the carrier space of $W^{\textsc{d}}$ is called a \textbf{Dirac state}.\\
\hspace*{6mm}
  The Euclidean covariant Dirac position operator $X^{\textsc{d}}$    is related to   $W^{\textsc{d}}$  by 
 \begin{equation} \label{NXH}
 N=\textnormal{\small{$\frac{1}{2}$}}\big(X^{\textsc{d}} H+H   X^{\textsc{d}} \big)
 \end{equation}
 In position space $L^2(\R^3,\C^4)$ the components of  $X^{\textsc{d}}$  
 are  the multiplication operators by the coordinate functions.  So one has a WL,  the \textbf{Dirac localization} $E^{\textsc{d}}$, with the   localization operators  in position space given  by 
 \begin{equation}\label{DWL}
E^{\textsc{d}\,pos}(\Delta)\psi=1_\Delta \psi, \quad \psi\in L^2(\R^3,\C^4)
\end{equation}

As known, the Dirac time evolution is  causal, i.e. 
\begin{equation}\label{CTEDS}
W^{\textsc{d}}(t)E^{\textsc{d}}(\Delta)W^{\textsc{d}}(t)^{-1}\le E^{\textsc{d}}(\Delta_t) 
\end{equation}
for $t\in \R$, $\Delta\subset \R^3$ measurable. See e.g. \cite[Remark on Theorem 1.2]{T92}. Here we infer this by  \cite[Theorem 10 (b)]{CL15} from the fact that the entire matrix-valued function $z\mapsto \operatorname{e}^{\operatorname{i}th(z)}$ on $\C^3$ is exponentially bounded\footnote{ An entire  matrix-valued function $f$ on $\C^d$ is called \textbf{exponentially bounded} or \textbf{of exponential type} with exponent $\delta\ge 0$ if there is a finite constant $C_\delta$ such that $\norm{f(z)}\le C_\delta \operatorname{e}^{\delta|z|}$ with $|z|^2=\sum_{j=1}^d|z_j|^2$ for  $z\in\C^d$. The \textbf{type} $\tau$ of $f$ is the infimum of all its exponents. If, in addition, 
$f|_{\R^3} \in L^2$ and $\tau'>0, \tau'\ge \tau$ then $\norm{f(z)}\le C_{\tau'} \operatorname{e}^{\tau'|\operatorname{Im} z|}$ follows (see \cite[Cor.\,10.10]{L14}).}\label{FNEB} 
 by $C_t\operatorname{e}^{|t|\,|z|}$ $\forall$ $z$ with $|z|^2=|z_1|^2+|z_2|^2+|z_3|^2$ and some finite constant $C_t$. \\
 \hspace*{6mm} 
      $E^{\textsc{d}}$   and $X^{\textsc{d}}$, in contrast to   $E^\textsc{nw}$ and $X^\textsc{nw}$,  do not commute with the sign of energy $\operatorname{sgn}(H)$, since otherwise the trace of 
$E^{\textsc{d}}$  on the positive energy subspace would yield  a WL with causal time evolution for the electron, which does not exist by (\ref{NLS}). Consequently the system $(W^\textsc{d}, E^{\textsc{d}})$ is irreducible. By the same reason,  Eq.\,(\ref{NXH}) cannot be solved for  $X^{\textsc{d}}$, as occurs  in case of the Bakamjian-Thomas-Foldy formula Eq.\,(\ref{BTFX}).
\\
\hspace*{6mm}
One observes that $E^{\textsc{d}}$ admits dilational covariance as $D_\lambda E^{\textsc{d}}(\Delta)D_\lambda^{-1}=E^{\textsc{d}}(\lambda \Delta)$, $\lambda>0$ holds for 
\begin{equation}\label{DDS}
(D_\lambda^{pos} \psi) (x):= \lambda^{-\frac{3}{2}}\psi(\textnormal{\tiny{$\frac{1}{\lambda}$}} x)
\end{equation}
\\
\hspace*{6mm}
Finally, by (\ref{PCPOL}),   we extend $E^{\textsc{d}}$  on $\mathfrak{S}$ by Poincar\'e covariance.\\


 \section{Dirac localization is causal}\label{DLIC}  In position space $L^2(\R^3,\C^4)$ Dirac's Hamiltonian $H$ is represented by   $H^{pos}=\mathcal{F}^{-1}H^{mom}\mathcal{F}=\sum_k\alpha_k\,\frac{1}{\operatorname{i}} \partial_k+ \beta m$.

\begin{Lem}\label{DRPS}
 In position space $L^2(\R^3,\C^4)$ the Dirac representation $W^{\textsc{d}}$   reads
 \begin{equation*}
 \big(W^{\textsc{d}\,pos}(g)\psi\big)(x)= s(A) \big(\operatorname{e}^{-\operatorname{i}y_0 H^{pos}}\psi\big)(y)
\end{equation*}
where $g=(t,b,A)\in\tilde{\mathcal{P}}$, 
and $(y_0,y):=g^{-1}\cdot (0,x)=A^{-1}\cdot (-t,x-b)$. 
\end{Lem}\\
{\it Proof.} By definition  $W^{\textsc{d}\,pos}\equiv \mathcal{F}^{-1}W^{\textsc{d}\,mom}\mathcal{F}$ with $W^{\textsc{d}\,mom}$ from Eq.\,(\ref{RDS}).
Apply first
Eq.\,(\ref{VSDPI}). Then $\big(W^{\textsc{d} \,mom}(g)\varphi\big)(p) =
s(A) \sum_{\eta =\pm 1}  
\operatorname{e}^{\operatorname{i}(t,b)\cdot (\eta \epsilon(p),p)}
 (\epsilon(q^\eta)/\epsilon(p))  \pi^{\eta}(q^\eta) \varphi(q^\eta)$. So  $\big( \mathcal{F}^{-1}W^{\textsc{d}\,mom}(g)\mathcal{F}\psi\big)(x)=s(A) (2\pi)^{-\frac{3}{2}}\sum_\eta\int \operatorname{e}^{\operatorname{i}xp}\operatorname{e}^{\operatorname{i}\mathfrak{a}\cdot \mathfrak{p}^\eta} \pi^{\eta}(q^\eta) \varphi(q^\eta)\, (\epsilon(q^\eta)/\epsilon(p)) \operatorname{d}^3p$  with $\mathfrak{a}=(t,b)$, $\varphi:=\mathcal{F}\psi$. It equals $s(A) (2\pi)^{-\frac{3}{2}}\sum_\eta\int \operatorname{e}^{\operatorname{i}xp} 
 \operatorname{e}^{\operatorname{i}\mathfrak{a}\cdot \mathfrak{p}^\eta}\pi^{\eta}(k) \varphi(k)\,  \operatorname{d}^3k$ by Lorentz-invariant integration and renaming the integration variable $q^\eta$ by $k$. Furthermore, using $\mathfrak{y}=(y_0,y)=g^{-1}\cdot (0,x)$ one has $xp+\mathfrak{a}\cdot \mathfrak{p}^\eta=-((0,x)-\mathfrak{a})\cdot \mathfrak{p}^\eta=-\mathfrak{y}\cdot \mathfrak{q}^\eta=-y_0\eta \epsilon(q^\eta)+yq^\eta=-y_0\eta \epsilon(k)+yk$. Hence $\sum_\eta\ \operatorname{e}^{\operatorname{i}xp} 
 \operatorname{e}^{\operatorname{i}\mathfrak{a}\cdot \mathfrak{p}^\eta} \pi^{\eta}(k)=  
 \operatorname{e}^{\operatorname{i}yk}\sum_\eta  \operatorname{e}^{-\operatorname{i}y_0\eta \epsilon(k)}\pi^{\eta}(k)= \operatorname{e}^{\operatorname{i}yk}\operatorname{e}^{-\operatorname{i}y_0 h(k)}$. Therefore, one concludes $\big( \mathcal{F}^{-1}W^{\textsc{d}\,mom}(g)\mathcal{F}\psi\big)(x)=s(A) (2\pi)^{-\frac{3}{2}}\int  \operatorname{e}^{\operatorname{i}yk}\operatorname{e}^{-\operatorname{i}y_0 h(k)}\varphi(k)\,  \operatorname{d}^3k=s(A) \big(\mathcal{F}^{-1}\operatorname{exp}(-\operatorname{i}y_0H^{mom})\varphi\big)(y)=s(A) \big(\operatorname{exp}(-\operatorname{i}y_0H^{pos})\psi\big)(y)$ as asserted.\qed\\

 More familiar than the position space wave-function $\psi$ in (\ref{DRPS}) is   the time-dependent Dirac wave-function $\Psi$. The latter is the solution of  the  initial-value problem for  the \textbf{Dirac equation} 
\begin{equation} 
i\partial_t\Psi = H^{pos}\Psi, \quad
\Psi(0,\cdot)=\psi \text{ on } \R^3 \text{ with } \psi\in \operatorname{dom}(H^{pos})
\end{equation}
Hence $\psi$ and $\Psi$ are related by $\Psi(t,x)=\big(\operatorname{e}^{-\operatorname{i}tH^{pos}}\psi\big)(x)= \big(W^{\textsc{d}\,pos}(-t)\psi\big)(x)$. From this relation and  (\ref{DRPS}) one  gets 
the transformation law under Poincar\'e transformations  for $\Psi$.  Indeed, let $\Psi_g$ be the wave-function with initial value $W^{\textsc{d}\,pos}(g)\psi$. Then 
\begin{equation}
 \Psi_g(\mathfrak{x})=s(A)\, \Psi(g^{-1}\cdot\mathfrak{x})
\end{equation}
 for $\mathfrak{x}=(t,x)$, since
$W^\textsc{d}(-t)W^\textsc{d}(g)= W^\textsc{d}(g')$ for $g':=(-t,0,I_2)g$ satisfying $g'^{-1}\cdot(0,x)=g^{-1}\cdot(t,x)$.


\begin{The}\label{CSDF} $(W^\textsc{d},E^\textsc{d})$ is a causal system \emph{(\ref{CS})}.
\end{The}

{\it Proof.} We are going to show $W^{\textsc{d}}(h)\,E^\textsc{d}(\Gamma)\,W^{\textsc{d}}(h)^{-1}\,\le \, E^\textsc{d}(\Gamma_h)$ for $h\in \tilde{\mathcal{P}}$ and $\Gamma$ a measurable subset of a spacelike hyperplane $\sigma$ (see (\ref{VEFCPOL})(c)).  \\
\hspace*{6mm} There are a subset $\Delta$ of the hyperplane $\varepsilon =\{0\}\times \R^3\equiv \R^3$ and $k\in  \tilde{\mathcal{P}}$ such that $\sigma=k\cdot \varepsilon$, $\Gamma=k\cdot \Delta$.
By Poincar\'e covariance Eq.\,(\ref{PCWL}), the  left hand side of the inequality reads $W^{\textsc{d}}(hk)\,E^\textsc{d}(\Delta)\,W^{\textsc{d}}(hk)^{-1}$. At the right hand side one has $(k\cdot \Delta)_h=k\cdot \Delta_{k^{-1}hk}$ and hence $E^\textsc{d}(\Gamma_h)=W^{\textsc{d}}(k)E^\textsc{d}(\Delta_{k^{-1}hk}) W^{\textsc{d}}(k)^{-1}$.  Put  $g:=k^{-1}hk$. Thus it suffices to show $W^{\textsc{d}}(g)\,E^\textsc{d}(\Delta)\,W^{\textsc{d}}(g)^{-1}\,\le \, E^\textsc{d}(\Delta_g)$, $g\in \tilde{\mathcal{P}}$. \\
\hspace*{6mm}
For this consider   any state $\psi$   localized in  $\Delta$, i.e. $E^{\textsc{d}\,pos}(\Delta)\psi=\psi$. By the causal time evolution Eq.\,(\ref{CTEDS}), $\operatorname{e}^{-\operatorname{i}tH^{pos}}\psi$ is localized in $\Delta_t$ for any $t\in\R$. Therefore and by (\ref{DWL}), according to (\ref{DRPS}),  $\big(W^{\textsc{d}\,pos}(g)\psi\big)(x)=0$ if $y\not\in \Delta_{y_0}$, where $\mathfrak{y}=g^{-1}\cdot \mathfrak{x}$ and $\mathfrak{x}=(0,x)$. Now, $y\not\in \Delta_{y_0}$  means $|y-a|>|y_0|$ $\forall$ $a\in\Delta$ or equivalently $(\mathfrak{y}-\mathfrak{a})^{\cdot 2}<0$ $\forall$ $\mathfrak{a}:=(0,a)$, $a\in\Delta$. Put $(z_0,z):=g\cdot (0,a)$. Then $(\mathfrak{y}-\mathfrak{a})^{\cdot 2} = (\mathfrak{x}-\mathfrak{z})^{\cdot 2}$. Therefore $W^{\textsc{d}\,pos}(g)\psi$ is localized in  
$\{\mathfrak{x}\in\varepsilon: (\mathfrak{x}-\mathfrak{z})^{\cdot 2}\ge 0 \textrm{ for some } \mathfrak{z}\in g\cdot \Delta\} = \Delta_g$. This implies the assertion.\qed\\

In  establishing  (\ref{CSDF}) we avoided the use of the cumbersome  propagator (see e.g. \cite[(1.87)]{T92}) but applied the position representation (\ref{DRPS}) and exploited causal time evolution Eq.\,(\ref{CTEDS}).\\
\hspace*{6mm}
We like to recall the equivalent descriptions of causality given in (\ref{VEFCPOL}).  In particular, causality is equivalent to local orthogonality,  viz. $E^\textsc{d}(\Gamma) \;E^\textsc{d}(\Gamma')=0$ if $\Gamma$ and $\Gamma'$ are spacelike separated.  Moreover, by  (\ref{FIDCWL}),  Dirac localization is frame-independent.\\


\section{Dirac localization and time evolution}\label{DTINC}  One is interested in  the temporal behavior of the carrier of a wavefunction. The following observation is instructive.
Let $R>0$ and let $\psi$ be a Dirac state localized in the  ball $B_R=\{x\in\R^3:|x|\le R\}$. Let $\psi_t$ denote the  state $W^{\textsc{d}\, pos}(t)\psi$ evolved in time $t\in\R$.
Fix $t_0\ne 0$. Then, due to causality,  $\psi_{t_0}$ is localized in $B_{R+|t_0|}$. This, however,  does not exclude that   $\psi_{t_0}$ actually is localized in a smaller ball $B_r$. Here  every $r>\big||t_0|-R\big|$  occurs.\\
\hspace*{6mm}
 Indeed,  let $0<\rho<R$ and let the state $\chi$ be localized in $B_\rho$.  Let $\varsigma$ denote the sign of $t_0$. Then $\psi:=W^{\textsc{d}\, pos}\big(\varsigma(\rho-R)\big)\chi$ is localized in $B_R$, and $\psi_{t}=W^{\textsc{d}\, pos}\big(t+\varsigma(\rho-R)\big)\chi$ is localized in $B_{r(t)}$ for $r(t):=\rho+|t+\varsigma(\rho-R)|$, whence the assertion for $r=r(t_0)=\rho+\big| |t_0|-R+\rho\big|$.\\
\hspace*{6mm} 
Actually a rather detailed picture of how in general the border of a wavefunction moves is available. It is drawn in the discussion  (\ref{DMBC}), which is based on the results (\ref{GGETIIED}), (\ref{SETIIED})  and the subsequent corollaries and lemma up to (\ref{LTSIE}).

\subsection{Mathematical preliminaries}\label{MPMC}
 The main mathematical tool is a theorem by  P\'olya and Plancherel  (see e.g. \cite{R86},  \cite{R74}), which is reported here in (\ref{TPP}). First recall the P-indicator  (i.e., the P\'olya-Plancherel indicator) $h_f$ of an entire function $f$ on $\C^d$ being 
\begin{equation}
 h_f(\lambda)=\sup\{h_f(\lambda,x):x\in\R^d\},\, \lambda\in\R^d \textrm{\, with\, } h_f(\lambda,x)=\varlimsup_{r\to \infty}\frac{1}{r}\ln |f(x+\operatorname{i}\lambda \,r)| 
 \end{equation}
  and the support function  $H_C$ for a convex set $C\subset \R^d$ being
 \begin{equation}\label{SFCS} 
  H_C(\lambda)=\sup\{x\lambda:x\in C\},\, \lambda\in\R^d
 \end{equation}

\begin{The}\label{TPP}   
A function $f:\C^d\to\C$ is entire and exponentially bounded  (see footnote in \emph{sec.\,\ref{DS}}) with $f|_{\R^d}\in L^2$ if and only if there is $g\in L^2(\R^d)$ vanishing outside a bounded set with
 \begin{equation*}  
 f(z)=(2\pi)^{-d/2}\int_{\R^d}\operatorname{e}^{-iqz}g(q)\operatorname{d}q 
\end{equation*}
 i.e., $f$ is the Fourier-Laplace transform of $g$.\\
\hspace*{6mm} 
  Then  $h_f=H_{C(g)}$, where $C(g)$ is the smallest convex set outside of which $g$ vanishes almost everywhere. \\
 \hspace*{6mm} 
  Moreover,  $f|_{\R^d}$  is bounded  by $(2\pi)^{-d/2}\int_{\R^d}|g(q)|\operatorname{d}q$ and, by the Riemann-Lebesgue lemma, 
  it  vanishes at infinity. Also,
  for each $\lambda$ one has $h_f(\lambda,x)=h_f(\lambda)$ for almost all $x$, and $h_{kf}=h_k+h_f$ holds for  any exponentially bounded entire function $k$. 
\end{The}

The Fourier-Laplace transform  of $\psi_t$ localized in $B_{R+|t|}$ is an entire function $\varphi_t$,  which 
 is exponentially bounded with exponent $R+|t|$, i.e., $|\varphi_t(z)|\le C\operatorname{e}^{(R+|t|)|z|}$, $z\in\C^3$.  Recall (\ref{DHMR}). Due to $h(z)^2=(z^2_1+z^2_2+z^2_3+m^2)I_4$, the time evolution yields 
 \begin{equation}\label{TED}
\varphi_t(z)=\operatorname{e}^{\operatorname{i}th(z)}\varphi(z)=\cos\big(t\epsilon(z)\big)\,\varphi(z)+\operatorname{i}t\operatorname{sinc}\big(t\epsilon(z)\big)\,h(z)\varphi(z)
\end{equation}
for every $z\in\C^3$.  Here $\epsilon$ satisfies $\epsilon(z)^2=z_1^2+z_2^2+z_3^2+m^2$, and 
$\operatorname{sinc}(w)=\sin(w)/w$ for $w\ne 0$, $\operatorname{sinc}(0)=1$.  From Eq.\,(\ref{TED}) one obtains 
\begin{equation}\label{STED}
\varphi_t+\varphi_{-t}=2\cos(t\epsilon)\,\varphi, \quad \varphi_t-\varphi_{-t}=2\operatorname{i}t\operatorname{sinc}(t\epsilon)\,h\varphi
\end{equation}
and
$(\varphi_t)_k\varphi_l=\cos(t\epsilon)\varphi_k\varphi_l+\operatorname{i}t\operatorname{sinc}(t\epsilon)(h\varphi)_k\varphi_l$ and hence
$\phi_{kl}=\operatorname{i}t\operatorname{sinc}(t\epsilon)\chi_{kl}$
  for  $\phi_{kl}:=(\varphi_t)_k\varphi_l-(\varphi_t)_l\varphi_k$ and  $\chi_{kl}:=(h\varphi)_k \varphi_l-(h\varphi)_l \varphi_k$, where $k,l=1,\dots,4$ enumerate the spinor components. \\
 \hspace*{6mm}
 There are  $k,l$  such that $\chi_{kl}\ne 0$. Indeed, assume $\chi_{kl}=0$ for all $k,l$. Then $\varphi_k\, h\varphi=(h\varphi)_k\, \varphi$ and hence $\epsilon^2\,\varphi_k\,\varphi=(h\varphi)_k\, h\varphi$. Then $(h\varphi)_k^2-\epsilon^2\,\varphi_k^2=0$. Fix $k,z_2,z_3$ such that $f(\zeta):=\varphi_k(\zeta,z_2,z_3)$ is not the null function. Set $g(\zeta):=(h\varphi)_k(\zeta,z_2,z_3)$. Choose the square root $\mu(\zeta)$ of $\zeta^2+z_2^2+z_3^2$ such that $g=\mu f$. This, however, is impossible as $g/f$ is meromorphic whereas $\mu$ is not.
Thus $\chi:=\chi_{kl}$, $\phi:=\phi_{kl}$ are non-zero entire exponentially bounded functions with exponents $2R$ and $2R+|t|$, respectively, satisfying
\begin{equation}\label{DRDTE}
\phi=\operatorname{i}t\operatorname{sinc}(t\epsilon)\,\chi
 \end{equation} 
 We are going to exploit the relations Eqs.\,(\ref{STED}) and (\ref{DRDTE}).

\begin{Lem}\label{EFSINC} Let $\mu, t,u,v$ be real, $\mu\ge 0$. Then there are  finite constants $A_t>0,B_t,C_t$ independent of $u,v$ such that 
\begin{itemize}
\item[(a)] $A_t \operatorname{e}^{|t v|}\le\big| \cos\big(t\sqrt{\mu^2+(u+\operatorname{i}v)^2}\,\big)\big|\le B_t\operatorname{e}^{|t v|}$
\item[(b)] $A_t  |u+\operatorname{i}v|^{-1}\operatorname{e}^{|t v|}\le\big| \operatorname{sinc}\big(t\sqrt{\mu^2+(u+\operatorname{i}v)^2}\,\big)\big|\le B_t  |u+\operatorname{i}v|^{-1} \operatorname{e}^{|t v|}$
\end{itemize}
for all $u$ and  $|v|>  C_t$. 
\end{Lem}

{\it Proof.}
 First we show
\begin{equation*} 
z\in \C, |z|\le \frac{1}{2}\; \Rightarrow \;\sqrt{1+z}=1+\zeta \textrm{\; with\; }  |\zeta|\le \frac{3}{4}|z|\tag{$\star$}
\end{equation*}
Indeed, let $f:[0,1]\to \C$, $f(r):=(1+r z)^{1/2}$. As $f'(r)=\frac{1}{2}(1+r z)^{-1/2}z$ one has $f(1)=1+\zeta$
with  $\zeta:=\int_0^1f'(r)\operatorname{d}r$  and $|\zeta|\le 1\cdot \frac{1}{2}(1-\frac{1}{2})^{-1/2}|z|$, whence ($\star$).\\

\hspace*{6mm}
Now  assume at once $t\ne 0$ and let in the following $|v|> \sqrt{2}\mu$. Put $s(u,v):=\sqrt{\mu^2+(u+\operatorname{i}v)^2}$. More precisely, 
$s(u,v):= (u+\operatorname{i}v)\sqrt{1+z}$ for $z:=\mu^2(u+\operatorname{i}v)^{-2}$ with $|z|=\mu^2(u^2+v^2)^{-1}\le\mu^2v^{-2}\le\frac{1}{2}$. \\

\hspace*{6mm}
By ($\star$), 
$s(u,v)=
(u+\operatorname{i}v)(1+\zeta)$ with $|u+\operatorname{i}v|\,|\zeta|\le \frac{3}{4}|u+\operatorname{i}v|\,|z|\le
\mu^2|v|^{-1}\le\mu$. This implies $|\operatorname{e}^{\pm\operatorname{i}ts(u,v)}|=
|\operatorname{e}^{\pm\operatorname{i}t(u+\operatorname{i}v)(1+\zeta)}|=\operatorname{e}^{\mp tv}|\operatorname{e}^{\pm\operatorname{i}t(u+\operatorname{i}v)\zeta}|$. One concludes
\begin{equation*}\operatorname{e}^{-|t|\mu}\operatorname{e}^{-tv}\le |\operatorname{e}^{\operatorname{i}ts(u,v)}|\le \operatorname{e}^{|t|\mu}\operatorname{e}^{-tv}, \quad \operatorname{e}^{-|t|\mu}\operatorname{e}^{tv}\le |\operatorname{e}^{-\operatorname{i}ts(u,v)}|\le \operatorname{e}^{|t|\mu}\operatorname{e}^{tv}\tag{$\star\star$}
\end{equation*}
for  all $v\in\R$ with $|v|> \sqrt{2}\mu$.  

\hspace*{6mm}
(a) $|\cos(w)|=\frac{1}{2}|\operatorname{e}^{\operatorname{i}w}+\operatorname{e}^{-\operatorname{i}w}|\le \frac{1}{2}(|\operatorname{e}^{\operatorname{i}w}|+|\operatorname{e}^{-\operatorname{i}w}|)$. For $w=ts(u,v)$ this yields by ($\star\star$) $|\cos(ts(u,v))|\le\frac{1}{2}( \operatorname{e}^{|t|\mu}\operatorname{e}^{-tv}+ \operatorname{e}^{|t|\mu}\operatorname{e}^{tv})\le \operatorname{e}^{|t|\mu}\operatorname{e}^{|tv|}$. Hence the right part of the  inequality of (a) holds for $B_t:=\operatorname{e}^{|t|\mu}$ and $C'_t=\sqrt{2}\mu$.\\
\hspace*{6mm}
For the left part of the inequality use $|\cos(w)|=\frac{1}{2}|\operatorname{e}^{\operatorname{i}w}+\operatorname{e}^{-\operatorname{i}w}|\ge \frac{1}{2}\big||\operatorname{e}^{\operatorname{i}w}|-|\operatorname{e}^{-\operatorname{i}w}|\big|$. Then for $w=ts(u,v)$ one gets by ($\star\star$) $|\cos(ts(u,v))|\ge\frac{1}{2}( \operatorname{e}^{-|t|\mu}\operatorname{e}^{|tv|}- \operatorname{e}^{|t|\mu}\operatorname{e}^{-|tv|})=\sinh\big(|t|(|v|-\mu)\big)$. Check $\sinh(x)\ge\frac{1}{4}\operatorname{e}^x$ for $x\ge \frac{\ln(2)}{2}$. Thus we conclude that the left part of the inequality holds for $A_t:=\frac{1}{4}\operatorname{e}^{-|t|\mu}$ and $C''_t:=\sqrt{2}\mu+\frac{\ln(2)}{2}\frac{1}{|t|}$.\\
\hspace*{6mm}
(b) Check first $|t\,s(u,v)|\ge \frac{|t|}{\sqrt{2}}|u+\operatorname{i}v|$, using $|\sqrt{1+z}|\ge \frac{1}{\sqrt{2}}$ for $|z|\le\frac{1}{2}$. Furthermore, $ |\sin(w)|=\frac{1}{2}|\operatorname{e}^{\operatorname{i}w}-\operatorname{e}^{-\operatorname{i}w}|\le \frac{1}{2}(|\operatorname{e}^{\operatorname{i}w}|+|\operatorname{e}^{-\operatorname{i}w}|)$. Hence,  as for (a),  the right part of the inequality holds for $B_t:=\frac{\sqrt{2}}{|t|}\operatorname{e}^{|t|\mu}$ and $C'_t=\sqrt{2}\mu$.\\
\hspace*{6mm}
Regarding the left part of the inequality of (b), we estimate $|ts(u,v)|^{-1}\ge (\frac{2}{3})^{1/2} |t(u+\operatorname{i}v)|^{-1}$, as $|s(u,v)|=|u+\operatorname{i}v|\,|\sqrt{1+z}\,|\le(\frac{3}{2})^{1/2} |u+\operatorname{i}v|$.  Furthermore, one has $|\sin(w)|=\frac{1}{2}|\operatorname{e}^{\operatorname{i}w}-\operatorname{e}^{-\operatorname{i}w}|\ge \frac{1}{2}\big||\operatorname{e}^{\operatorname{i}w}|-|\operatorname{e}^{-\operatorname{i}w}|\big|$.  Hence, proceeding as in (a), it follows that the left part of the inequality holds for $A_t:=(\frac{1}{24})^{1/2}\frac{1}{|t|}\operatorname{e}^{-|t|\mu}$ and $C''_t:=\sqrt{2}\mu+\frac{\ln(2)}{2}\frac{1}{|t|}$.
\qed\\

We use now (\ref{EFSINC}) to compute the P-indicator for $\cos(t\epsilon)$ and $\operatorname{sinc}(t\epsilon)$.

\begin{Lem}\label{PICESCE} For $t\in\R$ the functions $z\mapsto \cos\big(t\epsilon(z)\big)$ and $z\mapsto \operatorname{sinc}\big(t\epsilon(z)\big)$ are bounded on $\R^3$ and entire on $\C^3$ with exponent $|t|$, which is smallest. Moreover, $h_{\cos(t\epsilon)}(\lambda)=h_{\operatorname{sinc}(t\epsilon)}(\lambda)=|t|\,|\lambda|$ holds for $\lambda\in\R^3$. More precisely, $|t|\,|\lambda|=\lim_{r\to\infty}\frac{1}{r}\ln|f(p+\operatorname{i}\lambda r)|$, $p\in\R^3$ for  $f\in\{\cos(t\epsilon),\operatorname{sinc}(t\epsilon)\}$. These statements are equally valid if $\epsilon(z)^2$ is replaced with $z^2$, i.e., if the mass $m=0$.
 \end{Lem}

{\it Proof.} 
In the following the case $\epsilon(z)^2=z^2$, i.e. $m=0$, is included. 
We show the assertion for  $\operatorname{sinc}(t\epsilon)$. Regarding $\cos(t\epsilon)$ the proof is analogous. Assume at once $t\ne 0$.\\
\hspace*{6mm}
Obviously, $\operatorname{sinc}(t\epsilon)$ is bounded on $\R^3$ and entire on $\C^3$. Also, there is an entire function $s$ satisfying $s(z^2)=\operatorname{sinc}\big(t\epsilon(z)\big)$ with $z^2=z_1^2+z_2^2+z_3^2$ for all $z\in\C^3$. ---
Now $|\epsilon(z)|^2=|z^2+m^2|\le |z^2|+m^2 = |z|^2+m^2\le(|z|+m)^2$, whence $|\epsilon(z)|\le |z|+m$. Therefore, $|\sin\big(t\epsilon(z)\big)|=
\frac{1}{2}|\operatorname{e}^{\operatorname{i}t\epsilon(z)}-\operatorname{e}^{-\operatorname{i}t\epsilon(z)}|\le \operatorname{e}^{|t|\,|\epsilon(z)|}\le \operatorname{e}^{|t|m} \operatorname{e}^{|t|\,|z|}$ for all $z$.\\
\hspace*{6mm}
If $|z^2|\le 2m^2+1$ then  $|\operatorname{sinc}\big(t\epsilon(z)\big)|=|s(z^2)|\le C$ for some finite constant $C$. For $|z^2|> 2m^2+1$ one has $|\epsilon(z)|^2=|z^2+m^2|\ge |z^2|-m^2>m^2+1$, whence $|\epsilon(z)|^{-1}<1$.
Hence $|\operatorname{sinc}\big(t\epsilon(z)\big)|\le C'\operatorname{e}^{|t|\,|z|}$ for all $z$, where $C':=C+\frac{\operatorname{e}^{|t|m}}{|t|}$. So $|t|$ is an exponent for $\operatorname{sinc}(t\epsilon)$.\\ 
\hspace*{6mm}
In order to show that  $|t|$ is smallest assume that $0\le\delta <|t| $ is an exponent for  
$\operatorname{sinc}(t\epsilon)$. Let $\delta<\delta'<|t|$. Then obviously $|\sin\big(t\epsilon(z)\big)|\le C\operatorname{e}^{\delta'|z|}$, $z\in\C^3$ for some finite constant $C$. Let $w\in\C$. Choose $\zeta\in\C $ with $\zeta^2=w^2-m^2$. Then $w\in\{\pm\epsilon(0,0,\zeta)\}$ and $|\zeta|\le |w|+m$. Hence $|\sin(tw)|\le C\operatorname{e}^{\delta'|\zeta|}\le C'\operatorname{e}^{\delta'|w|}$ with $C':=C\operatorname{e}^{\delta'm}$.
Therefore also $|\cos(tw)|=|\sin(tw+\frac{\pi}{2})|\le C'\operatorname{e}^{\frac{\pi\delta'}{2|t|}}\operatorname{e}^{\delta'|w|}$, whence finally
 $|\operatorname{e}^{tw}|\le C''\operatorname{e}^{\delta'|w|}$, $w\in\C$ for some finite constant $C''$. This implies  the contradiction $\operatorname{e}^{(|t|-\delta')r}\le C''$ for all $r>0$.\\
\hspace*{6mm} 
 We turn to the P-indicator of $\operatorname{sinc}(t\epsilon)$. Assume at once $\lambda\ne 0$. Then $\epsilon(p+\operatorname{i}\lambda r)=\big(\mu^2+(\frac{p\lambda}{|\lambda|}+\operatorname{i}|\lambda|r)^2\big)^{1/2}$ with $\mu^2:=m^2+p^2-\big(\frac{p\lambda}{|\lambda |}\big)^2\ge 0$ independent of $r$. Hence by 
(\ref{EFSINC}) there are finite constants $A_t>0$, $B_t$ independent of $r$ that such $A_t\big|\frac{p\lambda}{|\lambda|}+\operatorname{i} |\lambda| \,r\big|^{-1}\operatorname{e}^{|t||\lambda|r}\le \operatorname{sinc}\big(t\epsilon(p+\operatorname{i}\lambda r)\big)\le B_t\big|\frac{p\lambda}{|\lambda|}+\operatorname{i} |\lambda| \,r\big|^{-1}\operatorname{e}^{|t||\lambda|r}
$, whence the assertion.\qed\\

\subsection{Motion of the border of a bounded localized wave function}\label{PBWF} The location of the carrier of a wavefunction is described by the positions of the supporting planes in every direction of the space. We are going to study the movements of these supporting planes.

\begin{Def} The unit vector $e\in\R^3$   and $\alpha\in[-\infty,\infty]$ determine the half-space 
$\{x\in\R^3: x\,e\le\alpha\}$ (which equals  $\emptyset$ or $\R^3$ if $\alpha\in\{-\infty,\infty\}$). For every $\psi\in L^2(\R^3,\C^4)\setminus\{0\}$ let $e(\psi)\in[-\infty,\infty]$ denote the largest $\alpha$ satisfying $E^{\textsc{d}}(\{x\in\R^3: x\,e\le\alpha\})\psi=0$. Similarly  for $\eta\in L^2(\R^3,\C)$
let $e(\eta)$ be the largest $\alpha$ with $1_{\{xe\le \alpha\}}\eta=0$ up to a null set. 
\end{Def}

 The meaning of $e(\psi)$ is best elucidated by 
the following result. Recall $E^{\textsc{d}}(\{x\in\R^3: x\operatorname{e}=\alpha\})=0$.  Let $$\overline{e}:=-e$$

\begin{Lem}\label{EEP}   $\psi$   is localized in
$\{x\in\R^3: e(\psi)\le xe \le -\overline{e}(\psi)\}$
with $]e(\psi), -\overline{e}(\psi)[$  the smallest interval with this property.
\end{Lem}

{\it Proof.} By definition $E^{\textsc{d}}(\{x\in\R^3: x\,e \le\alpha\})\psi=0$, $E^{\textsc{d}}(\{x\in\R^3: x\,\overline{e}\le \beta\})\psi=0$ exactly for all $\alpha\le e(\psi)$ and $\beta \le  \overline{e}(\psi)$. From this it follows $E^{\textsc{d}}\big(
\{x\in\R^3: e(\psi)\le xe \le -\overline{e}(\psi)\}\big)\psi=\psi$, whence the assertion.\qed\\

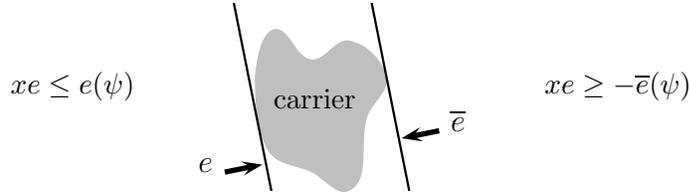
\begin{center}
\begin{pspicture}(3.5,2.5)
\pscurve*[linecolor=lightgray](0.25,1.5)(0.5,2.15)(1,1.75)(1.5,2)(2,1.5)(1.75,1)(1.5,0)(1,0.25)(0.45,0.5)(0.25,1.5)

\put(0.5,1.1){\small{carrier}}
\put(-3,1.3){\small{$xe\le e(\psi)$}}
\put(4.1,1.3){\small{$xe\ge -\overline{e}(\psi)$}}
\psline[linewidth=2pt]{->}(-0.15,0.25)(0.35,0.35)
\put(-0.5,0.25){$e$}

\psline[linewidth=2pt]{<-}(2.2,0.7)(2.7,0.8)
\put(2.85,0.8){$\overline{e}$}

\linethickness{0.3mm}
\put(0.47,0){\line(-1,5){0.5}}
\put(2.3,0){\line(-1,5){0.5}}
\end{pspicture}
\end{center}

Clearly,  $e(\psi)=\min_le(\psi_l)$. --- Let $C(\psi)$ denote the smallest convex set outside of which $\psi$ vanishes almost everywhere. Clearly, $\{x\in\R^3: x\,e\le e(\psi)\}\cap C(\psi)=\emptyset$ and $\{x\in\R^3: x\,e\le \alpha\}\cap C(\psi)$ is not a null set if $\alpha>e(\psi)$. Hence $e(\psi)=\inf\{xe:x\in C(\psi)\}$. These considerations are applicable as well  to every component $\psi_l$ of $\psi$. Therefore (see Eq.\,(\ref{SFCS}))
\begin{equation}\label{REH}
e(\psi)=-H_{C(\psi)}(-e), \quad e(\psi_l)=-H_{C(\psi_l)}(-e)
\end{equation}

\begin{Lem}\label{CDFE} Let $\psi$ be a Dirac state.  
Then $\R\to\R$, $t\mapsto e(\psi_t)$ is continuous.
\end{Lem}

{\it Proof.} Let $t,t_0\in\R$. By causality $e(\psi_t)\ge e(\psi_{t_0})-|t-t_0|$. This implies $\varliminf_{t\to t_0} e(\psi_t)\ge e(\psi_{t_0})$. Furthermore, for $P_t:= E^{\textsc{d}}(\{x:x\,e\le e(\psi_t)\})$ one has $0=P_t\psi_t=P_t\psi_{t_0} +P_t(\psi_t-\psi_{t_0})$, whence $\lim_{t\to t_0}P_t\psi_{t_0}=0$ as $\psi_t\to \psi_{t_0}$. This implies $\varlimsup_{t\to t_0}e(\psi_t)\le e(\psi_{t_0})$. Thus continuity of $t\to e(\psi_t)$  at $t_0$ holds.\qed\\

\begin{Lem}\label{SMF} Let $\psi$ be a Dirac state localized in a bounded region. Then 
\begin{equation*}
 \min\{e(\psi_t),e(\psi_{-t})\}= e(\psi) -|t| 
\end{equation*}
holds for every direction $e$  and  all times $t\in\R$.
\end{Lem}\\
 {\it Proof.} 
 By causality $e(\psi_t)\ge e(\psi) -|t|$ for all $t$, whence $\min\{e(\psi_t),e(\psi_{-t})\} \ge e(\psi)-|t|$. \\
\hspace*{6mm} 
 We prove now  the reverse inequality. Recall   $\phi= 2\cos(t\epsilon)\,\varphi$ for $\phi:=\varphi_t+\varphi_{-t}$ from Eq.\,(\ref{STED}).  Let $\theta:=\mathcal{F}^{-1}\phi|_{\R^3}$. Theorem (\ref{TPP}) applies to the components of $\varphi$ and, due to (\ref{PICESCE}), also to those of $\cos(t\epsilon)\,\varphi$.
Hence, using Eq.\,(\ref{REH}) and by (\ref{PICESCE}), $e(\theta_l)=-H_{C(\theta_l)}(-e)=-h_{\cos(t\epsilon)\,\varphi_l}(-e)=-h_{\cos(t\epsilon)}(-e)-h_{\varphi_l}(-e)=-|t|-H_{C(\psi_l)}(-e)=-|t|+e(\psi_l)$. Therefore $e(\theta)=\min_le(\theta_l)=-|t|+\min_le(\psi_l)=-|t|+e(\psi)$.\\
\hspace*{6mm}
It remains to show $\alpha:=\min\{e(\psi_t),e(\psi_{-t})\}\le e(\psi_t+\psi_{-t})$. Put $E_\alpha:= E^{\textsc{d}}(\{x\in\R^3: x\operatorname{e}\le\alpha\})$. Then $E_\alpha\psi_t=0$ and $E_\alpha\psi_{-t}=0$. Hence  $E_\alpha(\psi_t+\psi_{-t})=0$, whence the claim. \qed\\

\begin{The}\label{GGETIIED} Let $\psi$ be a Dirac state localized in a bounded region. Then 
\begin{equation*} e(\psi_t)\le -2\,\overline{e}(\psi)-e(\psi) -|t| 
\end{equation*}
holds for all directions $e$ and all times $t$. Recall \emph{(\ref{DHMR})}. If  $\psi\in \operatorname{dom}(H^{pos})$ or if more generally
$h\,\mathcal{F}\psi$ is bounded
on $\R^3$ then the inequality holds even with $<$  in place of $\le$.
\end{The}

{\it Proof.} We start from Eq.\,(\ref{DRDTE}) $\phi=\operatorname{i}t\operatorname{sinc}(t\epsilon)\,\chi$. Put here $\phi_{kl}:=(\varphi_t)_k\varphi_l$, $\chi_{kl}:=(h\varphi)_k\varphi_l$, whence $\phi=\phi_{kl}-\phi_{lk}$ and $\chi=\chi_{kl}-\chi_{lk}$.\\
\hspace*{6mm}
As $\varphi_l|_{\R^3} \in L^2$ and $(\varphi_t)_k|_{\R^3}$ is bounded,  $\phi_{kl}|_{\R^3}\in L^2$ so that 
(\ref{TPP}) applies to  $\phi_{kl}$. Let $\theta:=\mathcal{F}^{-1}\phi|_{\R^3}$, $\theta_{kl}:=\mathcal{F}^{-1}\phi_{kl}|_{\R^3}$. Obviously, $e(\theta)\ge \min\{e(\theta_{kl}),e(\theta_{lk})\}$.
Using  Eq.\,(\ref{REH}) one gets $e(\theta_{kl})=-H_{C(\theta_{kl})}(-e)=-h_{\phi_{kl}}(-e)=-h_{(\varphi_t)_k}(-e)-h_{\varphi_l}(-e)=-H_{C((\psi_t)_k)}(-e)-H_{C(\psi_l)}(-e)=e((\psi_t)_k)+e(\psi_l)\ge e(\psi_t)+e(\psi)$. It follows $e(\theta)\ge e(\psi_t)+e(\psi)$.\\
\hspace*{6mm}
We turn to the right hand side\, $\operatorname{i}t\operatorname{sinc}(t\epsilon)\,\chi$ of Eq.\,(\ref{DRDTE}).  Recall  $\operatorname{i}t\operatorname{sinc}(t\epsilon)\,\chi_{kl}=\phi_{kl}-\cos(t\epsilon)\varphi_k\varphi_l$ by Eq.\,(\ref{TED}). Note that $\cos(t\epsilon)\varphi_k|_{\R^3}$ is bounded. Hence $\operatorname{sinc}(t\epsilon)\,\chi_{kl}|_{\R^3}\in L^2$. However, $\chi_{kl}|_{\R^3}$ need not be square-integrable. Therefore  we consider instead $\chi'_{kl}:=s_\delta \chi_{kl}$ with $s_\delta:=\operatorname{sinc}(\delta\epsilon)$  for $\delta>0$. Then $\phi'=\operatorname{i}t\operatorname{sinc}(t\epsilon)\,\chi'$ for $\phi':=s_\delta\phi$ holds. As $h_{s_\delta}(e)=\delta$  by (\ref{PICESCE}), the analogous computation for 
$\theta':=\mathcal{F}^{-1}\phi'|_{\R^3}$ in place of $\theta$ yields $e(\theta')\ge -\delta+e(\psi_t)+e(\psi)$.
Moreover, (\ref{TPP}) applies to $\chi'$. Let $\xi':=\mathcal{F}^{-1}\chi'|_{\R^3}$. Then again, in the same way $e(\theta')=-|t|+e(\xi')$ follows.\\
\hspace*{6mm}
Next we examine  $-\overline{e}(\xi')$. Obviously $-\overline{e}(\xi')\le\max\{-\overline{e}(\xi'_{kl}),-\overline{e}(\xi'_{lk})\}$. By Eq.\,(\ref{REH}) and (\ref{TPP}) one has $-\overline{e}(\xi'_{kl})=H_{C(\xi'_{kl})}(e)=h_{\chi'_{kl}}(e)=h_{s_\delta(h\varphi)_k}(e)+h_{\varphi_l}(e)$ as $s_\delta(h\varphi)_k$ is exponentially bounded. Note  $|(h\varphi)_k(z)|\le q(z)\max_m|\varphi_m(z)|$ with
$q(z)^2:=4\sum_{m=1}^4|h(z)_{km}|^2$, where $h(z)_{km}$ is linear. Therefore $h_{s_\delta(h\varphi)_k}(e,x)=\varlimsup_{r\to\infty}\frac{1}{r}\big\{\ln|s_\delta(x+\operatorname{i}e\,r))|+ \ln|(h\varphi)_k(x+\operatorname{i}e\,r)|\big\}=\delta+\varlimsup_{r\to\infty}\frac{1}{r} \ln|(h\varphi)_k(x+\operatorname{i}e\,r)|\}$ (by (\ref{PICESCE})) $\le \delta+\varlimsup_{r\to\infty}\frac{1}{r}\big\{\ln|q(x+\operatorname{i}e\,r)|+\ln(\max_m|\varphi_m(x+\operatorname{i}e\,r)|)\big\}=\delta+0+\max_m\varlimsup_{r\to\infty}\frac{1}{r}\ln|\varphi_m(x+\operatorname{i}e\,r)|=\delta+\max_m h_{\varphi_m}(e,x)$. Furthermore, $\max_m h_{\varphi_m}(e)=\max_mH_{C(\psi_m)}(e)=\max_m\{-\overline{e}(\psi_m)\}=-\overline{e}(\psi)$. Also $h_{\varphi_l}(e)\le -\overline{e}(\psi)$. It follows  $-\overline{e}(\xi')\le \delta-2\,\overline{e}(\psi)$.\\
\hspace*{6mm}  
 Now, using $e(\xi')<- \overline{e}(\xi')$, one has the chain of inequalities $ -\delta+e(\psi_t)+e(\psi)\le e(\theta')=-|t|+e(\xi')<-|t| -\overline{e}(\xi')\le -|t|+ \delta-2\,\overline{e}(\psi)$ for $\delta>0$. The limit $\delta\to 0$ yields the final result $ e(\psi_t)+e(\psi)\le -|t|+-2\,\overline{e}(\psi)$. It remains to note that if $h\varphi$ is bounded on $\R^3$ one has $\chi_{kl}|_{\R^3}\in L^2$ so that $e(\xi)<- \overline{e}(\xi)$, and the chain holds even for $\delta=0$.
\qed\\

In other words,  suppose that, after or before time $|t|$, $\psi_t$ is localized in the half-space $\{x\in\R^3: x\operatorname{e}\ge\alpha\}$. Then, by (\ref{GGETIIED}), $|t|\le -\alpha-2\,\overline{e}(\psi)-e(\psi)$.  --- If $\psi$ is localized in $B_R$, then obviously  $e(\psi)\ge -R$ for every direction $e$. So,  if $\psi_t$ is localized in $B_r$, then by (\ref{GGETIIED}), $|t|\le r +3R$. \\

\begin{The}\label{SETIIED} Let $\psi$ be a Dirac state localized in a bounded region. Then there exists a unique time $t_e=t_e(\psi)\in\R$ such that 
\begin{equation*}  e(\psi_t)= e(\psi)+|t_e|-|t-t_e|
\end{equation*}
for all times $t\in\R$ and directions $e$. 
\end{The}\\
{\it Proof.}  
Since $t\to e(\psi_t)$ is  continuous by  (\ref{CDFE}) and bounded  above by (\ref{GGETIIED}) there is $t_e\in\R$ with $e(\psi_{t_e})=\sup_{t\in\R}e(\psi_t)$. Fix $t>0$.\\
\hspace*{6mm}
Now we apply (\ref{SMF})
to $\psi':=\psi_{t_e-t/2}$. Then $\min\{e(\psi'_{t'}),e(\psi'_{-t'})\} =e(\psi') -|t'|$ for all $t'\in\R$. As $e(\psi'_{t/2})=e(\psi_{t_e})\ge e(\psi_{t_e-t})=e(\psi'_{-t/2})$ it follows $e(\psi_{t_e-t})=e(\psi_{t_e-t/2})-t/2$. For $t/2$ in place of $t$ this reads  $e(\psi_{t_e-t/2})=e(\psi_{t_e-t/4})-t/4$. Hence $e(\psi_{t_e-t})=e(\psi_{t_e-t/4})-t/2-t/4$. From this one obtains in the same way $e(\psi_{t_e-t})=e(\psi_{t_e-t/8})-t/2-t/4-t/8$ and  finally $e(\psi_{t_e-t})=e(\psi_{t_e-t/2^n})-\sum_{k=1}^n t/2^k$ after  $n$ steps. Then by continuity (\ref{CDFE}) the limit $n\to \infty$ yields  $e(\psi_{t_e-t})=e(\psi_{t_e})-t$. --- 
Analogously, applying  (\ref{SMF}) to $\psi':=\psi_{t_e+t/2}$, one obtains  $e(\psi_{t_e+t})=e(\psi_{t_e})-t$.\\
\hspace*{6mm}
Thus  $e(\psi_t)= e(\psi_{t_e})-|t-t_e|$ holds for all $t\in\R$. In particular $e(\psi)= e(\psi_{t_e})-|t_e|$, whence the formula.
 Uniqueness of $t_e$ is obvious as $t\to e(\psi_t)$ has just one maximum at $t=t_e$.\qed

\begin{Cor}\label{MWTI} Let $\psi$ be a bounded localized Dirac state. Then for every direction $e$
\begin{itemize}
\item[(a)] $t\to e(\psi_t)$ assumes the  maximum, which is the only extremum,  at $t=t_e$
\item[(b)] the width $-\overline{e}(\psi_t)-e(\psi_t)$ of the carrier   in direction $e$  assumes its minimum $-\overline{e}(\psi_{t_{\overline{e}}})-e(\psi_{t_e}) + |t_{\overline{e}}-t_e|$ at the  times $t$ between $t_e$ and $t_{\overline{e}}$
\item[(c)]   $t_e(\psi_t)=t_e(\psi)-t$ and $e\big((\psi_t)_{t_e(\psi_t)}\big)=e(\psi_{t_e(\psi)})$        for $t\in\R$
\item[(d)] $e(\psi_{t_e})\le -\overline{e}(\psi_{t_{\overline{e}}})$ 
\end{itemize}
\end{Cor}
{\it Proof.} (a) and (b) follow easily from  (\ref{SETIIED}). \\
\hspace*{6mm} 
(c) Let $\tau\in\R$, $\psi':=\psi_\tau$, and $t'_e:=t_e(\psi')$. Apply (\ref{SETIIED}). Then $e(\psi'_t)= e(\psi')+|t'_e|-|t-t'_e|$ and $e(\psi')=e(\psi)+|t_e|-|\tau-t_e|$. As $\psi'_t=\psi_{t+\tau}$ also $e(\psi'_t)=e(\psi)+|t_e|-|t+\tau -t_e|$ holds. Therefore $|t+\tau-t_e|-|t-t'_e|=|\tau-t_e|-|t'_e|$ for all $t$, whence $t'_e=t_e-\tau$. Using this result, $e(\psi'_t)= e(\psi')+|t'_e|-|t-t'_e|=e(\psi)+|t_e|-|t-t'_e|=e(\psi_{t_e})-|t-t'_e|$, whence the claim putting $t=t'_e$.\\
\hspace*{6mm} 
(d) Apply (\ref{GGETIIED}) to $\eta:=\psi_{t_{\overline{e}}}$. For $t=t_e(\eta)$  one obtains $e(\eta_{t_e(\eta)})\le -2\overline{e}(\eta)-e(\eta)-|t_{e(\eta)}|$. 
Since 
$e(\eta_{t_e(\eta)})=e(\eta)+|t_{e(\eta)}|$ by (\ref{SETIIED}), one has $e(\eta_{t_e(\eta)})\le -\overline{e}(\eta)$. By  
(c), $e(\eta_{t_e(\eta)})=e(\psi_{t_e})$. The result follows.\qed

The result (\ref{MWTI})(d)  will be improved  by (\ref{ILCS})(b).\\

\begin{Cor}\label{CSETIIED} Let $\psi$ be a bounded localized Dirac state. Then for every direction $e$

\begin{itemize}
\item[(a)] $e(\psi_t)\le -\overline{e}(\psi)-|t-t_e|$
\item[(b)] $|t_e|+|t_{\overline{e}}|\le -\overline{e}(\psi)-e(\psi)$ 
\item[(c)] $2|t_e|<-\overline{e}(\psi)-e(\psi)$ if $t_e=t_{\overline{e}}$
\item[(d)]  If $h\,\mathcal{F}\psi$ is bounded
on $\R^3$ for $\psi$ in position representation, then the inequalities in \emph{(a), (b)} hold even with $<$  in place of $\le$.
\end{itemize}
\end{Cor}

{\it Proof.} (a) By (\ref{SETIIED})  and  (\ref{GGETIIED}) one has $e(\psi_t)=e(\psi) +|t_e|-|t-t_e|\le -2\overline{e}(\psi)-e(\psi)-|t|$. For $t=t_e$ this yields $|t_e|\le -\overline{e}(\psi)-e(\psi)$ and consequently $e(\psi_t)\le -\overline{e}(\psi)-|t-t_e|$. 
\\
\hspace*{6mm}
(b) Let $s,t\in\R$ and consider $e(\psi_{t+s})$. One the one hand, by (\ref{SETIIED}), $e(\psi_{t+s})=e(\psi)+|t_e|-|t+s-t_e|$. On the other hand, first using (\ref{GGETIIED}) and then applying (\ref{SETIIED}), one has $e(\psi_{t+s})\le -2\overline{e}(\psi_t)-e(\psi_t)-|s|= 
-2\big(\overline{e}(\psi)+|t_{\overline{e}}|-|t-t_{\overline{e}}|\big)-e(\psi)-|t_e|+|t-t_e|-|s|$. Hence 
$-2\big(\overline{e}(\psi)+e(\psi)\big)\ge 2|t_e|+2|t_{\overline{e}}|-2|t-t_{\overline{e}}|-|t-t_e| +|s|-|t+s-t_e|$. For $s=t_e-t$ this yields $-\overline{e}(\psi)-e(\psi)\ge |t_e|+|t_{\overline{e}}|-|t-t_{\overline{e}}|$. Then  $|t_e|+|t_{\overline{e}}|\le -\overline{e}(\psi)-e(\psi)$ follows for $t=t_e$. \\
\hspace*{6mm}
(c) By (\ref{EEP}), $0<-\overline{e}(\psi_t)-e(\psi_t)$ for all $t$.   Hence (\ref{SETIIED}) yields $|t_{\overline{e}}|+|t_e|-|t-t_{\overline{e}}|-|t-t_e|<-\overline{e}(\psi)-e(\psi)$.
This implies (c).\\
\hspace*{6mm}
(d) follows from the last part of  (\ref{GGETIIED}).
 \qed
 
The result (\ref{CSETIIED})(b)  will be improved  by (\ref{ILCS})(a).\\

Rather obvious are the transformations of $e(\psi)$ and $t_e(\psi)$ when $\psi$ is time-shifted (\ref{MWTI})(c) ore space-shifted (\ref{TCTTE}). They are less obvious if $\psi$ is boosted. Recall $A_{\rho e}$ in (\ref{ERIUB})(c) representing the boost in direction $e$ with rapidity $\rho\in\R$. 
 
\begin{Lem}\label{BCF} Let $\rho\in\R$ and $e$ be any direction. Put $\psi':=W^{\textsc{d}}(A_{\rho e})\psi$. One has
\begin{itemize}
\item[(a)] $e(\psi')=-\sinh(\rho)\,t_e+\cosh(\rho)\,e(\psi_{t_e})-\big|\cosh(\rho)\,t_e-\sinh(\rho)\,e(\psi_{t_e})\big|$
\item[(b)] $t_e(\psi')=\cosh(\rho)\,t_e-\sinh(\rho)\,e(\psi_{t_e})$
\item[(c)] $e(\psi'_{t_e(\psi')})=-\sinh(\rho)\,t_e+\cosh(\rho)\,e(\psi_{t_e})$
\end{itemize}
\end{Lem}
{\it Proof.} 
Assume without restriction $e=(0,0,1)$, whence $A_{\rho e}=A_\rho$\,, and use the position representation.  By (\ref{DRPS}) one has $\psi'_t(x)= s(A_\rho) \big(\operatorname{e}^{-\operatorname{i}y_0 H^{pos}}\psi\big)(y)$ for $t\in\R$ with $y_0:=-\cosh(\rho)\,t-\sinh(\rho)\,x_3$, $y_1:=x_1$, $y_2:=x_2$, $y_3:=\sinh(\rho)\,t+\cosh(\rho)\,x_3$. Now, $e(\psi_{-y_0})$ is the largest $\alpha\in\R$ such that $1_{\{x'_3\le\alpha\}}\psi_{-y_0}(x'_1,x'_2,x'_3)=0$ a.e., whence $\psi'_t(x)=0$ for $y_3\le \alpha$. Use $e(\psi_{-y_0})= e(\psi_{t_e})-|-y_0-t_e(\psi)|$.  Then $y_3\le \alpha$ means 
$$x_3\le e(\psi_{t_e})/\cosh(\rho)-\tanh(\rho)\,t-|\tanh(\rho)x_3-t_e/\cosh(\rho)+t|$$
Elementary computation shows that the set $J:=\{x\in\R:x\le a-|\lambda x-b|\}$ for $a,b\in\R$ and $\lambda\in]-1,1[$ equals $J=]-\infty, \beta]$ for $\beta:=\frac{1}{1-\lambda^2}(a-\lambda b-|b-\lambda a|)$. Thus, in the specific case $\beta$ equals $e(\psi'_t)$ and 
$$e(\psi'_t)=-\sinh(\rho)\,t_e+\cosh(\rho)\,e(\psi_{t_e})-|t-\cosh(\rho)\,t_e+\sinh(\rho)\,e(\psi_{t_e})|$$
for all $t$. This implies (b) and (c), and hence (a).\qed

As an application of (\ref{BCF}) we improve the results in (\ref{CSETIIED})(b) and (\ref{MWTI})(d).

\begin{Cor}\label{ILCS} Let $\psi$ be a bounded localized Dirac state. Then 
\begin{itemize}
\item[(a)] $|t_{\overline{e}}|\le \frac{1}{2}\big(-\overline{e}(\psi)-e(\psi)\big)$, \quad $|t_e|\le \frac{1}{2}\big(-\overline{e}(\psi)-e(\psi)\big)$
\item[(b)] $|t_{\overline{e}}-t_e|\le -\overline{e}(\psi_{t_{\overline{e}}})-e(\psi_{t_e})$
\end{itemize}
for every direction $e$.
\end{Cor}\\
{\it Proof.} (a) Obviously it suffices to prove the first relation. Perform the translation (\ref{TCTTE}) by $b:=-e(\psi)\,e$, whence without restriction $e(\psi)=0$ and hence $-\overline{e}(\psi)>0$.  Now assume the contrary $|t_{\overline{e}}|>-\frac{1}{2}\overline{e}(\psi)$. Consider  the boosted state $\psi':=W^{\textsc{d}}(A_{\rho e})\psi$. (\ref{BCF})(a) shows $e(\psi')=0$ since $e(\psi_{t_e})=|t_e|$, and
by  (\ref{BCF})(b) one has $t_{\overline{e}}(\psi')=\cosh(\rho)\,t_{\overline{e}}+\sinh(\rho)\,\overline{e}(\psi_{t_{\overline{e}}})=\cosh(\rho)\,t_{\overline{e}}+\sinh(\rho)\,\overline{e}(\psi)+\sinh(\rho)\,|t_{\overline{e}}|$.\\
\hspace*{6mm} 
Treat first the case $t_{\overline{e}}\ge 0$. Then $t_{\overline{e}}(\psi')=\e^\rho\big(t_{\overline{e}}+\frac{1}{2}\overline{e}(\psi)\big)-\frac{1}{2}\e^{-\rho}\overline{e}(\psi)>0$. Therefore, by (\ref{BCF})(a), $\overline{e}(\psi')=\sinh(\rho)\,t_{\overline{e}}+\cosh(\rho)\,\overline{e}(\psi_{t_{\overline{e}}})-t_{\overline{e}}(\psi')=-\e^{-\rho}\,t_{\overline{e}}+\e^{-\rho}\,\overline{e}(\psi_{t_{\overline{e}}})=\e^{-\rho}\,\overline{e}(\psi)$. Since $e(\psi')=0$ and $t_{\overline{e}}(\psi')\to \infty$, $\overline{e}(\psi')\to 0$ for $\rho\to \infty$, this yields a contradiction to (\ref{CSETIIED})(b). --- Now let $t_{\overline{e}}\le 0$. Similarly, one gets $t_{\overline{e}}(\psi')=\e^{-\rho}\big(t_{\overline{e}}-\frac{1}{2}\overline{e}(\psi)\big)-\frac{1}{2}\e^\rho\overline{e}(\psi)$ and $\overline{e}(\psi')=\e^{\rho}\,\overline{e}(\psi)$. Then $e(\psi')=0$, $t_{\overline{e}}(\psi')\to -\infty$, $\overline{e}(\psi')\to 0$ for $\rho\to -\infty$  contradicts again (\ref{CSETIIED})(b).\\
\hspace*{6mm}
(b) Consider the first relation in (a) for $\psi_{t_e}$ in place of $\psi$. By (\ref{MWTI})(c),(b) it reads $2|t_{\overline{e}}-t_e|\le -\overline{e}(\psi_{t_{\overline{e}}})-e(\psi_{t_e}) + |t_{\overline{e}}-t_e|$, whence the claim.
\qed\\

By (\ref{MWTI})(a), (\ref{ILCS})(b)
$$\emptyset\ne\; ]e(\psi_{t_e}),  -\overline{e}(\psi_{t_{\overline{e}}})[\; \subset [e(\psi_t), -\overline{e}(\psi_{t'})]\quad\forall \,t,t'\in\R$$
 This means  that in  direction $e$ the carrier of $\psi_t$ shrinks and expands around the nonempty  time fixed strip $\{x\in\R^3: e(\psi_{t_e}) < xe < -\overline{e}(\psi_{t_{\overline{e}}})\}$.  \\

\begin{Cor}\label{LTSIE} Let $\psi$ be a Dirac state localized in $B_R$ for some $R>0$.  Then one has
$$ e(\psi_t)=e(\psi_{R})+R-t \quad \forall\; t\ge R, \quad e(\psi_t)=e(\psi_{-R})+R+t \quad \forall\; t\le -R$$
 independently of the direction $e$.
\end{Cor}\\
{\it Proof.} From (\ref{SETIIED}) and (\ref{ILCS})(a) it follows for $t\ge R$ that $ e(\psi_t)=e(\psi_{t_e})+t_e-t$ and in particular $ e(\psi_{R})=e(\psi_{t_e})+t_e-R$, whence $ e(\psi_t)=e(\psi_{R})+R-t $. Similarly, for $t\le -R$ one has $ e(\psi_t)=e(\psi_{t_e})-t_e+t$ and in particular $ e(\psi_{-R})=e(\psi_{t_e})-t_e-R$, whence $ e(\psi_t)=e(\psi_{-R})+R+t $.\qed\\

\begin{DMBC}\label{DMBC} The results achieved so far provide the following picture. As long as $t<t_e$, 
$e(\psi_t)=e(\psi_{t_e})-t_e +t$  holds by (\ref{SETIIED}), which means that   the carrier of $\psi_t$ retreats  in  direction $e$ at the speed of light. Only after time $t_e$ the carrier advances in direction $-e$ at the speed of light as $e(\psi_t)=e(\psi_{t_e})+t_e -t$. So only then the wave function expands in the direction $-e$ as expected. The abrupt change at the time $t_e$ of the direction of  the motion with light velocity  to the opposite direction is like a bounce. It reminds of the phenomenon of the zitterbewegung. But this behavior is easy to understand. Let $\psi':=\psi_{t_e}$. Then due to homogeneity of time, i.e., the translational symmetry of time evolution, $\psi'$ satisfies $e(\psi'_t)=e(\psi')-|t|$ by (\ref{SETIIED}) and (\ref{MWTI})(c). So, as maximal permissible by causality, $\psi'$ expands in the future as well in the past  in direction $-e$ at the speed of light. In particular the result in (\ref{SETIIED}) does not single out some direction of time. On the contrary,  the reversal of motion is required by time reversal symmetry. Nevertheless for a short period the picture is complicated as  the time of change $t_e$ depends in general on the direction $e$ (see the example (\ref{NISE})). Therefore the carrier of the wave function performs the change from shrinking to expanding not isotropicly. According to  (\ref{CSETIIED})(a), in every direction $e$ the retreat equals at most the width of the carrier. 
Furthermore, the  times $t_e$  and $t_{\overline{e}}$, which pass until  the motion changes, are  limited  
by (\ref{ILCS})(a).
After, respectively before, the time corresponding to the diameter of the carrier a simultaneous  isotropic expansion   of the wave function with light velocity takes place in the future respectively in the past  (\ref{LTSIE}). So the phase of rebound is limited in time and space in the order of the diameter of the carrier at its minimum extension. 
\end{DMBC}

\subsection{Late-change states}
For the following construction (\ref{NISE}) we use the easily verifiable formulae 
\begin{equation}\label{TCTTE}
e\big(W^{\textsc{d}}(b)\psi\big)=e(\psi)+eb, \quad t_e\big(W^{\textsc{d}}(b)\psi\big)=t_e(\psi)
\end{equation} 
for all directions $e$ and translations  $b\in\R^3$. As to the first use translational  covariance of $E^{\textsc{d}}$ and for the latter recall that time and space translations commute.

\begin{Con}\label{NISE} of a  Dirac state $\psi$ with given values of $t_e$ and $t_{\overline{e}}$. Let $e\in\R^3$, $|e|=1$ and  $\overline{e}=-e$, and let $\tau\in\R\setminus\{0\}$ and $\delta>0$. 
Let $\psi^{(1)}$ be any  Dirac state   localized in a bounded region.  
 Set $\psi^{(2)}:=W^{\textsc{d}\, pos}(\delta e)\psi^{(1)}_\tau$. Finally, put
 $$\psi:=\psi^{(1)}+\psi^{(2)}$$
  It is not necessary to normalize $\psi$ here. In the following we  express the characteristic dates  $e(\psi),t_e$,  $\overline{e}(\psi), t_{\overline{e}}$ referring to $\psi$ by the input  dates 
$e(\psi^{(1)}),t^{(1)}_e$,  $\overline{e}(\psi^{(1)}), t^{(1)}_{\overline{e}}$ and the parameters $\tau, \delta$.

\hspace*{6mm} 
By  (\ref{SETIIED}), (\ref{MWTI})(c) and  Eq.\,(\ref{TCTTE}) one has
$t^{(2)}_e=t^{(1)}_e-\tau$ and $t^{(2)}_{\overline{e}}=t^{(1)}_{\overline{e}}-\tau$,
and  $e(\psi^{(1)}_t)=e(\psi^{(1)})+|t^{(1)}_e|-|t-t^{(1)}_e|$,  $e(\psi^{(2)}_t)=e(\psi^{(1)}_{t+\tau }) +\delta=e(\psi^{(1)})+|t^{(1)}_e|-|t+\tau-t^{(1)}_e|+\delta$ and similarly $\overline{e}(\psi^{(1)}_t)=\overline{e}(\psi^{(1)})+|t^{(1)}_{\overline{e}}|-|t-t^{(1)}_{\overline{e}}|$, $\overline{e}(\psi^{(2)}_t)=\overline{e}(\psi^{(1)})+|t^{(1)}_{\overline{e}}|-|t+\tau-t^{(1)}_{\overline{e}}|-\delta$.\\
 \hspace*{6mm} 
 Now, obviously $e(\psi_t)=\min\{e(\psi^{(1)}_t),e(\psi^{(2)}_t)\}$ and  $\overline{e}(\psi_t)=\min\{\overline{e}(\psi^{(1)}_t),\overline{e}(\psi^{(2)}_t)\}$. Hence $t_e$ and $t_{\overline{e}}$ are determined by  (\ref{SETIIED}). Write $e(\psi^{(2)}_t)-e(\psi^{(1)}_t) =
 d(t-t_e^{(1)})$ with $d(x):=|x|-|x+\tau|+\delta$  and $\overline{e}(\psi^{(2)}_t)-\overline{e}(\psi^{(1)}_t) =
 \overline{d}(t-t^{(1)}_{\overline{e}})$ with $\overline{d}(x):=|x|-|x+\tau|-\delta$. \\
\hspace*{6mm}
 We will distinguish the  cases (i) $|\tau|\le\delta$,  (ii) $\tau>\delta$, and  (iii) $-\tau>\delta$.  Note first
 \begin{equation*}
 |\tau|\le\delta\, \Leftrightarrow\,  d(t-t_e^{(1)})\ge 0 \; \forall t \, \Leftrightarrow\,  \overline{d}(t-t^{(1)}_{\overline{e}})\le 0 \; \forall t\tag{$\star$}
\end{equation*}
Indeed, $ d(t-t_e^{(1)})\ge 0$ is equivalent to $ |\tau|\le\delta$ as $d$ takes
 its minimum $-|\tau|+\delta$ at 
 $x=0$. Similarly, $\overline{d}$ takes its maximum $|\tau|-\delta$ at $x=-\tau$.\\
 
\hspace*{6mm}
(i) Consider  the case $|\tau|\le\delta$. By ($\star$) one has $e(\psi_t)=e(\psi^{(1)}_t)$, $\overline{e}(\psi_t)= \overline{e}(\psi^{(2)}_t)$,  whence $t_e=t^{(1)}_e$ and $t_{\overline{e}}=t^{(2)}_{\overline{e}}=t^{(1)}_{\overline{e}}-\tau$.  
So one obtains the given value of
$ t_{\overline{e}}-t_e$  by choosing $\tau =(t^{(1)}_{\overline{e}}-t^{(1)}_e)-(t_{\overline{e}}-t_e)$. By a subsequent time translation according to (\ref{MWTI})(c) one gets the prescribed values of $t_e$ and $t_{\overline{e}}$. 

\hspace*{6mm}
We examine the remaining cases (ii) and (iii) in order to verify that, as mentioned  in (\ref{L-CS}), the simple construction (\ref{NISE})  does not yield a Dirac state like (\ref{GPTEB}).

\hspace*{6mm}
(ii) Now let $\tau>\delta$. Then $d(x)=\tau+\delta>0$ for $x\le-\tau$, $d(x)=-2x-\tau+\delta$ for $-\tau\le x\le 0$, and $d(x)=-\tau+\delta$ for $x\ge 0$. Hence $d\big(\frac{1}{2}(-\tau+\delta)\big)=0$.  One infers  $t_e=t_e^{(1)}+\frac{1}{2}(-\tau+\delta)$. It follows $e(\psi)=e(\psi^{(1)})$ if $t_e\ge 0$ and $e(\psi)=e(\psi^{(2)})$ if $t_e\le 0$. In other words, $e(\psi)=e(\psi^{(1)})$ if 
$t^{(1)}_e\ge \frac{1}{2}(\tau-\delta)$ and $e(\psi)=e(\psi^{(1)})+|t^{(1)}_e|-|\tau-t^{(1)}_e|+\delta$ if $t^{(1)}_e\le \frac{1}{2}(\tau-\delta)$.\\
\hspace*{6mm}
Similarly, $\overline{d}(x)=\tau-\delta>0 $ for $x\le-\tau$, $\overline{d}(x)=-2x-\tau-\delta$ for $-\tau\le x\le 0$, and $\overline{d}(x)=-\tau-\delta$ for $x\ge 0$. Hence $\overline{d}\big(\frac{1}{2}(-\tau-\delta)\big)=0$. One infers $t_{\overline{e}}=t_{\overline{e}}^{(1)}-\frac{1}{2}(\tau+\delta)$.     It follows $\overline{e}(\psi)=\overline{e}(\psi^{(1)})$  if $t_{\overline{e}}\ge 0$ and 
$\overline{e}(\psi)=\overline{e}(\psi^{(2)})$  if  $t_{\overline{e}}\le 0$. In other words, $\overline{e}(\psi)=\overline{e}(\psi^{(1)})$  if $t^{(1)}_{\overline{e}}\ge \frac{1}{2}(\tau+\delta)$ and $\overline{e}(\psi)=\overline{e}(\psi^{(1)})+|t^{(1)}_{\overline{e}}|-|\tau-t^{(1)}_{\overline{e}}|-\delta$  if  $t^{(1)}_{\overline{e}}\le  \frac{1}{2}(\tau+\delta)$.

 \hspace*{6mm}
 (iii) Finally assume $-\tau>\delta$. Then $d(x)=\tau+\delta<0$ for $x\le 0$, $d(x)=2x+\tau+\delta$ for $0\le x \le -\tau$, and $d(x)=-\tau+\delta$ for $x\ge -\tau$. Hence $d\big(\frac{1}{2}(-\tau-\delta)\big)=0$. One infers $t_e=t_e^{(1)}-\frac{1}{2}(\tau+\delta)$.  It follows $e(\psi)=e(\psi^{(2)})$ if $t_e\ge 0$ and $e(\psi)=e(\psi^{(1)})$ if $t_e\le 0$. In other words, $e(\psi)=e(\psi^{(1)})+|t^{(1)}_e|-|\tau-t^{(1)}_e|+\delta$   if 
$t^{(1)}_e\ge \frac{1}{2}(\tau+\delta)$ and $e(\psi)=e(\psi^{(1)})$ if $t^{(1)}_e\le \frac{1}{2}(\tau+\delta)$.\\
\hspace*{6mm}
Similarly, $\overline{d}(x)=\tau-\delta<0$ for $x\le 0$, $\overline{d}(x)=2x+\tau-\delta$ for $0\le x \le -\tau$, and $\overline{d}(x)=-\tau-\delta$ for $x\ge -\tau$. Hence $\overline{d}\big(\frac{1}{2}(-\tau+\delta)\big)=0$. One infers $t_{\overline{e}}=t_{\overline{e}}^{(1)}+\frac{1}{2}(-\tau+\delta)$.   It follows $\overline{e}(\psi)=\overline{e}(\psi^{(2)})$  if $t_{\overline{e}}\ge 0$ and 
$\overline{e}(\psi)=\overline{e}(\psi^{(1)})$  if  $t_{\overline{e}}\le 0$. In other words, $\overline{e}(\psi)=\overline{e}(\psi^{(1)})+|t^{(1)}_{\overline{e}}|-|\tau-t^{(1)}_{\overline{e}}|-\delta$  if  $t^{(1)}_{\overline{e}}\ge  \frac{1}{2}(\tau-\delta)$ and $\overline{e}(\psi)=\overline{e}(\psi^{(1)})$  if $t^{(1)}_{\overline{e}}\le \frac{1}{2}(\tau-\delta)$.\qed
\end{Con}

 As an example, the construction  in (\ref{NISE}) for $\tau:=t^{(1)}_{\overline{e}}-t^{(1)}_e$, $\delta:=|\tau|$ with subsequent time translation by $t_e=t_{\overline{e}}$ yields a Dirac state  $\psi$ satisfying
 \begin{equation}\label{TETEOEZ} 
t_e=t_{\overline{e}}=0 \textrm{\, and } -\overline{e}(\psi)-e(\psi)= -\overline{e}(\psi^{(1)})-e(\psi^{(1)}) +
|t^{(1)}_{\overline{e}}-t^{(1)}_e |-|t^{(1)}_{\overline{e}}|- | t^{(1)}_e |
\end{equation}
The width of the carrier in direction $e$ in not increased since $|t^{(1)}_{\overline{e}}-t^{(1)}_e |-|t^{(1)}_{\overline{e}}|- | t^{(1)}_e |\le 0$.

\begin{Lem}\label{DSABTP}
For every $a,b\in\R$ with $a<b$ and $ |\tau| <\frac{1}{2}(b-a)$ 
 there is a Dirac state $\psi$ localized in a bounded region such that $a\le e(\psi)<-\overline{e}(\psi)\le b$  and $t_e=t_{\overline{e}}=\tau$.
\end{Lem}\\
{\it Proof.} Due to Eq.\,(\ref{TCTTE}) it is no restriction to assume $a=-b$. Let $0<\rho<b$. By Eq.\,(\ref{TETEOEZ}) there is  a Dirac state $\eta$ localized in a bounded region contained in $\{-\rho\le xe\le \rho\}$ with $t_e(\eta)=t_{\overline{e}}(\eta)=0$. Let $\varsigma$ denote the sign of $\tau$. Then, by causality, $\psi:=\eta_{\varsigma(-b+\rho)}$ is localized in $\{-b\le xe\le  b\}$. Moreover, $t_e=t_{\overline{e}}= \varsigma(b-\rho)$ holds by (\ref{MWTI})(c). The assertion follows for $\rho:=b-|\tau|$.\qed\\

\hspace*{6mm}
Before proceeding  we remind  of the symmetry of

\begin{Tir}\label{TIR}
For the Dirac system time reversal is represented by the antiunitary operator $\mathcal{T}$, which in position representation Eq.\,(\ref{DRPS}) using the Weyl matrices reads 
$$\mathcal{T}^{pos}\psi=\omega\,\overline{\psi}$$
with $\omega:=  -\operatorname{i}\alpha_1\alpha_3=-\operatorname{diag}(\sigma_2,\sigma_2)$. Note $\mathcal{T}^2=-I$. One finds $(\mathcal{T}^{mom}\varphi)(p)=\omega\overline{\varphi(-p)}$. Moreover $\big(\mathcal{T}^{pos}\,W^{\textsc{d}\,pos}(g)\,\mathcal{T}^{pos\,-1}\psi\big)(x)=  s(\sigma_2\overline{A}\sigma_2)\,\big(\operatorname{e}^{\operatorname{i}y_0 H^{pos}}\psi\big)(y)$ (cf.\,Eq.\,(\ref{DRPS})). Therefore $W^{\textsc{d}}(b,B)=\mathcal{T}\,W^{\textsc{d}}(b,B)\, \mathcal{T}^{-1}$ and 
\begin{equation}\label{RDBTTRO}
W^{\textsc{d}}(A_{\,-\rho e})=\mathcal{T}\,W^{\textsc{d}}(A_{\rho e})\, \mathcal{T}^{-1},\quad 
W^{\textsc{d}}(-t)=\mathcal{T}\,W^{\textsc{d}}(t)\, \mathcal{T}^{-1}
\end{equation}
i.e., Euclidean transformations are time-reversal invariant, boost and time direction are reversed. Recall that $A_{\rho e}=\operatorname{exp}(\frac{\rho}{2}\sum_{k=1}^3e_k\sigma_k)$  represents the boost in direction $e\in\R^3$, $|e|=1$ with rapidity $\rho$.\\
\hspace*{6mm}
 The crucial relations for this are $\sigma_2^2=I_2$ and $-\sigma_k=\sigma_2\,\overline{\sigma}_k\,\sigma_2$ for $k=1,2,3$, and hence $\overline{h(-\overline{z})}= \omega\, h(z)\, \omega$ for $z\in\C^3$. \\
 \hspace*{6mm}
 Obviously $\mathcal{T}E^{\textsc{d}}\mathcal{T}^{-1}=E^{\textsc{d}}$, i.e., $\mathcal{T}$ commutes with the Dirac localization, and  $\mathcal{T}\psi$ is localized in the same region as $\psi$ for every $\psi$.  It follows for every $e$
 \begin{equation}\label{BETIR}
 e(\psi)=e(\mathcal{T}\psi) \textrm{\; and\; }   t_e(\mathcal{T}\psi)= -t_e(\psi)
 \end{equation}
  Indeed, the first formula is obvious. As 
 $(\mathcal{T}\psi)_t=\mathcal{T}\psi_{-t}$, on the one hand one has $e\big((\mathcal{T}\psi)_t\big)=e(\psi)+|t_e(\mathcal{T}\psi)| +|t-t_e(\mathcal{T}\psi)|$ and on the other hand $e\big((\mathcal{T}\psi)_t\big)=e(\psi_{-t})=e(\psi)+ |t_e(\psi)|+|t+t_e(\psi)|$ for all $t$, whence the second formula.\qed
 \end{Tir}

\begin{L-CS}\label{L-CS} In (\ref{GPTEB}) we construct a bounded localized Dirac state $\psi$ satisfying $\varsigma t_{\overline{e}}= \frac{1}{2}\big(-\overline{e}(\psi)-e(\psi)\big)$ for a given direction $e$ and $\varsigma\in\{+,-\}$. (One verifies explicitly that it is not possible  to obtain such a state  by (\ref{NISE}), except in the trivial case that the initial  state $\psi^{(1)}$ already satisfies this condition.) By the spatial translation $W^{\textsc{d}}(b)$ for $b:=-e(\psi)e$ according to Eq.\,(\ref{TCTTE}) one achieves   
\begin{equation}\label{LTES}
e(\psi)=0, \quad -\overline{e}(\psi)=2 |t_{\overline{e}}|
\end{equation}
or equivalently by (\ref{SETIIED}) 
$$e(\psi_{t_e})=|t_e|, \quad-\overline{e}(\psi_{t_{\overline{e}}})= |t_{\overline{e}}|$$ 

Bounded localized states $\psi$ satisfying these relations
are particularly interesting. In view of (\ref{ILCS})(a) we call them \textbf{large -\,$t_{\overline{e}}$\,- states} or, less specifically, \textbf{late-change states}.\\
\hspace*{6mm}
 These states are characterized in (\ref{XCTLRCL}) by equivalent properties. Those of them, which are localized in $\{x\in\R^3:0\le xe\le 2\alpha\}$, $\alpha>0$, form an infinite dimensional subspace.
This is proved (for $\varsigma=+$) in sec.\,\ref{PPCS} (see (\ref{CCIDES}),\,(\ref{ACCIDES})) within the general frame of causal systems. What is more, every Dirac  state localized in $\{x\in\R^3:xe\ge0\}$ can be approximated by late-change states with $\varsigma=+$ (as well with $\varsigma=-$),  
see (\ref{CCPDS}).
This property is essential for the proof of the Lorentz contraction in (\ref{LCDWF}). Moreover,  for the existence of these states there are no reps of the lattice of causally complete regions in sec.\,\ref{LOCCR}, and there were no reps of  the lattice generated by the non-timelike relation in sec.\,\ref{LGNTLR} without these states.
\end{L-CS}

As  a corollary of (\ref{BCF}) one has  the particularly simple formulae for boosted late-change states.

\begin{Cor}\label{BLCS} Let $e$ be a direction. Let $\psi$ be a large -\,$t_{\overline{e}}$\,- state  \emph{(\ref{LTES})}. Put $\varsigma:=\operatorname{sign}t_{\overline{e}}$. Consider the boosted state $\psi':=W^{\textsc{d}}(A_{\rho e})\psi$ in direction $e$ with rapidity $\rho\in\R$. Then 
\begin{itemize}
\item $|t_e|<|t_{\overline{e}}|$, $\operatorname{sign}t_e\in\{0,\varsigma\}$, $t_e(\psi')=\e^{-\varsigma \rho} t_e$
\item $|t_{\overline{e}}-t_e|=-\overline{e}(\psi_{t_{\overline{e}}})-e(\psi_{t_e})$
\item $e(\psi')=0$
\item $t_{\overline{e}}(\psi')=\e^{-\varsigma \rho} t_{\overline{e}}$
\item $\overline{e}(\psi')=\e^{-\varsigma \rho}\overline{e}(\psi)$
\item $\psi'$ is a large -\,$t_{\overline{e}}$\,- state
\end{itemize}
\end{Cor}
{\it Proof.} The second claim follows from the first one by (\ref{LTES}). ---
By (\ref{BCF})(b),(a) one has   $t_e(\psi')=e^{-\operatorname{sign}(t_e) \rho}\,t_e$ and the items  three,  four, and five. --- Then (\ref{ILCS})(a) yields $|t_e(\psi')|\le -\frac{1}{2}\overline{e}(\psi')$ for all $\rho$, whence  either $t_e=0$ or $\operatorname{sign}t_e=\varsigma$. Hence $2|t_{\overline{e}}(\psi')|=2\e^{-\varsigma \rho} |t_{\overline{e}}|=-\e^{-\varsigma \rho}\overline{e}(\psi)=-\overline{e}(\psi')$ and the last item follows. --- Finally, $|t_e|<|t_{\overline{e}}|$ holds because of (\ref{CSETIIED})(c).\qed

It follows the 

\begin{Constr}\label{GPTEB} Let $a,b\in\R$ with $a<b$. Let $\varsigma\in\{+,-\}$.  Then there is a bounded localized Dirac state $\psi$  such that $a\le e(\psi)<-\overline{e}(\psi)\le b$  and $\varsigma t_{\overline{e}}= \frac{1}{2}\big(-\overline{e}(\psi)-e(\psi)\big)$. The spatial translation by $-e(\psi)e$ yields \emph{(\ref{LTES})}.
\end{Constr}\\
{\it Proof.}  By Eq.\,(\ref{BETIR}) and due to (\ref{ILCS})(a) it suffices to  exhibit $\psi$ with $t_{\overline{e}}\ge \frac{1}{2}\big(-\overline{e}(\psi)-e(\psi)\big)$.
As constructed in (\ref{DSABTP}),  there is  a bounded localized Dirac state $\eta$ with $a\le e(\eta)<-\overline{e}(\eta)\le b$  and $\upsilon:=t_{\overline{e}}(\eta)>0$. By (\ref{CSETIIED})(b),  $\upsilon\le-\overline{e}(\eta)-e(\eta)$. Let $0<\delta<\upsilon$. Set $\psi:=E^{\textsc{d}}\big(\{-\overline{e}(\eta)-\delta\le xe\le-\overline{e}(\eta)\}\big)\,\eta$ (it is not necessary to normalize $\psi$ here), $\tau:=t_{\overline{e}}$ and define $\eta':=\eta-\psi$. Obviously $a\le e(\psi)<-\overline{e}(\psi)\le b$ and
\begin{equation*}
 -\overline{e}(\eta)=-\overline{e}(\psi), \quad -\overline{e}(\eta')\le -\overline{e}(\eta)-\delta \le e(\psi)<-\overline{e}(\psi)\tag{$\star$}
 \end{equation*}
By (\ref{SETIIED}) and by ($\star$), $-\overline{e}(\psi_\tau)= -\overline{e}(\psi)-|\tau|$ and $-\overline{e}(\eta'_\tau)=-\overline{e}(\eta')-|\upsilon'|+|\tau-\upsilon'|\le -\overline{e}(\eta')+|\tau|\le e(\psi)+|\tau|$.\\
\hspace*{6mm}
 Assume 
$|\tau| < \frac{1}{2}\big(-\overline{e}(\psi)-e(\psi)\big)$, which means $ e(\psi)+|\tau|< -\overline{e}(\psi)-|\tau|$.
Hence $-\overline{e}(\eta'_\tau)<-\overline{e}(\psi_\tau)$. By (\ref{SETIIED}), 
$-\overline{e}(\psi_\tau)\le -\overline{e}(\psi_t)$ for all $t$. Still  by (\ref{SETIIED}), this implies 
$-\overline{e}(\eta'_t)<-\overline{e}(\psi_t)$ for all $t$. Now note $\eta_t=\eta'_t+\psi_t$. One infers $-\overline{e}(\eta_t)=-\overline{e}(\psi_t)$ for all $t$. In particular, $\upsilon=\tau$, whence by ($\star$) the contradiction $\tau >\delta \ge -\overline{e}(\psi)-e(\psi)$. 
\\
\hspace*{6mm}
Thus $|\tau| \ge \frac{1}{2}\big(-\overline{e}(\psi)-e(\psi)\big)$ holds. It remains to show $\tau\ge 0$. Assume the contrary. We start from $-\overline{e}(\eta')<-\overline{e}(\psi)$ by ($\star$). Then by (\ref{SETIIED}), for all $t\ge 0$ one has $-\overline{e}(\eta'_t)=-\overline{e}(\eta')-|\upsilon' | + |t-\upsilon' |\le -\overline{e}(\eta')+t< -\overline{e}(\psi)+t = -\overline{e}(\psi)-|\tau|+|t-\tau|=-\overline{e}(\psi_t)$. So 
$-\overline{e}(\eta'_t)<-\overline{e}(\psi_t)$ for all $t\ge 0$. This implies $-\overline{e}(\eta_t)=-\overline{e}(\psi_t)$ for all $t\ge 0$. For $t=\upsilon>0$ this means $-\overline{e}(\eta)-\upsilon=-\overline{e}(\psi)+\upsilon$. Since  $-\overline{e}(\eta)=-\overline{e}(\psi)$ by ($\star$), it contradicts $\upsilon\ne 0$.
\qed\\


\section{Long-term behavior of the probability of localization of Dirac states}\label{LTBPL} Of course, the probability of localization within the carrier of the  wave function evolving in time  stays at $1$. Insomuch the foregoing results on the movement of the border of the wave function  yield also an information about the time dependence of the probability of localization. However the probability stays not equally distributed across the carrier. We will show that for every Dirac state (not necessarily bounded localized) in the long term the probability of localization concentrates up to $1$ in the spherical  shell $B_{|t|}\setminus B_{r}$ for every radius $r>0$. \\
\hspace*{6mm}
The main mathematical tool is an application of the non-stationary phase method as shown in \cite[Theorem 1.8.]{T92} estimating Eq.\,(\ref{ENSPM})  for large  $|x|+|t|$. The part of the result, according to which the spatial probability in $B_{r}$ tends to zero (see (\ref{PRBG})),
 is a corollary to  \cite[Corollary 1.9.]{T92} by the fact that $\mathcal{C}^\infty_c(\R^3\setminus \{0\})$, the set of $\mathcal{C}^\infty$-functions with compact support in $\R^3\setminus \{0\}$, is dense in $L^2(\R^3)$.
 Rather analogously we prove in (\ref{VPOBT})  the fact  that  asymptotically  the spatial probability vanishes outside $B_{|t|}$.\\
\hspace*{6mm}
For the following proofs  the  obvious reduction to scalar-valued wave functions  in momentum representation is used. Let $\psi\in L^2(\R^3,\C^4)$ be a Dirac state in position representation and let $\varphi=\mathcal{F}\psi$ be its momentum representation. Regarding the time evolution one has $\varphi_t=\operatorname{e}^{\operatorname{i}th}\varphi$, i.e., $\varphi_t(p)=\operatorname{e}^{\operatorname{i}th(p)}\varphi(p) \,\forall\; p$.  Let $\eta\in\{+,-\}$ and recall that $\pi^\eta(p)=\frac{1}{2}(I+\frac{\eta}{\epsilon(p)}h(p))$ with $\epsilon(p)=\sqrt{|p|^2+m^2}$ is the projection in $\C^4$ onto the $2$-dimensional eigenspace of $h(p)$ with eigenvalue $\eta\, \epsilon(p)$.  Then  $\varphi^\eta:=\pi^\eta\varphi$
is the projection  of $\varphi$ onto the positive, respectively negative, energy eigenspace. Analogously 
 $(\varphi_t)^\eta:=\pi^\eta\varphi_t$. Note that $(\varphi_t)^\eta=(\varphi^\eta)_t=\operatorname{e}^{\operatorname{i}t\eta \epsilon}\varphi^\eta$, as $\operatorname{e}^{\operatorname{i}th}$ and $\pi^\eta$ commute. One concludes $(\psi_t)_l=\sum_\eta  (\psi^\eta_t)_l$,   $ (\psi^\eta_t)_l:= \big(\mathcal{F}^{-1}\varphi^\eta_t\big)_l=
  \mathcal{F}^{-1}\big(\operatorname{e}^{\operatorname{i}t\eta \epsilon}(\varphi^\eta)_l\big)$ for the $l$-th component of $\psi_t$, $l=1,\dots,4$. If $\varphi$  is also  integrable, then  $\varphi^\eta$ is so  and for each $l$ one has
 \begin{equation}\label{ENSPM}
 (\psi^\eta_t(x))_l=(2\pi)^{-3/2}\int \operatorname{e}^{\operatorname{i}(px+t\eta\epsilon(p))}(\varphi^\eta(p))_l\,\operatorname{d}^3p
 \end{equation}

\begin{The}\label{PRBG} Let $\psi$ be a Dirac state. Let $\varepsilon> 0$. Then there are $v\in\,]0,1[$ and $\tau>0$ such that $\norm{E^{\textsc{d}}(B_{v |t|})\psi_t}\le \varepsilon$ for all $|t|\ge \tau$. In particular
$$E^{\textsc{d}}(B_r)\psi_t\to 0, \quad |t|\to \infty$$ holds for every radius $r>0$.
\end{The}\\
{\it Proof.} Recall $\varphi=\mathcal{F}\psi$ and choose $\varphi'\in\mathcal{C}^\infty_c(\R^3\setminus \{0\},\C^4)$ with $\norm{\varphi-\varphi'}\le\varepsilon/2$. Hence $\norm{\psi-\psi'}\le\varepsilon/2$ for $\psi'=\mathcal{F}^{-1}\varphi'$. Choose $0<v<\inf\{\frac{|p|}{\epsilon(p)}:p\in \operatorname{supp}(\varphi')\}$.  Let $P_t := E^{\textsc{d}}(B_{v|t|})$. Now, according to \cite[Corollary 1.9.]{T92},  there is a constant $C_1$ such that $\norm{P_t\psi'_t}\le
C_1(1+|t|)^{-1}$ for all $t$. Let $\tau:=2C_1/\varepsilon$. Then $\norm{P_t\psi_t}\le \norm{P_t(\psi_t-\psi'_t)} + \norm{P_t\psi'_t}  \le$ $\norm{\psi-\psi'} +\,     C_1(1+|t|)^{-1} \le\varepsilon$ for $|t|\ge\tau$. --- Now fix $r>0$. Then for $|t|\ge \max\{\tau, \frac{r}{v}\}$ one has $\norm{E^{\textsc{d}}(B_r)\psi_t}\le \norm{E^{\textsc{d}}(B_{v |t|})\psi_t}\le \varepsilon$.
\qed\\

\begin{The}\label{VPOBT}  Let $\psi$ be a Dirac state. Then 
$$ E^{\textsc{d}}(\R^3\setminus B_{|t|})\psi_t\to 0, \;\;|t|\to \infty$$
If $\mathcal{F}\psi\in \mathcal{C}^\infty_c(\R^3,\C^4)$ holds then for every $N>0$ there is a finite constant $C_N$ such that $\norm{E^{\textsc{d}}(\R^3\setminus B_{|t|})\psi_t}\le C_N(1+|t|)^{-N}$ for all $t\in\R$.
\end{The}\\
{\it Proof.} Suppose first $\varphi:=\mathcal{F}\psi\in \mathcal{C}^\infty_c(\R^3,\C^4)$. Let $K:=\operatorname{supp}(\varphi)$. Set $\gamma:=\max\{\frac{|p|}{\epsilon(p)}:p\in K\}$. Clearly $0<\gamma<1$. For the estimation of the integral in Eq.\,(\ref{ENSPM})  consider $\phi^\eta(p):= (|x|+|t|)^{-1}\big(px-t\eta\epsilon(p)\big)$. Then $\nabla \phi^\eta(p)=  (|x|+|t|)^{-1}\big(x-\frac{t\eta}{\epsilon(p)}p\big)$ and $|\nabla \phi^\eta(p)|\ge  (|x|+|t|)^{-1}\big(|x|-|t|\frac{|p|}{\epsilon(p)}\big)\ge \frac{|x|-\gamma|t|}{|x|+|t|}$ for $p\in K$. Now suppose $|x|\ge|t|$. Then $|\nabla \phi^\eta(p)|\ge  \frac{|x|-\gamma|x|}{|x|+|x|}=\frac{1-\gamma}{2}>0$. This implies  (cf.\,\cite[(1.209)]{T92}) for $\eta\in\{+,-\}$, $l=1,\dots,4$, and for every $N>0$ that there is a finite constant $A_N$ with
$$\big|\big(\psi^\eta_t(x)\big)_l\big|\le A_N(1+|x|+|t|)^{-N}\textrm{ \;if\; } |x|\ge |t|$$
Hence $\norm{E^{\textsc{d}}(\R^3\setminus B_{|t|})\psi_t}\le \sum_\eta \norm{E^{\textsc{d}}(\R^3\setminus B_{|t|})\psi^\eta_t}$ and  $\norm{E^{\textsc{d}}(\R^3\setminus B_{|t|})\psi^\eta_t}^2=\int_{\R^3\setminus B_{|t|}}\norm{\psi^\eta_t(x)}^2\operatorname{d}x^3=\sum_l\int_{\R^3\setminus B_{|t|}}\big|\big(\psi^\eta_t(x)\big)_l\big|^2\operatorname{d}x^3\le 16\pi A_N^2\int_{|t|}^\infty (1+r+|t|)^{-2N} r^2\operatorname{d}r\le 16\pi A_N^2\int_{|t|}^\infty (1+r)^{-2N+2}\operatorname{d}r=\frac{16\pi}{2N-3}A_N^2(1+|t|)^{-2N+3}$ if $N>\frac{3}{2}$. Hence $\norm{E^{\textsc{d}}(\R^3\setminus B_{|t|})\psi_t}\le C_N(1+|t|)^{-N}$ for $N>0$ and  $C_N:=(32\pi/N)^{\frac{1}{2}}A_{N+\frac{3}{2}}$. \\
\hspace*{6mm}
Now consider a general Dirac state $\psi$. Let $\varepsilon>0$. Set $\varphi:=\mathcal{F}\psi$ and choose $\varphi'\in\mathcal{C}^\infty_c(\R^3,\C^4)$ with $\norm{\varphi-\varphi'}\le\varepsilon/2$. Hence $\norm{\psi-\psi'}\le\varepsilon/2$ for $\psi':=\mathcal{F}^{-1}\varphi'$. Let $P_t:=E^{\textsc{d}}(\R^3\setminus B_{|t|})$. By the foregoing result  there is a constant $C_1$ such that $\norm{P_t\psi'_t}\le
C_1(1+|t|)^{-1}$ for all $t$. Let $\tau:=2C_1/\varepsilon$. Then $\norm{P_t\psi_t} \le \norm{P_tW^{\textsc{d}\, pos}(t)(\psi-\psi')}  +  \norm{P_t\psi'_t} \le$ $\norm{\psi-\psi'}+C_1(1+|t|)^{-1}\le\varepsilon $ for $|t|\ge\tau$.\qed\\

\begin{AC}\label{AC} The results (\ref{PRBG}), (\ref{VPOBT}) hold also with respect to the Newton-Wigner localization of Dirac states $\psi$. In particular one has the asymptotic causality
\begin{equation}\label{ASC}
 E^{\textsc{nw}}(\R^3\setminus B_{|t|})\psi_t\to 0, \;\;|t|\to \infty
 \end{equation}
More generally  these results are valid  for all massive systems $[m,j,\eta]$ ($m>0,j\in\mathbb{Z}/2,\eta=+,-$) endowed with the Newton-Wigner localization.  Indeed, by (\ref{RIMS}) the evolution of the state  $\psi$  in position representation is
$ \psi_t(x)=(2\pi)^{-3/2}\int \operatorname{e}^{\operatorname{i}(px+t\eta\epsilon(p))}\big(\mathcal{F}\psi\big)(p)\,\operatorname{d}^3p$.
It equals Eq.\,(\ref{ENSPM}). Hence it suffices to apply the proofs of   (\ref{PRBG}) and (\ref{VPOBT}).\\
\hspace*{6mm}
Asymptotic causality  is  shown in   \cite[Proposition]{R81}. 
Note that (\ref{ASC}) holds for every state, even if the initial state $\psi_0$ is localized far away from the origin. In \cite{R81} it is also pointed out that (\ref{ASC}) is false for the massless system $[0,0,\eta]$.
But the failure of (\ref{ASC}) must not mean at all an  acausal behavior. We think of the fact that for some states of the Weyl systems, which are causal,   one has $\lim_{|t|\to\infty}\norm{E^{\textsc{w}}(\R^3\setminus B_{|t|})\psi_t}\ge 1/2$ by (\ref{GTERSWF})(b).
\qed
\end{AC}

Let $\psi$ be a Dirac state (not necessarily bounded localized) and let $e\in \R^3$, $|e|=1$ be a direction. Suppose that $e(\psi)$ is finite. This implies that $\psi$ is localized in the half-space $\{x\in\R^3: xe>e(\psi)\}$. By causality $\psi_t$ is localized in the half-space $\{x\in\R^3: xe>e(\psi)-|t|\}$. Therefore $e(\psi_t)\ge e(\psi)-|t|$. (Clearly, we imagine  that even (\ref{SETIIED})  holds for $e(\psi_t)$.) The following result on $e(\psi_t)$ is a trivial consequence of  (\ref{VPOBT}). Nevertheless, because of its importance  in view of (\ref{DSLSLHPTLC})(f), we display it by  

\begin{Cor}\label{CDSLSLHPTLC} Let $e$ be a direction. There is no Dirac state $\psi$  such that $e(\psi_{t_n})\ge |t_n|$ for some sequence $|t_n|\to \infty$.
\end{Cor}


Therefore,
as already announced in the comment following (\ref{DSLSLHPTLC}), one has
$P^>_\sigma=E(\sigma^>)$. More precisely one has (\ref{CCPDS}). Recall that $W^{\textsc{d}}(\alpha)$ represents the time shift by $\alpha\in\R$  and that $\psi_\alpha=W^{\textsc{d}}(\alpha)\psi$.

\begin{The}\label{CCPDS}
Let $\psi$ be a Dirac state  localized in $\{x\in\R^3:xe\ge0\}$. Then for 
 $\alpha>0$ and $\varsigma\in\{1,-1\}$
$$\psi=\lim_{\alpha\to\infty}\psi^{\alpha} \textnormal{\; for \;
} \psi^{\alpha}:= W^{\textsc{d}}(\varsigma\alpha)E^{\textsc{d}}(\{xe\le\alpha\})W^{\textsc{d}}(\varsigma\alpha)^{-1}\psi$$
 and $\psi^{\alpha}$ is localized in $\{x\in\R^3:xe\ge0\}$. In the definition of $\psi^\alpha$, $xe$ can be replaced with $|xe|$. Furthermore, $(\psi^\alpha)_{-\varsigma\alpha}$ is localized in $\{x\in\R^3:xe\le\alpha\}$ and $\psi^\alpha$ is localized in $\{x\in\R^3: xe\le 2\alpha\}$. If $\psi$ is localized in the (bounded) region $\Delta$ then $\psi^\alpha$ is localized in the (bounded) region $\Delta_{2\alpha}$.
 \end{The}
 
 {\it Proof.} Recall the definition of $P^>_\sigma$. Then due to (\ref{CDSLSLHPTLC})  the first part of the assertion follows from (\ref{DSLSLHPTLC})(f),(g)  for $\sigma=\{0\}\times\R^3\equiv\R^3$.  As $\psi_{-\varsigma\alpha}$ is localized in $\{x\in\R^3: xe \ge -\alpha \}$ by causality, $xe$ can be replaced with $|xe|$. 
 Obviously,  $(\psi^\alpha)_{-\varsigma\alpha}$ is localized in $\{x\in\R^3:xe\le\alpha\}$. So  $\psi^\alpha$ is localized in $\{x\in\R^3: xe\le 2\alpha\}$ by causality. Also the last part of the assertion holds by causality. \qed

\begin{Cor}\label{ECCPDS} $I=\lim_{|\alpha|\to\infty} W^{\textsc{d}}(\alpha)E^{\textsc{d}}(\{|xe|\le|\alpha|\})W^{\textsc{d}}(\alpha)^{-1}$.
\end{Cor}

The result (\ref{CCPDS}) is essential in proving the Lorentz contraction for Dirac wave functions in (\ref{LCDWF}).  
An alternative proof of   (\ref{ECCPDS}) and  (\ref{ECCPWS}) (see (\ref{CNGENTLNSLHP})) and
farther-reaching results are obtained 
 in sec.\,\ref{ECLNTLHP}  by group-theoretical methods.

\section{Dirac localization and boosts}\label{DLB} Recall that $A_{\rho e}=\operatorname{exp}(\frac{\rho}{2}\sum_{k=1}^3e_k\sigma_k)$  represents the boost in direction $e\in\R^3$, $|e|=1$ with rapidity $\rho$.
The formula from (\ref{ERIUB})(c)
$$\{x\in\R^3: 0\le xe \le  b\}_{A_{\rho e}}=\{x\in\R^3: 0\le xe \le b\operatorname{e}^{|\rho|}\}$$
 determines the region of influence of the boosted region $\{x\in\R^3:0\le xe \le  b\}$, $b>0$ in $\R^3$. By causality this implies for a Dirac state $\psi$ localized in $\{x\in\R^3: 0\le xe \le  b\}$ that the boosted state $W^{\textsc{d}\, pos}(A_{\rho e})\psi$ is localized in $\{x\in\R^3: 0\le xe \le b\operatorname{e}^{|\rho|}\}$ for all $\rho$. Moreover, in case that $e(\psi)<0<-\overline{e}(\psi)$ one has $\psi=\psi^{l}+\psi^{r}$, where $\psi^{l}:=1_{\{x\overline{e}\ge 0\}}\psi$, $\psi^{r}:=1_{\{xe\ge 0\}}\psi$ are not zero. As the foregoing consideration shows, the boosted wave functions $W^{\textsc{d}\, pos}(A_{\rho e})\psi^{l}$ and $W^{\textsc{d}\, pos}(A_{\rho e})\psi^{r}$ are still localized in $\{x\overline{e}\ge 0\}$ and $\{xe\ge 0\}$, respectively. So they do not interfere. Therefore for the study of  the carrier of a boosted Dirac state $W^{\textsc{d}\, pos}(A_{\rho e})\psi$ it is no restriction to assume that $\psi$ is localized in the half-space $\{xe\ge 0\}$.

\begin{The}\label{XCTLRCL} Let $e\in\R^3$ with $|e|=1$,  $\alpha>0$, and $\varsigma\in\{+,-\}$. Let $\psi$ be a Dirac state localized in a bounded region contained  in  $\{x\in\R^3: 0\le xe \le  2\alpha\}$. Then the  statements \emph{(a)\,--\,(d)} are equivalent.
\begin{itemize}
\item[(a)]  $\psi$ is a late-change state, i.e. $2\,\varsigma\,t_{\overline{e}}(\psi)= -\,\overline{e}(\psi)$ 
\item[(b)] $W^{\textsc{d}}(A_{\rho e})\psi$ is localized in $\{x\in\R^3: 0\le xe\le 2\alpha\,\operatorname{e}^{-\varsigma\rho}\}$ for all \, $\varsigma\,\rho\ge 0$.
\item[(c)] There is $c>0$ such that $W^{\textsc{d}}(A_{\rho e})\psi$ is localized in $\{x\in\R^3: 0\le xe\le c\}$ for all \,$\varsigma\,\rho\ge 0$.
\item[(d)] $\psi_{t}$  for $t=\varsigma \alpha$ 
is localized in $\{x\in\R^3:xe\le \alpha\}$.
\end{itemize}
\end{The}
{\it Proof.} We prove the assertion for the case $\varsigma=+$. Then  the case $\varsigma=-$ follows by Eq.\,(\ref{RDBTTRO}), Eq.\,(\ref{BETIR}) applying the former to $\mathcal{T}\psi$. Moreover,  due to rotational symmetry it suffices to consider the special case $e=(0,0,1)$, cf. the  proof of (\ref{ERIUB})(c).\\
\hspace*{6mm}
We start from (b). Then (c) holds trivially. We turn to (c) $\Rightarrow$ (a). By (\ref{DRPS}) and the footnote in sec.\,\ref{FDNWL},
$s(\operatorname{e}^{ \rho\, \sigma_3/2})^{-1} \big(W^{\textsc{d}\,pos}(A_\rho)\psi\big)(x)=\big(\operatorname{e}^{\operatorname{i}\sinh(\rho)x_3H^{pos}}\psi\big)\big(x_1,x_2,\cosh(\rho)x_3\big)$. Introduce $t:=\sinh(\rho) x_3$ for $\rho>0$. Since $x_3\ge c \Leftrightarrow t\ge \sinh(\rho) \,c$, so $\psi_t\big(x_1,x_2, \coth(\rho)\, t\big)=0$ for $t\ge \sinh(\rho) \,c$. Fix $t>0$. Then $J:=\{\coth(\rho)\,t: \sinh(\rho)\le t/c, \rho>0\}=
\{\sqrt{1+x^{-2}}\,t: 0<x\le t/c\}$. As $\{\sqrt{1+x^{-2}}: 0<x\le t/c\}=[\sqrt{1+c^2/t^2},\infty[$, one has $J=[\sqrt{c^2+t^2},\infty[$. Hence one finds $\psi_t(x)=0$ if $x_3\ge \sqrt{c^2+t^2}$, $t\ge0$. In other words, $-\overline{e}(\psi_t)\le \sqrt{c^2+t^2}$\, $\forall$ $t\ge0$. --- 
Apply (\ref{SETIIED}) to $\overline{e}(\psi_t)$ for $t\ge \max\{0,t_{\overline{e}},\gamma\}$ with $\gamma:=-\overline{e}(\psi)-|t_{\overline{e}}|-t_{\overline{e}}$. This yields $t+\gamma\le  \sqrt{c^2+t^2}$ and hence $t^2+2t\gamma +\gamma^2 \le t^2+c^2$. This implies $\gamma\le 0$. As $-\overline{e}(\psi)>0$ it follows 
$t_{\overline{e}}>0$ and then  $2\,t_{\overline{e}}\ge -\overline{e}(\psi)$. Thus (a) holds by (\ref{ILCS})(a).\\
\hspace*{6mm}
We show (a) $\Rightarrow$ (d). By (\ref{SETIIED}) one has $-\overline{e}(\psi_\alpha)=-\overline{e}(\psi)-|t_{\overline{e}}|+|\alpha-t_{\overline{e}}|$. Since $t_{\overline{e}}=-\overline{e}(\psi)/2\le \alpha$,  it follows $-\overline{e}(\psi_\alpha)= \alpha$, whence (d).\\ 
\hspace*{6mm}
It remains to prove (d) $\Rightarrow$ (b). Since $\psi_\alpha$ is localized in $\{x_3\le \alpha\}$ and $\psi$ is localized in $\{0\le x_3\}$, by causality $\psi_t=\psi_{\alpha+t-\alpha}$ is localized in $\{-|t|\le x_3\le \alpha+|t-\alpha|\}$ for every $t$. Therefore $\psi_t(x)=0$ for $t\le 0$, $x_3\le t$ or $t\ge 0$, $x_3\ge \alpha+|t-\alpha|$. This implies that $\psi_t\big(x_1,x_2, \coth(\rho)t\big)=0$ if $t<0$, $\rho>0$ or $t\ge (1-\operatorname{e}^{-2\rho})\,\alpha$, $\rho>0$ (since the latter is equivalent to $\coth(\rho) t\ge 2\alpha-t$, $\rho>0$). Now introducing $x_3:=t/\sinh(\rho)$ one has $\psi_{\sinh(\rho)x_3}\big(x_1,x_2,\cosh(\rho)x_3\big)=0$ for $x_3<0$, $\rho>0$ or $x_3\ge 2\alpha\,\operatorname{e}^{-\rho}$, $\rho>0$. Hence (b) follows.\qed\\

 Recall (\ref{L-CS}) for the introduction and the existence of late-change states. Furthermore
 see the  comment following (\ref{LBSLSET}) and the corollary (\ref{DLTES}). \\
\hspace*{6mm}
 Most remarkable in (\ref{XCTLRCL})  is the equivalence (b) $\Leftrightarrow$ (d) establishing a close relationship between the localization of the boosted state and  the state  evolved in time. It gives rise to the formula in (\ref{LBSLSET}). Recall also (\ref{BLCS}) by which (\ref{XCTLRCL})(b) can be replaced by
 $e(W^{\textsc{d}}(A_\rho)\psi)=0$, $\overline{e}(W^{\textsc{d}}(A_\rho)\psi)=\e^{-\varsigma \rho}\overline{e}(\psi)$ for all $\rho\in\R$.
 
 \begin{Cor}\label{LBSLSET} Let $e\in\R^3$ with  $|e|=1$,  $\alpha>0$,  and $\varsigma\in\{+,-\}$. Then, for $W=W^{\textsc{d}}, E=E^{\textsc{d}}$, one has
 $$ W(\varsigma\alpha)E(\{xe\le \alpha\})W(\varsigma\alpha)^{-1}E(\{xe\ge 0\})= \lim_{\varsigma\rho\to\infty} W(A_{\rho e})E(\{ 0\le xe\le 2\alpha\,\operatorname{e}^{-\varsigma\rho}\}) W(A_{\rho e})^{-1}$$
 where  the limit exists  since the projections decrease as $\varsigma\rho$ increases.
 \end{Cor}\\
  {\it Proof.} Like for the proof of (\ref{XCTLRCL}) we may restrict ourselves to the case $\varsigma=+$ and $e=(0,0,1)$. Then the projections on the right hand side read $E(\Gamma_\rho)$ for $\Gamma_\rho:=A_\rho\cdot \{0\le x_3\le2\alpha\operatorname{e}^{-\rho}\}$. Check $\hat{\Gamma}_\rho=\{-2\alpha\operatorname{e}^{-2\rho}\le x_0-x_3\le 0,0\le x_0+x_3\le 2\alpha\}$. Hence $\hat{\Gamma}_{\rho'}\subset \hat{\Gamma}_\rho$ for $\rho<\rho'$, whence $E(\Gamma_{\rho'})\le E(\Gamma_\rho)$ by (\ref{BOD}). Therefore the limit $\lim_{\rho\to\infty}E(\Gamma_\rho)$ exists yielding a projection $P_r$. Furthermore, according to Eq.\,(\ref{ALHPTLC}), Eq.\,(\ref{BLHPTLC}), the left hand side, denoted by $P_l$, is a projection. In particular, $E(\{x_3\ge 0\})$ commutes with $W(\alpha)E(\{x_3\le \alpha\})W(\alpha)^{-1}$.\\ 
\hspace*{6mm}  
 Suppose $P_l\varphi=\varphi$. The claim is $P_r\varphi=\varphi$. It is no restriction to assume that $\varphi$ is  bounded localized. Indeed, there is a bounded localized  $\varphi'$ with $\norm{\varphi-\varphi'}<\varepsilon$. By assumption $P_rP_l\varphi'=P_l\varphi'$ since $P_l\varphi'$ is bounded localized  by (\ref{CCPDS}). Then $\norm{P_r\varphi-\varphi}\le\norm{P_r\varphi-P_rP_l\varphi'}+\norm{P_rP_l\varphi'-\varphi}\le \norm{\varphi-P_l\varphi'}+\norm{P_l\varphi'-\varphi} =2\norm{P_l\varphi'-P_l\varphi}\le2\varepsilon$. --- Thus suppose $P_l\varphi=\varphi$ and $\varphi$ bounded localized.
Then $\varphi$ is localized in  $\{x_3\ge 0\}$ and $\varphi_{-\alpha}$ is localized in $\{x_3\le \alpha\}$. The latter implies by causality that $\varphi$ is localized  also in  $\{x_3\le 2\alpha\}$. Hence we may apply (\ref{XCTLRCL})(d)$\Rightarrow$(b) for $\varsigma=-1$. This yields   $E(\{ 0\le x_3\le 2\alpha\,\operatorname{e}^{\rho}\})W(A_\rho)\varphi=W(A_\rho)\varphi$ for all \, $-\rho\ge 0$. Therefore $E(\Gamma_\rho)\varphi=\varphi$ for all $\rho>0$. \\
\hspace*{6mm}
 Conversely, suppose now  $P_r\varphi=\varphi$. Recall $E(\Gamma_{\rho'})\le E(\Gamma_\rho)$ for $\rho<\rho'$. Then, $\varphi$ is localized in $\{0\le x_3\le 2\alpha\}$ since $\varphi=E(\Gamma_0)\varphi$, and 
$W(A_\rho)^{-1}\varphi=W(A_{-\rho})\varphi$ is localized in $\{ 0\le x_3\le 2\alpha\,\operatorname{e}^{-\rho}\}$ since  $\varphi=E(\Gamma_\rho)\varphi$.  In particular $P_r\varphi$ is bounded localized if  $\varphi$ is so.  Thus assume as above without restriction that $\varphi$ is bounded localized.
Hence we may apply (\ref{XCTLRCL})(b)$\Rightarrow$(d) for $\varsigma=-1$. Then $W(\alpha)E(\{x_3\le \alpha\})W(\alpha)^{-1}\varphi=\varphi$. Now  recall that $\varphi$ is localized in $\{x_3\ge 0\}$. $P_l\varphi=\varphi$ follows.\qed\\

  Applying  Eq.\,(\ref{BLHPTLC}), Eq.\,(\ref{ALHPTLC}) to the projection $P_l$ at the left hand side of (\ref{LBSLSET}) it follows from  (\ref{CCIDES}) that  $P_l$ is non-trivial and has infinite dimensional eigenspaces. 
So, only now we know that (the scalar multiples of) the large\,-\,$t_{\overline{e}}$\,-\,states  for $t_{\overline{e}}>0$ (as well for $t_{\overline{e}}<0$) localized in $\{ 0\le xe\le 2\alpha\}$, $\alpha>0$ actually  form 
 an infinite dimensional subspace, namely  the range of $P_l$.
 
\begin{Cor}\label{DLTES} For every direction $e$ the large\,-\,$t_{\overline{e}}$\,-\,states for $t_{\overline{e}}> 0$ (as well for $t_{\overline{e}}< 0$) are dense in the set of states localized in $\{x\in\R^3:xe\ge 0\}$.
\end{Cor}\\
{\it Proof.} Let the Dirac state $\psi$ be localized in $\{x\in\R^3:xe\ge 0\}$ and let $\varepsilon>0$. Choose a state $\psi'$ localized in a bounded region contained in $\{xe\ge 0\}$ with $\norm{\psi-\psi'}\le \varepsilon/2$. For $\varsigma\in\{1,-1\}$  choose $\alpha>0$ according to (\ref{CCPDS}) such that $\norm{\psi'-\psi'^\alpha}\le \varepsilon/2$. Note that by (\ref{CCPDS}) and by (\ref{XCTLRCL}),(d) $\Rightarrow$ (a) the state  $\psi'^\alpha$ is a 
large\,-\,$t_{\overline{e}}$\,-\,state.\qed

\subsection{Lorentz contraction}\label{LCDS} 
By (\ref{XCTLRCL})(a) it is clear that the cases $\varsigma=+$ and $\varsigma=-$ exclude each other. Consequently by (\ref{XCTLRCL})(c), there is no Dirac state $\psi$ localized in a bounded region such that the boosted states $W^{\textsc{d}\, pos}(A_{\rho e})\psi$, $\rho\in\R$ stay confined in any strip $\{x\in\R^3:a\le xe\le b\}$. If $|t_{\overline{e}}|<\frac{1}{2}(-\overline{e}(\psi)-e(\psi))$ then  $W^{\textsc{d}\, pos}(A_{\rho e})\psi$ is confined neither for $\rho\to \infty$ nor for $\rho\to - \infty$.\\
\hspace*{6mm}
Nevertheless, a series of numerical computations performed for the Dirac equation in one-dimensional discrete space by Mutze 2016 \cite{M16} supports  the idea that, for every Dirac state $\psi$, the  probability of localization of the boosted states $W^{\textsc{d}\, pos}(A_{\rho e})\psi$  in a strip $\{-\delta\le xe\le \delta\}$  around the origin tends to $1$ if the rapidity $\rho$ tends  to $\infty$ or $-\infty$. We like to call  this behavior  the Lorentz contraction of  the Dirac wave functions, which indeed is proven to hold true  by the following

\begin{The}\label{LCDWF} Let $\psi$ be any Dirac state, in particular an electron or positron state. Let $e\in\R^3$, $|e|=1$ be a direction. Then for every   $\delta>0$ 
$$\norm{E^{\textsc{d}}\big(\{x\in\R^3: -\delta\le xe \le \delta\}\big)\,W^{\textsc{d}}(A_{\rho e})\psi}\, \to 1, \;\; |\rho|\to\infty$$
\end{The}\\
{\it Proof.}  As argued at the beginning of sec.\,\ref{DLB} it is no restriction to assume that $\psi$ is localized in 
$\{xe\ge 0\}$. Then $W^{\textsc{d}}(A_{\rho e})\psi$ stays localized there for all $\rho$. Let $\varepsilon>0$. 
According to (\ref{DLTES}) choose a  large\,-\,$t_{\overline{e}}$\,-\,state $\psi'$ localized in $\{0\le  xe\le c\}$ for some $c>0$ with $\norm{\psi-\psi'}\le \varepsilon$. Let $Q_\rho:=E^{\textsc{d}}\big(\{ xe > \delta\}\big)\,W^{\textsc{d}}(A_{\rho e})$. 
Then, for all $\rho\le\rho_-:=\min\{0,\ln\frac{\delta}{c}\}$, one has $Q_\rho\psi'=0$ and hence $\norm{Q_\rho\psi} \le   \norm{Q_\rho \psi- Q_\rho\psi'}+\norm{Q_\rho\psi'}\le     \norm{\psi-\psi'}\le \varepsilon$.\\
\hspace*{6mm}
Now apply this result to $\mathcal{T}\psi$ with $\mathcal{T}$ the time reversal operator (see (\ref{TIR})). Accordingly  
$ \norm{Q_\rho\mathcal{T}\psi}\le\varepsilon$ for all $\rho\le -\rho_+$ for some $\rho_+\ge 0$. 
Since $Q_\rho\mathcal{T}=\mathcal{T}^{-1}Q_{-\rho}$ by Eqs.\,(\ref{RDBTTRO}), (\ref{BETIR}), 
 it follows $ \norm{Q_\rho\psi}\le\varepsilon$ for $\rho\ge \rho_+$, thus accomplishing the proof.\qed\\

Let $\psi$ be any Dirac state localized in the half-space $\{xe\ge 0\}$. As shown in the proof of (\ref{LCDWF}) there is an appropriately prepared  large\,-\,$t_{\overline{e}}$\,-\,state $\chi^\alpha$  localized in a bounded region  contained in the strip $\{0\le xe\le 2\alpha\}$, $\alpha>0$, which approximates $\psi$ arbitrarily well such that the probability of localization of $\psi$ in the carrier of $\chi^\alpha$ is almost one.  Due to the unitarity of the transformation   the latter  holds true also for the respective boosted states. The particular feature of $\chi^\alpha$ is that for every $\rho\ge 0$ the boosted state $W^{\textsc{d}\, pos}(A_{\rho e})\chi^\alpha$ is localized in the strip $\{0\le xe\le 2\alpha \operatorname{e}^{-\rho}\}$. This shows a Lorentz factor $\operatorname{e}^{\rho}$ which for large $\rho$ is almost twice the classical one $\gamma=\cosh(\rho)$.\\
\hspace*{6mm}
 What happens to the carrier of $\chi^\alpha$ in directions perpendicular to that of the boost? 
 Let $c>0$ such that $\chi^\alpha$ is localized in the cylinder $\Gamma:=\{x\in\R^3:0\le xe \le 2\alpha, |x-(xe)e|\le c\}$ with axis $\R e$, length $2\alpha$ and radius $c$. Then, due to causality, $W^{\textsc{d}\, pos}(A_{\rho e})\chi^\alpha$ is localized in $\{0\le xe\le 2\alpha \operatorname{e}^{-\rho}\}\cap \Gamma_{A_{\rho e}}$, which by (\ref{ERIUB})(d) is a truncated cone with axis $\R e$, length $2\alpha\operatorname{e}^{-\rho}$,   and radii  $c$ and $c+
\alpha(1-\operatorname{e}^{-2\rho})$ at $0$ and $ 2\alpha \operatorname{e}^{-\rho}e$, respectively.   Hence for $\rho \to \infty$ the carrier shrinks to the disc $\{x\in\R^3:xe=0, |x|\le c+\alpha\}$.

\subsection{Frame-dependence of Lorentz contraction}\label{FDOLC}
In classical mechanics it is widely discussed   whether the  length contraction  really exists or not. According to Einstein 1911 \cite{E11} it is a misleading question as  length contraction  does not \textquotedblleft really\textquotedblright\,exist, in so far as it does not exist for a comoving observer; though it \textquotedblleft really\textquotedblright\,exists, i.e. in such a way that it could be demonstrated in principle by physical means by a non-comoving observer.\\
\hspace*{6mm}
 We turn to the question about the reality of  the Lorentz contraction of a Dirac wave function. In order to be clear let us first recall briefly some well-known general facts (see e.g. \cite[sec.\,VIII]{A69}). Let $\mathfrak{R}$ denote the origin and the standard basis $\big(\mathfrak{o},(\mathfrak{e}^0,\mathfrak{e}^1,\mathfrak{e}^2,\mathfrak{e}^3)\big)$ of $\R^4$ as well the reference frame related to it. 
  The  description of the physical system will refer to  $\mathfrak{R}$
 in determining the representation of its states $S$ and observables $O$   by unit vectors $\psi$ and self-adjoint operators $A$, respectively,  in the Hilbert space of states.  The outcomes of measurements are the expectation values 
 $\langle \psi, A \psi\rangle$. The state space is unique up to Hilbert space isomorphisms $\iota$, as $\iota\psi$ and  $\iota A \iota^{-1}$ yield the same expectation values as $\psi$ and $A$, i.e.,  $ \langle \iota\psi, (\iota A \iota^{-1})\,\iota \psi\rangle=  \langle \psi, A \psi\rangle$. So every $\iota$ determines another representation of the states $S$ and observables $O$ of the system with 
  $\iota\psi$ and  $\iota A \iota^{-1}$ denoting the same state and observable, respectively, as  $\psi$ and $A$ do. This freedom plays a  decisive role in the description of symmetries.  In particular \textbf{relativistic symmetry} gives rise to a representation $W$ of $\tilde{\mathcal{P}}$ in the state space.\\
\hspace*{6mm}
Let $\mathfrak{R}'\equiv g^{-1}\cdot\mathfrak{R}$
 with $g\in \tilde{\mathcal{P}}$ be any other  reference frame.   $\mathfrak{R}'$ is the frame for which $g\cdot \mathfrak{x}$ denotes the same event as $\mathfrak{x}$ for 
$\mathfrak{R}$. (Recall that  
 $g\in \tilde{\mathcal{P}}$ acts as the Poincar\'e transformation Eq.\,(\ref{PTUCH}). Its coordinates refer to $\mathfrak{R}$ and are the same with respect to $\mathfrak{R}'$.) Consequently a Poincar\'e transformation described by $h\in \tilde{\mathcal{P}}$ with respect to 
 $\mathfrak{R}$ reads $h':=ghg^{-1}$ with respect to  
 $\mathfrak{R}'$. If a state $S$ and an observable $O$ are described by $\psi$ and $A$ with respect to  
 $\mathfrak{R}$, then they are described by $\psi':=W(g) \psi$ and $A':=W(g)\,A\,W(g)^{-1}$ with respect to  
 $\mathfrak{R}'$. In this way invariance  of the expectation values  under change of the reference frame is ensured. Similarly, since the Poincar\'e transformation $h$ is represented by $W(h)$ with respect to 
 $\mathfrak{R}$, its description is $W(g)W(h)W(g)^{-1}=W(ghg^{-1})$ with respect to 
 $\mathfrak{R}'$. Finally, what looks like $W'$ denoting the representation $W$ when described with respect to 
$\mathfrak{R}'$? Recall that $h'=ghg^{-1}$ denotes the Poincar\'e transformation $h$ when described in $\mathfrak{R}'$. Hence, by the foregoing result, $W'(h')$ has to equal $W(ghg^{-1})$. So one ends up with
\begin{equation} \label{STPR}
W'(k)=W(k)\quad \forall\; k\in\tilde{\mathcal{P}}
\end{equation} 
where $k$ is expressed on both sides by the coordinates of $\mathfrak{R}'$.
 
\hspace*{6mm}
Back to the initial  question,  note first that the Lorentz contraction (\ref{LCDWF}) can be observed. Indeed, let $S$ be any Dirac state   represented by the state vector $\psi$. For given $\varepsilon>0$ consider the observable $O$ described by $A:=E^{\textsc{d}}(\{|xe|\le\delta\})$ with $\delta> 0$ so small that  $\norm{E^{\textsc{d}}(\{|xe|\le\delta\})\psi}^2 \,\le \varepsilon$. According to (\ref{LCDWF}) there is $\tilde{\rho}>0$ such that $||E^{\textsc{d}}\big(\{  |xe| \le \delta\}\big)\,W^{\textsc{d}}(A_{\rho e})\psi||^2\ge 1-\varepsilon$ for all $\rho\ge\tilde{\rho}$. Let $\tilde{S}$ be the boosted state described  by $\tilde{\psi}:=W^{\textsc{d}}(A_{\tilde{\rho}e})\psi$. Then
\begin{equation}\label{LCDSM} 
\langle \psi,A \psi\rangle\le \varepsilon\;\textrm{ and }\;\langle \tilde{\psi},A \tilde{\psi}\rangle\ge 1-\varepsilon
\end{equation}
 Hence the observable $O$ distinguishes the state $S$ from the boosted state $\tilde{S}$. So the observer related to $\mathfrak{R}$ can ascertain the Lorentz contraction of the Dirac system.\\
\hspace*{6mm} 
What are the ascertainments  of an observer related to $\mathfrak{R}'$ provided with the  localization $(E^{\textsc{d}})'$? 
 By (\ref{PRWL}),\,(\ref{STPR}) one has $A'=E^{\textsc{d}}(g\cdot\{x_0=0,|xe|\le\delta\})$,  $\tilde{\psi}'=W^{\textsc{d}}(h')\psi'$ for $h'=ghg^{-1}$, $h:=A_{\tilde{\rho}e}$. Then $\langle \psi',A'\psi'\rangle\le\varepsilon$, $\langle \tilde{\psi}',A'\tilde{\psi}'\rangle\ge 1-\varepsilon$. Hence, observed from $\mathfrak{R}'$, the Dirac system in the state $S$ is highly localized in the spacelike region $g\cdot\{\mathfrak{x}: x_0=0, |xe|>\delta\}$, whereas in the boosted state $\tilde{S}$ it is highly localized in $g\cdot\{\mathfrak{x}:x_0=0, |xe|\le\delta\}$. The expected conclusion is that, due to relativistic symmetry,  the  Lorentz contraction of the Dirac system can be ascertained  in the same way and with the same result by any Lorentz observer.\\
\hspace*{6mm}
 On the other hand there is the \textbf{frame-dependence} of the Lorentz contraction, which is discussed now. By virtue of the Poincar\'e covariance of the localization no reference to a moving observer is needed.   Let a four vector $\mathfrak{e}$ be called a spacelike direction if 
$\mathfrak{e}\cdot \mathfrak{e}=-1$.

\begin{Cor} \label{FDLCDL} Let $\psi$ be a  Dirac state. Let $\sigma$ be a spacelike hyperplane and $\mathfrak{e}$ a spacelike direction parallel to $\sigma$. Boost them along $\mathfrak{e}$ with rapidity $\rho$ obtaining 
$\sigma_\rho$ and $\mathfrak{e}_\rho$. Then $$||E^{\textsc{d}}\big(\{\mathfrak{x}\in\sigma_\rho: |(\mathfrak{x}-\mathfrak{o})\cdot \mathfrak{e}_\rho|\le \delta\}\big)\psi ||\to 1 \textrm{ for } |\rho|\to \infty$$ where  $\mathfrak{o}\in\sigma$ is the fixed point of the boost.
 \end{Cor}\\
{\it Proof.}  Due to relativistic symmetry we may identify $\sigma$ with $\{0\}\times \R^3\subset \R^4$, $\mathfrak{o}$ with the origin $0$, and $\mathfrak{e}$ with $(0,e)$. Then the claim follows immediately from (\ref{LCDWF}) due to the covariance of $E^{\textsc{d}}$, by which $||E^{\textsc{d}}\big(A_{\rho e}\cdot\{|xe| \le \delta\}\big)\,\psi||^2 =||E^{\textsc{d}}\big(\{  |xe| \le \delta\}\big)\,W^{\textsc{d}}(A_{-\rho e})\psi||^2$.
\qed
 
 Thus, if the frame is moving fast enough depending on the state, then  the Dirac system is highly localized in a narrow strip perpendicular to the direction of  motion.
 One likes to remember the frame-dependence of the NWL (\ref{CFDNWL}): As soon as the frame of the bounded localized Newton-Wigner system moves, the latter  is no longer localized in any bounded region. Nevertheless, presumably the Newton-Wigner system undergoes the same Lorentz contraction (\ref{FDLCDL}) as the Dirac system. \\
\hspace*{6mm}
 The frame-dependence of the Lorentz contraction in classical  mechanics is striking by the fact that for the comoving observer it does not even exist. The same holds true for the Lorentz contraction of the Dirac wavefunctions. Moreover, due to  the Poincar\'e covariance of the localization no reference to a moving observer is needed,  but the fact refers to the expectation value of the corresponding localization observable. Indeed, recalling (\ref{LCDSM}), let $\mathcal{A}$ be an apparatus realizing  the localization observable $O$ described by $A$. Boost $\mathcal{A}$ according to  $A_{\tilde{\rho}e}$ thus obtaining the comoving apparatus $\tilde{\mathcal{A}}$. It realizes the localization observable $\tilde{O}$ for the space-like region $A_{\rho e}\cdot\{|xe| \le \delta\}$ (to which a comoving observer refers) described by $\tilde{A}:=E^{\textsc{d}}\big(A_{\rho e}\cdot\{|xe| \le \delta\}\big)$. Then due to the covariance of localization
 \begin{equation} \label{NLCFCM}
\langle \tilde{\psi},A \tilde{\psi}\rangle\ge 1-\varepsilon\;\textrm{ and }\;  \langle \tilde{\psi},\tilde{A} \tilde{\psi}\rangle\le \varepsilon
\end{equation} 
holds. This means that the non-comoving apparatus $\mathcal{A}$ ascertains the Lorentz contraction of the Dirac system whereas the comoving apparatus $\tilde{\mathcal{A}}$ ascertains non-contraction.

\hspace*{6mm}
In view of these clear results it is worth to consider also the particular reference frame 
$\mathfrak{R}':=A_{\tilde{\rho}e}\cdot \mathfrak{R}$. Imagine an object resting at $0$ with respect to 
 $\mathfrak{R}$, i.e., with world line  $t\mapsto t(1,0,0,0)$. When boosted according to $A_{\tilde{\rho}e}$ it moves with velocity  $\tilde{v}=\tanh(\tilde{\rho})$. Its world line is $t\mapsto t \cosh(\tilde{\rho})(1,\tilde{v}e)$. Hence 
  $\mathfrak{R}'$ is the rest frame of this boosted object. \\
\hspace*{6mm}
Now, with respect to 
 $\mathfrak{R}'$ the boosted state 
 $\tilde{S}$ is represented by $\tilde{\psi}'=W^{\textsc{d}}(A_{-\tilde{\rho}}) \tilde{\psi}=\psi$. Accordingly one is tempted to consider  
  $\mathfrak{R}'$ to be the rest frame of $S$. But  this interpretation seems to make sense only in the context of the history of the system. Anyway, the discerning observable $O$ is no longer represented by $A$ but by $A'=W^{\textsc{d}}(A_{-\tilde{\rho}e}) A W^{\textsc{d}}(A_{-\tilde{\rho}e})^{-1}$ so that $\langle \psi,A' \psi\rangle\ge 1-\varepsilon$. By the Poincar\'e covariance of the localization one has $A'=E^{\textsc{d}}(A_{-\tilde{\rho}e}\cdot \{x_0=0, |xe|\le \delta\})$. 
 Thus  $\langle \psi,A' \psi\rangle\ge 1-\varepsilon$ means also that the Dirac system in the state $\tilde{S}$ is almost localized  in the spacelike region  $\{\mathfrak{x}'\in\R^4:-\delta\le -x'_0/\sinh \tilde{\rho}=x'e/\cosh \tilde{\rho}\le\delta\}$ in Minkowski space.

\section{Causal   PO-Localization for the Electron } \label{CEL} 

 By (\ref{NLS}) there is no causal WL for the electron. But the trace  $T^e$ of  $E^{\textsc{d}}$  on the positive energy subspace of  $W^{\textsc{d}}$ yields a causal POL  for the Dirac electron by  (\ref{TCWL}).  Let $P^+$ denote the projection on  the subspace of positive energy. Then
 \begin{equation}\label{EPOLDE} 
T^e(\Gamma)\varphi= P^+E^\textsc{d}(\Gamma)\varphi
\end{equation} 
for all measurable subsets $\Gamma$ of spacelike hyperplanes and electron states $\varphi$. 
\\

  \subsection{Properties of $T^e$}\label{PTE} 
  For the following compare \cite[Corollary 1.7]{T92}. Call $\operatorname{supp}(\psi)=\bigcap_{\psi'\in[\psi]}\overline{\{\psi'\ne 0\}}$ the support of $\psi$. There is $\psi'\in[\psi]$ satisfying $\operatorname{supp}(\psi)=\overline{\{\psi'\ne 0\}}$.\footnote{ Generally, let $\Delta\subset \R^d$ be measurable and set $C:=\bigcap\{\overline{\Delta\setminus N}: N \textrm{ null set}\}$, 
 $C':=\{x\in\R^d: \lambda(\Delta\cap U)>0\,\forall \, U \textrm{ open, } x\in U\}$, and $C'':=\R^d\setminus \bigcup\{U \textrm{ open}: \lambda(U\cap \Delta)=0\}$ with $\lambda$ the Lebesgue measure. Then $C=C'=C''$. Moreover, $N:=\Delta\setminus C$ is a null set and $C=\overline{\Delta\setminus N}$. --- Indeed, one easily checks $C'=C''$.  If $x\in C'$ then $\lambda(\Delta \cap U)>0$ for all  open $U$ with $x\in U$. This implies that $(\Delta\setminus N) \cap U\ne \emptyset$ for every null set $N$. Hence $x\in \overline{\Delta\setminus N}$ for all $N$. Therefore $x\in C$. Thus $C'\subset C$.  For the proof of the reverse inclusion consider any countable base $\mathcal{B}$ of $\R^d$ and put $V:=\bigcup\{U\in\mathcal{B}: \lambda(U\cap \Delta)=0\}$. Obviously $\Delta \cap V$ is a null set. Let $N\supset \Delta \cap V$  be any null set and let $x\in\overline{\Delta\setminus N}$. Then  for every open $U$ with $x\in U$ there is $x'\in (\Delta\setminus V)\cap U$. One infers $\lambda(\Delta\cap U)>0$. Hence $x\in C'$. Thus $C\subset C'$. Finally,  note $V=\R^d\setminus C$. So $N:=\Delta\setminus C$ is a null set. Therefore $C\subset \overline{\Delta\setminus N}=\overline{C\cap N}\subset C$, whence the last part of the assertion.  --- The question remains open whether  $C\setminus \Delta$ is a null set.}

\begin{Cor} \label{SES} Let $\psi$ be an electron state in position space representation and  let $\Delta\ne \R^3$  be closed. Then $T^{e \,pos}(\Delta)\psi \ne \psi$ and
 the support of $\psi$ is $\R^3$.
  \end{Cor}\\
{\it Proof.}  $T^{e \,pos}(\Delta)\psi \ne \psi$ holds
by (\ref{NLS}). 
Hence $E^{\textsc{d}pos}(\Delta)\psi\ne\psi$ since $P^{+pos}\psi =\psi$. On the other hand $1_{ \overline{\{\psi'\ne 0\}}} \psi'=\psi'$ for every $\psi'\in[\psi]$, whence $E^{\textsc{d}pos}(\overline{\{\psi'\ne 0\}})\psi=\psi$. Therefore $\overline{\{\psi'\ne 0\}}=\R^3$, which ends the proof.\qed\\

We show now by (\ref{SPOL})  that $T^e$  is separated. The result (\ref{POLDE}) is a particular case of \cite[Theorem 8]{CL15}. In momentum space  
 one has
 $$\big(P^{+mom}\varphi\big)(p) = \pi^+(p)\varphi(p), \quad\: 
   \pi^+ (p):=
 \frac{1}{2}\left( \text{{\it \small{I}}}_4+ 
\frac{1}{\epsilon(p)} h(p)\right)$$
 with $\epsilon(p):=\sqrt{|p|^2+m^2}$. Since 
$(D_\lambda^{mom} \varphi)(p)= \lambda^\frac{3}{2} \varphi(\lambda p)$, for   $P^+_\lambda := D_\lambda P^+ D_\lambda^{-1}$ one finds $\big(P_ \lambda^{+mom}\varphi\big)(p)=
 \pi^+(\lambda p)\varphi( p)$.
 For every $p\in\R^3$, the matrices $\pi^+(\lambda p)$ and 
  $$\pi_0 (p):=
 \frac{1}{2}\left( \text{{\it \small{I}}}_4+ 
\frac{1}{|p|} \sum_k\alpha_k p_k\right)$$ 
  are orthogonal projection 
with $\pi^+(\lambda p)\to \pi_0(p)$ for $\lambda\to\infty$. Then  $\big(Q^{mom}\varphi\big)(p) :=\pi_0 (p)\varphi(p)$ defines a non-zero orthogonal projection operator on $L^2(\R^3,\C^4)$.
By dominated convergence $P^+_ \lambda \to Q$ strongly as $\lambda\to\infty$. 
Thus the premises of  (\ref{SPOL})  hold true and one has

\begin{The}\label{POLDE} The trace  $T^e$ of  $E^{\textsc{d}}$  on the positive energy subspace is a causal separated POL of the Dirac electron.  If  $\phi$ is any Dirac state with $Q\,\phi\ne 0$, then $P^+D_n^{-1}\phi\ne 0$ up to finitely many $n$ and 
$$\varphi_n:=\frac{1}{||P^+D_n^{-1}\phi ||} \, P^+D_n^{-1}\phi $$
 yields a sequence of electron states localized at $0$.
\end{The}

One may imagine a point-localized sequence $(\varphi_n)$ of electron states  to be a progressive preparation of the system in order to get an increasing localization around a point. Regarding  the sequences in (\ref{POLDE}), the result  in (\ref{IEPLS}), which is a special case of 
\cite[Theorem 9]{CL15}, shows  that this process needs unlimited energy. --- Let $|P|$ denote, in  momentum representation,  the multiplication operator by $|p|$.  

\begin{Lem}\label{IEPLS}
Let $\phi$ be a Dirac state with $Q\,\phi\ne 0$. Consider  $\varphi_n$  from \emph{(\ref{POLDE})}. Then, for every $n$,   
$\varphi_n\in \operatorname{dom}(H)$ if and only if $Q\,\phi \in \operatorname{dom}(|P|)$. Let $Q\,\phi \in \operatorname{dom}(|P|)$. Then
$$\frac{1}{n}\langle \varphi_n, H\varphi_n\rangle \to \frac{1}{ ||Q\,\phi ||^{2}}  \langle Q\,\phi,|P| \,Q\,\phi\rangle  > 0   \quad  \text{ for }\; n\to \infty $$
\end{Lem}\\
{\it Proof.}  We work in momentum representation omitting the superscript {\it mom}. --- Let $p\in\R^3$.  One easily verifies that the positive matrices $\text{{\it \small{M}}}_0(p):=
|p| \pi_0(p)$ and  $\text{{\it \small{M}}}_n(p):=\frac{\epsilon(np)}{n}\pi^+(np) $  satisfy 
$$\text{{\it \small{M}}}_0(p)\le \text{{\it \small{M}}}_n(p) +mI_4,\quad \text{{\it \small{M}}}_n(p)\le \text{{\it \small{M}}}_0(p)+mI_4$$
Now note that $|P|Q$ is the matrix multiplication operator by $ \text{{\it \small{M}}}_0$. Since  $HP^+$ is the matrix multiplication operator by $ \epsilon(\cdot) \pi^+$, it follows that 
$\frac{1}{n}D_nH P^+D_n^{-1}$ is the matrix multiplication operator by $ \text{{\it \small{M}}}_n $. The first part of the assertion follows. ---  As to the second part, $\frac{1}{n}\langle \varphi_n, H\varphi_n\rangle=\frac{1}{n} \langle D_n\varphi_n, D_nH\varphi_n\rangle =  ||P^+_n\phi ||^{-2}  \int \frac{\epsilon(np)}{n}   \big|\pi^+(np)\phi(p) \big|^2 \operatorname{d}^3p$, which  by dominated convergence tends to $||Q\phi ||^{-2}  \int  |p| \big| \pi_0(p)\phi(p) \big|^2 \operatorname{d}^3p=||Q\phi ||^{-2}  \langle Q\phi,|P|Q\phi\rangle $. Finally, $ \langle Q\,\phi,|P|Q\,\phi\rangle >0$ as   $|P|Q\,\phi \ne 0$.\qed\\

As to the relevance of (\ref{IEPLS}) one notes that  the assumption $Q\,\phi \in \operatorname{dom}(|P|)\setminus \{0\}$ is  satisfied by all $\phi =(I+|P|)^{-1}\phi'$ with $Q\phi'\ne 0$, since $Q$ commutes with $|P|$. 
Presumably, the result $\lim_{n\to\infty} \langle \varphi_n, H\varphi_n\rangle =\infty$ obtained in (\ref{IEPLS}) holds for every point-localized sequence $(\varphi_n)$ of electron states.\\


\subsection{Point-localized sequences of electron states}
 As to the behavior of point-localized  sequences of  electron states with respect to $T^e$ under Poincar\'e transformations we recall the general result  (\ref{PCCPLSS}).
Certain point-localized sequences of electron states (see Eq.\,(\ref{PLSES}) below)   have already been  constructed and studied in great detail  in   \cite{BM99} 1999 and     \cite{BFM05} 2005. The concept is further formalized in   \cite{M02} 2001. Essentially this section  is taken from \cite[sec.\,I]{CL15}.

For the construction of  point-localized  sequences of electron states we provide
\begin{equation}\label{DHO}
   h(p)  =    \epsilon(p)\;              y(p)^{-1}  \, \gamma_5   \,    y(p), \quad y(p):= \frac{1}{2}\sqrt{\frac{\epsilon(p)}{\epsilon(p)+m}}(I_4+\gamma_5\beta)      \left(\beta+\frac{1}{\epsilon(p)} h(p)   \right)
\end{equation}
with $\gamma_5=-\operatorname{i}\alpha_1\alpha_2\alpha_3$. Note that $\gamma_5$ equals $\operatorname{diag}(I_2,-I_2)$ in the Weyl representation. Moreover, $y(p)$ is  self-adjoint unitary, whence $y(p)^{-1}=y(p)$. Hence, if $c(p)\in\C^4$ with $||c(p)||=1$ and $c(p)=\gamma_5 c(p)$, then $u(p):=  y(p)^{-1}c(p)$ satisfies 
$h(p)  u(p)=\epsilon(p) u(p)$ and $\pi^+(p)  u(p)=u(p)$.\\
\hspace*{6mm} Now,  in addition,  let $c(p)$ be constant or, more generally,  choose $c(p),\; p\in\R^3$ such that $c_0(p):=\lim_{\lambda\to \infty}c(\lambda p)$ exists and satisfies $c_0(\lambda p)=c_0(p)$ for  $\lambda>0$. Then 
\begin{equation}\label{DIG}
  u(\lambda p) \to u_0(p):=\frac{1}{2} (I_4+\gamma_5\beta)\left(\beta+\frac{1}{|p|}\sum_k\alpha_k p_k\right)\,c_0(p) 
\end{equation}
and
\begin{equation}\label{DIEV}
 \pi_0 (p)u_0(p)=u_0(p),\quad u_0(\lambda p)=u_0(p), \,\lambda>0
\end{equation}

\begin{Exa} \label{EPLSES}   
The formula for $(\varphi_n)$  in (\ref{POLDE}) yields rather general  point-localized sequences of electron states. Multiplying $\varphi_n$ in momentum representation by $ \operatorname{e}^{-\operatorname{i}bp}$, one obtains a sequence   localized at $b$.
For some concrete examples in momentum representation 
we specialize $\phi:=f\,u_0$ with $f\in L^2(\R^3)$,  $||f||_2=1$  and $u_0:\R^3\to \C^4$ measurable, bounded, and
 $\pi_0(p)\,u_0(p) = 0$ not a.e. The latter guarantees  $Q^{mom} \phi\ne 0$. \\
 \hspace*{6mm} The simplest choice is $u_0\in \C^4\setminus \{0\}$ constant. Another useful  choice is a dilation invariant   unit eigenvector $u_0(p)$ of  
$\pi_0(p)$ (cf.\;Eq.\,(\ref{DIEV})).
Then  (\ref{POLDE})  yields, up to the normalizing constant,   the localized sequence at $b$
\begin{equation}\label{EPLS}
\varphi'_n(p) =n^{-\frac{3}{2} } \operatorname{e}^{-\operatorname{i}bp}f( \textnormal{\tiny{$\frac{1}{n}$}}  p)\,  \pi^+ (p)\,u_0(p)
\end{equation}
Closely related to Eq.\,(\ref{EPLS}) is  the  amplitude for a wave packet Eq.\,(\ref{PLSES}). Following  \cite[Eq.\,(14)]{BM99}, \cite[Eq.\,(2.1)]{BFM05}, \cite[Eq.\,(29)]{M02} we consider 
\begin{equation}\label{PLSES}
\varphi_n(p) :=n^{-\frac{3}{2} } \operatorname{e}^{-\operatorname{i}bp}f(\textnormal{\tiny{$\frac{1}{n}$}}  p)\,u(p)
\end{equation}
with $u(p)$ a unit eigenvector of  $\pi^+ (p)$. $\varphi_n$ in Eq.\,(\ref{PLSES})  is already normalized. In the corollary below it is supposed that the dilational limit $u_0(p):=\lim_{\lambda \to \infty} u(\lambda p)$ exists. Obviously this is the case if $u(p) :=  y(p)^{-1}c$ with $y(p)$ in Eq.\,(\ref{DHO}) and $c\in \C^4$ a unit vector with $c=\gamma_5 c$, see Eq.\,(\ref{DIEV}).

\begin{Cor}  The amplitude for a wave packet $(\varphi_n)$   in \emph{Eq.\,(\ref{PLSES}) }  is  a sequence of electron states localized at $b$ if  the dilational limit $u_0(p):=\lim_{\lambda \to \infty} u(\lambda p)$ exists.
\end{Cor}\\
{\it Proof.}
Clearly  $u_0(p)$ is a dilation invariant unit eigenvector of $\pi_0(p)$. So
$(\varphi_n)$    equals asymptotically   $(\varphi'_n)$ in Eq.\,(\ref{EPLS}), since obviously $||\varphi_n-\varphi'_n|| =   ||D_n\varphi_n- D_n\varphi'_n||  \to 0$. The result follows. \qed

\vspace{0,5mm}
Section 4 of \cite{BM99} deals with a proof of a weaker result  than this for Schwartz functions $f$ in Eq.\,(\ref{PLSES}). The choice $f(p):=\pi^{-3/4} \operatorname{e}^{-|p|^2/2}$ is discussed in detail in \cite{BM99}, \cite{BFM05}, \cite{M02}.
\end{Exa}

\begin{Exa} In view of Hegerfeldt's result reported in  sec.\,\ref{SDLD}
 it is worth verifying that there are sequences $(\varphi_n)$ of electron states localized at $0$, for which 
  the spatial probability decays exponentially   
 \begin{equation}\label{ESDLD}
  \langle \varphi_n,T^e(\{|x|>r\})\,\varphi_n\rangle   \in  \mathcal{O}(\operatorname{e}^{-Kr}), \quad r\to \infty
 \end{equation}
 for  $K< 2m$ ($c=\hbar=1$). 
For  the following   example,  inspired by  \cite[sec.\,2]{BK03}, choose in (\ref{EPLSES}) $u_0\in\C^4\setminus\{0\}$ constant and $f:=\mathcal{F}g$ with $g\in L^2(\R^3)\setminus \{0\}$ satisfying $\operatorname{e}^{b|\cdot|}g\in L^2(\R^3)$,  $b\in[0,a[$ for some $a>0$. E.g. $g(x):=\sqrt{\frac{a^3}{\pi}}\operatorname{e}^{-a|x|}$ and $f(p)=\frac{\sqrt{8a^5}}{\pi} \big(a^2+|p|^2\big)^{-2}$. By  (\ref{POLDE})  the sequence
 \begin{equation}
\varphi_n(p) = C_n      f(\textnormal{\tiny{$ \frac{1}{n}$}}  p) \, \pi^+ (p)\,u_0
\end{equation}
 with $C_n$  a normalizing constant,
is localized at $0$. The claim is that, if $n\ge\frac{m}{a}$, then $\varphi_n$ satisfies  Eq.\,(\ref{ESDLD}) for all $K<2m$.

 {\it Proof.}  For the proof  a result on Fourier transformation (see e.g. \cite[Theorem IX.13]{RS75}) is used. By this result, $f$ has an analytic continuation to the strip $\{w\in \C^3: |\operatorname{Im}w|<a\}$
such that  $f_q\in L^2(\R^3)$, $f_q(p):=  f(p+\operatorname{i}q)$,  for $q\in\R^3$, $|q|<a$ with $\sup_{|q|\le b}||f_q||<\infty $ for every $b\in[0,a[$. Then, for $n\ge\frac{m}{a}$, the strip of analyticity of $p\to f(\frac{1}{n}p)$ has the half-width $na\ge m$. Thus, because of the factor $\frac{1}{\epsilon(p)}$ present  in 
$\pi^+ (p)$,
 the strip of analyticity for  every component of $\varphi_n$ is limited to 
$\{w\in \C^3: |\operatorname{Im}w|<m\}$. This is the crucial point. The remaining premises in \cite[Theorem IX.13]{RS75}) on $\varphi_n$ concerning square-integrability are easily checked.  Put  $\psi_n:=\mathcal{F}^{-1}\varphi_n$.
It follows $\operatorname{e}^{c|\cdot|}\psi_n\in L^2(\R^3)$ for all $c\in[0,m[$. As a consequence $\langle \varphi_n,T^{e\,mom}(\{|x|>r\})\,\varphi_n\rangle = \langle \psi_n,E^{\textsc{d}\,pos}(\{|x|>r\})\,\psi_n\rangle=\int_{\{|x|>r\}}|\psi_n(x)|^2\operatorname{d}^3x= 
\int_{\{|x|>r\}}    \operatorname{e}^{-2c|x|}|\operatorname{e}^{c|x|}\psi_n(x)|^2\operatorname{d}^3x\le \operatorname{e}^{-2cr}\int_{\{|x|>r\}}|\operatorname{e}^{c|x|} \psi_n(x)|^2 \operatorname{d}^3x\le ||\operatorname{e}^{c|\cdot|}\psi_n||^2 \operatorname{e}^{-2cr}$. \hfill{$\Box$}
 \end{Exa}

There are as well point-localized sequences of electron states with only  power-like decay  of the spatial probability, as e.g.,    $\langle \varphi_n,T^ {e\,mom}(\{|x|>r\})\,\varphi_n\rangle \in\Theta(r^{-5})$ for $\varphi_n(p) = C_n      \operatorname{e}^{- |p|/n} \pi^+ (p)\,u_0$. What is more, this example  suggests, contrary to some literature,  that the exponential decay of the spatial probability  is {\it not}  typical. The electron wave functions with non-exponentially decaying  tails in the position representation of the Dirac system are even predominant, at least from a mathematical point of view.  The reason is that exponential decay requires analyticity properties  of the electron wave function in momentum representation, like in the example on  Eq.\,(\ref{ESDLD}).  This is clearly explained in   \cite[sec.\,2]{BK03}. 
\\

 Summarizing, the formalism of POL yields an  arbitrarily good localization obeying causality.  More precisely,   $T^ e$ is separated so that there are sequences of electron states localized at any point, and  $T^ e$  is causal in that the probability of localization  in the region of influence does not decrease. Moreover,
good localization is not related to the spatial decay of the  spatial probability.  But generally, due to causality, massive relativistic particles  have tails of the  spatial probability  which do not 
decay  faster than $\operatorname{e}^{-4\pi r /\lambda_C}$. The preparation of a point-localized sequence of electron states requires unlimited energy as the expectation values   of the energy in electron states of a point-localized sequence tend to infinity.\\

\section{Negative Energy States} \label{NES}

Let $\Delta$ be a closed region $\ne\R^3$. As we know the localization of an electron in $\Delta$ is not possible. For every electron state $\psi$ the probability of localization  $\langle\psi,T^e(\Delta)\psi\rangle$ is less than $1$. The reason one imagines is that attempting the localization causes such a large  uncertainty in energy that
pair-production occurs. Therefore, as it stands, a one-particle theory cannot furnish a complete description.  We share this view. At the same time  we think that  PO-localization  takes account of this situation in an effective manner.  We will argue that applying the localization operator $E(\Delta)$ of a causal system  to a particle  state does not create a pure state but the mixed state (\ref{SO}) of particle and antiparticle states. As such  their energies are positive for stability. Thus there is no  negative energy problem. On the other hand one recalls that these states continue to be  not localized in $\Delta$, despite the attempt in this regard.  But  we will see (\ref{LPL}) that there a particle state is created with an increased probability of localization in $\Delta$ with respect to the original one.\\ 
 \hspace*{6mm}
 This scenario appears quite reasonable. Of course its validity  has to be confirmed at concrete situations. For instance the situation underlying  the Klein paradox could be examined  as a device for the localization in a half-space. For this purpose one can employ  the thorough investigations on the former available in the literature (see \cite{BR76} and its references). ---
In the following we simply write $E,T$ for $E^{\textsc{d}},T^e$ and $P$ for $P^+$. We follow  \cite[sec.\,J]{CL15}.

Let the electron be in the  state $\varphi_1$, consider a region $\Delta\subset \R^3$, and suppose $T(\Delta)\varphi_1\ne 0$. 
Since $\varphi_1$ is a state of the Dirac system, $ \langle \varphi_1,E(\Delta) \varphi_1\rangle$ is the probability to get an affirmative answer at the position measurement to the question whether the Dirac system is in the region $\Delta$.  Recall that
$\langle \varphi_1,T(\Delta)\varphi_1\rangle =\langle \varphi_1,E(\Delta) \varphi_1\rangle$ as $P\varphi_1=\varphi_1$. Hence 
$\langle \varphi_1,T(\Delta)\varphi_1\rangle$ is the probability $\omega_1>0$ of localization in $\Delta$ of the electron in the state $\varphi_1$. It is also the probability  that after the measurement the Dirac system is in the state 
\begin{equation*}
\phi_1 := \norm{E(\Delta)\varphi_1 }^{-1} E(\Delta)\varphi_1
\end{equation*}
This state is regarded to be a non-observable virtual superposition $\phi_1=P\phi_1+ \bar{P}\phi_1$, $\bar{P} := I-P$, of an electron and a positron state.  Positrons are described by \textquotedblleft negative energy\textquotedblright\,states since, being antiparticles, they travel \textquotedblleft backwards\textquotedblright\,in time.   More precisely one has

\begin{Lem}\label{AUEMR}
Let $\mathcal{O}^{m,\eta}$ for $m>0$, $\eta=\pm$ denote the mass shell $\{\mathfrak{p}\in\R^4: p_0=\eta \epsilon(p)\}$.  Let  $j\in\N_0/2$. Then the antiunitary operator $\mathcal{C}:L^2(\mathcal{O}^{m,+},\C^{2j+1})\to L^2(\mathcal{O}^{m,-},\C^{2j+1})$, $(\mathcal{C}F)(\mathfrak{p}):=D^{(j)}(\operatorname{i}\sigma_2)\overline{F(-\mathfrak{p}})$ satisfies $$U^{m,j,-}=\mathcal{C}\,U^{m,j,+}\,\mathcal{C}^{-1}$$ where $\big(U^{m,j,\eta}(\mathfrak{a},A)F\big)(\mathfrak{p})=\operatorname{e}^{\operatorname{i}\mathfrak{a}\cdot\mathfrak{p}}D^{(j)}\big(Q(\mathfrak{p})^{-1}A\,Q(A^{-1}\cdot\mathfrak{p})\big)\,F(A^{-1}\cdot \mathfrak{p})$ is the irreducible massive representation $[m,j,\eta]$ of $\tilde{\mathcal{P}}$. $\mathcal{C}$ is uniquely determined up to a constant phase.
\end{Lem}\\
{\it Proof.} Note that $\operatorname{i}\sigma_2=\left( \begin{array}{cc} 0 & 1\\ -1 & 0 \end{array}\right)$ satisfies $\overline{B}=(\operatorname{i}\sigma_2)\,B\,(\operatorname{i}\sigma_2)^{-1}$  for $B\in SU(2)$. Further recall that $U^{m,j,\eta}$ is irreducible. Then the assertion is easily verified.\qed\\

So, by (\ref{AUEMR}),  the representation $[m,j,-]$ is antiunitarily equivalent to $[m,j,+]$. Hence, like $[m,j,+]$, it also represents  a particle with mass $m$ and spin $j$. For \textbf{stability} its energy has to be positive, too.

 \begin{Int}\label{IMRNE} $[m,j,+]$ represents a  particle with  \textbf{positive energy}, mass $m>0$, and spin $j\in \mathbb{N}_0/2$, and 
$[m,j,-]$ represents its antiparticle, which has \textbf{positive energy}, too,  the same mass $m$ and spin $j$.  
\end{Int} 

\hspace*{6mm} 
At this stage one need not worry about the conservation  of charge as it concerns the compound system consisting of the Dirac system and the measuring apparatus.  Accordingly, after the measurement the electron state
\begin{equation*}
\varphi_2 :=   \norm{ P\phi_1 }^{-1}P\phi_1= \norm{ T(\Delta)\varphi_1 }^{-1} T(\Delta)\varphi_1
\end{equation*}
occurs with probability  $\sigma_2$ given by the product $\langle \varphi_1,T(\Delta)\varphi_1\rangle \norm{ P\phi_1}^2=\norm{ T(\Delta)\varphi_1 }^2$. \\
\hspace*{6mm} 
As $I-E(\Delta)= E(\Delta')$ with $\Delta' := \R^3\setminus \Delta$, one has at the same time the information that there is the electron state $\varphi_2':=  \norm{ T(\Delta')\varphi_1 }^{-1} T(\Delta')\varphi_1$ with probability $\sigma_2' := \norm{ T(\Delta')\varphi_1 }^2$. Thus the probability of the occurrence of an electron  after the position measurement  is
\begin{equation*}
\sigma_2 + \sigma_2'=1-2\langle \varphi_1,T(\Delta)T(\Delta') \varphi_1\rangle
\end{equation*}
The  state $\phi_1$ also gives rise to the positron state $\bar{\varphi}_2 =  \norm{ \bar{P}E(\Delta)\varphi_1 }^{-1} \bar{P}E(\Delta)\varphi_1$ with probability $\bar{\sigma}_2 =  \norm{ \bar{P}E(\Delta)\varphi_1 }^2$. Also, there is the positron state  $\bar{\varphi}'_2 =  \norm{ \bar{P}E(\Delta')\varphi_1 }^{-1} \bar{P}E(\Delta')\varphi_1$ with probability $\bar{\sigma}'_2 =  \norm{ \bar{P}E(\Delta')\varphi_1 }^2$. One easily checks that $\bar{\sigma}_2+\bar{\sigma}'_2=2\langle \varphi_1,T(\Delta)T(\Delta') \varphi_1\rangle = 1-\sigma_2 - \sigma_2'$.\\
\hspace*{6mm} 
These considerations suggest that the influence of the apparatus represented by $E(\Delta)$ does not create eigenstates of $E(\Delta)$ as being unobservable for a super-selection rule, but a mixed state given by the density operator 
\begin{equation}\label{SO}
 W := PEP_{\varphi_1} EP +PE'P_{\varphi_1} E'P +\bar{P}EP_{\varphi_1} E\bar{P} +\bar{P}E'P_{\varphi_1} E'\bar{P}  
\end{equation}
Here, for short, $E=E(\Delta)$, $E'=E(\Delta')$, and $P_{\varphi_1}$ the projection on the subspace $\C\varphi_1$. Clearly $W\ge 0$ and, using the formula $QP_\phi Q=\norm{Q\phi}^2\norm{\phi}^{-2}P_{Q\phi}$ for any projection $Q$, one easily verifies $\operatorname{tr}(W)=1$. Note that the sum in Eq.\,\eqref{SO} is not orthogonal.

\begin{The}\label{PELPM}  One has $ \frac{1}{2}\le \sigma_2 + \sigma_2'\le 1$.  If $\Delta$ and  $\Delta'$ contain each an open ball up to a Lebesgue null set, then $\sigma_2 + \sigma_2' < 1$.
\end{The}

{\it Proof.} By the Spectral Theorem there is a probability measure $\mu$  on $[0,1]$ such that $\langle \varphi_1,T(\Delta)^k\varphi_1\rangle=\int t^k \operatorname{d} \mu(t)$ for $k\ge 0$.
Hence $\langle \varphi_1,T(\Delta)T(\Delta') \varphi_1\rangle=\int (t-t^2)\operatorname{d}\mu(t) \le \int \frac{1}{4}\operatorname{d}\mu=\frac{1}{4}$. Clearly $\sigma_2 + \sigma_2'\le 1$. ---
As to the second part of the assertion, one notes first that, due to the translational covariance, $T$ vanishes at the Lebesgue null sets.  Hence without restriction  $\Delta$ and  $\Delta'$ have a non-empty interior. Then it follows from (\ref{NLS})
 that the kernels of $T(\Delta)$ and $T(\Delta')$ are trivial, whence the assertion.\qed\\

 In conclusion, one infers that the missing probability $1-\sigma_2 - \sigma_2'$ for the occurrence of an electron state is due to a possible positron state. As to the interpretation of this result, we imagine that at the measurement of the position of the electron in $\Delta$  there is a probability not greater than $\frac{1}{2}$ (and arbitrarily small for a good localized electron as (\ref{NEPOM}) will show) that a positron state is created by the energy released 
 at the separation of  the measuring apparatus from the Dirac system.\\

 We ask now if the probability of localization $\omega_1$  increases by performing repeatedly the  position measurement. One finds for $n\in\N$
\begin{equation*}
\omega_n=  \langle \varphi_n,T(\Delta)\varphi_n\rangle = \frac{ \langle \varphi_1,T(\Delta)^{2n-1}\varphi_1\rangle }{ \langle \varphi_1,T(\Delta)^{2n-2}\varphi_1\rangle}
\end{equation*}
where the electron state $\varphi_n\,=\,  \norm{ T(\Delta)^{n-1}\varphi_1 }^{-1} T(\Delta)^{n-1} \varphi_1$ occurs with probability $\sigma_n=$ $\norm{T(\Delta)^{n-1}\varphi_1}^2$.

\begin{The}\label{LPL}  $\omega_n \nearrow \omega := \sup_m\omega_m$ and $\sigma_n \searrow \sigma := \inf_m\sigma_m$.  Moreover,  $\sigma>0$ if and only if  
$\varphi_1$ is not orthogonal to the kernel of      $I-T(\Delta)$. If  $\sigma>0$ than $\omega=1$.
\end{The}

{\it Proof.}  Let $\mu$ be the measure introduced in the proof of (\ref{PELPM}). Set  $\gamma_k := \int t^k \operatorname{d} \mu(t)$. Then $\sigma_n=\gamma_{2n-2}$, $\omega_n=\gamma_{2n-1}/\gamma_{2n-2}$. Since $t^k\searrow 0$ as $k\to\infty$  for $t\in[0,1[$, by dominated convergence $\gamma_k\searrow \mu(\{1\}) =\langle \varphi_1,P_1\varphi_1\rangle$, where $P_1$ denotes the orthogonal projection onto the kernel  of $I-T(\Delta)$. The assertions on $(\sigma_n)$ follow.\\
\hspace*{6mm} 
Now we prove $\gamma_{k-1}\gamma_{k+1} - \gamma_{k}  \gamma_{k}\ge 0$ for $k\in \N$, which obviously  implies monotony $\omega_n\le \omega_{n+1}$ for $n\in\N$. 
Let $m := \mu\otimes \mu$ be the product measure on $[0,1]^2$. Then $0\le  \int (t-s)^2(st)^{k-1} \operatorname{d} m(s,t) = \int (s^{k-1}t^{k+1}+s^{k+1}t^{k-1} -2s^kt^k) \operatorname{d} m(s,t) = \gamma_{k-1}\gamma_{k+1} + \gamma_{k+1}\gamma_{k-1} -2 \gamma_{k}  \gamma_{k}$. \qed\\

\hspace*{6mm} We add

\begin{Lem}\label{LPLZ} $\omega >0$, $\mu(]\omega,1])=0$  and   $\mu(]t,1])>0$ for $t<\omega$.
\end{Lem}

{\it Proof.} Set $\mu_k := \rho_k \mu$ with the density $\rho_k(t) := \gamma_k^{-1}t^k$. Then $\mu_k$ is a probability measure on $[0,1]$ and  $\omega_n=\int t\operatorname{d} \mu_{2n-2}(t)$, $n\in\N$.\\
\hspace*{6mm}Let  $H := \bigcup \big\{ \,]a,1]:a\in[0,1]\cap \mathbb{Q},\mu(]a,1])=0\big\}$. Obviously there is $b\in [0,1]$ such that $H=]b,1]$, and $\mu(H)=0$. One has $b>0$ since otherwise $\mu=\delta_0$ contradicting $T(\Delta)\varphi_1\ne 0$. The claim is $\omega =b$.\\
\hspace*{6mm} Let $0\le c<c'<b$ with $c'\in\mathbb{Q}$. Then $\mu_k([0,c])=\gamma_k^{-1}\int_{[0,c]} t^k \operatorname{d} \mu(t)  \le \gamma_k^{-1}c^k\mu([0.c])$, and $\gamma_kc^{-k}=\int (\frac{t}{c})^k \operatorname{d} \mu(t)\ge \int_{]c',1] } (\frac{t}{c})^k \operatorname{d} \mu(t)  \ge (\frac{c'}{c})^k \mu(]c',1]) \to \infty$ for $k\to \infty$ since $ \mu(]c',1]) >0$ by definition of $H$.
Hence  $\mu_k([0,c]) \to 0$ for $k\to \infty$. 
Then $c-c\,\mu_k([0,c]) = c\,\mu_k(]c,1]) \le \int t\operatorname{d}\mu_k(t) = \int_{[0,c]}\cdots +\int_{]c,1]}\cdots \le c\,\mu_k([0,c])+  b\,\mu_k(]c,1])=b+(c-b)\mu_k([0,c])$ implies $c\le \liminf_k \int t\operatorname{d}\mu_k(t) \le \limsup_k\int t\operatorname{d}\mu_k(t) \le b$ for all $c<b$. Hence $\lim_n\omega_n=b$. \qed\\

 So, indeed, according to (\ref{LPL}) one gets the localization  improved by the position measurements. But, at the same time, the observation of the localized states becomes seldom as the probabilities $\sigma_n$ in general tend to zero. Moreover (\ref{LPLZ}) shows  that, depending on the initial state $\varphi_1$,  $\lim_n\omega_n$ may be less than $1$ even if $\norm{ T(\Delta) }=1$. Hence the repeated measurement of the position  in general is not suited to prepare a point-localized sequence of states. On the other hand, as shown in the following theorem, the measurement of the position for electron states of a point-localized  sequence generates positron states only with vanishing probability.



\begin{The}\label{NEPOM} Let $(\varphi_n)$ be a sequence of electron states localized at $b\in\R^3$. Then there is a sequence of Dirac states $\phi_m$ satisfying $E(B)\phi_m=\phi_m$ up to finitely many $m$ for every ball $B$ around $b$ and such that $(I-P)\phi_m\to 0, \; m\to \infty$.
\end{The}  

{\it Proof.}  Fix the ball $B^m$ with radius $1/m$. As $P\varphi_n=\varphi_n$,  $\norm{E(B^m)\varphi_n}\ge \norm{PE(B^m)P\varphi_n} 
= ||T(B^m)\varphi_n||\to 1$, whence $C^m_n:=$ $||E(B^m)\varphi_n||> 0$ up to finitely many $n$. Put $\phi^m_n:=(1/C^m_n)E(B^m)\varphi_n$. Then $C^m_n||P\phi^m_n|| = 
||PE(B^m)P\varphi_n||=||T(B^m)\varphi_n|| \to 1$, whence $\norm{P\phi^m_n}\to 1$ since $C^m_n\le 1$. Hence $\norm{(I-P)\phi^m_n}^2=1-\norm{P\phi^m_n}^2\to 0$ for $n\to\infty$. This holds for every $m$. Now choose a subsequence $n_m$ such that $1-\norm{P\phi_m}^2<\frac{1}{m}$ for $\phi_m:=\phi^m_{n_m}$.\qed

  Hence according to (\ref{NEPOM}) an arbitrarily small  negative energy component may suffice to localize the Dirac system in  a bounded region.
We examine more closely the decomposition into positive energy and negative energy part of a Dirac state $\phi$. Let us say that
 the spatial probability \textbf{decays fast}  if    $\langle \phi, E(\{|x|>r\})\phi\rangle\in \mathcal{O}(\operatorname{e}^{-Kr})$  as $r\to \infty$ for some $K>2m$.

\begin{Lem}\label{DCLDS}  Let $\phi$ and $\phi'$ be Dirac states with fast decaying spatial probabilities. If $P\phi=P\phi'$ then $\phi=\phi'$ holds.
\end{Lem}

{\it Proof.} Check the  estimate $\langle \phi-\phi', E(\{|x|>r\}) \,(\phi-\phi')\rangle \le2 \langle \phi,  E(\{|x|>r\})\,\phi\rangle+ 2 \langle \phi', E (\{|x|>r\})\,\phi'\rangle$. Assume now $\phi\ne\phi'$. Let $\chi$ denote the normalized $\phi-\phi'$. Then  $\chi$ has a fast decaying spatial probability.  
 By assumption  $\chi$ is  a negative energy state. This contradicts the result by Hegerfeldt reported in sec.\,\ref{SDLD}. \qed\\

In particular, the bounded localized Dirac states $\phi$ are uniquely determined by their positive energy components $P\phi$. 
Since, due to causality (\ref{VEFCPOL})(c), the dense subspace of these states $\phi$ is invariant under   $W^\textsc{d}$  and since the subspaces of positive and negative energy are also invariant,  $W^\textsc{d}$ is already determined  by its action on the  subspace of positive energy.\\

\section{Dependence on  positive mass  and the case of zero mass}\label{DPMZM}   The  results on the Dirac system are equally 
valid for every mass $m>0$. They depend on the mass due to the time evolution $t\mapsto W^{\textsc{d}\,m}(t)=\operatorname{exp}(itH)$ with the Dirac Hamiltonian $H=\sum_k\alpha_kP_k +\beta m$. The superscript $m$ is introduced here in order to remind of this dependence. It is worthwhile to reflect a moment about the dependence  on  $m>0$.\\
\hspace*{6mm}
By means of the dilations  $D_\lambda$, $\lambda>0$ (see Eq.\,(\ref{DDS})) one establishes  the following relationship between the time evolutions for different masses
\begin{equation*}
W^{\textsc{d}\,\lambda m}(t)=D_\lambda^{-1}W^{\textsc{d}\,m}(\lambda t)D_\lambda
\end{equation*}
The question is about  the dependence of $e(\psi_t)=e(W^{\textsc{d}m}(t)\psi)$ on $m>0$. Using the relation $E^{\textsc{d}}(\lambda \Delta)=D_\lambda E^{\textsc{d}}(\Delta) D_\lambda^{-1}$
one finds $e(D^{pos}_\lambda \psi)=\lambda e(\psi)$ and 
\begin{equation*}
t^{\lambda m}_e(\psi)=\textnormal{\tiny{$\frac{1}{\lambda}$}}t^{m}_e(D^{pos}_\lambda \psi)
\end{equation*}
Hence the maximum of $e(\psi_t)$ and   its position $t^m_e(\psi)$ may depend on $m>0$.
Independent of $m>0$ are $\min\{e(\psi_t),e(\psi_{-t})\}$, since equal to $e(\psi) -|t|$ by (\ref{SMF}), and the upper bound $e(\psi_t)\le -2\,\overline{e}(\psi)-e(\psi) -|t|$  (\ref{GGETIIED}). This subject would merit more analysis.

\hspace*{6mm}
In view of the study of the causal Weyl systems in sec.\,\ref{OWL} the question is which results in sec.\,\ref{DTINC}, \ref{DLB} persist for mass zero. Indeed, for $m=0$, the Dirac Hamiltonian $H$ becomes the orthogonal sum $\bigoplus_{\chi=\pm} H^\chi$ of the Weyl Hamiltonians so that by (\ref{WRPS}) and Eq.\,(\ref{DRPS})  the  Dirac system 
$(W^{\textsc{d}\,m},E^{\textsc{d}})$ is reduced to the orthogonal sum of the Weyl systems, i.e.,
\begin{equation*}
(W^{\textsc{d}\,0},E^{\textsc{d}})=\bigoplus_{\chi=\pm} (W^{\textsc{w}\chi},E^{\textsc{w}})
\end{equation*}
By (\ref{CSWF})  the latter is still a causal system. Note  that also the time reversal operator $\mathcal{T}$ (\ref{TIR}) is the orthogonal sum of the respective operators for the Weyl systems (see before sec.\,\ref{FIWL}).\\
\hspace*{6mm}
 Therefore and since (\ref{PICESCE}) holds also for $\epsilon(z)^2=z^2$, as a consequence, all results in sec.\,\ref{PBWF} hold, by the same proofs, also for the Weyl systems.\\
\hspace*{6mm}
The results in sec.\,\ref{LTBPL} about the long term behavior of the probability of localization of  Dirac states are not equally valid for Weyl states. In  sec.\,\ref{LOCCRS} they are replaced  with results, which are stronger or necessarily weaker than the analogous ones for Dirac states.\\
\hspace*{6mm}
 However, it is important that (\ref{ECCPWS}) holds which is the analogous result (\ref{ECCPDS}) for Dirac states. As a consequence, all results for Dirac states in sec.\,\ref{DLB} equally hold true  for Weyl states.


\section{The Weyl Systems}\label{OWL}
We study now the prominent \textbf{massless} causal systems (\ref{CHWS}), the Weyl systems  $(W^{\textsc{w}\chi}, E^{\textsc{w}})$, $\chi\in\{+,-\}$. They are derived in   (\ref{RCL}). In momentum representation on $L^2(\R^3,\C^2)$ they read
\begin{equation}\label{2WS}
\big(W^{\textsc{w}\chi \,mom}(t,b,A)\varphi\big)(p) =\operatorname{e}^{-\operatorname{i}bp}
\sum_{\eta =\pm 1} \operatorname{e}^{\operatorname{i}t \eta |p|} \pi^{\chi\eta}(p)s^\chi(A)^{*-1}     
\varphi(q^\eta),\quad E^{\textsc{w}mom}=\mathcal{F}E^{can}\mathcal{F}^{-1}
\end{equation}

where $s^+(A)=A$, $s^-(A)=A^{*-1}$, 
$h^\chi(p)=\chi(\sigma_1p_1+\sigma_2p_2+\sigma_3p_3)$ and 
$\pi^{\chi\eta}(p)=\frac{1}{2}\big(I_2+\frac{\eta}{|p|}h^\chi(p)\big)$. For the energy and the cone representation of the Weyl systems see (\ref{A:ESRWS}).\\

\hspace*{6mm}
They may turn out to be relevant for the description of the recently discovered \textbf{Weyl fermions} \cite{S et al15}, \cite{L et al15}. By Eq.\,(\ref{DWTL}) the Weyl representation $W^{\textsc{w} \chi}$ is the orthogonal sum 
$$[0,\chi
\textnormal{\tiny{$\frac{1}{2}$}},\chi] \oplus [0,\chi
\textnormal{\tiny{$\frac{1}{2}$}},-\chi]$$ The case $\chi=+$ means  a massless fermion with helicity $+\frac{1}{2}$, which is called  the right-handed Weyl fermion as its spin points in the  direction of its motion. It is
accompanied by its left-handed  antiparticle. There is also a left-handed  Weyl fermion accompanied by its right-handed antiparticle in the case $\chi=-$. We like to refer to the interpretation (\ref{IRNE}). Note that, in accordance with experimental evidence, other combinations of a Weyl fermion with its antiparticle are excluded by the requirement of a causal localization, see  (\ref{DDE}). This is a further fact in favor of causal localization.

\begin{POLWP}\label{POLWP} $E^{\textsc{w}}$ admits dilational covariance quite as $E^{\textsc{d}}$ due to $D^{pos}_\lambda$  in Eq.\,(\ref{DDS}).
Since $\pi^{\chi\eta}(\lambda p)=\pi^{\chi\eta}(p)$ for  $\lambda>0$ the projection $P^{\chi\eta}$ onto the state space of the Weyl particle $[0,\chi/2,\eta]$ commutes with $D_\lambda$. Let $T^{\chi\eta}$ denote the trace of $E^{\textsc{w}}$ by $P^{\chi\eta}$, cf.\,(\ref{EPOLDE}). Obviously  the results in sec.\,\ref{PTE} on the electron carry over on the Weyl particle. In particular, $T^{\chi\eta}$ is separated and the formula analogous to that in (\ref{POLDE})  holds, which here results to be particularly simple, i.e., $\varphi_n:=D_n^{-1}\varphi$ ($n\in\N$) for every state $\varphi$ in the range of $P^{\chi\eta}$.  Finally 
recall that  $E^{\textsc{w}}$  extends on $\mathfrak{S}$ by Poincar\'e covariance (\ref{PCPOL}). So $T^{\chi\eta}$ is defined on $\mathfrak{S}$ and, by  (\ref{TCWL}) and (\ref{CSWF}) below, $T^{\chi\eta}$ is causal. 
\end{POLWP}

Like (\ref{NXH}) for the Dirac system one has $N=\textnormal{\small{$\frac{1}{2}$}}\{X^{\textsc{w}},H\}$. Note that,
 as the proof of  (\ref{RCLX}) shows, for a massless SCT $(W_E,E)$ the booster $N$ satisfies
\begin{equation*}\label{NCPOW}
N=\textrm{\SMALL{$\frac{1}{2}$}}\{X,H\}
\end{equation*}
 if and only if $(W_E,E)$ is unitarily equivalent to the orthogonal sum for $\chi\in\{+,-\}$ of multiples of $(W^{\textsc{w}\chi},E^{\textsc{w}})$.\\
 \hspace*{6mm} 
Since time reversal operator $\mathcal{T}$ 
is used in the proof of some results, let us note that for the Weyl systems  it is given by $\mathcal{T}^{pos}\psi=-\sigma_2\overline{\psi}$ (cf.\,(\ref{TIR})).
   
\subsection{Weyl systems are causal}\label{FIWL} Compare this section  with sec.\,\ref{DLIC}. As to the notations recall  (\ref{DRPS}) and put $H^{\chi pos}:=\chi \sum_k\sigma_k\,\frac{1}{\operatorname{i}} \partial_k$, $\chi\in\{+,-\}$.\\

\begin{Lem}\label{WRPS}
 In position space $L^2(\R^3,\C^2)$ the Weyl representation $W^{\textsc{w} \chi}$   reads
 \begin{equation*}
\big(W^{\textsc{w}\chi\,pos}(g)\psi\big)(x)= s^\chi(A) \big(\operatorname{e}^{-\operatorname{i} y_0 H^{\chi pos}}\psi\big)(y)
\end{equation*}
\end{Lem}\\
{\it Proof.} By definition  $W^{\textsc{w}\chi\,pos}\equiv \mathcal{F}^{-1}W^{\textsc{w}\chi\, mom}\mathcal{F}$ with $W^{\textsc{w}\chi\,mom}$ from Eq.\,(\ref{2WS}). 
Apply first
Eq.\,(\ref{VSPI}). Then $\big(W^{\textsc{w}\chi \,mom}(g)\varphi\big)(p) =
s^\chi(A) \sum_{\eta =\pm 1}  
\operatorname{e}^{\operatorname{i}(t,b)\cdot (\eta |p|,p)}
 (|q^\eta|/|p|)  \pi^{\chi\eta}(q^\eta) \varphi(q^\eta)$. Now proceed as in the proof of (\ref{DRPS}).\qed\\

As in the case of the Dirac representation (\ref{DRPS}) one introduces   the time-dependent  wave-function $\Psi$ related to the position space wave-function $\psi$ by $\Psi(t,x)=\big(\operatorname{e}^{-\operatorname{i}t H^{\chi pos}}\psi\big)(x)$.
Then the former is the solution of  the  initial-value problem for  the \textbf{Weyl equation} 
\begin{equation} 
\operatorname{i}\partial_t\Psi = H^{\chi pos}\Psi, \quad
\Psi(0,\cdot)=\psi  \text{ with } \psi\in \operatorname{dom}(H^{\chi pos})
\end{equation}
 See also  
 \cite[(5.234), (5.235)]{FG70}.  Under Poincar\'e transformations $g$  it transforms like   
$ \Psi_g(\mathfrak{x})=s^\chi(A)\, \Psi(g^{-1}\cdot\mathfrak{x})$ 
 for $\mathfrak{x}=(t,x)$.
 
 \begin{Cor}\label{CSWF} $(W^{\textsc{w}\chi},E^\textsc{w})$, $\chi\in\{+,-\}$ are causal systems \emph{(\ref{CS})}. 
 \end{Cor}\\
 {\it Proof.}  By (\ref{RCL}) $(W^{\textsc{w}\chi},E^\textsc{w})$ is an SCT. Here we  may  infer this also by  \cite[Theorem 10 (b)]{CL15} from the fact that  $z\mapsto \operatorname{e}^{\operatorname{i}th^\chi(z)}$  is exponentially bounded 
 (cf.\,after Eq.\,(\ref{CTEDS})). The assertion  follows from  (\ref{WRPS}) imitating the proof for (\ref{CSDF}).\qed\\

So causality \begin{equation}\label{CWS} 
  W^{\textsc{w}\chi}(g)\,E^\textsc{w}(\Gamma)\,W^{\textsc{w}\chi}(g)^{-1}\,\le \, E^\textsc{w}(\Gamma_g)
  \end{equation}
    ($g\in \tilde{\mathcal{P}}$, $\Gamma\in \mathfrak{S}$) holds. See (\ref{VEFCPOL}) for equivalent formulations of causality.
\\

 \begin{Sum}\label{SRWS} 
 As already checked in  sec.\,\ref{DPMZM}, the results on the propagation of the border of a Dirac wave function in sec.\,\ref{PBWF} hold equally, by the same proofs,  for Weyl wave functions. In particular, for a Weyl state $\psi$ localized in a bounded region one has (cf.\,(\ref{SETIIED}))
\begin{equation}\label{PBWWF} 
e\big(W^{\textsc{w}\chi}(t)\psi\big)=e(\psi) +|t_e^\chi|-|t-t_e^\chi|
  \end{equation}

Furthermore, still examined in  sec.\,\ref{DPMZM}, the results in sec.\,\ref{DLB} regarding the effect of a boost on the localization stay valid for the Weyl systems. The proofs  are the same except for using (\ref{ECCPWS}), which it is harder to prove than the analog (\ref{ECCPDS}) for the Dirac system. This is due to a remarkable difference between the long term behavior of the probability of localization of the Dirac system and the Weyl systems, compare sec.\,\ref{LTBPL} with sec.\,\ref{LOCCRS}. The key result in this context is (\ref{TERSWF}) providing a rather explicit description of the time evolution of the radially symmetric Weyl wave functions.
 So an important result is 
\begin{equation}\label{WLBSLSET}
 W(\varsigma\alpha)E(\{xe\le \alpha\})W(\varsigma\alpha)^{-1}E(\{xe\ge 0\})= \lim_{\varsigma\rho\to\infty} W(A_{\rho e})(\{ 0\le xe\le 2\alpha\,\operatorname{e}^{-\varsigma\rho}\}) W(A_{\rho e})^{-1}
\end{equation}
for $W=W^{\textsc{w}\chi}, E=E^{\textsc{w}}$ (cf.\,(\ref{LBSLSET})),  which relates the effects on  localization of boost and time evolution. The left hand side tends to $E(\{xe\ge 0\})$ for $\alpha\to \infty$, which is one of the  completeness results, see (\ref{ECCPWS}) and the comments on it.
Hence last not least 
   the Lorentz contraction holds also  for Weyl states $\psi$ (cf.\,(\ref{LCDWF})), i.e., for every   $\delta>0$ one has
\begin{equation}\label{LCWWF} 
 \norm{E^{\textsc{w}}\big(\{x\in\R^3: -\delta\le xe \le \delta\}\big)\,W^{\textsc{w}\chi}(A_{\rho e})\psi}\, \to 1, \;\; |\rho|\to\infty
\end{equation}
\end{Sum}

 \section{Long-term behavior of the probability of localization of Weyl states}\label{LOCCRS}
Let $\psi_t:=W^{\textsc{w}\chi}(t)\psi$, $t\in\R$ denote the (right- respectively left-handed) Weyl state $\psi$ evolved in time.
As for the Dirac system in sec.\,\ref{LTBPL} we will study the behavior of the probability of localization  for large $|t|$. The following result (\ref{WSVPOBT}) is stronger than the analogous  (\ref{PRBG}) for the Dirac system. (Occasionally we will omit the superscript {\it pos}.)

\begin{The}\label{WSVPOBT}  Let $\psi\in L^2(\R^3,\C^2)$ be a  Weyl state.  Then for every $v\in]0,1[$ 
$$ E^{\textsc{w}}(B_{v|t|})\,\psi_t\to 0, \;\;|t|\to \infty$$
If $\mathcal{F}\psi\in \mathcal{C}^\infty_c(\R^3\setminus\{0\},\C^2)$, then for every $N>0$ there is a finite constant $C_N$ such that $\norm{E^{\textsc{w}}(B_{v|t|})\psi_t}\le C_N(1+|t|)^{-N}$ for all $t\in\R$.
\end{The}\\
{\it Proof.} Suppose first $\varphi:=\mathcal{F}\psi\in \mathcal{C}^\infty_c(\R^3\setminus\{0\},\C^2)$. For the estimation of the integral 
 \begin{equation}\label{WENSPM}
 (\psi^\eta_t(x))_l=(2\pi)^{-3/2}\int \operatorname{e}^{\operatorname{i}(px+t\eta\chi\, |p|)}(\varphi^\eta(p))_l\,\operatorname{d}^3p
 \end{equation}
for $l=1,2$ (cf. Eq.\,(\ref{ENSPM}))  consider $\phi^{\eta\chi}(p):= (|x|+|t|)^{-1}\big(px-t\eta\chi\,|p|\big)$. Then $\nabla \phi^{\eta\chi}(p)=  (|x|+|t|)^{-1}\big(x-\frac{t\eta\chi}{|p|}p\big)$ and $|\nabla \phi^{\eta\chi}(p)|\ge  (|x|+|t|)^{-1}(|t|-|x|)$ for $p\ne 0$. Now suppose $|x|\le v|t|$. Then $|\nabla \phi^{\eta\chi}(p)|\ge (v|t|+|t|)^{-1}(|t|-v|t|)=\frac{1-v}{1+v}>0$.  This implies  (cf.\,\cite[(1.209)]{T92}) for $\eta,\chi\in\{+,-\}$, $l=1,2$, and for every $N>0$ that there is a finite constant $A_N$ with
$$\big|\big(\psi^\eta_t(x)\big)_l\big|\le A_N(1+|x|+|t|)^{-N}\textrm{ \;if\; } |x|\le v |t|$$
Hence $\norm{E^{\textsc{w}}( B_{v|t|})\psi_t}\le \sum_\eta \norm{E^{\textsc{w}}( B_{v|t|})\psi^\eta_t}$ and  $\norm{E^{\textsc{w}}( B_{v|t|})\psi^\eta_t}^2=\int_{ B_{v|t|}}|\psi^\eta_t(x)|^2\operatorname{d}x^3=\sum_l\int_{ B_{v|t|}}\big|\big(\psi^\eta_t(x)\big)_l\big|^2\operatorname{d}x^3\le 8\pi A_N^2\int_{0}^{v|t|} (1+r+|t|)^{-2N}r^2\operatorname{d}r\le 8\pi A_N^2(1+|t|)^{-2N}\int_{0}^{v|t|} r^2\operatorname{d}r\le \frac{8\pi}{3}v^3A_N^2(1+|t|)^{-2N+3}$ if $N>\frac{3}{2}$. 
 Hence $\norm{E^{\textsc{w}}(B_{v|t|})\psi_t}\le C_N(1+|t|)^{-N}$ for $N>0$ and  
 $C_N:=(16\pi/3)^{\frac{1}{2}}A_{N+\frac{3}{2}}$. \\
\hspace*{6mm}
Now consider a general Weyl state $\psi$. Let $\varepsilon>0$. Set $\varphi:=\mathcal{F}\psi$ and choose $\varphi'\in\mathcal{C}^\infty_c(\R^3\setminus\{0\},\C^2)$ with $\norm{\varphi-\varphi'}\le\varepsilon/2$. Hence $\norm{\psi-\psi'}\le\varepsilon/2$ for $\psi':=\mathcal{F}^{-1}\varphi'$. Let $P_t:=E^{\textsc{w}}( B_{v|t|})$. By the foregoing result  there is a constant $C_1$ such that $\norm{P_t\psi'_t}\le
C_1(1+|t|)^{-1}$ for all $t$. Let $\tau:=2C_1/\varepsilon$. Then $\norm{P_t\psi_t} \le \norm{P_t(\psi_t-\psi'_t)}  +  \norm{P_t\psi'_t} \le$ $\norm{\psi-\psi'}+C_1(1+|t|)^{-1}\le\varepsilon $ for $|t|\ge\tau$.\qed\\

So roughly speaking, in the long term the Weyl fermion tends to infinity with any subluminal velocity. This however does not imply that $ E^{\textsc{w}}(B_{|t|})\,\psi_t\to 0$ as $|t|\to \infty$. The result (\ref{GTERSWF})(b) also shows  that in general neither the other extremum $E^{\textsc{w}}(\R^3\setminus B_{|t|})\,\psi_t\to 0$ occurs. This is one of the main differences  between  Weyl systems and Dirac system (\ref{VPOBT}).\\

\hspace*{6mm} 
A  function $\psi\in L^2(\R^3, \C^n)$ is called radially symmetric if $\psi(x)=g(|x|)$, $x\in\R^3$, where $g:[0,\infty[\to \C^n$ is measurable and $jg$ with the identity map $j(r):=r$ is square integrable. Note that $\norm{\psi}^2=4\pi\int_0^\infty r^2|g(r)|^2\operatorname{d}r$.

\begin{The}\label{TERSWF} Let $\psi=g(|\cdot|)$ be a radially symmetric Weyl state in the position representation. Then for $t\in\R$ and $x\in\R^3$,  $\varsigma\in\{+,-\}$ one has
$$ \psi_t= A^+_t+A^-_t+R_t$$
with 
\begin{itemize}
\item $A^\varsigma_t(x):=\pi^{\chi\varsigma}(\frac{1}{|x|}x)\frac{|x|+\varsigma t}{|x|}g\big(\big||x|+\varsigma t\big|\big)$
\end{itemize}
and the remainder 
\begin{itemize}
\item $R_t(x)=h^\chi(\frac{1}{|x|}x)\frac{1}{2|x|^2}\big(G\big(\big||x|+t\big|\big)-G\big(\big||x|-t\big|\big)\big)$ with $G(r):=-\int_0^r \rho g(\rho)\operatorname{d}\rho$
\end{itemize}
Furthermore, $A^\varsigma_t,R_t\in L^2(\R^3,\C^2)$ and
\begin{itemize}
\item[(a)]\, $\langle A^+_t(x), A^-_t(x)\rangle=0$
\item[(b)]\, $2\norm{A^\varsigma_t}^2=1-\varsigma \operatorname{sgn}(t)\norm{E^{\textsc{w}}(B_{|t|})\psi}^2$, in particular $A^\varsigma_t\to 0$ for $\varsigma t\to \infty$
\item[(c)]\, $\norm{ E^{\textsc{w}}( B_{|t|}) A^\varsigma_t}^2=\frac{1}{2}\norm{ E^{\textsc{w}}( B_{|t|}) \psi}^2$ if $ -\varsigma t\ge 0$
\item[(d)]\, $R_t\to 0$ for $|t|\to \infty$
\end{itemize}
and finally
\begin{itemize}
\item[(e)]\, $\norm{E^{\textsc{w}}(B_{|t|})\psi_t}^2\to \frac{1}{2}$ for $|t|\to \infty$
\end{itemize}
\end{The}

{\it Proof.} Let $\mathcal{S}$ denote the Fourier-sine transformation on $L^2([0,\infty[)$ given by $(\mathcal{S}h)(s):=\sqrt{\frac{2}{\pi}}\int_0^\infty \sin(rs)\,h(r)\operatorname{d}r$ if $h$ is integrable. Recall that 
$\mathcal{S}$ is unitary with $\mathcal{S}^{-1}=\mathcal{S}$. One verifies that  $\varphi:=\mathcal{F}\psi$ is radially symmetric $\varphi=f(|p|)$ with
\begin{equation*}
 jf=\mathcal{S}(jg)\tag{1}
\end{equation*}
Indeed, for $R>0$ and $B\in SU(2)$, consider $(2\pi)^{3/2}\varphi^R(p):=\int_{B_R}\operatorname{e}^{-\operatorname{i}xp}\psi(x)\operatorname{d}^3x=\int_{B_R}\operatorname{e}^{-\operatorname{i}B\cdot x B\cdot p}\psi(x)\operatorname{d}^3x=\int_{B_R}\operatorname{e}^{-\operatorname{i}xB\cdot p}g(|x|)\operatorname{d}^3x$. Choose $B=B(p)^{-1}$ (see Eq.\,(\ref{CCSEF})) and use spherical coordinates $r=|x|, x_1=r\cos(\phi) \sin(\theta), x_2=r\sin(\phi) \sin(\theta), x_3=r\cos(\theta)$. Then,  substituting  $\cos(\theta)$ by the variable $\xi$,   $\varphi^R(p)=(2\pi)^{-3/2}\int_{B_R}\operatorname{e}^{-\operatorname{i}x_3|p|}g(|x|)\operatorname{d}^3x=(2\pi)^{-1/2}\int_{0}^R\int_{-1}^1 \operatorname{e}^{-\operatorname{i}r|p| \xi}\operatorname{d}\xi\,g(r) r^2\operatorname{d}r =\frac{1}{|p|}\sqrt{\frac{2}{\pi}}\int_0^R\sin(|p|r)rg(r)\operatorname{d}r$. The limit in the mean for $R\to\infty$ yields $\varphi(p)=\frac{1}{|p|}(\mathcal{S}(jg))(|p|)$, whence (1).

\hspace*{6mm}
$\bullet$ \,Similarly  compute $\psi_t=\mathcal{F}^{-1}(\operatorname{e}^{\operatorname{i}th^\chi(\cdot)}\varphi)$. 
Let $(2\pi)^{3/2}\psi_t^R(x):=\int_{B_R}\operatorname{e}^{\operatorname{i}px}\operatorname{e}^{\operatorname{i}th^\chi(p)}\varphi(p)\operatorname{d}^3p=\int_{B_R}\operatorname{e}^{\operatorname{i}pB\cdot x}\operatorname{e}^{\operatorname{i}th^\chi(B^{-1\cdot}p)}f(|p|)\operatorname{d}^3p=\int_{B_R}\operatorname{e}^{\operatorname{i}pB\cdot x}B^{-1}\operatorname{e}^{\operatorname{i}th^\chi(p)}Bf(|p|)\operatorname{d}^3p$ by Eq.\,(\ref{RIH}). \\
\hspace*{6mm}
Note that
 $h^\chi(p)^2=|p|^2I_2$. Hence
 $$\operatorname{e}^{\operatorname{i}th^\chi(p)}
=\cos(t|p|)I_2+\operatorname{i}\sin(t|p|)\textnormal{\tiny{$\frac{1}{|p|}$}}h^\chi(p)$$
In spherical coordinates $s=|p|, p_1=s\cos(\phi) \sin(\theta), p_2=s\sin(\phi) \sin(\theta), p_3=s\cos(\theta)$ one has 
$$h^\chi(p)=\chi s \big(\cos(\phi)\sin(\theta)\sigma_1+\sin(\phi)\sin(\theta)\sigma_2+\cos(\theta)\sigma_3\big)$$ 
For $B=B(x)^{-1}$ the exponent  $p\,B\cdot x$ becomes $p_3|x|=s|x| \cos(\theta)$, which does not depend on the variable $\phi$. So the integrations $\int_0^\pi\int_0^{2\pi}\sin(\theta)\operatorname{d}\theta \operatorname{d}\phi$ can be easily performed. Using Eq.\,(\ref{CRCS}), $\psi_t^R(x)=(2\pi)^{-1/2}\int_0^R\int_{-1}^1\operatorname{e}^{\operatorname{i}s|x|\xi} \big(\cos(ts)I_2+\operatorname{i} \sin(ts)\xi h^\chi(\frac{1}{|x|}x)
 \big)f(s)\operatorname{d}\xi \,s^2\operatorname{d}s=\frac{1}{|x|}\sqrt{\frac{2}{\pi}}\int_0^R \sin(s|x|)\cos(ts)\,sf(s)\operatorname{d}s\;\,+\,\;h^\chi(\textnormal{\tiny{$\frac{1}{|x|}$}}x) \frac{1}{|x|}\sqrt{\frac{2}{\pi}}
\int_0^R\cos(s|x|)\sin(ts)\,sf(s)\operatorname{d}s\,\;+\,\, R^{\textsc{r}}_t(x)$,  
\begin{equation*}
R^{\textsc{r}}_t(x):=-h^\chi(\textnormal{\tiny{$\frac{1}{|x|}$}}x) \textnormal{\footnotesize{$\frac{1}{|x|^2}$}}\textnormal{\tiny{$\sqrt{\frac{2}{\pi}}$}}
\int_0^R \sin(s|x|)\sin(ts)f(s)\operatorname{d}s\tag{2}
\end{equation*}
\hspace*{6mm}
Next we use $2\sin(s|x|)\cos(ts)= \sin(s(|x|+t))+ \sin(s(|x|-t))$, $2\cos(s|x|)\sin(ts)= \sin(s(|x|+t))- \sin(s(|x|-t))$, and $u=\operatorname{sgn}(u)\,|u|$ for $u\in\R$. Furthermore, the limit in the mean of $\sqrt{\frac{2}{\pi}}\int_0^R \sin(s\,|u|)sf(s)\operatorname{d}s$ for $R\to \infty$ is $(\mathcal{S}(jf))(|u|)=|u|g(|u|)$ by (1). Thus, putting things together for $u=|x|\pm t$, one confirms the formula for $A_t^\varsigma$. \\
\hspace*{6mm}
$\bullet$ \,As to the formula for $R_t$, according to (2) the claim is
\begin{equation*}
(\mathcal{S}(f_t))(r)=-\textnormal{\tiny{$\frac{1}{2}$}}\big(G(|r+t|)-G(|r-t|)\big)\tag{3}
\end{equation*}
for $f_t(s):=\sin(ts)f(s)$. Indeed, as $|f_t(s)|\le s|f(s)|$, one has $f_t\in L^2$ and in (2) the limit in the mean for $R\to \infty$ exists. Moreover, $jg$ is locally integrable as $jg\in L^2$, whence $G$ is well-defined. \\
\hspace*{6mm}
In  (2) use $-2\sin(sr)\sin(ts)=\cos(s(r+t))-\cos(s(r-t))$. In proving (3) first assume $f\in \mathcal{C}_c$, i.e., $f$ continuous with compact support. Then, for $u\ge 0$, $G(u)=-\int_0^u\rho g(\rho)\operatorname{d}\rho=-\int_0^u (\mathcal{S}(jf))(\rho)\operatorname{d}\rho=-\int_0^u \textnormal{\tiny{$\sqrt{\frac{2}{\pi}}$}}\int_0^\infty \sin(\rho s)sf(s)\operatorname{d}s\operatorname{d}\rho$, which by Fubini's theorem equals $\textnormal{\tiny{$\sqrt{\frac{2}{\pi}}$}}\int_0^\infty\cos(su)f(s)\operatorname{d}s-\textnormal{\tiny{$\sqrt{\frac{2}{\pi}}$}}\int_0^\infty f(s)\operatorname{d}s$ . By this formula for $G(u)$ with $u=|r\pm t|$, (3) follows. \\
\hspace*{6mm}
Now consider a general $f$. Fix $t\ne 0$. Denote the right hand side of (3) by $G_t(r)$. Let $\varepsilon >0$. Choose $f'\in \mathcal{C}_c$ with $\norm{jf-jf'}\le \frac{\varepsilon}{1+2|t|}$. Then $\norm{\mathcal{S}f_t-G_t}\le \norm{\mathcal{S}f_t-\mathcal{S}f'_t}+$ $\norm{\mathcal{S}f'_t-G'_t}+\norm{G'_t-G_t}$. As just shown, $\norm{\mathcal{S}f'_t-G'_t}=0$. Next,
$\norm{\mathcal{S}f_t-\mathcal{S}f'_t}=$ $\norm{f_t-f'_t}=\norm{ h_t(jf-jf')}\le |t|\norm{ jf-jf'}\le \varepsilon/2$ with $h_t(s):=\sin{ts}/s$, as $|h_t(s)|\le|t|$. Finally, $G'_t-G_t=\frac{1}{2}\int_{|r-t|}^{|r+t|}\big(\mathcal{S}(jf)-  \mathcal{S}(jf')\big)(\rho) \operatorname{d}\rho$, whence $\norm{G'_t-G_t}\le \frac{1}{2}\int_{|r-t|}^{|r+t|}$ $\norm{\big(\mathcal{S}(jf)-  \mathcal{S}(jf')\big)(\rho)} \operatorname{d}\rho\le \frac{1}{2}\big||r+t|-|r-t|\big|^{1/2}\norm{\mathcal{S}(jf)-\mathcal{S}(jf')}\le\frac{1}{2}(2|t|)^{1/2}\norm{jf-jf'}\le \varepsilon/2$. Hence $\norm{\mathcal{S}f_t-G_t}\le \varepsilon$, which proves the claim.\\

We turn to the proof of (a) -- (e).\\
\hspace*{6mm}
(a) Recall that $\pi^{\chi\varsigma}(x)$ are projections in $\C^2$ with $\pi^{\chi-}(x)=I_2-\pi^{\chi+}(x)$.\\
\hspace*{6mm}
(b) $2\int|A_t^\varsigma(x)|^2\operatorname{d}^3x= \int \big(\frac{|x|+\varsigma t}{|x|}\big)^2\big\langle \big(I_2+\varsigma 
h^\chi(\frac{1}{|x|}x)\big) g\big(\big||x|+\varsigma t\big|\big),g\big(\big||x|+\varsigma t\big|\big)\big\rangle \operatorname{d}^3x=2\pi\int_0^\infty\int_{-1}^1(r+\varsigma t)^2\langle (I_2+\varsigma\chi \xi\sigma_3)g(|r+\varsigma t|),g(|r+\varsigma t|)\rangle\operatorname{d}\xi\operatorname{d}r=4\pi\int_0^\infty(r+\varsigma t)^2|g(|r+\varsigma t|)|^2\operatorname{d}r=4\pi \int_{\varsigma t}^\infty r'^2|g(|r'|^2\operatorname{d}r'<\infty$, whence $A^\varsigma_t\in L^2$ and hence also $R_t\in L^2$. Moreover, as $\norm{E^{\textsc{w}}(B_{|t|})\psi}^2=4\pi \int_0^{|t|} r^2|g(r)|^2\operatorname{d}r$, (b) holds.\\
\hspace*{6mm}
(c) As in (b) one computes $2\norm{ E^{\textsc{w}}( B_{|t|}) A^\varsigma_t}^2=2\int_{B_{|t|}}|A_t^\varsigma(x)|^2\operatorname{d}^3x=4\pi\int_{0}^{|t|}(r+\varsigma t)^2|g(|r+\varsigma t|)|^2\operatorname{d}r=4\pi \int_{-|t|}^{0} r'^2|g(|r'|)|^2\operatorname{d}r'=\norm{ E^{\textsc{w}}( B_{|t|}) \psi}^2$ as $|t|+\varsigma t=0$ by assumption.\\
\hspace*{6mm}
(d) First assume $f\in\mathcal{C}_c$. Then $R_t(x) =\lim_{R\to \infty}R^{\textsc{r}}_t(x)$ in (2) for every $x\in\R^3\setminus\{0\}$. Since $h(s):= \frac{\sin(s|x|)}{|x|}f(s)$ is integrable,
$\int_0^\infty \sin(ts)h(s)\operatorname{d}s \to 0$ as $|t|\to \infty$  by the Riemann-Lebesgue lemma. Hence
$R_t(x)\to 0, \;|t|\to \infty$  for every  $x\ne 0$.\\
\hspace*{6mm}
 Next use the obvious estimation $\big|\sin(s|x|)/|x|\big|\le s$ if $|x|\le1$ and $\le 1/|x|$ if $|x|>1$. Therefore, as $|R_t(x)|\le \frac{1}{|x|} \textnormal{\tiny{$\sqrt{\frac{2}{\pi}}$}}\int_0^\infty\big|\frac{\sin(s|x|)}{|x|}\big|\norm{f(s)}\operatorname{d}s$, one has   $|R_t(x)|\le M(x)$ with $M(x):= \frac{C}{|x|}$ if $|x|\le1$ and $M(x)\le \frac{C}{|x|^2}$ if $|x| >1$ for some finite constant $C$. Note $\int M(x)^2\operatorname{d}^3x=8\pi C^2<\infty$. So by dominated convergence $\int |R_t(x)|^2\operatorname{d}x^3 \to 0$ for $|t|\to \infty$, as claimed.\\
\hspace*{6mm}
Now consider a general $f$.  Let $\varepsilon >0$. There exists a radially symmetric $\varphi\in\mathcal{C}_c$ with
$\norm{\varphi-\varphi'}\le \varepsilon/6$. Plainly $f'\in\mathcal{C}_c$. Let $\psi':=\mathcal{F}^{-1}\varphi'$. Then $\norm{\psi-\psi'}\le\varepsilon/6$, whence $\norm{\psi_t-\psi'_t}\le\varepsilon/6$ for all $t$. As in (b) one computes 
$\norm{A_t^\varsigma-A_t'^{\,\varsigma}}^2=2\pi \int_{\varsigma t}^\infty r'^2|g(|r'|)-g'(|r'|)|^2\operatorname{d}r'$, whence $\norm{A_t^\varsigma-A_t'^{\,\varsigma}}\to 0$ for $\varsigma t\to \infty$ and $\norm{A_t^\varsigma-A_t'^{\,\varsigma}}\to\norm{\psi-\psi'}$
  for $\varsigma t\to -\infty$.\\
\hspace*{6mm}  
   Hence for all $|t|$ large enough one has $\norm{R'_t}\le \varepsilon/6$  and $\norm{A_t^\varsigma-A_t'^{\,\varsigma}}\le \varepsilon/3$. Therefore $\norm{R_t}\le \norm{R'_t}+\norm{R_t-R'_t} = \norm{R'_t} +\norm{\psi_t-A^+_t-A_t^- -(\psi'_t-A_t'^+-A_t'^-)}\le \norm{R'_t} +\norm{\psi-\psi'}+$ $\norm{A_t^+-A_t'^-}+\norm{A_t^- -A_t'^-}\le \varepsilon$.\\
\hspace*{6mm}   
(e)   Due to (d) and (a) one has $\lim_{|t|\to\infty} \norm{E^{\textsc{w}}(B_{|t|})\psi_t}^2=\lim_{|t|\to\infty} \norm{E^{\textsc{w}}(B_{|t|})(A_t^++A_t^-)}^2=\lim_{|t|\to\infty} \norm{E^{\textsc{w}}(B_{|t|})A_t^+}^2+\lim_{|t|\to\infty} \norm{E^{\textsc{w}}(B_{|t|})A_t^-}^2$.  Hence the assertion follows by (b) and (c).\qed

\begin{Cor}\label{GTERSWF} Let $\psi$ be radially symmetric around some point $b\in\R^3$, i.e., $\psi=g(|\cdot -b|)$. Then 
\begin{itemize}
\item[(a)] $\lim_{t\to\pm\infty}\norm{E^{\textsc{w}}(B_{|t|})\psi_t}^2=\frac{1}{2}\pm \,2\pi\int_0^1 \int_0^{|b|\xi} \xi\, r^2\,
\langle g(r),h^\chi(\frac{1}{|b|}b)g(r)\rangle\operatorname{d}r\operatorname{d}\xi$\\

\item[(b)] $\frac{1}{4}\le \lim_{t\to\pm\infty}\norm{E^{\textsc{w}}(B_{|t|})\psi_t}^2\le\frac{3}{4}$\\

\item[(c)]  $\lim_{|t|\to\infty}\norm{E^{\textsc{w}}(\{x\in\R^3:|xe|\le|t|\})\psi_t}= 1$
\end{itemize}
\end{Cor}
{\it Proof.}  (a) In order to make easier the computation we first perform the Euclidean transformation $h:=(\beta e_3,B(-b)^{-1})$, $\beta:=|b|$ getting $W^{\textsc{w}}(h)E^{\textsc{w}}(B_{|t|})\psi_t= E^{\textsc{w}}(h\cdot B_{|t|})W^{\textsc{w}}(h)\psi_t=E^{\textsc{w}}(B_{\beta,|t|})\tilde{g}_t(|\cdot|)$
with $B_{\beta,|t|}:=\beta e_3+B_{|t|}$, $\tilde{g}:=B(-b)^{-1}g$, and $\langle g(r),h^\chi(\frac{1}{|b|}b)g(r)\rangle=-\chi \langle \tilde{g}(r),\sigma_3 \tilde{g}(r)\rangle$ by Eq.\,(\ref{CRCS}). So we have to compute $\norm{E^{\textsc{w}}(B_{\beta,|t|})\tilde{\psi}_t}^2$ for $\tilde{\psi}: =\tilde{g}(|\cdot |)$. Henceforth we  may omit the tilde.\\
\hspace*{6mm}
According to (\ref{TERSWF}), $E^{\textsc{w}}(B_{\beta,|t|})\psi_t=\sum_\varsigma E^{\textsc{w}}(B_{\beta,|t|})A^\varsigma_t+E^{\textsc{w}}(B_{\beta,|t|})R_t$ with $R_t\to 0$ for $|t|\to\infty$ and $\norm{\sum_\varsigma E^{\textsc{w}}(B_{\beta |t|})A^\varsigma_t}^2=\sum_\varsigma\norm{E^{\textsc{w}}(B_{\beta,|t|})A^\varsigma_t}^2$. Hence  by (\ref{TERSWF})(b) it suffices to compute
$\lim_{ \varsigma t\to -\infty}\norm{E^{\textsc{w}}(B_{\beta |t|})A^\varsigma_t}^2$. \\
\hspace*{6mm}
Suppose at once $|t|>\beta$. Then a point in $\R^3$ with polar angle $\theta\in[0,\pi]$ is on the surface of $B_{\beta,|t|}$ if and only if, according to the cosine rule, $|t|^2=\beta^2+ r(\theta)^2-2r(\theta)\beta \cos(\theta)$ or, equivalently, $r(\theta)=\sqrt{|t|^2-\beta^2\sin^2(\theta)}+\beta \cos(\theta)$. Let $\tilde{r}(\xi):=\sqrt{|t|^2-\beta^2(1-\xi^2)}+\beta \xi$. It satisfies $\tilde{r}\big(\cos(\theta)\big)=r(\theta)$.\\
\hspace*{6mm}
 Now,
$\norm{E^{\textsc{w}}(B_{\beta,|t|})A^\varsigma_t}^2=\int 1_{B_{\beta,|t|}}(x)\big(\frac{|x|+\varsigma t}{|x|}\big)^2\big\langle\pi^{\chi\varsigma}(\frac{1}{|x|}x)g\big(\big||x|+\varsigma t\big|\big),g\big(\big||x|+\varsigma t\big|\big)\big\rangle\operatorname{d}x^3$. It becomes $2\pi\int_{-1}^1\int_{-\infty}^\infty h_{t}(r',\xi)\operatorname{d}r'\operatorname{d}\xi$ with $h_{t}(r',\xi):=1_{[\varsigma t,\tilde{r}(\xi)+\varsigma t]}(r')k(r',\xi)$ and $k(r',\xi):= r'^2\big\langle \frac{1}{2}(I_2+\varsigma\chi \xi\sigma_3) g(|r'|),g(|r'|)\big\rangle$ by elementary computation. Obviously 
$$|h_{t}(r',\xi)|\le r'^2|g(|r'|)|^21_{[-1,1]}(\xi)$$ 
which is integrable on $]-\infty.\infty[\times[-1,1]$. We are going to apply dominated convergence for $\varsigma t\to -\infty$. \\
 \hspace*{6mm}
 For $\varsigma t<0$ one has $\varsigma t=-|t|$ and hence $\tilde{r}(\xi)+\varsigma t=\sqrt{|t|^2-\beta^2(1-\xi^2)}+\beta \xi- |t|\to \beta \xi$. So 
$h_t(r',\xi)\to  1_{]-\infty,\beta \xi]}(r')k(r',\xi)$.
 Therefore $\lim_{\varsigma t \to -\infty}\norm{E^{\textsc{w}}(B_{\beta,|t|})A^\varsigma_t}^2=2\pi\int_{-1}^1\int_{-\infty}^0 k(r',\xi)\operatorname{d}r' \operatorname{d}\xi+ 2\pi\int_{-1}^1\int_{0}^{\beta \xi} k(r',\xi)\operatorname{d}r' \operatorname{d}\xi$. One easily does the $\xi$-integration for the first summand obtaining $2\pi\int_{-\infty}^0r'^2\norm{g(|r'|)}^2\operatorname{d}r'=\frac{1}{2}$. As to the second summand one finds $\int_{-1}^0\int_{0}^{\beta \xi} k(r',\xi)\operatorname{d}r' \operatorname{d}\xi=\int_{-1}^0\int_{-\beta\xi}^{0} k(-r',\xi)\operatorname{d}r' \operatorname{d}\xi=\int_{0}^1\int_{0}^{\beta \xi} -k(-r,-\xi)\operatorname{d}r \operatorname{d}\xi$. Since $k(r,\xi)-k(-r,-\xi)=\varsigma\chi \xi r^2\langle \sigma_3g(r),g(r)\rangle$, the proof of (a) is accomplished.
 
\hspace*{6mm}
(b)  Estimate $\big|\,2\pi\int_0^1 \int_0^{|b|\xi} \xi\, r^2\,
\langle g(r),h^\chi(\frac{1}{|b|}b)g(r)\rangle\operatorname{d}r\operatorname{d}\xi\big|\le  \,2\pi\int_0^1 \int_0^{\infty} \xi\, r^2\,|g(r)|^2\operatorname{d}r\operatorname{d}\xi=\int_0^1\frac{1}{2}\xi\operatorname{d}\xi=\frac{1}{4}$, whence the assertion.

\hspace*{6mm}
(c) Arguing as  for (a), due to  the Euclidean motion $h=(-B(\varsigma e)^{-1}b, B(\varsigma e)^{-1})$ with $\varsigma =\pm$ such that $\beta:= \varsigma be \ge 0$, it suffices to consider the case $b=0$,  to replace $\{|xe|\le|t|\}$ with $S_{\beta,|t|}:=\{|x_3-\beta|\le|t|\}$, $\beta\ge 0$, and to show
$\lim_{\varsigma t\to -\infty}\norm{E^{\textsc{w}}(S_{\beta,t})A^\varsigma_t} =1$.\\
\hspace*{6mm}
 Suppose at once $|t|>\beta$. By elementary trigonometry a point in $\R^3$ with polar angle $\theta\in[0,\pi]$ is on the boundary of $\{|x_3-\beta|\le|t|\}$ if and only if $r(\theta)=\frac{|t|+\beta}{\cos(\theta)}$ for $0\le \theta<\frac{\pi}{2}$ and $r(\theta)=\frac{|t|-\beta}{\cos(\pi-\theta)}$ for $\frac{\pi}{2}<\theta\le\pi$. \\
\hspace*{6mm}
Hence,  as in (a),
$\norm{E^{\textsc{w}}(S_{\beta,|t|})A^\varsigma_t}^2=
2\pi\int_{-1}^1\int_{-\infty}^\infty h_{t}(r',\xi)\operatorname{d}r'\operatorname{d}\xi$ with $\tilde{r}(\xi)=\frac{\beta}{\xi}+\frac{|t|}{|\xi|}$,   $\xi\ne 0$. If $\xi\not\in \{-1,0,1\}$, then $\tilde{r}(\xi)+\varsigma t 
\to \infty $ and $h_{t}(r',\xi)\to k(r',\xi)$ for $\varsigma t\to -\infty$. 
Finally, one has $ 2\pi\int_{-1}^1\int_{-\infty}^\infty k(r',\xi)\operatorname{d}r'\operatorname{d}\xi=
2\pi  \big(\int_{-1}^1 \operatorname{d}\xi \big)\big(    \int_{0}^\infty r^2\norm{g(r)}^2\operatorname{d}r\big)+\varsigma \chi \,   \pi \big(\int_{-1}^1\xi \operatorname{d}\xi\big)  \big( \int_0^\infty r^2 \langle \sigma_3g(r),g(r)\rangle \operatorname{d}r\big)=1+0$, thus accomplishing the proof. \qed\\

For the  Weyl states $\psi$ in (\ref{GTERSWF}) one has $\norm{E^{\textsc{w}}(\R^3\setminus B_{|t|})\psi_t}\ge \gamma$ for large $|t|$, for every $\gamma<1/2$. 
So the proof of  (\ref{ECCPDS}) via (\ref{CDSLSLHPTLC})  for  the Dirac system cannot be adopted to the Weyl systems since (\ref{CDSLSLHPTLC}) for Weyl states is not available at this stage. However,   the result for the Weyl systems analogous to (\ref{ECCPDS})  holds true and is derived in (\ref{ECCPWS}) by means of   (\ref{GTERSWF})(c) due to appendix (\ref{TSLPS}). As (\ref{ECCPWS}) shows,  (\ref{GTERSWF})(c) actually is valid for all Weyl states.\\


\begin{The}\label{ECCPWS}  $I=\lim_{|\alpha|\to\infty} W^{\textsc{W}\chi}(\alpha)E^{\textsc{W}}(\{|xe|\le |\alpha|\})W^{\textsc{W}\chi}(\alpha)^{-1}$.
\end{The}\\
{\it Proof.}  $P(\alpha):=W^{\textsc{W}\chi}(\alpha)E^{\textsc{W}}(\{|xe|\le|\alpha|\})W^{\textsc{W}\chi}(\alpha)^{-1}$ is an orthogonal projection. It suffices to show $P(\alpha)\psi\to\psi$ for all elements $\psi$ of a total set $L$. Indeed, given $\varepsilon>0$ and a state $\psi$ there is $\psi'$ in the linear hull $\langle L\rangle$ with $\norm{\psi-\psi'}\le \varepsilon/3$. Then, by assumption, $\norm{\psi-P(\alpha)\psi}\le$  $\norm{\psi-\psi'}+\norm{\psi'-P(\alpha)\psi'}+$ $\norm{P(\alpha)\psi'-P(\alpha)\psi}\le 2\norm{\psi-\psi'}+\norm{\psi'-P(\alpha)\psi'}\le \varepsilon$ for all $|\alpha|$ large enough. --- Furthermore 
it suffices to show $\norm{P(\alpha)\psi}\to \norm{\psi}$ for every $\psi\in L$. Indeed,  $\langle (I-P(\alpha))\psi,P(\alpha)\psi\rangle =0$, whence  $\norm{\psi}^2=\norm{(I-P(\alpha))\psi}^2+\norm{P(\alpha)\psi}^2$. By assumption $\norm{(I-P(\alpha))\psi}^2=\norm{\psi}^2-\norm{P(\alpha)\psi}^2\to 0$. This means $(I-P(\alpha))\psi\to 0$ as claimed. --- Finally,  $\norm{P(\alpha)\psi}=\norm{E^{\textsc{W}}(\{|xe|\le|\alpha|\})\psi_{-\alpha}}$. Thus the result holds by appendix  (\ref{TSLPS}) and (\ref{GTERSWF})(c).\qed\\

 Arguing by means of  time reversal it is clear that  it suffices to prove (\ref{ECCPWS})  for  $\alpha\to \infty$ (or $\alpha\to -\infty$). We like to recall that, due to (\ref{DSLSLHPTLC})(f),  (\ref{ECCPWS}) is equivalent to $I=P^>_\sigma +P^<_\sigma$ and the other similar equalities. 
 Hence, for the Weyl systems,  also the version analogous to
 (\ref{CCPDS}) is equivalent to (\ref{ECCPWS}). Therefore the density result analogous to (\ref{DLTES}) concerning late-change states is valid, too.

\begin{Cor}\label{GECCPWS} Consider finitely many directions $e^1,\dots, e^m$, i.e., $e^j\in\R^3$, $|e^j|=1$. Then  $I=$\,s-$\lim_{|\alpha|\to\infty} W^{\textsc{W}\chi}(\alpha)E^{\textsc{W}}(\{x\in\R^3: |xe^1|\le|\alpha|,\dots,|xe^m|\le |\alpha| \})W^{\textsc{W}\chi}(\alpha)^{-1}$.
\end{Cor}\\
{\it Proof.} Let $P(\alpha,m)$ denote the orthogonal projection on the right hand side. We proceed by induction on $m$. The base case $m=1$ holds by (\ref{ECCPWS}). It follows the inductive step $m\to m+1$. Let $P(\alpha):=W^{\textsc{W}\chi}(\alpha)E^{\textsc{W}}(\{|xe^{m+1}|\le|\alpha|\})W^{\textsc{W}\chi}(\alpha)^{-1}$. Check $P(\alpha,m+1)=P(\alpha)P(\alpha,m)=P(\alpha,m)P(\alpha)$. Then for every state $\psi$, $\norm{P(\alpha,m)\psi}^2=$ $\norm{P(\alpha)P(\alpha,m)\psi}^2+\norm{\big(I-P(\alpha)\big)P(\alpha,m)\psi}^2$, whence $\norm{P(\alpha,m+1)\psi}^2=\norm{P(\alpha,m)\psi}^2-\norm{P(\alpha,m)\big(I-P(\alpha)\big)\psi}^2\to 1+0$ by assumption and since $\norm{P(\alpha,m)\big(I-P(\alpha)\big)\psi}\le$ $ \norm{\big(I-P(\alpha)\big)\psi}\to 0$ by (\ref{ECCPWS}). Since $\norm{\big(I-P(\alpha,m+1)\big)\psi}^2=1-\norm{P(\alpha,m+1)\psi}^2$, this implies $\big(I-P(\alpha,m+1)\big)\psi\to 0$, finishing the proof.\qed\\

For $e^1,e^2,e^3$ the standard basis of $\R^3$, $C_{|t|}:=\{x\in\R^3: |xe^1|\le|t|,\dots,|xe^3|\le |t|\}$ is the cube with sides parallel to the axes $\R^3e^j$ circumscribed about the ball $B_{|t|}$. Choosing more directions $e^1,\dots, e^m$ one gets a figure $F_{|t|}$ which is even much more close to $B_{|t|}$. Nevertheless, by (\ref{GECCPWS}), $\lim_{|t|\to\infty}\norm{E^{\textsc{w}}(F_{|t|})\psi_t}^2=1$ for every Weyl state $\psi$. This is remarkable in view of the result (\ref{GTERSWF})(b).\\


\section{No  Causal  Spin Zero Systems}\label{SD1}

The results in sec.\,\ref{CMSCT}, \ref{MLST} imply that there are  no  (massive or massless) spin--$0$ systems with causal time evolution. Actually, this is a rather general fact not restricted to relativistic systems, see \cite[Corollary 2]{CL15}. One may ask about its relevance  for the
  \textbf{Klein-Gordon equation}  
\begin{equation} \label{KGE} 
 - \partial^2_t  \Psi(t,x)  =-\left( \nabla^2 -m^2\right)\Psi(t,x)
\end{equation} 
which is a well-established relativistic wave equation for the description of a massive  spinless particle. It is manifestly relativistic as for every solution $\Psi$  of  Eq.\,(\ref{KGE}) and every $g\in\tilde{\mathcal{P}}$ the transformed wave function $\Psi_g(\mathfrak{x}):=\Psi(g^{-1}\cdot\mathfrak{x})$ with $\mathfrak{x}=(t,x)$ is still a solution. (For a detailed discussion of Eq.\,(\ref{KGE}) 
see \cite{FV58}). We  will show explicitly  that, in contrast to what  holds for the Dirac and Weyl equations, the multiplication of $\Psi$ by the space variables  $x_j$, $j=1,2,3$  and by the indicator functions $1_\Delta$ of measurable sets $\Delta\subset \R^3$ does not define observables as these operators turn out to be not self-adjoint.  
This failure is expected due to the fact that otherwise the canonical PM  $E(\Delta)\Psi:=1_\Delta \Psi$ would (as in the Dirac and Weyl cases) constitute a causal WL, which as we know does not exist.\\

The starting point for the discussion of Eq.\,(\ref{KGE}) are their plane wave solutions  $(t,p)\mapsto \operatorname{e}^{-\operatorname{i}\eta\epsilon(p)t+\operatorname{i}px}$, $p\in\R^3$,  $\eta\in\{-1,+1\}$ with $\epsilon(p)=\sqrt{|p|^2+m^2}$. Superposing these waves one gets the linear space $\mathcal{H}^\textsc{kg}$ of solutions of Eq.\,(\ref{KGE}) 

\begin{equation}\label{SLKG}
\Psi(t,x)=\sum_\eta \int_{\R^3}\, \varphi_\eta(p)  \operatorname{e}^{-\operatorname{i}\eta\epsilon(p)t} \operatorname{e}^{\operatorname{i}px}  \, \frac{\operatorname{d}^3p}{\sqrt{2\epsilon(p)}}, \quad  \varphi_\eta\in L^2(\R^3)
\end{equation}

More precisely, $\Psi(t,\cdot)=\sum_\eta \mathcal{F}^{-1}\left( (2\epsilon)^{-\frac{1}{2}}    \operatorname{e}^{-\operatorname{i}\eta\epsilon t} \varphi_\eta \right)$, which is well-defined as $\frac{|\varphi_\eta|}{\sqrt{\epsilon}} \in L^2$. The factor $(2\epsilon)^{-\frac{1}{2}}$ contributes to the Lorentz invariant measure 
$\frac{\operatorname{d}^3p}{2\epsilon(p)}$ so that
$ \Psi_g(\mathfrak{x})=\sum_\eta\int_{\R^3}(W^\eta(g)\varphi_\eta)(p)\operatorname{e}^{-\operatorname{i}\eta\epsilon(p)t} \operatorname{e}^{\operatorname{i}px}  \, \frac{\operatorname{d}^3p}{\sqrt{2\epsilon(p)}} 
$ holds. Here
\begin{equation}
\big(W^\eta(\mathfrak{a},A)\varphi\big)(p)=\sqrt{\epsilon(q^\eta)/\epsilon(p)}\, \operatorname{e}^{\operatorname{i} \mathfrak{p}^{\eta}\cdot \,\mathfrak{a}} \,\varphi(q^\eta)
\end{equation}
is the  representation  $[m,0,\eta]$ of $\tilde{\mathcal{P}}$  on $L^2(\R^3)$ from Eq.\,(\ref{RIMS}). Recall that $W^\eta$ is unitary. \\
\hspace*{6mm} Let $\tau\equiv ((\tau,0),I_2)\in\tilde{\mathcal{P}}$ denote the time translation by $\tau\in\R$. Then one easily checks that consistently $\Psi_\tau(t,x)=\Psi(t-\tau,x)$. In the following $\widehat{\Psi_t}:= \mathcal{F}\Psi(t,\cdot)$.


\begin{Lem} The space $ \mathcal{H}^\textsc{kg}$ of solutions $\Psi$ of the Klein-Gordon equation given by Eq.\,\emph{(\ref{SLKG})} provided with the inner product
$$\langle \Psi,\Psi'\rangle :=\int_{\R^3}\left( \epsilon(p)\,  \overline{\widehat{\Psi_t}(p)}\,\widehat{\Psi'_t}(p)+ \frac{1}{\epsilon(p)} \, \overline{\partial_t\widehat{\Psi_t}(p)}\,\partial_t\widehat{\Psi'_t}(p)\right)\operatorname{d}^3p$$
which does not depend on time $t$, is a Hilbert space, on which $W^\textsc{kg}(g)\Psi:=\Psi_g$ is a representation $[m,0,+]\oplus [m,0,-]$ of  $\tilde{\mathcal{P}}$. In particular,  the time evolution of the wave functions determined by the Klein-Gordon equation is unitary.
\end{Lem}

{\it Proof.} Let $\iota: L^2(\R^3)\oplus L^2(\R^2)\to \mathcal{H}^\textsc{kg}$, $\iota(\varphi_+,\varphi_-):= \Psi$ be  the linear surjection given by Eq.\,(\ref{SLKG}). Then, as shown,  $W^\textsc{kg}\,\iota=\iota\, (W^+\oplus W^-)$ holds. Since  $\widehat{\Psi_t}=(2\epsilon)^{-\frac{1}{2}}\sum_\eta\operatorname{e}^{-\operatorname{i}\eta\epsilon t}\varphi_\eta$ and $\partial_t\widehat{\Psi_t}=-\operatorname{i}(\epsilon/2)^{\frac{1}{2}}\sum_\eta \eta\operatorname{e}^{-\operatorname{i}\eta\epsilon t}\varphi_\eta$ it follows $\varphi_\eta= \operatorname{e}^{\operatorname{i}\eta\epsilon t} \left(  (\epsilon/2)^\frac{1}{2} \widehat{\Psi_t}+\operatorname{i}\eta (2\epsilon)^{-\frac{1}{2}}       \partial_t\widehat{\Psi_t}    \right)$, whence $\iota$ is also injective with $\iota^{-1}\Psi= (\varphi_+,\varphi_-)$. Then $\langle \Psi,\Psi'\rangle :=\langle \iota^{-1}\Psi,\iota^{-1}\Psi'\rangle$ is a inner product on $ \mathcal{H}^\textsc{kg}$ making $ \mathcal{H}^\textsc{kg}$  a Hilbert space and $W^\textsc{kg}$ unitary. It is easy to check the above formula for $\langle \Psi,\Psi'\rangle$.  \hfill{$\Box$}\\

Now, as to the posed question about the self-adjointness of the multiplication operator $(X_j\Psi)(t,x)=x_j\Psi(t,x)$, one easily verifies that  in momentum space, which is the carrier space of $W^+\oplus W^-$, the latter equals 
\begin{equation}\label{KGPO}
X^{mom}_j=-\frac{\operatorname{i}}{2}\frac{P_j}{P^2+m^2I}+\operatorname{i}\partial_{p_j} 
\end{equation}
revealing a non-vanishing bounded skew-adjoint part.\\

\section{Lattice Generated by the Non-Timelike Relation}\label{LGNTLR}

As shown in (\ref{MFEMSLS}), (\ref{EMTVMH})
there are  maximal spacelike sets which are not  bases, i.e., for which the sets of dependency are not the whole spacetime $\R^4$. One may feel this fact to be an unsatisfactory property of the spacelike relation $\perp$. It turns out that the weaker orthogonality $\perp'$ meaning non-equal and non-timelike, i.e. 
$$\mathfrak{x}\perp'\mathfrak{y}\quad  \Leftrightarrow\quad  \mathfrak{x}\ne\mathfrak{y} \textnormal{ and } (\mathfrak{x}-\mathfrak{y})^{\cdot 2}\le 0$$ 
generates a lattice $\mathcal{M}'$, which from a mathematical point of view has a richer structure than  $\mathcal{M}$. In Cegla, Jadczyk 1977 \cite[Theorem 1]{CJ77} it is shown that $(\mathcal{M}',\subset, \perp')$ is a complete orthomodular lattice and, what is more, that $(\R^4,\perp')$ is a complete $D$-space, which means that every $\perp'$-complete set $M=M^{\land'}$ is the $\perp'$-completion of any maximal $\perp'$-set contained in $M$. In
\cite[Corollary 1]{CJ77} it is also shown that every timelike line intersects every maximal $\perp'$-set in $\R^4$. Hence, every maximal $\perp'$-set  is a $\perp'$-base. More generally, as a consequence noted in (\ref{SIMLAND}), for every $\perp'$-set its set  of $\perp'$-determinacy equals its $\perp'$-completion. \\
\hspace*{6mm}  
Therefore in addition to the well-known spacelike hyperplanes there are also the non-timelike not spacelike hyperplanes, which are  $\perp'$-bases as well. An example of such a hyperplane is $\{\mathfrak{x}: x_0=x_3\}$. The question is about their meaning. Obviously they do not constitute a reference frame. In fact  there is no direction $\mathfrak{e}\in\R^4$,  
$\mathfrak{e}\cdot \mathfrak{e}=1$ with $\mathfrak{e}\cdot (\mathfrak{x}-\mathfrak{x}')=0$ \;$\forall$ $\mathfrak{x},\mathfrak{x}'\in\sigma$, which would determine the time coordinate of the frame. One may  attempt to consider such a hyperplane to be some high boost limit of a spacelike hyperplane. This idea is concretized  in sec.\,\ref{ECLNTLHP}, when localization operators are attributed to $\perp'$-complete regions. Thereby rather  interesting properties of causal systems are predicted, which are confirmed for the Dirac and the Weyl systems.

 $\perp'$-orthogonality, in contrast to $\perp$-orthogonality, considers lightlike separated events to be independent. This is reasonable within classical mechanics where all objects are macroscopic and  move slower than light. Let us remark that Casini 2002 \cite{C02} studies $\perp'$-orthogonality on general spacetime manifolds with dimension $\ge2$ showing that the respective lattice $\mathcal{M}'$ is complete orthomodular and atomic. \cite[sec.\,3]{C02} proposes an interpretation of $\mathcal{M}'$  in terms of propositions for classical particles. In the first instance,  to every $M\subset \R^4$ a proposition is attributed, namely the proposition that the particle passes through $M$. But, as observed by \cite[sec.\,3]{C02}, in general  the logical opposite of the proposition attributed to $M$ (i.e., the particle does not meet $M$)  is not a proposition represented by some $L\subset \R^4$. To be clear note first  that a classical particle in Minkowski space  is described by a timelike  line. Let $\mathcal{T}_M$ denote the set of all  timelike  lines meeting $M$, which represents the set of all particles  which pass through $M$. The equality $\mathcal{T}_{M_1}=\mathcal{T}_{M_2}$ means that $M_1$ and $M_2$ represent the same proposition. One easily verifies  $\mathcal{T}_{M_1}\cap\mathcal{T}_{M_2}=\emptyset$ $\Leftrightarrow$ $M_1\perp' M_2$. This means that the propositions attributed to $M_1$ and $M_2$ exclude each other if and only if they are $\perp'$-orthogonal.  Put   $\mathcal{T}:=\mathcal{T}_{\R^4}$. There is the following result.

 \begin{Lem}\label{JLGNTLR} $ \mathcal{T}\setminus \mathcal{T}_M= \mathcal{T}_L$ if and only if $ \mathcal{T}_M=\mathcal{T}_{M^{\land'}}$   and $ \mathcal{T}_L=\mathcal{T}_{M^{\perp'}}$.
 \end{Lem}\\
 {\it Proof.} $ \mathcal{T}\setminus \mathcal{T}_M= \mathcal{T}_L$ implies $ \mathcal{T}_M\cap  \mathcal{T}_L=\emptyset$, whence $L\subset M^{\perp'}$ and hence $ \mathcal{T}\setminus \mathcal{T}_{M^{\land'}}\subset  \mathcal{T}\setminus \mathcal{T}_M=\mathcal{T}_L\subset 
 \mathcal{T}_{M^{\perp'}}$. --- Let $A$ be a maximal $\perp'$-set in $M^{\land'}$ and $B$ a maximal $\perp'$-set in $M^{\perp'}$. Then $A\cap B=\emptyset$ and $A\cup B$ is a maximal $\perp'$-set. Therefore $u\in  \mathcal{T}_{M^{\perp'}}$ $\Leftrightarrow$ $u\cap M^{\perp'}\ne \emptyset\Leftrightarrow u\cap B\ne \emptyset$  (by (\ref{SIMLAND}))\,$\Leftrightarrow u\cap A= \emptyset$ $\Leftrightarrow$ $u\cap  M^{\land'}=\emptyset$ (by (\ref{SIMLAND}))\,$\Leftrightarrow$ $u\in \mathcal{T}\setminus \mathcal{T}_{M^{\land'}}$. This proves $ \mathcal{T}\setminus \mathcal{T}_{M^{\land'}}=
 \mathcal{T}_{M^{\perp'}}$, whence the assertion.\qed\\
  
   Hence  (\ref{JLGNTLR}) clarifies that only the set of propositions represented by $\perp'$-complete sets is  negation-closed. This is one reason why consider  $\mathcal{M}'$ for an adequate   description of  spatial-temporal localization of classical particles. But, obviously, not every reasonable physical proposition concerning the 
 latter belongs to   $\mathcal{M}'$. For instance the proposition that the particle passes through $M$ \textbf{and} passes through $L$ for $M,L\in \mathcal{M}'$ in general does not correspond to the conjunction  $M \land' L$ representing the proposition that the particle passes through $M\cap L$. ($M \land' L$ is merely 
 the maximal proposition  in $\mathcal{M}'$ implying $M$ and  $L$.) \\
  \hspace*{6mm} 
 The question is about the
relevance   of $\perp'$ for quantum mechanics. 
The concept of a causal system does not discern $\perp$- and $\perp'$-orthogonality as shown in (\ref{ELOLSO}), despite the fact that motions at the speed of light really occur as the 
motion of the border of the Dirac and Weyl wave functions (see (\ref{LTSIE}),\,(\ref{SRWS})). Moreover, as already remarked above,  the respective lattices $\mathcal{M}$ and $\mathcal{M}'$ show significant differences, and just the latter proves to be particularly relevant for causal systems.\\
 \hspace*{6mm}
Before tackling the question of a possible rep of $\mathcal{M}^{'borel}$ (\ref{PPCBS}) and proving the implications for the Dirac and Weyl systems (see sec.\,\ref{SRLBSGNTLR},\,\ref{ECLNTLHP}) we like to study $\mathcal{M}'$ in some detail. The examples at the level of exercises mainly serve to become more familiar with $\mathcal{M}'$ and can be skipped.

\subsection{$\perp'$-sets and  non-timelike hyperplanes}\label{OSNTH}
Let $A$ be a subset of Minkowski space. $A$ is said to be non-timelike or $\perp'$-orthogonal or a $\perp'$-set, if $(\mathfrak{x}-\mathfrak{y})^{\cdot 2}\le 0$ for every $\mathfrak{x},\mathfrak{y} \in A$. Note that every one-point set is a $\perp'$-set. \\
\hspace*{6mm}
Let $M\subset\R^4$ and let $A\subset M$ be a $\perp'$-set. Then there is a maximal  $\perp'$-set in $M$ containing $A$. This holds true by an obvious application of Zorn's lemma. In particular, every set $M$ contains a maximal $\perp'$-set in $M$. \\
\hspace*{6mm}
By definition, a maximal $\perp'$-set is a maximal $\perp'$-set  in $\R^4$. Any maximal  $\perp'$-set $A$ is closed. (Indeed,  let $\mathfrak{x}_n\in A$ with $\mathfrak{x}_n\to \mathfrak{x}$ for some $\mathfrak{x}\in\R^4$. Then,  for all $\mathfrak{y}\in A$, $0\ge (\mathfrak{x}_n-\mathfrak{y})^{\cdot 2}\to (\mathfrak{x}-\mathfrak{y})^{\cdot 2}$, whence $\mathfrak{x}\in A$.) Every $\perp'$-set is contained in a maximal $\perp'$-set.  For the following lemma see \cite[(2)]{CJ79}.

\begin{Lem}\label{MPPS} $\Sigma$ is a maximal $\perp'$-set if and only if it is the graph $\{\big(\xi(x),x\big):x\in\R^3\}$    of a  map $\xi:\R^3\to \R$ satisfying $|\xi(x)-\xi(y)|\le |x-y|$. $\xi$ is uniquely determined. Clearly, $\xi$ is Lipschitz continuous.
\end{Lem}

{\it Proof.} Let $\Sigma$ be a maximal $\perp'$-set. Let $x\in\R^3$. By \cite[Corollary 1]{CJ77} the timelike line $(0,x)+\R(1,0)$ intersects $\Sigma$. The intersection consists of one point only as $\Sigma$ is $\perp'$-orthogonal. Hence there is a unique $\xi(x)\in\R$ with $(\xi(x),x)\in\Sigma$. Since $\Sigma$ is  $\perp'$-orthogonal, $|\xi(x)-\xi(y)|\le |x-y|$ holds. --- Conversely,  obviously $\Sigma:=\operatorname{graph}(\xi)$ is $\perp'$-orthogonal. Assume $\mathfrak{y}\in\Sigma^{\perp'}$. Then $(\xi(y),y)\in\Sigma$ and  $|y_0-\xi(y)|\le |y-y|=0$, whence the contradiction $\mathfrak{y}=(\xi(y),y)$. Hence $\Sigma$ is maximal. 
\qed\\

 Plainly, every non-timelike hyperplane $\kappa$ is a maximal $\perp'$-set. Also, $\kappa$ is a non-timelike hyperplane if and only if it is the graph of $\xi(x):=\rho +xe$, $x\in\R^3$, for unique $\rho\in\R$ and $e\in\R^3$ with $|e|\le1$ or, equivalently, if $\kappa=\{\mathfrak{x}: \mathfrak{x}\cdot \mathfrak{e}=\rho\}$ for unique $\rho\in\R$ and non-spacelike
$\mathfrak{e}=(1,e)$, and $\kappa$ is non-timelike not spacelike  if and only if $|e|=1$. 
 Every non-timelike not spacelike hyperplane $\kappa$ is given by $\kappa=g\cdot \{\mathfrak{x}: x_0=x_3\}$ for the Poincar\'e transformation
 $g=\big((\rho,0),B\big)h$ with $B\in SU(2)$ satisfying $B\cdot (0,0,1)=e$ and arbitrary $h$ 
leaving $\{\mathfrak{x}: x_0=x_3\}$ invariant (see (\ref{IGNSLHP})).

\hspace*{6mm}
If  $\kappa$ is a non-timelike not spacelike hyperplane  and if $\sigma$ is a spacelike hyperplane then  $\pi:=\kappa\cap\sigma$ is a spacelike plane. Let $M$ be a subset of one of the two open half-hyperplanes which constitute $\kappa\setminus\pi$. Then, according to the definition in  sec.\,\ref{PPCS}, one determines the open half-hyperplane $\sigma_M$ in $\sigma\setminus\pi$ such that $M$ does not lie at the same side of $\pi$ as $\sigma_M$.

\begin{Examp}\label{GMPPOS} Let $\sigma,\tau$ be two non-intersecting spacelike hyperplanes and let $\kappa$ be a non-timelike not spacelike hyperplane. Then $$\sigma_M\cup M \cup \tau_M$$ is a maximal $\perp'$-set.
Here $M$ denotes the closed subset of $\kappa$ between $\sigma$ and $\tau$.
 \end{Examp}
 
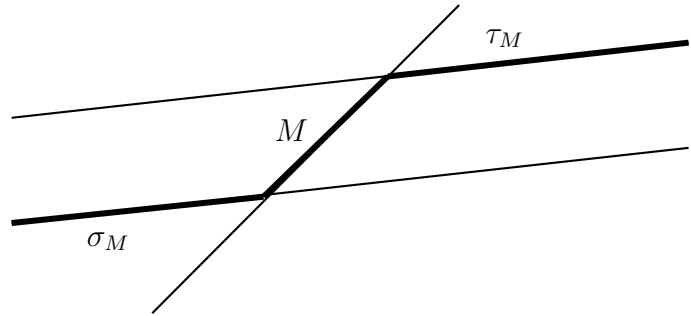
 \begin{pspicture}(-9,-1)(2,3)

\psline(-6,1.5)(3,2.5)
\psline(-6,0.1)(3,1.1)

\psline(-4.13,-1.1)(-0.05,3)
\psline[linewidth=0.8mm](-6,0.1)(-2.65,0.45)
\psline[linewidth=0.8mm](-2.65,0.45)(-1,2.05)
\psline[linewidth=0.8mm](-1,2.05)(3,2.5)

\put(-5,-0.2){$\sigma_M$}
\put(0.3,2.5){$\tau_M$}
\put(-2.5,1.2){$M$}

\end{pspicture}

  {\it Proof.} We anticipate the result (\ref{PCEPRD}). So it suffices to show that every timelike line $l=\mathfrak{y}+\R \mathfrak{z}$, $\mathfrak{z}=(1,z), |z|<1$ intersects $\sigma_M\cup M \cup \tau_M$ just once. \\
\hspace*{6mm}
Let $\sigma=\{\mathfrak{x}: \mathfrak{x}\cdot \mathfrak{e}=\rho_\sigma\}$, $\tau=\{\mathfrak{x}: \mathfrak{x}\cdot \mathfrak{e}=\rho_\tau\}$, $\kappa=\{\mathfrak{x}: \mathfrak{x}\cdot \mathfrak{e}_\kappa=\rho_\kappa\}$  with $\mathfrak{e}=(1,e)$, $|e|<1$ and $\mathfrak{e}_\kappa=(1,e_\kappa)$, $|e_\kappa|=1$ and assume without restriction  $\rho_\sigma<\rho_\tau$ so that $M=\{\mathfrak{x}\in\kappa:\rho_\sigma-\rho_\kappa\le \mathfrak{x}\cdot(\mathfrak{e}-\mathfrak{e}_\kappa)\le\rho_\tau-\rho_\kappa\}$,
 $\sigma_M=\{\mathfrak{x}\in\sigma: \mathfrak{x}\cdot(\mathfrak{e}-\mathfrak{e}_\kappa)<\rho_\sigma-\rho_\kappa\}$, and $\tau_M=\{\mathfrak{x}\in\tau: \mathfrak{x}\cdot(\mathfrak{e}-\mathfrak{e}_\kappa)>\rho_\tau-\rho_\kappa\}$. Moreover, $\kappa=\kappa_1\cup M\cup\kappa_2$ with $\kappa_1:=\{ \mathfrak{x}\in\kappa: \mathfrak{x}\cdot(\mathfrak{e}-\mathfrak{e}_\kappa)<\rho_\sigma-\rho_\kappa\}$ and $\kappa_2:=\{ \mathfrak{x}\in\kappa: \mathfrak{x}\cdot(\mathfrak{e}-\mathfrak{e}_\kappa)>\rho_\tau-\rho_\kappa\}$.\\
\hspace*{6mm} 
 Now one  notes $l\cap \kappa=\{\mathfrak{y}+s \mathfrak{z}\}$ for $s:=\frac{1}{\mathfrak{z}\cdot\mathfrak{e}_\kappa}(\rho_\kappa-\mathfrak{y}\cdot\mathfrak{e}_\kappa)$ and $l\cap \sigma=\{\mathfrak{y}+t \mathfrak{z}\}$ for $t:=\frac{1}{\mathfrak{z}\cdot\mathfrak{e}}(\rho_\sigma-\mathfrak{y}\cdot\mathfrak{e})$, due to $\mathfrak{z}\cdot\mathfrak{e}_\kappa>0$, $\mathfrak{z}\cdot\mathfrak{e}>0$. By a short computation one finds 
 $$(\mathfrak{y}+s \mathfrak{z})\cdot (\mathfrak{e}-\mathfrak{e}_\kappa)    <\rho_\sigma-\rho_\kappa\; \Leftrightarrow \;(\mathfrak{y}+t \mathfrak{z})\cdot (\mathfrak{e}-\mathfrak{e}_\kappa)<\rho_\sigma-\rho_\kappa$$
 This proves $l\cap \kappa_1\ne\emptyset$ $\Leftrightarrow$ $l\cap\sigma_M\ne \emptyset$.
 Analogously $l\cap \kappa_2\ne\emptyset$ $\Leftrightarrow$ $l\cap\tau_M\ne \emptyset$ holds. A moment of reflection shows the proof.\\
  \hspace*{6mm} 
Alternatively, because of Poincar\'e covariance, for the proof of (\ref{GMPPOS}) it  suffices to verify that for $e\in\R^3$, $|e|=1$ and $\alpha>0$
\begin{equation}\label{SGMPPOS}
 \{\mathfrak{x}:x_0=0, xe< 0\}\cup \{\mathfrak{x}: 0\le x_0=xe\le\alpha\}\cup\{x_0=\alpha, xe>\alpha\}
 \end{equation}  
 is a maximal $\perp'$-set.\footnote{ Let $P'$ denote the set in Eq.\,(\ref{SGMPPOS}) for $e=(0,0,1)$.\\
\hspace*{6mm}
(i) We show $\sigma_M\cup M\cup \tau_M=g\cdot P'$ for some $g\in\tilde{\mathcal{P}}$. Indeed, let $\sigma, \tau, \kappa$ be as in the proof of (\ref{GMPPOS}). Then $\sigma=g\cdot \{\mathfrak{x}: x_0=0\}$ and $\tau=g\cdot \{\mathfrak{x}: x_0=\alpha\}$  with $\alpha:=(1-e^2_\sigma)^{-1/2}(\rho_\tau-\rho_\sigma)$ for $g=g_0\,(b,B)$ with $g_0:=\big((\rho_\sigma,0),Q(\mathfrak{e})\big)$ and arbitrary $(b,B)\in ISU(2)$ (see the footnote in sec.\,\ref{PPCS}). Now, $\kappa':=g_0^{-1}\kappa$ is a non-timelike not spacelike hyperplane and hence $\kappa'=\{\mathfrak{x}: \mathfrak{x}\cdot \mathfrak{e}'=\rho'\}$ for some $\mathfrak{e}'=(1,e')$, $|e'|=1$ and $\rho'\in\R$.  Specify $(b,B)$ such that $e'=B\cdot (0,0,1)$ and $be'=\rho'$, whence $\kappa'=(b,B)\cdot\{\mathfrak{x}: x_0=x_3\}$. Thus $\kappa=g\cdot \{\mathfrak{x}: x_0=x_3\}$. Finally, following the definition of $M$, $\sigma_M$, and $\tau_M$, one verifies the assertion.\\
\hspace*{6mm}
 (ii) For the sake of completeness we show that the only $g\in\tilde{\mathcal{P}}$ leaving $P'$ invariant are $g=\big((0,a_1,a_2,0),\operatorname{e}^{\operatorname{i}\varphi\sigma_3/2})$, $a_1,a_2,\varphi\in\R$. Indeed, these $g$ obviously satisfy $g\cdot P'=P'$. Hence for a general $g=(\mathfrak{a},A)$ assume at once $\mathfrak{a}=(a_0,0,0,a_3)$. Observe first $\mathfrak{a}=g\cdot 0\in P'$. Next check $\{\mathfrak{x}\in P':\mathfrak{x}^{\cdot 2}=0\}=\{ s\mathfrak{e}: s\in[0,\alpha]\}$ for $\mathfrak{e}:=(1,0,0,1)$. Since $A$ acts orthocronously,  $(A\cdot \mathfrak{e})_0>0$. Therefore $g\cdot s\mathfrak{e}\in P'$ for $s\in [0,\alpha]$ excludes $a_0=0$ or $a_3=\alpha$. It follows $\mathfrak{a}=a_0\mathfrak{e}$ and $g\cdot s\mathfrak{e}\in M:= \{\mathfrak{x}:0\le x_0=x_3\le\alpha\}$. Hence $(A\cdot \mathfrak{e})_0=(A\cdot \mathfrak{e})_3$. Because of $(A\cdot \mathfrak{e})^{\cdot 2}=0$ one has $A\cdot \mathfrak{e}=\lambda \mathfrak{e}$ with $\lambda>0$.  
 Then (see \,(\ref{IGNSLHP}))
  \begin{displaymath}
A\cdot \mathfrak{e}=\lambda \mathfrak{e}\quad \Leftrightarrow \quad  A\in A_{\ln(\lambda)}\,E(2) \quad \Leftrightarrow \quad A=\,
\operatorname{e}^{\ln(\lambda)\sigma_3/2} \,   \operatorname{e}^{\operatorname{i}\varphi\sigma_3/2}  \,
\left(\begin{array}{cc} 1& w\\ 0 & 1\end{array}\right)
\end{displaymath}
with $\varphi\in\R$, $w=u+\operatorname{i} v\in\C$. So $A$ leaves $M$ invariant and hence $(A\cdot\mathfrak{x})_0=\operatorname{e}^\rho(x_0+ux_1-vx_2) \in[0,\alpha]$ $\forall$ $\mathfrak{x}\in M$, whence $u=v=0$ and $\rho=0$, thus accomplishing the proof.}\qed

\subsection{$\perp'$-completion and  set of $\perp'$-determinacy}
 Let $M\subset \R^4$. Then $M^{\land '}:=M^{\perp'\perp'}$ denotes the $\perp'$-completion of $M$. Moreover, $M$ is said to be $\perp'$-\textbf{complete} if $M=M^{\land '}$. 
The set of $\perp'$-determinacy of $M$  is defined as 
$$M^{\sim'}:=\{\mathfrak{x}:\forall\;\mathfrak{z} \textnormal{ with }\mathfrak{z}^{\cdot 2}>0\;\exists\, s\in\R \textnormal{ with } \mathfrak{x}+s\mathfrak{z}\in M\}$$
It consists of  all points $\mathfrak{x}$ such that every timelike  line through $\mathfrak{x}$  meets $M$.
\\

 Obviously,  $M\subset M^{\sim'}$. Further, $M^{\sim'\sim'}= M^{\sim'}$ holds. (Indeed, let $\mathfrak{x}  \in M^{\sim'\sim'}$. For $ \mathfrak{z}^{\cdot2}> 0$ there is $s\in \R$ with $\mathfrak{x}+s\mathfrak{z}\in M^{\sim'}$. Hence, there is $s'\in\R$ with $(\mathfrak{x}+s\mathfrak{z})+s'\mathfrak{z}\in M$, whence $\mathfrak{x}\in M^{\sim'}$.) Clearly $M^{\sim'}_1\subset M^{\sim'}_2$ if $M_1\subset M_2$.

\begin{Lem}\label{PCEPRD} 
$M^{\sim'}\subset M^{\land '}$. 
\end{Lem}\\
{\it Proof.} Fix $\mathfrak{x}\in M^{\sim'}$. Then there is a timelike $\mathfrak{w}$ with $\mathfrak{x}+\mathfrak{w} \in M$. Hence $\big((\mathfrak{x}+\mathfrak{w})-\mathfrak{x}\big)^{\cdot 2}=\mathfrak{w}^{\cdot 2}>0$,  whence $\mathfrak{x}\not\in M^{\perp'}$. Now let 
 $\mathfrak{y}\in M^{\perp'}$. Then $\mathfrak{z}:=\mathfrak{x}-\mathfrak{y}\ne 0$. Assume $\mathfrak{z}^{\cdot 2}>0$. Then there is $s\in\R$ with $\mathfrak{x}-s\mathfrak{z}\in M$. Clearly, $s\ne 1$ as $\mathfrak{y}\not\in M$. It follows the contradiction $0\ge \big((\mathfrak{x}-s\mathfrak{z})-\mathfrak{y}\big)^{\cdot 2}=(\mathfrak{z}-s\mathfrak{z})^{\cdot 2}=(1-s)^2\mathfrak{z}^{\cdot 2}>0$. Thus $\mathfrak{z}^{\cdot 2}\le 0$ contrary to the assumption. This implies $\mathfrak{x}\in M^{\perp'\perp'}=M^{\land '}$.\qed \\

 Thus, by  (\ref{PCEPRD}), one has  $M\subset M^{\sim'}=M^{\sim'\sim'} \subset M^{\land'}$, whence $M=M^{\sim'}$ if $M=M^{\land'}$. But $M=M^{\sim'}$ does not imply that $M$ is $\perp'$-complete.  A simple example is $M:=\R^4\setminus \R\mathfrak{e}$ for $\mathfrak{e}=(1,0,0,0)$ with $M=M^{\sim'}$ and $M^{\land'}=\R^4$. (Indeed, $M=M^{\sim'}$ is obvious.  Next note $M^{\perp'}\subset \R\mathfrak{e}$, since generally $M^{\perp'}\cap M=\emptyset$, and $s  \mathfrak{e} \not\perp' (2+s,0,0,1)$ with $(2+s,0,0,1)\in M$. Hence $M^{\perp'}=\emptyset$, whence $M^{\land'}=\R^4$.)

  \begin{Lem}\label{AOPSS}  The one-point sets  are $\perp'$-complete.
   \end{Lem}\\
{\it Proof.} This is easily checked by (\ref{SIMLAND}).\qed\\

 \begin{Exampl}\label{CCTPSS} for the $\perp'$-completion of a two-point set.  Let $s\in\R\setminus\{0\}$. Then $$\{0,(s,0,0,0)\}^{\land'}=\{\mathfrak{x}:|x|< |\textnormal{\tiny{$\frac{s}{2}$}}|, \, |x_0-\textnormal{\tiny{$\frac{s}{2}$}}|+|x|\le |\textnormal{\tiny{$\frac{s}{2}$}}|\}=\{\mathfrak{x}: x_0=\textnormal{\tiny{$\frac{s}{2}$}}, |x|< |\textnormal{\tiny{$\frac{s}{2}$}}|\}^{\land'}$$
\end{Exampl}\\
{\it Proof.} We proceed as for (\ref{CCTPS}) and may be brief. $\{(t,0,0,0)\}^{\perp'}=\{\mathfrak{y}:|y_0-t|\le|y|\}\setminus \{(t,0,0,0)\}$ for every $t$. Hence $D:=\{0\}^{\perp'}\cap \{(s,0,0,0)\}^{\perp'}=\{\mathfrak{y}: |y_0-\frac{s}{2}|+|\frac{s}{2}|\le |y|\}$ and $\{0,(s,0,0,0)\}^{\land'}=D^{\perp'}=\{\mathfrak{x}:|x_0-y_0|\le |x-y|\; \forall\, \mathfrak{y}\in D\}\setminus D=\{\mathfrak{x}: |x_0-\frac{s}{2}|+|y_0-\frac{s}{2}|\le \big||x|-|y|\big|\; \forall \;  \mathfrak{y}\in D\}\setminus D$. Hence $|x|<|\frac{s}{2}|$ for $\mathfrak{x}\in D^{\perp'}$, whence one infers the first =. The second = follows as for (\ref{CCTPS}) applying (\ref{PCEPRD}).\qed\\

\begin{The}\label{SIMLAND}  Let $A$ be a $\perp'$-set. Then $A^{\sim'}=A^{\land'}$.
\end{The}\\
{\it Proof.} Because of (\ref{PCEPRD}) it remains to show $A^{\land'}\subset A^{\sim'}$. Let $\Lambda$ be a maximal $\perp'$-set with $A\subset\Lambda$, see sec.\,\ref{OSNTH}. Put $B:=\Lambda\setminus A$. Then $B\perp' A$, whence $B\perp' A^{\land}$.
 \\
\hspace*{6mm}
 Now fix $\mathfrak{x}\in A^{\land'}$ and let $\mathfrak{z}^{\cdot 2}>0$. Then, by \cite[Corollary 1]{CJ77}, there is $s\in\R$ with $\mathfrak{x}+s\mathfrak{z}\in \Lambda$. Assume $\mathfrak{x}+s\mathfrak{z}\in B$. Then $(\mathfrak{x}+s\mathfrak{z})\perp'  \mathfrak{x}$. This implies $s\ne 0$ and the contradiction $0\ge \big((\mathfrak{x}+s\mathfrak{z})-\mathfrak{x}\big)^{\cdot 2}=s^2\mathfrak{z}^{\cdot 2}>0$. Thus $\mathfrak{x}+s\mathfrak{z}\in A$. This proves $\mathfrak{x}\in A^{\sim'}$.\qed\\

\begin{Exa}  Let $\mathfrak{a}\in\R^4$, $r>0$ and $\Delta:=\{\mathfrak{x}:x_0=a_0, |x-a|<r\}$. Then $$\Delta^{\land'}=\{\mathfrak{x}:|x_0-a_0| + |x-a|\le r\}\setminus\{\mathfrak{x}: x_0=a_0, |x-a|=r\}$$ 
Indeed, without restriction assume $\mathfrak{a}=0$. Then, by (\ref{SIMLAND}), $\mathfrak{x}\in \Delta^{\land'}$ if and only if for every $|z|<1$ there is $s\in\R$ with $\mathfrak{x}+s(1,z)\in\Delta$. This means that $s=-x_0$ and $|x-x_0z|<r$ for all $|z|<1$. Then either $x_0=0$ and $|x|<r$ or $x_0\ne 0$ and $|y|+(1-\varepsilon)|x_0|<r$ for all $\varepsilon>0$, whence the claim.
\end{Exa}\\

Let $M\subset \R^4$ be $\perp'$-complete. Then a set $A\subset M$ is said to be a \textbf{$\perp'$-base of $M$} if $A$ is a $\perp'$-set with $A^{\sim'}=M$. (Recall $A^{\sim'}=A^{\land'}$  by (\ref{SIMLAND}).)
Every \textbf{$\perp'$-base}  $A$ of $M$ is a maximal  $\perp'$-set in $M$, since $M=A^{\sim'}\subset A^{\land'}=A^{\perp'\perp'}$, whence $A^{\perp'}\cap M\subset M^{\perp'}\cap M=\emptyset$. Conversely, recall that by \cite[Theorem 1]{CJ77} every maximal $\perp'$-set $A$ in $M$  is a $\perp'$-base of $M$. Hence, in particular, every $\perp'$-complete
set has a $\perp'$-base. ---
By definition, a set is a  \textbf{$\perp'$-base} if it is a $\perp'$-base for $\R^4$. So for $A\subset \R^4$ we point out 
\begin{equation}\label{EPBMPS} 
A  \perp'\textnormal{-base } \Leftrightarrow\; A \textnormal{ maximal $\perp'$-set}
\end{equation}

\begin{Lem}\label{SDAUCS} Let $(M_\iota)_\iota$ be a family in $\mathcal{M}'$ such that $M_\iota\perp' M_\kappa$ if $\iota\ne \kappa$.  Then $\big(\cup_\iota M_\iota\big)^{\sim'}=\big(\cup_\iota M_\iota\big)^{\land'}$.
\end{Lem}\\
{\it Proof.} Let $A_\iota$ be a $\perp'$-base of $M_\iota$. Put $A:=\cup_\iota A_\iota$. Obviously $A$ is a $\perp'$-set. Hence, by (\ref{SIMLAND}), $\big(\cup_\iota M_\iota\big)^{\sim'}\supset \big(\cup_\iota A_\iota\big)^{\sim'}=A^{\sim'}=A^{\land'}=\lor'_\iota A^{\land'}_\iota=\lor'_\iota M_\iota=\big(\cup_\iota M_\iota\big)^{\land'}$. The proof is accomplished by (\ref{PCEPRD}).\qed\\

\begin{Lem}\label{ETHSSHPS} Let $A$ be a subset of a $\perp'$-base $\Sigma$. Then \emph{(i)} $\Sigma \setminus A=A^{\perp'}\cap \Sigma$, \emph{(ii)} $(\Sigma \setminus A)^{\perp'}=A^{\land'}$, \emph{(iii)} $(\Sigma \setminus A)^{\land'}=A^{\perp'}$,  \emph{(iv)} $A=A^{\land'}\cap \Sigma$, and finally \emph{(v)}  $A\supset\bigcup_{\mathfrak{y} \in A^{\land'}}\{\mathfrak{x}\in\Sigma: (\mathfrak{x}-\mathfrak{y})^{\cdot 2}> 0\}$.
\end{Lem}

{\it Proof.}  Recall first $A\subset \Sigma$, $A\cap A^{\perp'} =\emptyset$, and $\Sigma\setminus A\subset A^{\perp'}$ as $\Sigma$ is a $\perp'$-set. Hence (i).
--- Then obviously $A^{\land'}\subset (\Sigma \setminus A)^{\perp'}$. We show now $(\Sigma \setminus A)^{\perp'}\subset A^{\sim'}$. Then (ii) follows by (\ref{PCEPRD}). So let $\mathfrak{x}\in (\Sigma \setminus A)^{\perp'}$. This means $(\mathfrak{x}-\mathfrak{y})^{\cdot 2}\le 0$, $\mathfrak{x}\ne \mathfrak{y}$ for all $\mathfrak{y}\in \Sigma \setminus A$. Let  $\mathfrak{z}^{\cdot 2}> 0$. As $\Sigma^{\sim'}=\R^4$ there is $s\in\R$ with $\mathfrak{x}-s\mathfrak{z} \in\Sigma$. Assume $\mathfrak{x}-s\mathfrak{z} \in\Sigma\setminus A$. Then $s\ne 0$ and the contradiction $0\ge(\mathfrak{x}-\mathfrak{x}+s \mathfrak{z})^{\cdot 2}=s^2\mathfrak{z}^{\cdot 2}> 0$ follows. Therefore $\mathfrak{x}-s\mathfrak{z} \in A$ showing $\mathfrak{x}\in\tilde{A}$. --- By orthocomplementation  (ii) implies (iii). --- Further, (iv) equals (i) for $\Sigma\setminus A$ in place of $A$ because of (ii). 
---  Finally, as to  (v), consider  $\mathfrak{x}\in\Sigma\setminus A$. This implies $\mathfrak{x}\in A^{\perp'}$ and hence, $\mathfrak{x}\perp' A^{\land'}$, which implies $\mathfrak{x}\in \bigcap_{\mathfrak{y}\in A^{\land'}}\{\mathfrak{x'}\in\Sigma:(\mathfrak{x'}-\mathfrak{y})^{\cdot 2}\le 0\}$ as $\mathfrak{x}\in\Sigma$.
  \qed\\

\subsection{$\perp'$-diamonds}
A set $M\subset \R^4$ is called a  $\perp'$-diamond if there is a subset $\Delta$  of a non-timelike hyperplane $\sigma$ with $M=\Delta^{\land'}$. $\Delta$ is called a \textbf{flat} ($\perp'$-)\textbf{base} of $M$.  By (\ref{ETHSSHPS})(iii), $M^{\perp'}=(\sigma\setminus \Delta)^{\land'}$ is a $\perp'$-diamond, too.

\begin{Exam}\label{PPD} for $\perp'$-diamonds. {\it Clearly, $\emptyset$ and $\R^4$ are $\perp'$-diamonds. Let $0<a\le b$ and $\Gamma:=\{\mathfrak{x}: (x_0,x_3)=(r\,a,r\,b) \textrm{ for } r\in [0,1]\,\}$. Then $\Gamma^{\perp'}=\{\mathfrak{x}:|x_0|\le-x_3\}\cup \{\mathfrak{x}:|x_0-a|\le x_3-b\}$,  $\Gamma^{\land'}=\{\mathfrak{x}:a-b\le x_0-x_3\le 0,\, 0\le x_0+x_3\le a+b\}$, and  $(\Gamma\setminus\{\mathfrak{x}:x_0= x_3=0\})^{\perp'}=\Gamma^{\perp'}\setminus\{\mathfrak{x}:x_0=a, x_3=b\}$, $(\Gamma\setminus\{\mathfrak{x}:x_0= x_3=0\})^{\land'}=\Gamma^{\land'}\setminus\{\mathfrak{x}:x_0= x_3=0\}$.}\\
\hspace*{6mm}
We verify the claim for $(\Gamma\setminus\{\mathfrak{x}:x_0= x_3=0\})^{\land'}$ using (\ref{SIMLAND}). The remaining formulae  follow similarly. So, $(\Gamma\setminus\{\mathfrak{x}:x_0= x_3=0\})^{\sim'}$ if and only if for every $\mathfrak{z}=(1,z)$, $|z|<1$ there are $s\in\R$, $r\in]0,1]$ such that $x_0+s=ra$, $x_3+sz_3=rb$. Eliminating $r$ and $s$ one finds the equivalent condition $0<x_3-x_0z_3\le b-az_3$ for all $|z_3|< 1$. This means
$|x_0|\le x_3$, $x_3>0$ and $x_3-b\le-|x_0-a|$, whence the assertion.\qed
\end{Exam}

For $a<b$ the bases of the  $\perp'$-diamonds in (\ref{PPD}) are spacelike. For $a=b$ the $\perp'$-diamond $\Gamma^{\land'}$ coincides with is own base $\Gamma$, which is not spacelike. More generally, for $\perp'$-diamonds with not spacelike bases see (\ref{NTLSD}).

\begin{Lem}\label{NTLSD}   Let $\kappa$ be a non-timelike  not spacelike hyperplane. This means $\kappa=\{\mathfrak{x}: \mathfrak{x}\cdot \mathfrak{e}=\rho\}$ with some $\rho\in\R$ and some 
    lightlike  $\mathfrak{e}=(1,e)\in\R^4$ with $|e|=1$.  Let $\Delta\subset \kappa$. Then $\mathfrak{y}\in\Delta^{\land'}$ if and only if
  \begin{equation*}
 \mathfrak{y}'+\{\mathfrak{x}:\mathfrak{x}\cdot \mathfrak{e}=0, \;\big(x-(xe)e\big)^2<\lambda\,xe\}\subset \Delta\tag{1}
 \end{equation*}
with $\lambda:=\rho- \mathfrak{y}\cdot \mathfrak{e}$ and $ \mathfrak{y}':= \mathfrak{y}+\frac{1}{2}\lambda (1,-e)\in\kappa$. In particular one has
  \begin{equation*}
\Delta=\Delta^{\land'}\tag{2}
\end{equation*}
 if $\Delta$ is bounded or, more generally, if for every $\mathfrak{y}'\in\kappa$ there is $(\xi_n)$ with  $\liminf \xi_n=-\infty$, $\limsup \xi_n=\infty$ such that $\mathfrak{y}'+\xi_n\mathfrak{e}\not\in \Delta$ .
\end{Lem}\\
{\it Proof.} By (\ref{SIMLAND}) one has to determine the intersection $C$ of $\kappa$ with the interior of the light-cone with apex $\mathfrak{y}$.  Check  $\mathfrak{z}\cdot \mathfrak{e}\ne 0$. So for every timelike $\mathfrak{z}$ one finds  $(\mathfrak{y}-s\mathfrak{z})\cdot \mathfrak{e}=\rho$ with $s=\lambda/(\mathfrak{z}\cdot \mathfrak{e})$. Check also that $\mathfrak{x}=\mathfrak{z}/(\mathfrak{z}\cdot \mathfrak{e})$ for some timelike $\mathfrak{z}$ if and only if $\mathfrak{x}\cdot \mathfrak{e}=1$ and $|x|<|x_0|$. Hence $C=\mathfrak{y}+\lambda\{\mathfrak{x}:\mathfrak{x}\cdot \mathfrak{e}=1, |x|<|x_0|\}$, which equals the left hand side of (1) by the substitution $\mathfrak{x}=\frac{1}{\lambda}\mathfrak{x}'+\frac{1}{2}(1,-e)$. As to (2),
note that $\lambda=0$ if and only if $ \mathfrak{y}\in\Delta$. Hence, if
$\mathfrak{y}\in\Delta^{\land'}\setminus\Delta$, then the projection of $\Delta$ onto the space $\R^3\equiv \{0\}\times \R^3$ contains the interior of a circular paraboloid with axis $e$ and minimum point $y-\frac{1}{2}\lambda e$.
\qed\\

 With regards to sec.\,\ref{ECLNTLHP}  the $\perp'$-diamonds in (\ref{SNTLSD}) are particularly interesting, since they have   a spacelike flat base and a not spacelike one as well.\\
 \hspace*{6mm} 
  Let $\pi$ be a spacelike plane, let $\sigma$ be any spacelike hyperplane containing $\pi$ and let $\sigma^>$ be one of the two open half-hyperplanes constituting $\sigma\setminus\pi$. 
  For every  non-timelike hyperplane $\tau$ containing $\pi$  let $\tau^>$ denote the half-hyperplane constituting $\tau\setminus\pi$, which is at the same side  of $\pi$ as $\sigma^>$ (cf.\,(\ref{EHHP})). 
  There are exactly two different non-timelike not spacelike hyperplanes $\tau$, denoted by $\kappa_\pm$, which contain $\pi$.\footnote{ Explicitly, let $\sigma=\{\mathfrak{x}: \mathfrak{x} \cdot \mathfrak{e}_\sigma=\rho_\sigma\}$,  
  $\pi =\{\mathfrak{x}\in\sigma: xe=\rho \}$ with
  $\mathfrak{e}_\sigma=(1,e_\sigma)$, $|e_\sigma|<1$, 
$e\in\R^3$, $ |e|=1$, $\rho_\sigma,\rho\in\R$. Then $\tau$ is a non-timelike hyperplane containing $\pi$ if and only if there is $\lambda\in [\lambda_-,\lambda_+]$ with  $\lambda_\pm :=-e_\sigma e \pm\sqrt{1-e_\sigma^2+(e_\sigma e)^2}$ such that
\begin{equation*}
\tau=\{\mathfrak{x}: \mathfrak{x} \cdot \mathfrak{e}_\tau=\rho_\tau\} \textnormal{ with }\mathfrak{e}_\tau=(1,e_\tau),\, e_\tau:=e_\sigma+\lambda e,\; \rho_\tau:=\rho_\sigma-\lambda \rho 
\end{equation*}
Only for $\lambda\in\{\lambda_-,\lambda_+\}$, $\tau\equiv \chi_\pm$ is not spacelike. --- This is easily checked using  $1\ge|e_\sigma+\lambda e|^2=e_\sigma^2+2\lambda e_\sigma e+\lambda^2$.
}

\begin{Lem}\label{SNTLSD}   $(\tau^>)^{\land'}=(\sigma^>)^{\land'}$ and $(\sigma^>)^{\land'}=\bigcup_\tau \tau^>$, $(\sigma^>)^{\perp'}=\bigcup_\tau \tau^\le$. In particular, $(\kappa_-^>)^{\land'}=(\kappa_+^>)^{\land'}=(\sigma^>)^{\land'}$.
\end{Lem}\\
{\it Proof.} Because of Poincar\'e symmetry it suffices to check the assertion for
$\sigma=\{\mathfrak{x}:x_0=0\}$, $\pi=\{\mathfrak{x}:x_0=0, x_3=0\}$, $\sigma^>=\{\mathfrak{x}:x_0=0, x_3>0\}$. Then, indeed, a short computation shows that the non-timelike hyperplanes containing $\pi$ are $\tau=\{\mathfrak{x}:x_0-\alpha x_3=0\}$ for $\alpha\in[-1,1]$. So, the not spacelike hyperplanes are  $\kappa_\pm=\{\mathfrak{x}:x_0\mp x_3=0\}$, and $\tau^>=\{\mathfrak{x}:x_0-\alpha x_3=0, x_3>0\}$.\\
\hspace*{6mm}
Now we compute $(\tau^>)^{\perp'}=\{\mathfrak{y}:(\mathfrak{y}-\mathfrak{x})^{\cdot 2}\le 0 \,\forall \mathfrak{x}\in\tau^>\}\setminus \tau^>=\{\mathfrak{y}:|y_0-\alpha x_3|\le|y_3-x_3|\,\forall x_3>0\}\setminus \tau^>$. Therefore   $\mathfrak{y}\not\in (\tau^>)^{\perp'}$ if $y_3>0$ and hence  $(\tau^>)^{\perp'}=\{\mathfrak{y}: y_3\le 0, |y_0-\alpha x_3|\le x_3-y_3 \,\forall x_3>0\}=\{\mathfrak{y}: y_3\le-|y_0|\}$ independently of $\alpha\in[-1,1]$. Similarly one verifies 
$\{\mathfrak{y}: y_3\le-|y_0|\}^{\perp'}=\{\mathfrak{y}: y_3\ge |y_0|\}\setminus\{0\}$. The assertion follows.\qed\\

\subsection{The lattices   $\mathcal{M}'$ and $\mathcal{M}^{'borel}$}\label{LMMBOR}
 The set $\mathcal{M}^{'}$ of all $\perp'$-complete subsets   of $\R^4$ is a complete orthomodular lattice  (cf. \cite[sec.\,4]{CJ77}). It is atomic without the covering property (see  (\ref{AOPSS}),  (\ref{CCTPSS})). Moreover, all its elements are measurable.\\
\hspace*{6mm} 
  As to the proof of the latter, 
 for $M\subset\R^4$ consider 
the set of $\perp'$-influence  
\begin{equation*}
M^{\lor'}:=M\cup \bigcup_{\mathfrak{y}\in M}\{\mathfrak{x}:(\mathfrak{x}-\mathfrak{y})^{\cdot 2}>0\}
\end{equation*}
 which obviously equals $\R^4\setminus M^{\perp'}$. Following  (\ref{MCCS}) one has $U(M)\subset M^{\lor'}\subset M^\lor$, whence $M^{\lor'}$ is measurable. It follows that the $\perp'$-complete sets are measurable.\\
\hspace*{6mm}
The above formula for $M^{\lor'}$ also implies that $M^{\lor'}$ is Borel if  $M$ is so. It follows that $M^{\land'}$ is Borel if   $M$ is Borel (cf.  \cite[sec.\,4]{CJ77}). Therefore the set 
\begin{equation}\label{PPCBS}
\mathcal{M}^{'borel}:=\{M\subset\R^4: M\; \perp'\textrm{-complete Borel set}\}
\end{equation}
 is a sublattice of $\mathcal{M}'$,  which   is $\sigma$-complete orthomodular atomic without the covering property.\\
\hspace*{6mm}
 Furthermore,  it is worth be noted that for every $M\in \mathcal{M}'$ there is  $M_0\in\mathcal{M}^{'borel}$, $M_0\subset M$ with $\lambda^4(M\setminus M_0)=0$. (Indeed, let $M\in \mathcal{M}'$. Since $M$ is measurable there is a Borel $B\subset \R^4$, $B\subset M$ with $\lambda^4(M\setminus B)=0$. Then $M_0:=B^{\land'}$ satisfies the assertion.)\\
\hspace*{6mm} 
Finally, by \cite[Lemma 4.1]{CJ77}, if $M$ is $\perp'$-complete and  $A$ is a maximal $\perp'$-set in $M$, then $M$ is Borel if and only if  $A$ is Borel.  \\

\section{A Rep  
of the  Lattice of Borel Sets 
Generated by  the Non-Timelike Relation}\label{SRLBSGNTLR}

The interest in the causal structure of spacetime continues to exist and many articles are dedicated  to this subject.  However despite the common conviction that causal logics should be relevant for quantum theory it seems that not many attempts have been made to study a possible relationship other than  the already mentioned assignment in local quantum theory.  We cite Borowiec, Cegla, Jadczyk, Jancewicz 1977-1979 \cite{J77},  \cite{CJn77}, \cite{CJ79}, \cite{BJ79}, where  reps of the causal logic are related to Poincar\'e covariant conserved (operator) density currents.

Let us start stating that there is no rep of the lattice $\mathcal{M}'$ generated by the non-timelike relation. This is due to the fact that all $\perp'$-complete sets are measurable. It is quite the situation concerning 
the lattice $\mathcal{M}$ of causally complete sets, and also the proof of  (\ref{NRLCCR}) applies quite literally. \\
\hspace*{6mm}
In contrast the lattice $\mathcal{M}^{'borel}$  has reps. Recall that a rep $(W,F')$ of $\mathcal{M}^{'borel}$ regards a relativistic 
system described by the representation $W$  of $\tilde{\mathcal{P}}$ with $F'$   a  projection valued  normalized monotone locally orthogonal $\sigma$-orthoadditive map on $\mathcal{M}^{'borel}$ being   Poincar\'e covariant with respect to $W$.  \\
\hspace*{6mm}
The existence of a rep of $\mathcal{M}^{'borel}$ is per se  a remarkable fact. However, the  reps  $(W,F')$ of $\mathcal{M}^{'borel}$, which we know and are going to present,  have a  mass-squared operator  with  spectrum $]-\infty,\mu^2]$ for every $\mu\ge 0$  and infinite spinor dimension  so that  we do not consider them   to represent a physical system (cf. the beginning of sec.\,\ref{DCRS}).

\hspace*{6mm} 
The following is based on the construction of a particular representation $W'$ of $\tilde{\mathcal{P}}$ in Doplicher, Regge, Singer 1968 \cite{DRS68}.\\
\hspace*{6mm}
 Recall that $\mathcal{T}$ denotes the set of all timelike straight lines in $\R^4$. Every $u\in\mathcal{T}$ is of the form $u=\mathfrak{x}+\R\mathfrak{v}$ with $\mathfrak{v}^{\cdot 2}>0$. Clearly $\tilde{\mathcal{P}}$ acts transitively on $\mathcal{T}$ in the natural way. For $u_0:=\R\mathfrak{e}$, $\mathfrak{e}:=(1,0,0,0)$ one has the stabilizer subgroup $\tilde{\mathcal{P}}_{u_0}=\R\mathfrak{e}\times SU(2)$.  Their irreducible representations are equivalent to $d^{\mu,j}(t,B):=\operatorname{e}^{\operatorname{i}\mu\, t} D^{j}(B)$
for $\mu\in\R$, $j\in \N_0/2$. \\
\hspace*{6mm}
Let $d$ be a representation of $\tilde{\mathcal{P}}_{u_0}$ acting on $H_d$. We build the induced representation $W':=d^{\tilde{\mathcal{P}}}$ of $\tilde{\mathcal{P}}$ as described  in sec.\,\ref{A:IT}. To this  identify 
$$\mathcal{T}\, \equiv\, \R^3\times O_1\;(\subset \R^6)$$
where $O_1:= \{v\in\R^3:|v|<1\}$  by the parametrization $u=(x,v)\equiv (0,x)+\R(1,v)$. Endow $\mathcal{T}$  with the Lebesgue measure $m$, the restriction of $\lambda^6$ to $\mathcal{T}$.\\
\hspace*{6mm}
 Then
the group action becomes $\mathfrak{a}\cdot (x,v)=(x+a-a_0v,v)$, $B\cdot (x,v)=(B\cdot x,B\cdot v)$, and $A_\rho\cdot (x,v) =(y,w)$ with
$ y:=\big(x_1,x_2,\cosh(\rho)x_3\big)-\frac{\sinh(\rho)x_3}{\cosh(\rho)+\sinh(\rho)v_3}\big(v_1,v_2,$ $ \sinh(\rho)+\cosh(\rho)v_3\big)$ 
 and $w:=\frac{1}{\cosh(\rho)+\sinh(\rho)v_3}\big(v_1,v_2,\sinh(\rho)+\cosh(\rho)v_3\big)$. \\
 \hspace*{6mm}
 For general $A\in SL(2,\C)$ one has  $A=A_{\rho e}B$ with $B\in SU(2)$ and  $A_{\rho e}:=(AA^*)^{1/2}$ representing  the pure velocity transformation along $e\in\R$, $|e|=1$ determined by $A_{\rho e}=B(p)A_\rho B(p)^{-1}$ (\ref{CRCS}).\\
\hspace*{6mm} 
 So a straight forward  computation yields for $g=(\mathfrak{a},A)$ 
\begin{equation*}
  \frac{\operatorname{d}m^g}{\operatorname{d}m}(x,v)=\big( A^{-1}\cdot (1,v)\big)_0^{-5}= \big(\cosh(\rho)-\sinh(\rho)\,ev\big)^{-5}=
  \big(1-\tanh^2(\rho)\big)^{5/2}\big(1-\tanh(\rho)\,ev\big)^{-5}\tag{1}
 \end{equation*}
Hence $m$ is quasi-invariant under $\tilde{\mathcal{P}}$.\\
\hspace*{6mm}
A Borel section for the surjection $\tilde{\mathcal{P}}\to \mathcal{T}$, $g\mapsto g\cdot u_0$ is given by $q(u):=\big(\mathfrak{x}-\frac{x_0}{v_0}\mathfrak{v},Q(\mathfrak{v})\big)$ for $u=\mathfrak{x}+\R\mathfrak{v}$, $\mathfrak{v}^{\cdot 2}>0$ with  the canonical cross section $Q$ (\ref{PCCS}). The Wigner rotation becomes 
\begin{equation*}
R(u,g)=\left(\Big(   \frac{x_0}{|v_0|}-\frac{(g^{-1}\cdot \mathfrak{x})_0}{|(A^{-1}\cdot \mathfrak{v})_0|}\Big) \sqrt{\mathfrak{v}^{\cdot 2}}\,\mathfrak{e}, \,Q(\mathfrak{v})^{-1}AQ(A^{-1}\cdot\mathfrak{v})\right)\tag{2}
 \end{equation*}
 
\hspace*{6mm}
We are going to study $W'$ given in sec.\,\ref{A:IT}\,(2) 
\begin{equation*}
(W'(g)\psi)(u)=\sqrt{(\operatorname{d}m^g/ \operatorname{d}m)(u)} \,d\big(R(u,g)\big)\psi(g^{-1}u)\tag{3}
\end{equation*}

$W'$ acts on $L^2(\mathcal{T},H_d)$.
$(W',E^{can})$ is a canonical system of imprimitivity on $(\tilde{\mathcal{P}},\mathcal{T})$, which is irreducible if $d$ is irreducible (cf.\;sec.\,\ref{A:IT}). It is  trivially extended to all (Lebesgue) measurable subsets of 
$\mathcal{T}$.\\
\hspace*{6mm}
Next let us determine the spectrum of the mass-squared operator $C$ in case that  $d$ is irreducible.
  To this we compute $K(t,b):=\mathcal{F}'W'(t,b)\mathcal{F}^{'-1}$, where $(\mathcal{F}'\psi)(p,v):=(2\pi)^{-3/2}\int \operatorname{e}^{-
 \operatorname{i}px}\psi(x,v)\operatorname{d}^3x$ for  $\psi\in\mathcal{C}_c$. For $g=(t,b)$ the derivative  (1) is constant $=1$ and $R(u,g)$  becomes $(t\sqrt{1-|v|^2}\,\mathfrak{e},I_2)$. Hence $\big(K(t,b)\,\varphi\big)(p,v)=\exp(-\operatorname{i}p\,b)\,\exp\big(\operatorname{i}t(p\,v+\mu\sqrt{1-|v|^2})\big)\varphi(p,v)$. So $C$ is the multiplication operator by $\gamma(p,v):=\big(p\,v+\mu\sqrt{1-|v|^2}\big)^2-|p|^2$, and its spectrum equals the essential numerical range of $\gamma$. As $\gamma$ is continuous the latter is the closure of its codomain. One finds spectrum$(C)=\,]-\infty,\mu^2]$.\\
 
\hspace*{6mm}
Now we turn to the  skillful  twist in \cite{DRS68} regarding $(W',E^{can})$. Recall
$$\mathcal{T}_M=\{u\in \mathcal{T}: u\cap M\ne \emptyset\}$$
for every $M\subset\R^4$. 
The following properties of   $\mathcal{T}_M$ (up to the last) are easy to verify:
\begin{itemize}
\item $\mathcal{T}_{\R^4}=\mathcal{T}$, $\mathcal{T}_\emptyset=\emptyset$
\item $M_1\subset M_2$ $\Rightarrow$ $\mathcal{T}_{M_1}\subset \mathcal{T}_{M_2}$
\item   $\cup_\iota \mathcal{T}_{M_\iota}=\mathcal{T}_{\cup_\iota M_\iota}$
\item $\mathcal{T}_{M_1}\cap\mathcal{T}_{M_2}=\emptyset$ $\Leftrightarrow$ $M_1\perp' M_2$
\item $\mathcal{T}_{M_1}=\mathcal{T}_{M_2}$ $\Leftrightarrow$ $M^{\sim'}_1= M^{\sim'}_2$
\item  $\mathcal{T}_M=\mathcal{T}_{M^{\land'}}$ if $M$ is $\perp'$-orthogonal (use (\ref{SIMLAND}))
\item  $\mathcal{T}=\mathcal{T}_M$ if $M$ is maximal $\perp'$-orthogonal (by (\ref{EPBMPS}))
\item $g\cdot \mathcal{T}_M= \mathcal{T}_{g\cdot M}\; \forall \,g\in \tilde{\mathcal{P}}$
\item $\mathcal{T}_M\equiv \bigcup_{\mathfrak{y}\in M}\{(y-y_0v,v):v\in O_1\}$
\item $ \mathcal{T}\setminus \mathcal{T}_{M^{\land'}}=\mathcal{T}_{M^{\perp'}}$ (see (\ref{JLGNTLR}))
\end{itemize}

In general $\mathcal{T}_M$ is not measurable. This holds true even if $M$ is $\perp'$-complete and hence measurable.

\begin{Exa} In order to provide a $\perp'$-complete $M$ with non-measurable $\mathcal{T}_M$, let $Y\subset [0,1]^3$ be any non-measurable set with respect to $\lambda^3$. Obviously there is a $\sigma$-compact $C\subset Y$ satisfying $\lambda^3(C)=\sup\{\lambda^3(K): K\subset Y \textrm{ compact}\}<\infty$. Then clearly $Y_0:=Y\setminus C$ is still not measurable. Moreover,
for every Borel set of $\R^3$ with $B\subset Y_0$ one has due to tightness 
$\lambda^3(B)=\sup\{\lambda^3(K): K\subset B \textrm{ compact}\}=0$. Now, put $M:=\{0\}\times Y_0\subset \R^4$. \\
\hspace*{6mm}
As $\lambda^4(\{0\}\times \R^3)=0$, $M$ is measurable with respect to $\lambda^4$.
 $M$ is even $\perp'$-complete. Indeed, let $\mathfrak{x}\in M^{\sim'}$.  Then one easily checks   $ \{y\in\R^3: |x-y|< |x_0|\} \subset Y_0$. This implies $x_0=0$, whence  $\mathfrak{x}\in M$. Hence $M^{\sim'}=M$. By (\ref{SIMLAND}) $M^{\sim'}=M^{\land'}$ since $M$ is a $\perp'$-set. Therefore $M$ is $\perp'$-complete.\\
\hspace*{6mm}
The claim is that $\mathcal{T}_M=Y_0\times O_1$ is not measurable with respect to $\lambda^6$. --- First let us exclude that $\mathcal{T}_M$ is a null set. Assume the contrary. Then  there is a Borel set $B$ of $\R^6$ with $B\supset \mathcal{T}_M$ and $\lambda^6(B)=0$. So, by  Tonelli's theorem, $0=\lambda^6(B)=\int_{\R^3}\lambda^3(B_v)\operatorname{d}\lambda^3(v)$.
Hence $\lambda^3(B_v)=0$ for some $v\in O_1$. This is a contradiction, since $B_v=\{y\in\R^3:(y,v)\in B\}\supset
\{y\in\R^3:(y,v)\in \mathcal{T}_M\}=Y_0$. --- Now assume that $\mathcal{T}_M$ is measurable. Then there is a Borel set $B$ of $\R^6$ with $B\subset \mathcal{T}_M$ and $\lambda^6( \mathcal{T}_M\setminus B)=0$. Clearly $\lambda^6(B)>0$ since $ \mathcal{T}_M$ is not a null set. So, by Tonelli's theorem, $0<\lambda^6(B)=\int_{\R^3}\lambda^3(B_v)\operatorname{d}\lambda^3(v)$. Hence $\lambda^3(B_v)>0$ for some $v\in O_1$. This is a contradiction, since $B_v=\{y\in\R^3:(y,v)\in B\}\subset
\{y\in\R^3:(y,v)\in \mathcal{T}_M\}=Y_0$.\qed
\end{Exa}

Fortunately one has

\begin{Lem}\label{BMMTM} If $M\subset \R^4$ is Borel then  $\mathcal{T}_M$ is measurable with respect to $\lambda^6$.
\end{Lem}\\
 {\it Proof.} Consider the map 
$ f:\R\times \R^3\times O_1\to \R^4\times O_1,\; f(\xi,x,v):=(\xi,x+\xi v,v)$
and the projection $\pi_{\R^3\times O_1}(\xi,x,v):=(x,v)$. One easily verifies
$\mathcal{T}_M\,=\,\pi_{\R^3\times O_1}\big(f^{-1}(M\times O_1)\big)$.
Now, if $M$ is Borel, then  $f^{-1}(M\times O_1)$ is Borel, whence $\mathcal{T}_M$ is measurable by the Measurable Projection Theorem. (More precisely, as $f^{-1}(M\times O_1)$ is Borel in the product space of the Suslin spaces $\R$ and $\R^3\times O_1$, the projection $\mathcal{T}_M$
is Suslin in $\R^3\times O_1$, which together with its Lebesgue measurable sets is a complete measurable space. Since $\R^3\times O_1$ is  metric, the assertion follows. See e.g. \cite{www. med}.)
\qed\\

\begin{The}\label{RLBSGNTLR} Recall $W'=d^{\tilde{\mathcal{P}}}$ \emph{(3)} and put $F'(M):=E^{can}(\mathcal{T}_M)$ for $M\in\mathcal{M}^{'borel}$. Then\\
\hspace*{6mm}
\emph{(a)} $(W',F')$ is a rep of $\mathcal{M}^{'borel}$.   

\hspace*{6mm}
 \emph{(b)} $(W',E)$ is a causal system with $E(\Delta):=E^{can}(\mathcal{T}_\Delta)$, $\Delta\in\mathfrak{S}$. In particular $E(\Delta )= F'( \Delta^{\land'})$ if $\Delta$ is Borel, and  $\big(E(\Delta)\psi\big)(x)=1_\Delta(x)\psi(x)$ for $\psi\in L^2(\R^3,\mathcal{S})\equiv L^2(\R^3\times O_1, H_d)$ for $\mathcal{S}:=L^2(O_1,H_d)$, 
if $\Delta\subset\R^3$ is measurable and $\Delta\equiv\{0\}\times\Delta$.   
 
 \hspace*{6mm}
  \emph{(c)} $W'|_{ISU(2)}$ is induced by the representation $D$ of $SU(2)$ in $\mathcal{S}$ given by $\big(D(B)f\big)(v):=d(0,B)f(B^{-1}\cdot v)$. If $d=d^{\mu,j}$ then the  spectrum of $C$ is $]-\infty,\mu^2]$ and the spinor space $\mathcal{S}$ is $L^2(O_1,\C^{2j+1})$. 
\end{The}\\ 
{\it Proof.} By (\ref{BMMTM}) $F'$ is well-defined. We use now tacitly the above  list of properties of 
$\mathcal{T}_M$. Normalization and monotony are clear. --- If $M_1\perp' M_2$ then $\mathcal{T}_{M_1}\cap \mathcal{T}_{M_2}=\emptyset$, whence $F'(M_1)F'(M_2)=E^{can}(\mathcal{T}_{M_1})E^{can}(\mathcal{T}_{M_2})=0$. --- If $M_n\perp' M_m$    for $n\ne m$,  then $\mathcal{T}_{M_n}\cap \mathcal{T}_{M_m}=\emptyset$ and hence $\sum_nF'(M_n)
=\sum_nE^{can}(\mathcal{T}_{M_n})=E^{can}(\cup_n\mathcal{T}_{M_n})=E^{can}(\mathcal{T}_{\cup M_n})=E^{can}(\mathcal{T}_{(\cup M_n)^{\sim'}})=E^{can}(\mathcal{T}_{(\cup M_n)^{\land'}})=F'\big((\cup M_n)^{\land'}\big)=F'(\lor'M_n)$ using (\ref{SDAUCS}). --- Since $W'(g)E^{can}(\mathcal{T}_M)W'(g)^{-1}=E^{can}(g\cdot \mathcal{T}_M)=E^{can} (\mathcal{T}_{g\cdot M})$, covariance of $F'$ follows.\\
\hspace*{6mm}
$(W',E)$ is a causal system by the following  (\ref{RLCCRRCL}).  The spectrum of $C$ is already determined above.
For the remainder see sec.\,\ref{DCRSPR}.\qed

The question remains open whether the causal system $(W',E)$ in (\ref{RLBSGNTLR})
is irreducible if $d$ is irreducible.

\begin{Pro}\label{RLCCRRCL} Let    $(W,F')$ be a rep  of  $\mathcal{M}^{'borel}$. Set $E(\Delta):=F'(\Delta^{\land'})$ for  Borel $\Delta\in\mathfrak{S}$. Then $E$ extends to all $\Delta\in\mathfrak{S}$, and  $(W,E)$ is a causal system.
\end{Pro} \\
{\it Proof.} Let $\sigma$ be a spacelike hyperplane. Let $\Delta_n\subset\sigma$, $n\in\N$ be Borel and  disjoint. Then  $M_n:=\Delta^{\land'}_n$ is Borel (see sec.\,\ref{LMMBOR}) and $M_n^{\perp'}=\Delta_n^{\perp'}=(\sigma\setminus\Delta_n)^{\land'}\supset \Delta^{\land'}_m=M_m$ for $n\ne m$ by (\ref{ETHSSHPS})(iii). Hence $M_n\in\mathcal{M}^{'borel}$ are mutually orthogonal, whence $\sum_n E(\Delta_n)=\sum_n F'(M_n)=F'(M)$ for $M:=\lor'_nM_n$ by $\sigma$-orthoadditvity of $F'$. For $\Delta:=\bigcup_n\Delta_n$ obviously one has $\Delta^{\land'}=(\bigcup_n M_n)^{\land'}=M$, whence $E(\Delta)=F'(\Delta^{\land'})=F'(M)$. This proves $\sigma$-additivity for $E$. --- Moreover, $E(\sigma)=F'(\R^4)=I$ and $E(\emptyset)=F'(\emptyset)=0$. ---
Poincar\'e covariance carries over from $F'$ to $E$ since $(g\cdot \Delta)^{\land'} =g\cdot \Delta^{\land'}$. --- Now extend $E$ to all of $\mathfrak{S}$, see (\ref{PCPOL}). ---
Spacelike separated $\Delta,\Gamma\in\mathfrak{S}$ obviously are $\perp'$- orthogonal. Let $\Delta_0\subset\Delta$ be Borel with $E(\Delta_0)=E(\Delta)$ and similarly be $\Gamma_0$.
So the completions $\Delta^{\land'}_0$ and $\Gamma^{\land'}_0$ are $\perp'$-orthogonal, whence $E(\Delta)\, E(\Gamma)=E(\Delta_0)\, E(\Gamma_0)=F'(\Delta^{\land'}_0)\,F'(\Gamma^{\land'}_0)=0$ by local $\perp'$-orthogonality of $F'$. Hence $E$ is causal by (\ref{VEFCPOL})(f). --- Thus $(W,E)$ is a causal system.\qed\\

\hspace*{6mm}
Coming back to (\ref{RLBSGNTLR}), there the map $F'$ on $\mathcal{M}^{'borel}$ into the lattice  of projections is not a lattice homomorphism. 
Indeed by monotony of $F'$ one has $F'(M\land' L)\le F'(M)\land F'(L)$ for all $M,L\in \mathcal{M}^{'borel}$. But  equality does not hold in general. For instance, $M:=\{\mathfrak{x}:|x_0-1|+|x|\le 1\}$ and $L:=\{\mathfrak{x}:|x_0+1|+|x|\le 1\}$ are $\perp'$-diamonds with $\mathcal{T}_{M\cap L}=\mathcal{T}_{\{0\}}=\{0\}\times O_1$ and  $\mathcal{T}_M\cap \mathcal{T}_L=\{(x,v):|v|\le 1-|x|,|x|\le 1\}$. Since $\lambda^6(\mathcal{T}_{M\cap L})=0$ and $\lambda^6(\mathcal{T}_M\cap \mathcal{T}_L)=4\pi^2/45> 0$, one has $F'(M\land' L)=0$ and $F'(M)\land F'(L)\ne 0$.\\
\hspace*{6mm}
Hence the  quantum proposition $ F'(M)\land F'(L)$ meaning that the particle passes through $M$ \textbf{and} passes through $L$  \textbf{does not} imply the proposition $F'(M\land' L)$ meaning that the particle passes through $M\cap L$.\\
\hspace*{6mm}
Plainly this fact is in full accordance with physical intuition and thus the latter does not require a lattice homomorphism from $\mathcal{M}^{'borel}$ into the quantum proposition system. Actually  mathematics does not even allow a reasonable homomorphism. More precisely one has

\begin{Lem}\label{HMPL} Let $h$ be a homomorphism from $\mathcal{M}^{'borel}$ into the lattice of projections of a separable Hilbert space, i.e., $h$ satisfies $h(M\land' L)=h(M)\land h(L)$ and $h(M\lor' L)=h(M)\lor h(L)$ for all $M,L $. Then $h$ vanishes at all bounded $M \in \mathcal{M}^{'borel}$.
\end{Lem}\\
{\it Proof.} Clearly, $h$ is monotone. (Indeed, if $L\subset M$ then $h(L)=h(L\land' M)=h(L)\land h(M)\le h(M)$.) Since $h(M)\le h(\R^4)$ for all $M$ it is no restriction to assume $h(\R^4)=I$. Now
remind of the fact that $h$ preserves orthocomplementation. (Indeed, $0=h(\emptyset)=
h(M\land' M^{\perp'})
=h(M)\land h(M^{\perp'})=h(M) h(M^{\perp'})$ and  $I=h(\R^4)=h(M\lor' M^{\perp'})=h(M)\lor h(M^{\perp'})$, whence $h(M^{\perp'})=I-h(M)=h(M)^\perp$.) Next consider the atoms 
$\{\mathfrak{a}\}$ for $\mathfrak{a}=(k,a)$, $k\in\mathbb{Z}$, $a\in\R^3$ (see (\ref{AOPSS})).  Fix $k$. Then they are mutually orthogonal and so are $h(\{\mathfrak{a}\})$. Since the Hilbert space is separable, $h(\{\mathfrak{a}\})=0$ up to countable many.  Therefore there are $a_k\in\R^3$ with $|a_{k}|<\frac{1}{|k|}$ and $h(\{\mathfrak{a}_k\})=0$ for $\mathfrak{a}_k:=(k,a_k)$.\\
\hspace*{6mm}
Now consider the diamond $M_n:=\{\mathfrak{a}_n\}\lor'\{\mathfrak{a}_{-n}\}$ for $n\in\N$ (cf. \cite{C02} at the end of sec.\,3). Clearly $h(M_n)=0$. Note that $\{(n,0)\}\lor'\{(-n,0)\}=\{\mathfrak{x}: |x|<0, |x_0|+|x|\le n\}$.  This makes clear that every bounded $M$ is contained in some $M_n$. Hence $h$ vanishes at all bounded $M\in\mathcal{M}^{'borel}$.\qed\\

Clearly $h$ in (\ref{HMPL}) is locally orthogonal and orthoadditive. If $h$ is $\sigma$-orthoadditive, then by (\ref{PLRPOR})(c) it follows $h=0$.\\

\section{Extension of Causal Localizations to Non-Timelike Hyperplanes }\label{ECLNTLHP}

We do not know whether there exists  a rep $(W,F')$  of  $\mathcal{M}^{'borel}$ with non-negative mass-squared operator and finite spinor dimension. But
in  (\ref{NRLCCRRCLP}) we derive some consequences  of a rep in general, which  refer to more profound properties of causal systems. Then  in (\ref{NGENTLNSLHP}) these are shown to be valid   for the Dirac and the Weyl systems. Hence, in addition to its mere mathematical convenience, the non-timelike relation $\perp'$ seems to have physical relevance. It draws attention to the non-timelike not spacelike hyperplanes, of which Poincar\'e covariant localizations did not take note so long. \\ 
\hspace*{6mm}
Of course, by physical grounds we stay  with the concept of causality as described  by a causal system $(W,E)$  defined in (\ref{CS}). So local orthogonality of $E$, which is equivalent to causality (see (\ref{VEFCPOL})(f)), refers to spacelike separation rather than to non-timelike separation. However, this does not discredit $\perp'$ since by (\ref{ELOLSO}) local orthogonality and local $\perp'$-orthogonality coincide.

\begin{Lem}\label{ELOLSO} Let $\Delta, \Gamma\in\mathfrak{S}$. Then\\ $$\Delta \perp \Gamma \Rightarrow \Delta \perp' \Gamma \Rightarrow (\Delta\setminus N)\perp \Gamma \textnormal{ for some null set } N\subset \Delta\Rightarrow E(\Delta)\,E(\Gamma)=0$$
\end{Lem}\\
{\it Proof.} The first arrow $\Rightarrow$ holds obviously, the last one due to (\ref{PCPOL})(a). As to the second arrow suppose $\Delta \perp' \Gamma$. Let $\sigma$ be a spacelike hyperplane with $\Delta\subset \sigma$. Then check $N:=\{\mathfrak{x}\in \Delta:\mathfrak{x}\not\perp \Gamma\}\subset\ \{\mathfrak{x}\in \sigma\cap \Gamma^{\perp'}:\mathfrak{x}\not\perp \Gamma\}\subset \{\mathfrak{x}\in \sigma:(\mathfrak{x}-\mathfrak{y})^{\cdot 2}\le 0 \textnormal{ for all } \mathfrak{y}\in\Gamma,  (\mathfrak{x}-\mathfrak{y})^{\cdot 2}\ge 0 \textnormal{ for some } \mathfrak{y}\in\Gamma\} \subset \{\mathfrak{x}\in \sigma:\sup_{\mathfrak{y}\in\Gamma}(\mathfrak{x}-\mathfrak{y})^{\cdot 2}=0\}$, which is the null set $N(\Gamma,\sigma)$ in (\ref{LMRI}).\qed\\

Moreover, the analogue of (\ref{BOD}) for $\perp'$-diamonds is valid. It shows in particular that  $E$ assigns to all measurable spacelike flat bases of a $\perp'$-diamond the same operator. 

\begin{Lem}\label{BODS} 
 Let $\sigma, \tau$ be  spacelike hyperplanes. Consider   a $\perp'$-diamond $M=\Delta^{\land'}$  with measurable flat base $\Delta \subset \sigma$. Let $\Gamma \subset M\cap \tau$ be a measurable set. Then $E(\Gamma)\le E(\Delta)$.
 If $\Gamma^{\land'}=M$ then $E(\Delta)=E(\Gamma)$.
  \end{Lem}
  
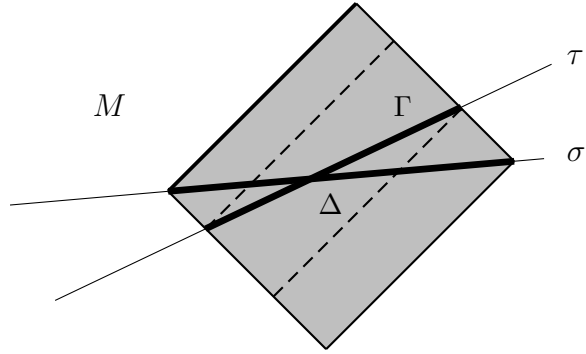
  \begin{pspicture}(-7,-1)(3,2.5)

\psline*[linecolor=lightgray]
(-2,0)(0.5,2.5)(2.6,0.4)(0.1,-2.1)(-2,0)
\psline(0.5,2.5)(2.6,0.4)
\psline(-2,0)(0.1,-2.1)

\psline(0.1,-2.1)(2.6,0.4)
\psline[linewidth=0.9mm]{-}%
(-2,0)(2.6,0.4)
\psline[linewidth=0.5mm]{-}%
(-2,0)(0.5,2.5)
\put(-1.5,-0.5){\line(2.1,1){-2}}
\put(-1.5,-0.5){\line(2.1,1){4.6}}
\put(-2,0){\line(4.6,0.4){5}}
\put(-2,0){\line(4.6,0.4){-2.1}}
\psline[linestyle=dashed](-1.5,-0.5)(1,2)
\psline[linestyle=dashed](-0.6,-1.4)(1.9,1.1)


\linethickness{0.9mm}
\put(-1.5,-0.5){\line(2.1,1){3.4}}

\put(1,1){\small{$\Gamma$}}
\put(0,-0.3){\small{$\Delta$}}
\put(3.3,1.7){$\tau$}
\put(3.3,0.4){$\sigma$}
\put(-3,1){$M$}
\end{pspicture}

\vspace{1cm}  
 {\it Proof.} By (\ref{LMRI}), $M(\Gamma,\sigma)=\bigcup_{\mathfrak{y}\in\Gamma}\{\mathfrak{x}\in\sigma: (\mathfrak{x}-\mathfrak{y})^{\cdot 2}>0\}\subset \Gamma_\sigma$ equals $\Gamma_\sigma$ up to a null set of $\sigma$. Recall $\Gamma\subset \Delta^{\land'}$. Then, by (\ref{ETHSSHPS})(v), $M(\Gamma,\sigma)\subset \Delta$. So causality  (\ref{VEFCPOL})(a)  implies $E(\Gamma)\le E(\Gamma_\sigma)=E(M(\Gamma,\sigma))\le E(\Delta)$.  \qed\\

\hspace*{6mm} 
We turn to the announced properties of a rep $F'$ of $\mathcal{M}^{'borel}$. To this end we need the results in (\ref{PLRPOR}) and (\ref{IGNSLHP}).

\begin{Lem}\label{PLRPOR} Let $F'$ be a rep of $\mathcal{M}^{'borel}$.
\begin{itemize}
\item[(a)]  Let $L,M\in \mathcal{M}^{'borel}$ with $L\subset M$. Then $F'(M\cap L^{\perp'})=F'(M)-F'(L)$. In particular, $F'$ is monotone. 
\item[(b)]
Let $M_n\in\mathcal{M}^{'borel}$, $n\in\N$ with $M_1\supset M_2\supset M_3\supset\dots$. Then $F'(\bigcap_nM_n)=\lim_nF'(M_n)$.
\item[(c)] Let $M_n\in\mathcal{M}^{'borel}$, $n\in\N$ with $M_1\subset M_2\subset M_3\subset\dots$. Then $F'(\bigvee_nM_n)=\lim_nF'(M_n)$.

\end{itemize}
\end{Lem}
{\it Proof.}  
(a) By orthomodularity on has the orthogonal decomposition $M=L\lor (M\cap L^{\perp'})$. Hence orthoadditivity of $F'$ yields $F'(M)=F'(L)+F'(M\cap L^{\perp'})\ge F'(L)$. --- (b) Put $M:=\bigcap_nM_n$. Apply (\ref{ODCDS}) and $\sigma$-orthoadditivity of $F'$ to $(M_n)$ as well to $(M_1,M_2, \dots,M_m,\emptyset,\emptyset,\dots)$ for $m\in\N$. Then $F'(M_1\cap M^{\perp'})=\sum_nF'(M_n\cap M_{n+1}^{\perp'})=\lim_m\sum_{n=1}^m F'(M_n\cap M_{n+1}^{\perp'})=\lim_mF'(M_1\cap M_m^{\perp'})$, whence $F'(M_1)-F'(M^{\perp'})=F'(M_1)-\lim_mF'(M_m)$ proving (b). --- (c) follows from (b) for $(M_n^{\perp'})$ as generally $F'(M^{\perp'})=I-F'(M)$ for $M\in\mathcal{M}^{'borel} $.\qed\\

\begin{Lem}\label{IGNSLHP} Let $\chi$ be the non-timelike not spacelike hyperplane $\{\mathfrak{x}:x_0=x_3\}$ and $\tilde{\mathcal{P}}_\chi:=\{g\in\tilde{\mathcal{P}}: g\cdot \chi\subset \chi\}$  the invariance subgroup. Then
\begin{displaymath}
 \tilde{\mathcal{P}}_\chi=\Big\{(\mathfrak{a},A)\in\tilde{\mathcal{P}}: \mathfrak{a}\in\chi, \;A=\left(\begin{array}{cc} z& w\\ 0 & 1/z\end{array}\right), z\in\C\setminus\{0\}, w\in\C\Big\}
\end{displaymath}

Hence  $\tilde{\mathcal{P}}_\chi=IST(2)$, i.e.,  the inhomogeneous  group of upper triangular $2\times 2$ matrices with determinant $1$.   So   $g\in\tilde{\mathcal{P}}_\chi$ if and only if 
\begin{displaymath}
g=\big(\mathfrak{a},\,\operatorname{e}^{\zeta\sigma_3/2}\big)\big(0,\,
\left(\begin{array}{cc} 1& w\\ 0 & 1\end{array}\right)\big)
\end{displaymath}
with $\mathfrak{a}\in\chi$ and  $\zeta=\rho+\operatorname{i}\varphi,w=u+\operatorname{i}v\in\C$. The first factor $\big(\mathfrak{a},\,\operatorname{e}^{\zeta\sigma_3/2}\big)$ represents the obvious part consisting in a translation $\mathfrak{a}$ along $\chi$, a boost  $\Lambda(\operatorname{e}^{\rho\,\sigma_3/2})$ along the third axis, and a rotation $\Lambda(\operatorname{e}^{\operatorname{i}\varphi\,\sigma_3/2})$ in the $x_1x_2$ plane.  Put  $r:=\frac{1}{2}|w|^2$. The second factor acts by
\begin{displaymath}
\Lambda\big(\left(\begin{array}{cc} 1& w\\ 0 & 1\end{array}\right)\big)=\left(\begin{array}{cccc} 1+r& u & -v & -r\\ u &1 & 0 & -u\\ -v & 0 & 1 & v\\ r & u & -v & 1-r\end{array}\right)
\end{displaymath}
\end{Lem}\\
{\it Proof.} Write $\chi=\{\mathfrak{x}: \mathfrak{x}\cdot \mathfrak{e}=0\}$ with $\mathfrak{e}:=(1,0,0,1)$. Then 
$g=(\mathfrak{a},A)\in\tilde{\mathcal{P}}_\chi$ if and only if $\chi=g\cdot \chi=\{\mathfrak{x}: (g^{-1}\cdot \mathfrak{x})\cdot\mathfrak{e}=0\}=\{\mathfrak{x}: \mathfrak{x}\cdot (A\cdot \mathfrak{e})=\mathfrak{a}\cdot(A\cdot \mathfrak{e})\}$, whence equivalently  $\mathfrak{a}\cdot(A\cdot \mathfrak{e})=0$ and $A\cdot \mathfrak{e}=\lambda \mathfrak{e}$ for some $\lambda\ne 0$, see after (\ref{MPPS}). Actually $\lambda>0$ as $A$ acts orthocronously. This means $\mathfrak{a}\in\chi$ and $A=A_{\ln(\lambda)}A'$ with $A'\in SL(2,\C)_\mathfrak{e}=E(2)$. Recall $A\in E(2) \Leftrightarrow A_{22}=\overline{A}_{11}$, $A_{21}=0$. The remainder follows by explicit  computation. 
\qed\\

 Now let $W(\alpha)$ denote the time shift by $\alpha\in\R$ and recall
that $A_{\rho e}=\operatorname{exp}(\frac{\rho}{2}\sum_{k=1}^3e_k\sigma_k)$  acts as the boost in direction $e\in\R^3$, $|e|=1$ with rapidity $\rho$. Let  $\varsigma\in\{1,-1\}$.\\
\hspace*{6mm}
 Let $\mathfrak{S}_\infty$ denote the set of all measurable subsets of non-timelike  not spacelike hyperplanes.

\begin{The}\label{NRLCCRRCLP} Let $(W,F')$ be a rep of  $\mathcal{M}^{'borel}$. Set $E(\Delta):=F'(\Delta^{\land'})$ for Borel $\Delta\in\mathfrak{S}$ and $P(\Delta):=F'(\Delta^{\land'})$ for Borel 
$\Delta\in\mathfrak{S}_\infty$. Then  $E$ and $P$ extend to all sets of $\mathfrak{S}$ and $\mathfrak{S}_\infty$, respectively, and
\begin{itemize}
\item[(a)] $(W,E)$ is a causal system\\
\item[(b)] $(W,P)$ is a Poincar\'e covariant  localization. More precisely, on every non-timelike not spacelike hyperplane $P$ is a PM and $W(g)P(\Delta)W(g)^{-1}=P(g\cdot\Delta)$ holds for all $g\in\tilde{\mathcal{P}}$ and $\Delta\in\mathfrak{S}_\infty$
\end{itemize}
There is a close relationship between $E$ and $P$. Let $\alpha>0$ and $e\in\R^3$, $|e|=1$. One has
\begin{itemize}
\item[(c)] $P(\{ 0\le \varsigma x_0=xe\le \alpha\})= \lim_{\varsigma\rho\to\infty} E(A_{\rho e}\cdot\{x_0=0,  0\le xe\le 2\alpha\,\operatorname{e}^{-\varsigma\rho}\})$,  where  the limit exists since the projections decrease as $\varsigma\rho$ increases\\
\item[(d)] $P(\{0\le  \varsigma x_0=xe\le \alpha\})=E(\{x_0=\varsigma\alpha, xe\le \alpha\})E(\{x_0=0, xe\ge 0\})$\\
\item[(e)] $P(\{0\le \varsigma x_0=xe\})
=E(\{\mathfrak{x}: x_0=0, xe\ge 0\})$
\end{itemize}

\end{The}

{\it Proof.} (a) $(W,E)$ is a causal system by (\ref{RLCCRRCL}). 

\hspace*{6mm}
(b) Poincar\'e covariance of $P$ follows immediately from the Poincar\'e covariance of $F'$ since $(g\cdot \Delta)^{\land'} =g\cdot \Delta^{\land'}$. --- As noted after (\ref{MPPS}) a non-timelike hyperplane is a maximal $\perp'$-set and hence by Eq.\,(\ref{EPBMPS}) a $\perp'$-base. Now $\sigma$-additivity of $P$ on every non-timelike not spacelike hyperplane follows as the $\sigma$-additivity of $E$ in (\ref{RLCCRRCL}).  --- Extend $P$ to all of $\mathfrak{S}_\infty$, see (\ref{PCPOL}).

\hspace*{6mm}
(c) Obviously the sets $M_\rho:=\{-2\alpha\operatorname{e}^{-2\varsigma\rho}\le \varsigma x_0-xe\le 0, 0\le \varsigma x_0+xe\le2\alpha\}$, $\rho\in\R$ converge to $M:=\{0\le \varsigma x_0=xe\le \alpha\}$ for $\varsigma\rho\to \infty$, as 
 $$M_{\rho'}\subset M_\rho \quad\textnormal{if\, } \varsigma\rho<\varsigma\rho'\quad\textnormal{and\; }  M=\cap_{\rho}\,M_\rho$$ 
 We check $M_\rho=\Gamma_\rho^{\land'}$ for $\Gamma_\rho:=A_{\rho e}\cdot\{x_0=0, 0\le xe\le 2\alpha \operatorname{e}^{-\varsigma\rho} \}$ ($=\{0\le-\sinh(\rho)\,x_0+\cosh(\rho)\,xe\le 2\alpha \operatorname{e}^{-\varsigma\rho} \}$). 
Indeed,  $\Gamma_\rho^{\land'}=A_{\rho e}\cdot\{x_0=0, 0\le xe\le 2\alpha \operatorname{e}^{-\varsigma\rho} \}^{\land'}=A_{\rho e}\cdot \{-2\alpha\operatorname{e}^{-\varsigma\rho}\le x_0-xe\le 0, 0\le x_0+xe\le 2\alpha\operatorname{e}^{-\varsigma\rho}\}=\{-2\alpha\operatorname{e}^{-\varsigma\rho-\rho}\le x_0-xe\le 0, 0\le x_0+xe\le 2\alpha\operatorname{e}^{-\varsigma\rho+\rho}\}=M_\rho$. ---
Hence $M_\rho$
 is a $\perp'$-diamond with the spacelike flat base $\Gamma_\rho$, and $M=M^{\land'}$, cf. also (\ref{NTLSD})(2). So using  (\ref{PLRPOR})(b)
one  infers $P(M)=F'(M)=\lim_{\varsigma\rho\to \infty}F'(M_\rho)=\lim_{\varsigma\rho\to \infty}E(\Gamma_\rho)$ showing (c).

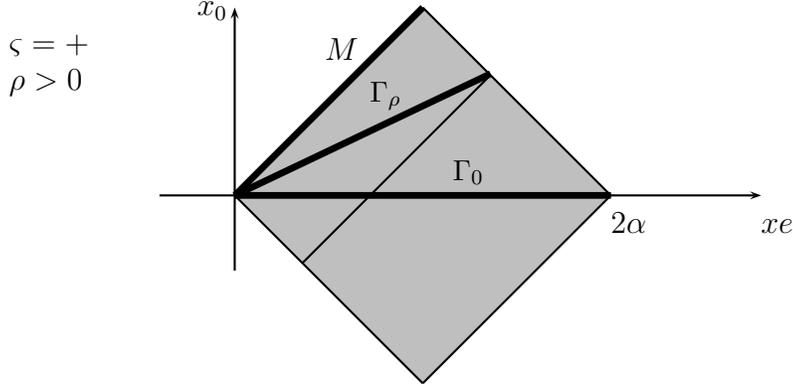
\begin{pspicture}(-6,-2.5)(3,3)
\psline{->}(-2,-1)(-2,2.5)
\psline{->}(-3,0)(5,0)


\put(-5,1.9){$\varsigma=+$}
\put(-5,1.4){$\rho>0$}
\psline*[linecolor=lightgray]
(-2,0)(0.5,2.5)(3,0)(0.5,-2.5)(-2,0)
\psline(0.5,2.5)(3,0)
\psline(-2,0)(0.5,-2.5)
\psline(0.5,-2.5)(3,0)
\psline(-1.1,-0.9)(1.4,1.6)
\linethickness{0,9mm}
\put(-2,0){\line(1,0){5}}
\put(-2,0){\line(2.1,1){3.4}}
\put(-2,0){\line(1,1){2.5}}
\put(-0.2,1.25){\small{$\Gamma_\rho$}}
\put(0.9,0.2){\small{$\Gamma_0$}}
\put(3,-0.5){$2\alpha$}
\put(5,-0,5){$xe$}
\put(-2.5,2.4){$x_0$}
\put(-0.8,1.8){$M$}
\end{pspicture}

\hspace*{6mm}
(d) By 
Eq.\,(\ref{ALHPTLC}) the right hand side of (d) equals $I-E(\{x_0=\varsigma \alpha, xe> \alpha\})-E(\{x_0=0, xe <0\})$ and hence by definition $=I-F'(\{ x_0=\varsigma \alpha, xe> \alpha\}^{\land'})-F'(\{x_0=0,xe <0\}^{\land'})$, and further $=I-F'(\{\varsigma x_0= xe> \alpha\}^{\land'})-F'(\{\varsigma x_0=xe <0\}^{\land'})$ due to (\ref{SNTLSD}), and finally $=F'(\{ 0\le \varsigma x_0=xe\le \alpha\}^{\land'})=P(\{0\le  \varsigma x_0=xe\le \alpha\})$ by orthoadditivity of $F'$.

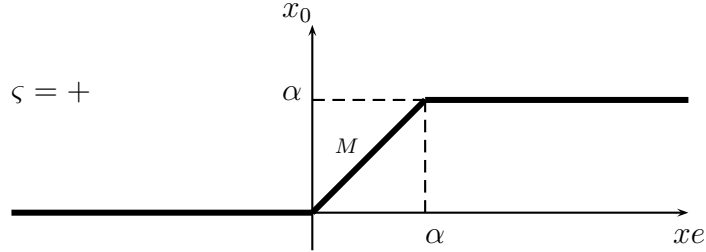
\begin{pspicture}(-8,-1)(2,3)
\psline{->}(-5,0)(3,0)
\put(2.8,-0.4){$xe$}
\psline{->}(-2,-0.5)(-2,2.5)
\put(-2.4,2.6){$x_0$}
\psline[linewidth=0.8mm](-6,0)(-2,0)
\psline[linewidth=0.8mm](-2,0)(-0.5,1.5)
\psline[linewidth=0.8mm](-0.5,1.5)(3,1.5)
\put(-2.4,1.5){$\alpha$}
\psline[linestyle=dashed](-2,1.5)(-0.5,1.5)
\put(-0.5,-0.4){$\alpha$}
\psline[linestyle=dashed](-0.5,0)(-0.5,1.5)
\put(-1.7,0.8){\tiny{$M$}}
\put(-6,1.5){$\varsigma=+$}

\end{pspicture}

\hspace*{6mm}
(e) By (\ref{SNTLSD}) one has $P(\{0\le \varsigma x_0=xe\})=F'(\{0\le \varsigma x_0=xe\}^{\land'})=F'(\{\mathfrak{x}: x_0=0, xe\ge 0\}^{\land'})=E(\{\mathfrak{x}: x_0=0, xe\ge 0\}$.\qed\\

Let us comment on (\ref{NRLCCRRCLP}). The question is what does it mean that   a state $\varphi$ is {\it localized} in $\{ 0\le \varsigma x_0=xe\le \alpha\}$, i.e. $P(\{0\le \varsigma x_0=xe\le \alpha\})\varphi=\varphi$. Clearly it is senseless saying that the spatial probability of the system   in  $\{ 0\le \varsigma x_0=xe\le \alpha\}$ in the state $\varphi$ is $1$, since $\{ \varsigma x_0=xe\}$ is not a reference frame. In the following one recognizes $\varphi$ to be a 
\textbf{ late-change state} 
in case of  a Dirac or Weyl state.\\
\hspace*{6mm}
According to (c), the answer is that $\varphi$ is just a state for which the boosted state $W(A_{\varsigma \rho e})\varphi$ is localized in $\{x_0=0, 0\le xe\le 2\alpha \operatorname{e}^{-\rho}\}$ for every rapidity $\rho$. 
(Recall Poincar\'e covariance (\ref{PCPOL})(b).)
This is particularly interesting for $\rho\to \infty$ as the region $\{x_0=0, 0\le xe\le 2\alpha \operatorname{e}^{-\rho}\}$ of localization shrinks to an arbitrarily narrow strip at the origin.  This is  pure Lorentz contraction shown  for Dirac and Weyl wave functions in sec.\,\ref{DLB} and (\ref{SRWS}), respectively.\\
\hspace*{6mm}
Furthermore, according to (d),  $\varphi$ is just a state which is localized in $\{x_0=0, xe\ge 0\}$
and which at time $\varsigma \alpha$, i.e., $W(\varsigma \alpha)\varphi$, is localized in $\{x_0=0, xe\le \alpha\}$. (Note that the right hand side of (d) is the product of two projection which commute due to causality, and recall Poincar\'e covariance (\ref{PCPOL})(b).) The equality of the right hand sides of (c) and (d) is shown to be valid  for Dirac and Weyl states in (\ref{LBSLSET}) and (\ref{SRWS}), respectively. Moreover, by Eq.\,(\ref{ALHPTLC}), Eq.\,(\ref{BLHPTLC}), and (\ref{CCIDES}), 
the right hand side of (d)
 is a  non-trivial projection with infinite dimensional eigenspaces.
\\
\hspace*{6mm}
Since $P(\{0\le  \varsigma x_0=xe\})=\cup_{\alpha>0}P(\{0\le  \varsigma x_0=xe\le \alpha\})$
by (b), actually  (e) is a completeness property. Every state localized in $\{x_0=0, x_3\ge 0\}$ can be approximated by the particular states described in  (c), (d). This is shown to be true for Dirac and Weyl states in (\ref{DLTES}) and (\ref{ECCPWS}), respectively.

\hspace*{6mm}
Hence  to every   rep $(W,F')$ of $\mathcal{M}^{'borel}$ there are related a causal system $(W,E)$ and a Poincar\'e covariant localization $(W,P)$ on $\mathfrak{S}_\infty$ satisfying (\ref{NRLCCRRCLP})(c),\,(d),\,(e).  These relations, which refer to more profound properties of causal systems, are easily elucidated by its geometrical meaning (cf.\,above). Now we are going to prove the existence of the localization $(W,P)$ for the Dirac and the Weyl systems. It is  obvious that much more efforts have to be made towards a possible construction of  a rep $(W,F')$ of $\mathcal{M}^{'borel}$. But, at least, the Dirac and the Weyl systems allow for a further Poincar\'e covariant extension of $E$ (recall  (\ref{PCPOL}) for the first extension) to PM on all non-timelike hyperplanes. More precisely on has (\ref{NGENTLNSLHP}).\\

\hspace*{6mm}
 Let $\chi$ denote the non-timelike not spacelike hyperplane $\{\mathfrak{x}: x_0=x_3\}$ and recall $\tilde{\mathcal{P}}_\chi$ in (\ref{IGNSLHP}).

\begin{The}\label{NGENTLNSLHP} Let $(W,E)$ be a causal system, which is the finite orthogonal sum of Dirac and Weyl systems.
Then there is a unique Poincar\'e covariant localization $(W,P)$ on $\mathfrak{S}_\infty$ such that
\begin{equation*}
P(\{\mathfrak{x}\in\chi: 0\le \varsigma x_3\le\alpha\})=\lim_{\rho\to\infty}
E\big(\operatorname{e}^{\,\rho\,\sigma_3/2}\cdot\{x_0=0, 0\le \varsigma x_3\le 2\alpha \operatorname{e}^{-\rho} \}\big)\tag{1}
\end{equation*}
for all $\alpha>0$, $\varsigma\in\{-1,1\}$. The limit in \emph{(1)} exists since the projections decrease as $\rho$ increases. 
\end{The}\\
{\it Proof.} Obviously it is no restriction assuming that $(W,E)$ is the Dirac system or a Weyl system. First,  in (a) and (b), we prove the existence of $P$. Put $\Delta_\rho:=\{x_0=0, 0\le\varsigma x_3\le 2\alpha \operatorname{e}^{-\rho} \}$. 

\hspace*{6mm}
(a) Let $(W,E)$ be the Dirac system. Then $E^{\textsc{d}\,mom}=\mathcal{F}E^{can}\mathcal{F}^{-1}$, $\big(E^{\textsc{d}\,mom}(\Delta_\rho)\varphi\big)(p)=\int 
k_\rho(s-p_3)
\varphi(p_1,p_2,s)\operatorname{d}s$ for integrable $\varphi\in L^2(\R^3,\C^4)$ with $k_\rho(u):=\frac{\exp(\,\operatorname{i} 2\alpha \operatorname{e}^{-\rho}\varsigma \,u\,)-1}{2\pi \operatorname{i} \varsigma \,u}$. Hence, using Poincar\'e covariance one finds according to Eq.\,(\ref{RDS}) 
 $$\big(E^{\textsc{d}\,mom}(A_\rho\cdot\Delta_\rho)\varphi\big)(p)=\sum_{\eta,\eta'}\int 
k_\rho(s-q^\eta_3) 
 \pi^\eta(p)s(A_\rho)^{-1}\pi^{\eta'}(p_1,p_2,s)s(A_\rho)\varphi(p_1,p_2,s')\operatorname{d}s$$
with $q_3^\eta=\cosh(\rho)p_3-\eta\sinh(\rho)\epsilon(p)$ and   $s':=\eta'\sinh(\rho)\epsilon(p_1,p_2,s)+\cosh(\rho) s$. Note that $s=\cosh(\rho)s'-\eta'\sinh(\rho)\epsilon(p_1,p_2,s')$. We apply Eq.\,(\ref{VSDPI}) in order to move $s(A_\rho)$ to the left and  substitute the variable of integration  $s$ by $s'$. One finds
\begin{equation*}
\big(E^{\textsc{d}\,mom}(A_\rho\cdot\Delta_\rho)\varphi\big)(p)=\sum_{\eta,\eta'}\int 
k_\rho(s-q^\eta_3)
\pi^\eta(p)s(A_\rho)^{-2}\pi^{\eta'}(p_1,p_2,s')\varphi(p_1,p_2,s')\operatorname{d}s'\tag{2}
\end{equation*}
Now it is easy to perform in (1) the limit $\rho\to\infty$. Check $|k_\rho(u)|\le\frac{1}{\pi}\alpha\operatorname{e}^{-\rho}$ and $s(A_\rho)^{-2}=\operatorname{diag}(\operatorname{e}^{-\rho},\operatorname{e}^{\rho},\operatorname{e}^{\rho},\operatorname{e}^{-\rho})$, whence $\norm{k_\rho(s-q^\eta_3)s(A_\rho)^{-2}}\le \frac{\alpha}{\pi}$ for $\rho\ge 0$. Due to dominated convergence one gets for every $p\in\R^3$
\begin{equation*} 
\lim_{\rho\to\infty}\big(E^{\textsc{d}\,mom}(A_\rho\cdot\Delta_\rho)\varphi\big)(p)=\sum_{\eta,\eta'}\int k(\mathfrak{p},\mathfrak{p}')\pi^\eta(p)S\pi^{\eta'}(p')\varphi(p')\operatorname{d}p_3'\tag{3}
\end{equation*}
where $\mathfrak{p}:=(\eta\epsilon(p),p)$ and $\mathfrak{p}'=(p_0',p')$ with $p'_1:=p_1$, $p'_2:=p_2$, $p'_0:=\eta'\epsilon(p')$,  and $k(\mathfrak{p},\mathfrak{p}'):=\frac{\exp (\operatorname{i}\alpha\varsigma(p_0-p_3-p'_0+p'_3))-1}{\operatorname{i}\pi\varsigma(p_0-p_3-p'_0+p'_3)}$, $S:=\operatorname{diag}(0,1,1,0)$.\\

\hspace*{6mm} 
The right hand side of (1) is a special case of the right hand side of (\ref{NRLCCRRCLP})(c). Hence, by (\ref{BODS}), the limit $\lim_{\rho\to\infty}E^{\textsc{d}\,mom}(A_\rho\cdot\Delta_\rho)\varphi$ exists, whence for integrable $\varphi$ its value at every $p\in\R^3$ is given by the right hand side of (3).\\
\hspace*{6mm}
One notes that $k(\mathfrak{p},\mathfrak{p}')$ is just the kernel for $P^{shell}(M)$ in (\ref{A:SI})(b). Actually the next step is to show that the shell representation of $P(M):=\lim_{\rho\to\infty}E^{\textsc{d}}(A_\rho\cdot\Delta_\rho)$ coincides with $P^{shell}(M)$ in (\ref{A:SI})(b). To this, according to (\ref{A:ESRDS}), we have to perform several equivalence transformations, namely  $P^{shell}(M)=(Y\xi Z)^{-1}P^{mom}(M)(Y\xi Z)$. We indicate some of the intermediate steps.\\
\hspace*{6mm}
  Straightforward computations yield the unwieldy $4\times 4$ matrices $V(p)$ and $2\times 4$ matrices $U_\eta(p)$, which satisfy
$ (Y\xi Z\phi)(p)=V(p)\big(\phi(\mathfrak{p}^+),(\phi(\mathfrak{p}^-)\big)$ for $p\in\R^3$, $\mathfrak{p}^\eta=(\eta\epsilon(p),p)$ and 
$((Y\xi Z)^{-1}\varphi)(\mathfrak{p})=U_\eta(p)\varphi(p)$ for $\mathfrak{p}\in\mathcal{O}^m$, 
$\mathfrak{p}=(\eta\epsilon(p),p)$. However, after repeated applications of the formulae regarding the canonical cross section in (\ref{PCCS}) the relevant expressions became clear:
\begin{eqnarray*}
S\pi^\eta(p)V(p) & = & \frac{1}{2\epsilon(p)}
\left(\begin{array}{cc} \Tiny{\big(\begin{array}{cc} 0 & 0\\ 0 & 1\end{array}\big)}\delta_{+,\eta} &  \Tiny{\big(\begin{array}{cc} 0 & 0\\ 0 & 1\end{array}\big)}\delta_{-,\eta}\\  \Tiny{\big(\begin{array}{cc} 1 & 0\\ 0 & 0\end{array}\big)}\delta_{+,\eta} &  -\Tiny{\big(\begin{array}{cc} 1 & 0\\ 0 & 0\end{array}\big)}\delta_{-,\eta}\end{array}\right)\left(\begin{array}{cc} \sqrt{\epsilon(p)-p_3}I_2& 0\\ 0 & \sqrt{\epsilon(p)+p_3}I_2 \end{array}\right)\\
U_\eta(p)\pi^{\eta'}(p) S &  =  & \delta_{\eta,\eta'}\sqrt{\epsilon(p)-\eta p_3}\left(\begin{array}{cc}\Tiny{\big(\begin{array}{cc} 0 & 0\\ 0 & 1\end{array}\big)} & \Tiny{\big(\begin{array}{cc} 1 & 0\\ 0 & 0\end{array}\big)}\end{array}\right)
\end{eqnarray*}
Now it is easy to verify that in the shell representation  the right hand side of (3) coincides for every $\mathfrak{p}\in\mathcal{O}^m$ with the right hand side of (\ref{A:SI})(b). This confirms the claim.\\
\hspace*{6mm}
The considerations so far show by (\ref{A:SI}), (\ref{A:ESRDS}) that $(W^{\textsc{d}}|_{\tilde{\mathcal{P}}_\chi},P)$ is a covariant PM on $\chi$ satisfying (1). Now $P$ is extended to $\mathfrak{S}_\infty$ in a Poincar\'e covariant manner. One achieves this unique extension arguing as in (\ref{PCPOL}). This concludes the proof of existence of $P$ in the Dirac case.

\hspace*{6mm}
(b) Now we consider the Weyl system $(W^{\textsc{w}\chi}, E^{\textsc{w}})$. (Here $\chi=\pm$ denotes the handedness (cf. after Eq.\,(\ref{2WS})) not to be confused with the hyperplane $\chi=\{x_0=x_3\}$.) One gets quite analogously to the Dirac case
\begin{equation*} 
\lim_{\rho\to\infty}\big(E^{\textsc{w}\,mom}(A_\rho\cdot\Delta_\rho)\varphi\big)(p)=\sum_{\eta,\eta'}\int k(\mathfrak{p},\mathfrak{p}')\pi^{\chi\eta}(p)S^\chi\pi^{\chi\eta'}(p')\varphi(p')\operatorname{d}p_3'\tag{4}
\end{equation*}
with $S^+:=\operatorname{diag}(0,1)$, $S^-:=\operatorname{diag}(1,0)$.\\
\hspace*{6mm}
It remains to check that the cone representation of $P(M):=\lim_{\rho\to\infty}E^{\textsc{w}}(A_\rho\cdot\Delta_\rho)$ 
in (4) coincides with $P^{cone}(M)$ in (\ref{A:SI}). According to (\ref{A:ESRWS}), the former is   $P^{cone}(M)=(Y^\chi\xi_0 Z^\chi)^{-1}P^{mom}(M)(Y^\chi\xi_0 Z^\chi)$. The explicit computations just use repeatedly Eq.\,(\ref{CCSEF}), Eq.\,(\ref{CRCS}).

This concludes the proof of the existence of $P$. We turn to  its uniqueness. Let $(W,E)$ be a causal system and let $P$, $P'$ be two PM on $\chi$ which are covariant with respect to 
$W|_{\tilde{\mathcal{P}}_\chi}$ and which satisfy (1). The assertion is $P=P'$.

\hspace*{6mm}
(c) Put $\overline{\R}:=\R\cup \{-\infty,\infty\}$. For  $a,b\in\overline{\R}^3$ with $a_k\le b_k$, $k=1,2,3$   define the \textbf{box} 
 \begin{equation*}  
\langle a, b[_\chi :=\{\mathfrak{x}\in\chi: a_1\le x_3-x_1<b_1, a_2\le x_3+x_2<b_2, a_3\le x_3<b_3\}
\end{equation*}
in $\chi$. Call  a \textbf{figure} the union of finitely many boxes, and let $\mathcal{Q}_\chi$ denote the set of all figures.
 The claim is that every figure is the union of finitely many disjoint boxes and that 
$\mathcal{Q}_\chi$ is an algebra.\\ 
\hspace*{6mm}
Indeed, it is well-known that the unions of finitely many disjoint boxes $\langle a, b[ :=\{x\in\R^3: a_k\le x_k<b_k, k=1,2,3\}$, $a,b\in\overline{\R}^3$   form an algebra $\mathcal{Q}^3$ in $\R^3$.\footnote{ $\mathcal{Q}^3$ is $\cap$-stable since the intersection of two boxes is a box. Check that the complement of a box $\R^3\setminus \langle a,b[$ is the union of  disjoint boxes. Then also the complement of a figure is the union of finitely many disjoint boxes. Therefore  $\mathcal{Q}^3$ is an algebra.  Finally, let $M=B_1\cup \dots \cup B_n$ be a figure with boxes $B_i$. Then 
$M=B_1\cup (B_2\setminus B_1)\cup\dots \cup \big(B_n\setminus(B_1\cup\dots\cup B_{n-1})\big)$ is the union of disjoint boxes.} The bijective linear map $l: \chi\to \R^3$, $l(\mathfrak{x}):= (x_3-x_1,x_2+x_3,x_3)$ (with $l^{-1}(x)=(x_3, x_3-x_1,x_2-x_3,x_3)$) satisfies 
$\mathcal{Q}_\chi=l^{-1}(\mathcal{Q}^3)$,  $\langle a, b[_\chi \,= l^{-1}(\langle a, b[\,)$. This confirms  the claim.
 
\hspace*{6mm}
(d) The claim now is $P|_{\mathcal{Q}_\chi}=P'|_{\mathcal{Q}_\chi}$. Indeed, due to (1), $P$ and $P'$ coincide at the boxes $\{0\le x_3<\alpha\}$ and $\{-\alpha\le x_3<0\}$, $\alpha>0$. (Note that $P$ and $P'$ vanish at the Lebesgue null sets $\{x_3=\gamma\}$, $\gamma\in\R$.) By continuity $P$, $P'$ are equal also at the boxes $\{0\le x_3<\infty\}$, $\{-\infty<x_3<0\}$. Then one easily infers from additivity that
$P$, $P'$ coincide at all boxes $B(\alpha,\beta):=\{\mathfrak{x}\in\chi: \alpha\le x_3<\beta\}$, $\alpha,\beta \in\overline{\R}$, $\alpha\le \beta$.\\
\hspace*{6mm}
 Next check $\{\mathfrak{x}\in\chi: a_1\le x_3-x_1<b_1\}=\kappa\cdot B(a_1,b_1)$ and  $\{\mathfrak{x}\in\chi: a_2\le x_3+x_2<b_2\}=\kappa'\cdot B(a_2, b_2)$ for $\kappa:=I_2+\frac{1}{2}(\sigma_1+\operatorname{i}\sigma_2)$, $\kappa':=I_2+\frac{1}{2}(\operatorname{i}\sigma_1-\sigma_2)$ in $ST(2)$, see (\ref{IGNSLHP}). Hence
$\langle a, b[_\chi =\big(\kappa\cdot  B (a_1,b_1)\big)\cap \big(\kappa'\cdot B(a_2, b_2)\big)\cap  B(a_3, b_3)$.
Therefore $P(\langle a, b[_\chi)=P\big(\kappa\cdot  B (a_1,b_1)\big)P\big(\kappa'\cdot B(a_2, b_2)\big)P\big(B(a_3,b_3)\big)$ and hence $P(\langle a, b[_\chi)=W(\kappa)P\big(B(a_1,b_1)\big)W(\kappa)^{-1}\,W(\kappa')P\big(B(a_2,b_2)\big)W(\kappa')^{-1}\,P\big(B(a_3,b_3)\big)$ by $ST(2)$-covariance. This equation holds equally for $P'$ in place of $P$. Since generally $P\big(B(\alpha,\beta)\big)=P'\big(B(\alpha,\beta)\big)$, it follows $P(\langle a, b[_\chi)=P'(\langle a, b[_\chi)$ for all boxes, whence the claim.

\hspace*{6mm}
(e) Now we are ready for the proof of $P=P'$.  For every vector $\varphi$, $M\to
\langle \varphi, P(M)\varphi\rangle=\langle \varphi, P'(M)\varphi\rangle$ defines a finite $\sigma$-additive content on $\mathcal{Q}_\chi$. It allows a unique extension to a measure $m$ on the $\sigma$-algebra of Borel sets of $\chi$ generated by $\mathcal{Q}_\chi$
(see e.g.\,\cite[5.7]{B68}). Hence $m$ coincides with $\langle \varphi, P(\cdot)\varphi\rangle$ and $\langle \varphi, P'(\cdot)\varphi\rangle$ at Borel sets.
One infers that $P$  and $P'$ coincide at Borel sets, whence $P=P'$ arguing as in (\ref{PCPOL}).\qed

\begin{Cor}\label{CNGENTLNSLHP} $P$ from \emph{(\ref{NGENTLNSLHP})} satisfies \emph{(c),\,(d),\,(e)} in \emph{(\ref{NRLCCRRCLP})}. One obtains 
an alternative proof of   \emph{(\ref{ECCPDS})} and  \emph{(\ref{ECCPWS})}.
  \end{Cor}

{\it Proof.} Note $B(\varsigma e)\cdot\{x_0=x_3,0\le \varsigma x_3<\alpha\}=\{0\le\varsigma x_0=xe<\alpha\}$. Hence by covariance one infers  that (c) and (1) are equivalent. So (c) holds.
Therefore, due  to (\ref{LBSLSET}) and Eq.\,(\ref{WLBSLSET})
also (d) holds. Hence, by 
(\ref{DSLSLHPTLC})(b), $P(\{0\le \varsigma x_0=xe\})
\le E(\{\mathfrak{x}: x_0=0, xe\ge 0\})$. Replacing $(e,\varsigma)$ by $(-e,-\varsigma)$ one gets
$P(\{0\ge \varsigma x_0=xe\})
\le E(\{\mathfrak{x}: x_0=0, xe\le 0\})$. The sum of the left hand sides is $I$. Also the sum of the right hand sides equals $I$. This proves that the inequalities actually are equalities. Hence (e) holds and $I=\lim_{|\alpha|\to\infty} E(\{x_0=\alpha, |xe|\le|\alpha|\})$ follows.\qed\\

As an example consider the maximal $\perp'$-set $\sigma_M\cup M\cup \tau_M$ from (\ref{GMPPOS}). One has 
$$ I= E(\sigma_M)  +   P(M)   +  E(\tau_M)$$
Indeed, it suffices to verify this equation in the special case Eq.\,(\ref{SGMPPOS}). This follows easily from 
$ P(\{\mathfrak{x}:0\le x_0=xe\le \alpha\})=
E(\{\mathfrak{x}: x_0=\alpha, xe\le \alpha\})-E(\{\mathfrak{x}: x_0=0, xe\le 0\})$, which holds by 
 (\ref{NRLCCRRCLP})(d)  because of  Eq.\,(\ref{BLHPTLC}), Eq.\,(\ref{ALHPTLC}). Furthermore, because of (\ref{SNTLSD}) and (\ref{BODS}), the spacelike hyperplanes $\sigma$ and $\tau$ may even intersect as in the figure below.\\  
 
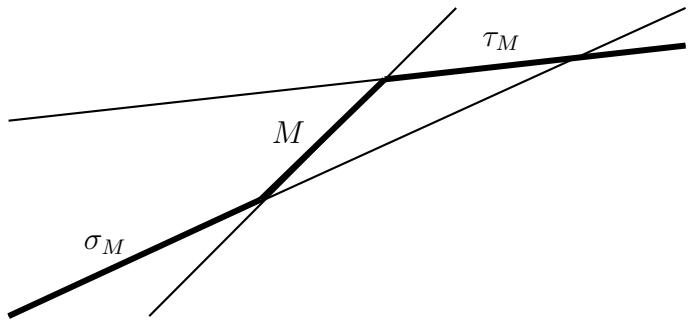
\begin{pspicture}(-9,-1)(2,3)
\psline(-6,-1.1)(3,3)
\psline(-6,1.5)(3,2.5)
\psline(-4.13,-1.1)(-0.05,3)
\psline[linewidth=0.8mm](-6,-1.1)(-2.65,0.45)
\psline[linewidth=0.8mm](-2.65,0.45)(-1,2.05)
\psline[linewidth=0.8mm](-1,2.05)(3,2.5)

\put(-5,-0.2){$\sigma_M$}
\put(0.3,2.5){$\tau_M$}
\put(-2.5,1.2){$M$}

\end{pspicture}

\section{Systems with Causal Time Evolution}\label{DCRS}

After having studied  in detail the massive and massless causal systems of fundamental importance, i.e., in sec.\,\ref{DS} 
up to sec.\,\ref{NES}  the Dirac system 
$(W^\textsc{d},E^\textsc{d})$ for every positive mass and in sec.\,\ref{OWL}, sec.\,\ref{LOCCRS} the two Weyl systems $(W^{\textsc{w}\chi}, E^{\textsc{w}})$, $\chi\in\{+,-\}$, now the goal is to determine all other causal systems. 
In so doing  we will consider  only \textbf{physical systems}, i.e., systems with non-negative mass-squared operator 
$$ C=H^2-P_1^2-P_2^2-P_3^2\,\,\ge \,0$$
and  finite spinor dimension  (sec.\,\ref{FDSB}). Then it will turn out that the Dirac and the Weyl systems are the only irreducible causal systems. Actually we will not succeed as yet to rule out  certain obviously unlikely irreducible massive and massless SCT, see the conclusions at the end of sec.\,\ref{MCSR} and sec.\,\ref{SMLCS}.  Recall that  there exist causal systems (\ref{RLBSGNTLR}), which do not satisfy the properties $C\ge 0$ and finite spinor dimension. 
 We start studying  a preliminary object, the SCT.  For this recall the definitions in sec.\,\ref{SPCPOL} and \ref{CPOL}, in particular (\ref{POLCTE}). 

\begin{Def}\label{SCTER} A  \textbf{system with causal time evolution}  (SCT) consists of a representation $W$ of $\tilde{\mathcal{P}}$ and, with respect to $W$, a Poincar\'e covariant  WL $E$ such that the time evolution is causal, i.e.,
\begin{equation*}\label{CTEWL}
W(t)E(\Delta)W(t)^{-1}\le E(\Delta_t)
\end{equation*}
 for all measurable $\Delta\subset\R^3$ and all $t\in\R$.  An SCT is called \textbf{massive} if $C>0$, and \textbf{massless} if $C=0$. If $(W,E)$ is an SCT then any SCT, which is unitarily equivalent to   $(SWS^{-1},E)$, where $S$ is  some unitary operator commuting with the representation $W|_{\R\otimes ISU(2)}$ of the little kinematical group,  is said to be \textbf{unitarily related}   to $(W,E)$.   
\end{Def}

Clearly,  causal systems are SCT. A complete explicit description of the SCT is provided in the following sections.  The massive ones are treated  in \cite{CL15} and here we summarize the relevant results in (\ref{DMST}). The massless SCT are derived in sec.\,\ref{SMLCS}. The main result on massless SCT is (\ref{CMLST}).

\subsection{Position representation}\label{DCRSPR}  In analyzing the SCT $(W,E)$
 the  imprimitivity theorem by Mackey 1949, 1955 \cite{M49},\cite{M55}, reported in (\ref{A:IT}), is essential. By this theorem, up to unitary equivalence,  $W|_{ ISU(2)}$  is  a representation  induced from a representation $D$ of $SU(2)$ and  $E$ is the related system of imprimitivity. Let $D$ act on the Hilbert space $\mathcal{S}$. Then $(W,E)$ is unitarily equivalent to the position representation  $(W^{pos},E^{can})$ (cf. \cite{W62},\cite{L69}), which is characterized by
\begin{equation}\label{PR}
 (W^{pos}(b,B)\psi)(x) :=D(B)\,\psi(B^{-1}\cdot (x-b)),  \quad E^{can}(\Delta)\psi :=1_\Delta\psi
 \end{equation}
 on $L^2(\R^3,\mathcal{S})$ with $(b,B)\in ISU(2)$ and $\Delta\subset \R^3$  a measurable set.  $E^{can}$ is called  the \textbf{canonical} PM. Note that $W$ is not determined by $W|_{ISU(2)}$. \\
  \hspace*{6mm} 
  The spin$-j$ components for $j\in\N_0/2$ of $W|_{ISU(2)}$ are the eigenspaces to the eigenvalues $j(j+1)$ of the square of the spin vector operator $S:=J-X\times P$. The orbital angular momentum $X\times P$ refers to the \textbf{causal} position operator $X$, which in position representation is given by $\big(X_k^{pos}\psi\big)(x)=x_k\psi(x)$.
  
\subsection{Spinor basis}\label{FDSB}  By Eq.\,(\ref{PR}), for $x\in\R^3$ the vector $\psi(x)$ is a spinor, whence  $\mathcal{S}$ is called the spinor space.  Its dimension  is called the  \textbf{spinor dimension} of $W$.  Clearly it is an invariant and characteristic for the SCT.  \\
 \hspace*{6mm}  We shall write $D$ as the orthogonal sum $\oplus_j\nu_jD^{(j)}$ of  the spin components, where  
 \begin{equation}\label{MSC}
 \nu_j\in\{0,1,2, \dots\}\cup\{\infty\}
 \end{equation}
  is the \textbf{multiplicity} of the irreducible representation $D^{(j)}$ for every spin  $j\in \N_0/2$. Then the spinor dimension is $\sum_j(2j+1)\nu_j$.
  The standard matrix  form of $D^{(j)}$   satisfies
 \begin{equation}\label{DRSG}
 D^{(j)}(\operatorname{diag}(c,\overline{c}))=\operatorname{diag}(c^{2j},c^{2j-2},\dots,c^{-2j+2},c^{-2j})
\end{equation}
for $c\in\C$, $|c|=1$.  Accordingly the standard basis for $\C^{2j+1}$ is denoted by $(|j,s\rangle)$, where the label $s$ runs through the helicity values $2j,2j-2,\dots,-2j+2,-2j$. In order to obtain a basis of the spinor space  $\mathcal{S}$ one has to take account of the multiplicities $\nu_j$ by an additional index $\iota$ indicating the $\iota$-th copy of $D^{(j)}$. One gets the orthogonal spinor basis
 \begin{equation}\label{OSB}
(|j,\iota,s\rangle) \text{ with } j\in |s|+\N_0,\; \iota = 1,2,\dots, \nu_j, \, s=2j,2j-2,\dots,-2j+2,-2j
\end{equation} 
 \\  
 From now on we will consider only  SCT with \textbf{finite spinor dimension}. So all $\nu_j$ are finite and only finitely many of them are non-zero. From a physical point of view this is a reasonable restriction as infinite dimensional spinor theories have not found  application so far. 
 
  \subsection{Momentum representation of time evolution}\label{MRTE}  The Fourier transformation $\mathcal{F}$  on $ L^2(\R^3,\mathcal{S})\simeq \mathcal{S}\otimes L^2(\R^3)$  acts as $I_\mathcal{S}\otimes \mathcal{F}_\C$. Then the momentum representation $W^{mom}:=\mathcal{F}\,W^{pos}\,\mathcal{F}^{-1}$ satisfies
  \begin{equation}
 \big(W^{mom}(b,B)\varphi\big)(p)=\operatorname{e}^{-\operatorname{i}bp}D(B)\,\varphi(B^{-1}\cdot p)
 \end{equation}
 
Let $H$ be  the generator  of the time evolution $W(t)=\operatorname{e}^{\operatorname{i} tH}$, $t\in\R$. Since $H$ is translational invariant
  there are self-adjoint  matrices $h(p)$ acting on $\mathcal{S}$  such that  
 \begin{equation}
(H^{mom}\varphi)(p)=h(p)\varphi(p)
\end{equation} 
 defined on its natural domain.  $H$ is rotational invariant, too.  So  $h$ is rotational covariant and  $h$ can be chosen such that $h(\rho\, e_3)$,  $\rho\ge 0$, commutes 
 with $D(B)$ if  $B$ is diagonal, and  $h(p) = D(B(p))    h(|p| e_3) D(B(p))^{-1}$, whence
 \begin{equation}\label{RIH}
 h(B\cdot p)=D(B)\,h(p)\,D(B)^{-1} 
\end{equation}
for all  $p\in \R^3$ without exception (see  \cite[Eq.\,(14)]{CL15}).\\
 \hspace*{6mm} 
 Here $B(p)$ denotes the \textbf{helicity cross section} satisfying $B(p)\in SU(2)$ and $|p|B(p)\cdot e_3=p$ for all $p\in\R^3$. Explicitly
\begin{equation}\label{CCSEF}
B(p) =\left( \begin{array}{cc} a_+ & -\overline{b}\,a_-\\ b\,a_- & a_+ \end{array}\right), \quad a_\pm:=\sqrt{\frac{|p|\pm p_3}{2|p|}}, \quad b:=\frac{p_1+ip_2}{|p_1+ip_2|} \textrm{ for } (p_1,p_2)\ne 0
\end{equation}
and $B(\alpha e_3)$  equals $I_2$ if $\alpha\ge 0$ and $-i\sigma_2$ if $\alpha <0$. Note  $B(\lambda p)=B(p)$ for all $\lambda>0$ and for $|p|=1$ the important relation
\begin{equation}\label{CRCS} 
B(p)\,\sigma_3\, B(p)^{-1}=\sum_{k=1}^3p_k\sigma_k
\end{equation}

  Analogous formulae hold for $W(t)$, $t\in\R$ in place of $H$ such that
 \begin{equation}\label{TEGR}
\big(W^{mom}(t)\varphi\big)(p)=v_t(p)\varphi (p),\quad\,   v_t(p)=\operatorname{e}^{\operatorname{i} t h(p)}    
\end{equation}

\subsection{Causal time evolution} As proven in  \cite[Theorem 1]{CL15} the causal time evolution of the SCT $(W,E)$ 
 implies  that $h$ is linear in momentum, i.e., for all $p\in \R^3$
\begin{equation}\label{GCTE}
h(p)=A_1p_1+A_2p_2+A_3p_3+M
\end{equation}
where $A_k$, $M$ are constant self-adjoint matrices in spinor space with $||\sum_k p_kA_k ||=\gamma |p|$ for some $\gamma\in [0,1]$. 
By rotational covariance Eq.\,(\ref{RIH})  of $h$ one has
\begin{equation}\label{RIA}
 D(B) \big(\sum_{k=1}^3 p_k A_k \big)  \,D(B)^{-1}=\sum_{k=1}^3  (B \cdot p)_k A_k, \quad D(B)MD(B)^{-1}=M
\end{equation}

Further, since the mass-squared operator $C$ is supposed to be non-negative,  by \cite[Lemma 16, Lemma 1]{CL15} one has  $\big(C^{mom}\varphi\big)(p)=M^2 \varphi(p)$. Hence $C$ commutes with $(W,E)$ and the spectrum of $C$ is finite non-negative.  This implies

\begin{Lem} \label{ECA} Every SCT is the finite orthogonal sum of SCT with $C=m^2I$ for pairwise different scalars $m\ge 0$. The  matrices $A_k$, $M$ satisfy $\{A_k,A_j\}=2\delta_{kj}I$, $\{A_k,M\}=0$.
\end{Lem}\\
In particular $\gamma=1$ follows. Further analysis will reveal the structure of the matrices $A_k$, $M$. \\

\section{Massive Causal Systems}\label{MCSR}

In \cite{CL15}  massive SCT  are treated in detail.\footnote{ In \cite{CL15} a massive SCT is called a massive causal system.} We  summarize the main results  from \cite{CL15} providing a complete description of the massive SCT. Then the aim  is to single out the massive causal systems. To this, in view of (\ref{FIDCWL}),  we  study the  dependence of their localizations on the  frame of reference.
\\

\subsection{Construction of the massive SCT}\label{CMSCT} One starts from the fundamental Dirac localization $(V^\textsc{d},U^\textsc{d},E^\textsc{d}):=(W^\textsc{d}|_{\R\otimes ISU(2)},E^\textsc{d})$ (see sec.\,\ref{DS}). By  a simple construction, which enlarges the spinor space, one gets for every positive mass $m$ and   every $n\in\N$  the $n$-th \textbf{Dirac tensor-localization}
 \begin{equation}\label{DTL}
 (V_n,U_n,E_n):=(V^\textsc{d},U^\textsc{d},E^\textsc{d})\otimes (I_n,U^{(\frac{n}{2}-\frac{1}{2})},I_n)
 \end{equation}
with $I_n$ the identity on $\C^n$ and $U^{(\frac{n}{2}-\frac{1}{2})}(b,B):= D^{(\frac{n}{2}-\frac{1}{2})}(B)$. In particular, $h_n(p)= h^\textsc{d}(p) \otimes I_n$ with $h^\textsc{d}(p)=\sum_{k=1}^3\alpha_k p_k + \beta m$, whence the Dirac tensor-matrices are $\alpha_k\otimes I_n$, $\beta\otimes I_n$. In the position representation $E_n$ is the canonical PM. The  time evolutions $V_n$  
are causal. The Dirac tensor-localizations are irreducible and mutually unitarily inequivalent.\\
\hspace*{6mm}
  Every representation $(V_n,U_n)$  for mass $m>0$ of the little kinematical group can be extended to a representation $W_n$ of $\tilde{\mathcal{P}}$ on  $L^2(\R^3,\C^4\times \C^n)$.  For $(t,b,A) \in \tilde{\mathcal{P}}$   the latter is given by
\begin{equation}\label{IMCS}
\big(W^{mom}_n(t,b,A)\varphi\big)(p) :=\operatorname{e}^{-\operatorname{i}bp}
\sum_{\eta =\pm 1}\left( \operatorname{e}^{\operatorname{i}t\eta\epsilon(p)}  \pi^\eta(p)s(A)^{*-1} \otimes D^{(\frac{n}{2}-\frac{1}{2})} \big(R(\mathfrak{p}^\eta,A) \big) \right)    \varphi(q^\eta)
\end{equation}
As to the  notations recall
Eq.\,(\ref{RDS}) and following. Furthermore recall the Wigner rotation   
$R(\mathfrak{p}^\eta,A)$
with respect to the canonical cross section $Q$ with
the important property $R(\mathfrak{p}^\eta,B)=B$ for $B\in SU(2)$ (see (\ref{PCCS})). Note that $W_1$ is just the Dirac representation $W^\textsc{d}$.
Hence for every mass $m>0$ a sequence $(W_n,E_n)$ of irreducible and mutually unitarily  inequivalent massive SCT  is constructed.\\
\hspace*{6mm}
 $(W_1,E_1)$ is the  Dirac system  $(W^\textsc{d},E^\textsc{d})$ studied in  sec.\,\ref{DS}.
 
\hspace*{6mm}
From \cite[sec.\,B,D,E]{CL15} we extract

\begin{The}\label{DMST} \emph{(i)} Every massive SCT $(W,E)$ is a  unique orthogonal sum of finitely many massive SCT for  different masses $m>0$.\\
 \hspace*{6mm} \emph{(ii)} If $(W,E)$ is a massive SCT for mass $m>0$, then  
it is unitarily related  to a  finite orthogonal sum  of   $(W_n,E_n)$ for mass $m$    with uniquely determined multiplicities.  \\
\hspace*{6mm} \emph{(iii)} Finally, a massive SCT $(W,E)$  is unitarily equivalent  to an orthogonal sum  of finitely many  $(W_n,E_n)$ with possibly various  masses if and only if the booster  of $W$ satisfies
  \begin{equation}\label{NCPOM}
N=\frac{1}{2}\{X,H\} + \frac{\operatorname{sgn}(H)}{C^{1/2}+|H|} P \times \left(J- X\times P -\frac{\operatorname{i}}{4} [X,H] \times  [X,H] \right)
\end{equation}
where $X=(X_1,X_2,X_3)$ denotes the causal position operator associated with $E$.
\end{The}

\subsection{Determination of the massive causal systems}\label{FDMCS}
By (\ref{CSDF}) the Dirac localization is causal and hence, by (\ref{FIDCWL}),  frame-independent. In contrast,  we recall,   there is no Dirac state   localized  by NWL in a bounded region  with respect to two frames moving relative to each other (see sec.\,\ref{FDNWL}). This may be not surprising as, with respect to NWL, the Dirac time evolution is not causal. But,  surprisingly,  as shown in (\ref{SCTLFD}), an SCT  need not localize frame-independently.  \\

Let $(W,E)$ be an irreducible massive SCT. By (\ref{DMST})(i),(ii) it is unitarily related to a  finite orthogonal sum  of   $(W_n,E_n)$ for some mass  $m>0$.  So, in the position representation, $E$ is the canonical PM and, in the momentum representation, 

\begin{equation}\label{MCSRST}
\big(W^{mom}(t,b,A)\varphi\big)(p) =\operatorname{e}^{-\operatorname{i}bp}s(p)^{-1}
\sum_{\eta =\pm 1} \operatorname{e}^{\operatorname{i}t\eta\epsilon(p)}w^\eta(A,p) 
 s(q^\eta)    \varphi(q^\eta) 
\end{equation}
for almost all $p\in\R^3$. Here, $w^\eta(A,p):= \pi^{\eta}(p)s(A)^{*-1} \otimes D' \big(R(\mathfrak{p}^\eta,A) \big)$ with some finite dimensional representation $D'$ of $ISU(2)$. Further, according to (\ref{SCTER}), $S$  is  some unitary operator  commuting with $W|_{\R\times ISU(2)}$ so that $(S^{mom}\varphi)(p)=s(p)\varphi(p)$ with unitary matrices $s(p)$ acting on spinor space, satisfying rotational covariance $s(B\cdot p)= D(B)s(p)D(B)^{-1}$ for $D:=(D^{\frac{1}{2}}\oplus D^{\frac{1}{2}})\otimes D'$, and  commuting with the projections on the energy eigenspaces $s(p)(\pi^\eta(p)\otimes I')=(\pi^\eta(p)\otimes I')s(p)$ with $I':=D'(I_2)$. Therefore  $s(p)=D(B(p))s(|p|e_3)D(B(p))^{-1}$,  and $s(re_3)$, $r\ge 0$ commutes with $D(B)$ for $B$ diagonal and with $(r\alpha_3+m\beta)\otimes I'$ (see Eq.\,(\ref{RIA}) with $h(p)= h^\textsc{d}(p) \otimes I'$ in Eq.\,(\ref{GCTE})).\\

\begin{The}\label{DMCS} Let $(W,E)$ be an irreducible  massive SCT, which  localizes frame-independently. Suppose that  $p\mapsto s(p)$ (see \emph{Eq.\,(\ref{MCSRST})}) has an entire extension. Then $(W,E)$ is unitarily related to a finite multiple of
the  Dirac system  $(W^\textsc{d},E^\textsc{d})$. 
\end{The}\\
{\it Proof.} 
Let $\varphi$ be localized in a bounded region. Then, by assumption, so is $\varphi_\rho:=W^{mom}(A_\rho)\varphi$, where $A_\rho:=\operatorname{e}^{\frac{\rho}{2}\sigma_3}$, $\rho\ne 0$. Hence, by the Paley-Wiener Theorem,  these functions are the restriction of entire functions on $\C^3$, still denoted by $\varphi$ and $\varphi_\rho$. 
 As $q^\eta=(p_1,p_2,-\eta\beta\epsilon+\alpha p_3)$ with  $\alpha:= \cosh\rho$, $\beta:=\sinh \rho$,  and $\epsilon=\sqrt{p_1^2+p_2^2+p_3^2+m^2}$, there are entire  functions $\phi_1$, $\phi_2$ such that $(s(A_\rho)^{*-1}\otimes I_n)s(q^\eta)\varphi(q^\eta)=\phi_1(p) -\eta\epsilon \phi_2(p)$. Furthermore, explicit computation yields 
\begin{equation}\label{WRDB}
R(\mathfrak{p}^\eta,A_\rho)=d(\eta p)^{-1/2}\left( \begin{array}{cc} \gamma  (m+\epsilon) -\delta \eta p_3 & \delta \eta (p_1-\operatorname{i}p_2)     \\ -\delta \eta (p_1+\operatorname{i}p_2)     &\gamma (m+\epsilon) -\delta \eta p_3 \end{array} \right)
\end{equation}
 with  $d(p):=(m+\epsilon) (m+\alpha\,\epsilon-\beta p_3)$ and constants $\gamma:=\cosh\frac{\rho}{2}$, $\delta:=\sinh\frac{\rho}{2}$.\\
 \hspace*{6mm}
  Note that each component $D^{(j)}_{kl}(M)$ is a homogeneous polynomial of degree $2j$ of the entries of $M$. Hence there are  matrix-valued functions $A^{(j)}$, $B^{(j)}$ with polynomial entries such that 
 $ D^{(j)} \big(R(\mathfrak{p}^\eta,A_\rho) \big)=\varDelta^{(j)}(\eta p) \big(A^{(j)}(\eta p)+\epsilon B^{(j)}(\eta p)\big)$ with $\varDelta^{(j)}(\eta p):=d(\eta p)^{-j}I_{2j+1}$. Recall that $D'$ is the orthogonal sum of some $D^{(j)} $. Accordingly we write $ D' \big(R(\mathfrak{p}^\eta,A_\rho) \big)=\varDelta'(\eta p) \big(A'(\eta p)+\epsilon B'(\eta p)\big)$. 
Hence,  by Eq.\,(\ref{MCSRST}), one has

 \begin{equation*}
2s(p) \varphi_\rho(p)= \sum_\eta  \left( \big( \text{{\it \small{I}}}_4+ 
\frac{\eta}{\epsilon} h^\textsc{d}(p)\big) \otimes \varDelta'(\eta p)\big(A'(\eta p)+\epsilon B'(\eta p)\big)\right)\big(\phi_1(p)-\eta\epsilon\phi_2(p)\big) 
\end{equation*}
 \hspace*{6mm}
Now we examine the point $p_*:=(\operatorname{i}m,0,0)$. There $\epsilon$ vanishes. Hence
one infers $\sum_\eta\big( \eta  I_4\otimes \varDelta'(\eta p_*) A'(\eta p_*)
\big)\,\xi_*=0$  for $\xi_*:=( h^\textsc{d}(p_*)\otimes I') \phi_1(p_*)=\big( h^\textsc{d}(p_*)s(A_\rho^{*-1})\otimes I'\big)s(p_*) \varphi(p_*)$. Here, plainly, $s(p_*)$ denotes the analytic continuation of $s$ at $p_*$. As $\varDelta'(\eta p_*) A'(\eta p_*)=D' \left(\begin{array}{cc} \gamma   &  \operatorname{i}\eta \delta     \\ -\operatorname{i} \eta\delta     &\gamma  \end{array} \right)$, one gets equivalently 
\begin{equation*}
\xi_*=\left(I_4\otimes D' \left(\begin{array}{cc} \alpha   &  \operatorname{i}\beta     \\ -\operatorname{i} \beta     &\alpha  \end{array} \right) \right)\,\xi_*\tag{$\star$}
\end{equation*}

This condition implies that $D'$ is  a multiple of the unit representation. Indeed, assume the contrary. Recall that $\varphi(p_*)= \int\operatorname{e}^{mx_1}\psi(x)\operatorname{d}^3x$ 
for some $\psi \in L^2(\R^3,\C^4\times \C^n)$ with bounded support. So one may alter $\psi$ a bit  within its support to guarantee $\varphi(p_*)\ne 0$. Further, note that $s(p_*)$ is invertible, as the equation $I=s(z)s(\overline{z})^*$ holds true by analytic continuation for all $z\in\C^3$. Hence one may multiply $\psi$  by a suitable constant spinor matrix in order that $\xi_*$ does not  satisfy Eq.\,$(\star)$. This accomplishes the proof.  \qed\\

\begin{Cor}\label{SCTLFD} 
The SCT    $(W_n,E_n)$ for $n>1$ do not localize frame-independently and hence are not causal.
\end{Cor}

{\it Proof.} The result in (\ref{DMCS}) applies with $s(p)=I_{4n}$ for all $p$.\qed

 \hspace*{6mm}
In conclusion we are  convinced that the assumption of analyticity of s in (\ref{DMCS})  can be dropped. 
Compare also  the analogous result (\ref{FDWTL}) for massless SCT going without such an assumption.
Anyway, due to  (\ref{FIDCWL}), the higher Dirac tensor-systems $(W_n,E_n)$, $n>1$ are revealed to be mathematical artefacts, interim results  in determining the massive causal systems.
 It is quite sure that the only irreducible massive SCT, which  are causal,  are unitarily equivalent to the Dirac systems  $(W^\textsc{d},E^\textsc{d})$ for masses $m>0$ studied in  sec.\,\ref{DS}.

\section{Massless Causal Systems}\label{SMLCS}

The analysis of the massless causal systems proceeds  analogously to that of the massive ones. However there is an important difference arising from  the matrices $A_1,A_2,A_3,M$   (\ref{ECA}). In the massive case one is dealing with the Clifford algebra with four basis elements, which has only one irreducible representation, the $4$-dimensional Dirac representation  (see \cite[Lemma 2]{CL15}). The massless case $M=0$, however, is determined by  the Clifford algebra generated by three basis elements, which has just two inequivalent irreducible representations, the  $2$-dimensional Weyl representations (see (\ref{CA3BE})).

\begin{Lem}\label{CA3BE} Let $A_1,A_2,A_3$ be any (respectively, self-adjoint) square matrices. Then, they satisfy $\{A_k,A_j\}=2\delta_{kj}I$ if and only if, up to (respectively, unitary) equivalence,
$$ A_k=(\sigma_k\otimes I') \oplus (-\sigma_k\otimes I'')$$
with $\sigma_k$, $k=1,2,3$ the Pauli matrices and $I',I''$ unit matrices, and $\oplus$ the direct (respectively, orthogonal) sum.
\end{Lem}

{\it Proof.} The group $G$ generated by the Clifford algebra $\{\alpha_k,\alpha_j\} =2\delta_{kj}1$, \,$k,j=1,2,3$ has order $16$. Its elements are $1,-1$ and all words in $\alpha_k$ which result from $\pm\, \alpha_1\alpha_2\alpha_3$ by canceling  at most $2$ of them. There are $10$ conjugacy classes, namely $\{1\}$, $\{-1\}$, $\{\alpha_1\alpha_2\alpha_3\},$ $\{-\alpha_1\alpha_2\alpha_3\}$, and $\{g,-g\}$ for all other group elements $g$. Every representation of $G$, which maps $-1$ to $I$, actually is a representation of the abelian group $G/ \{1,-1\}$. Hence there are  $8$ inequivalent one-dimensional representations of $G$. 
So, up to equivalence, there are just two inequivalent irreducible representation $\rho_1,\rho_2$ left. Let $d_1$ and $d_2$ be their dimensions. They satisfy $16=8 \cdot 1^2 +d_1^2+d_2^2$, whence $d_1=d_2=2$. One easily checks that $\rho_1(\pm 1):=\pm I _2$  and $\rho_1(\alpha_k):=\sigma_k$, $k=1,2,3$ defines an irreducible representation of $G$. Similarly does  $\rho_2(\pm 1):=\pm I _2$  and $\rho_2(\alpha_k):=-\sigma_k$, $k=1,2,3$. Since $\operatorname{tr}(\rho_1(\alpha_1\alpha_2\alpha_3)) =2\operatorname{i} \ne \operatorname{tr}(\rho_2(\alpha_1\alpha_2\alpha_3))=-2\operatorname{i}$, $\rho_1$ and $\rho_2$ are inequivalent.  By the same reason, $\rho_1$, $\rho_2$ are inequivalent even as representations of the Clifford algebra, known as Weyl representations.  Finally, if one limits oneself   to self-adjoint representations of the algebra, which means to unitary representations of the group, equivalence becomes unitary equivalence. Moreover, the direct sum must be replaced by an orthogonal sum. \qed

\begin{Lem}\label{RCRC} 
 Let $D$ be a finite dimensional representation of $SU(2)$ and let  $A_1,A_2,A_3$ be self-adjoint matrices.  They satisfy rotational covariance $D(B) \big(\sum_{k=1}^3 p_k A_k \big)  \,D(B)^{-1}=\sum_{k=1}^3  (B \cdot p)_k A_k$ (cf.\,\emph{Eq.\,(\ref{RIA})}) and $\{A_k,A_j\}=2\delta_{kj}I$ (cf.\,\emph{(\ref{ECA})}) if and only if, up to unitary equivalence,  $A_k=(\sigma_k\otimes I') \oplus (-\sigma_k\otimes I'')$ and 
 \begin{equation*}
 D=(D^{(\frac{1}{2})}\otimes D')   \oplus (D^{(\frac{1}{2})} \otimes D'')
 \end{equation*}
  for some representations $D'$, $D''$  of $SU(2)$.
\end{Lem}  

{\it Proof.}  By Lemma (\ref{CA3BE}) we may assume at once  $A_k=(\sigma_k\otimes I') \oplus (-\sigma_k\otimes I'')$. Recall the relation $B(\sum_kp_k\sigma_k)B^*=\sum_k(B\cdot p)_k\sigma_k$.  
So  the representation $L:=  (D^{(\frac{1}{2})} \otimes I') \oplus (D^{(\frac{1}{2})} \otimes I'') $ of $SU(2)$ satisfies  $L(B) \big(\sum_{k=1}^3 p_k A_k \big)  \,L(B)^{-1}=\sum_{k=1}^3  (B \cdot p)_k A_k$. This implies the \textquotedblleft if\textquotedblright part of the assertion. As to the \textquotedblleft only if\textquotedblright   part, note that $L(B)^{-1}D(B)$ commutes with $A_k$, $k=1,2,3$.
Note that $A_k=A'_k\oplus A''_k$ with $A'_k:=\sigma_k\otimes I'$ and $A''_k:=-\sigma_k\otimes I''$ determines the primary decomposition  of the representation of $G$ (see the proof of (\ref{CA3BE})) defined by $A_k$, $k=1,2,3$. Therefore  $L(B)^{-1}D(B) = R'(B)\oplus R''(B)$ for some $R'(B)$ and  $R''(B)$, which commute with $A'_k$ and $A''_k$, respectively.
 Hence $R'(B)=I_2\otimes D'(B)$ and $R''(B)=I_2\otimes D''(B)$ with uniquely determined matrices $D'(B)$, $D''(B)$ for every $B\in SU(2)$. Then $D(B)=L(B)\big( R'(B)\oplus R''(B)\big)$ yields the formula on $D$. The latter implies that $D'$, $D''$ are representations of $SU(2)$.\qed
 
 \subsection{The massless WL with causal time evolution}\label{MWLWCTE}
 
 The following results (\ref{DDE}), (\ref{ICLC}), and (\ref{CLC}) are an immediate consequence of the foregoing considerations (\ref{ECA}),  (\ref{CA3BE}), (\ref{RCRC}). Recall that $\big(U^{pos}(b,B)\psi\big)(x) =D(B)\,\psi(B^{-1}\cdot (x-b))$ and  $E^{pos}=E^{can}$, i.e., $E^{pos}(\Delta)\psi =1_\Delta\psi$. A  WL with causal time evolution (\ref{POLCTE}) is called massless if $C=0$.
 
\begin{The}\label{DDE} Let $(V,U,E)$ be a massless WL  with causal time evolution of smallest spinor dimension. Then the spinor dimension is $2$ and $D=D^{(\frac{1}{2})}$. Moreover, there is $\chi \in\{+,-\}$ such that, up to unitary equivalence,
\begin{equation}\label{WCHH}
H^{\chi pos}=\chi\left(\sigma_1\frac{1}{\operatorname{i}}\frac{\partial}{\partial x_1}+\sigma_2\frac{1}{\operatorname{i}}\frac{\partial}{\partial x_2}+\sigma_3\frac{1}{\operatorname{i}}\frac{\partial}{\partial x_3}\right)
\end{equation}
\end{The}

The result in  (\ref{DDE}) is a derivation of the Weyl equations as a mere consequence of the principle of causality in relativistic quantum mechanics. 

\begin{Def}\label{DMLWLCTE} The massless WL  with causal time evolution, for which in position representation $H$ is given by 
Eq.\,(\ref{WCHH}) and $D=D^{(\frac{1}{2})}$, 
 is called the \textbf{right-handed}, respectively  \textbf{left-handed} \textbf{Weyl localization}. It is denoted by $(V^{\textsc{w}\chi},U^\textsc{w},E^\textsc{w})$ with $\chi=+$ and $\chi=-$, respectively.  
\end{Def}

\begin{The}\label{ICLC} Let $(V,U,E)$ be an irreducible massless  WL with causal time evolution. Then there exists   and $n\in\N$ and $\chi \in\{+,-\}$ such that $(V,U,E)$ is unitarily equivalent to the  massless WL with causal time evolution $(V_n^\chi,U_n,E_n)$ given by
\begin{align*}
V_n^\chi(t) &= V^{\textsc{w}\chi} (t)\otimes I_n     &  \text{time translations}\\
U_n(b)  &= U^\textsc{w} (b)\otimes I_n  & \text{space translations}  \\
U_n(B) &= U^\textsc{w}(B)\otimes  D^{(\frac{n}{2}-\frac{1}{2})}(B)  & \text{rotations} \\
 E_n(\Delta)&= E^\textsc{w} (\Delta)\otimes I_n  & \text{localization operators}
\end{align*}
where $I_n$ is the identity on $\C^n$.
\end{The}

\begin{Def} 
Let the massless  WL  with causal time evolution $(V_n^\chi,U_n,E_n)$, $n\in\N$,  $\chi \in \{+,-\}$ be  called   \textbf{Weyl tensor-localizations}.
\end{Def}

Note that  the   Weyl tensor-localizations for $n=1$  are just the Weyl localizations. The spinor space of the  Weyl tensor-localizations for $n$ is $\C^2 \otimes \C^n \simeq\C^{2n}$. Hence the spinor dimension is $2n$.  As $D_n \simeq D^{(\frac{1}{2})}   \otimes D^{(\frac{n}{2}-\frac{1}{2})} \simeq D^{(\frac{n}{2}-1)}\oplus D^{(\frac{n}{2})}$, the helicity spectrum is $\{-\frac{n}{2}, -\frac{n}{2}+1,\dots,\frac{n}{2}\}$ with multiplicity $1$ for helicity $\frac{n}{2},-\frac{n}{2}$ and $2$ else.

\begin{The}\label{CLC}  Every massless  WL with causal time evolution is unitarily equivalent to a finite orthogonal sum  of Weyl tensor-localizations with uniquely determined multiplicities.
\end{The}

We will show that each representation $(V_n^\chi ,U_n)$ of the little kinematical group can be extended to an, up to unitary equivalence, unique representation $W_n^\chi$ of $\tilde{\mathcal{P}}$. The following insertion sec.\,\ref{IIMRPG} provides some  facts used for this objective.  

\subsection{On  irreducible massless representations of $\tilde{\mathcal{P}}$}\label{IIMRPG} 
Let $\mathcal{O}^{0,\eta}$ for  $\eta=\pm$ denote the light half-cone $\{\mathfrak{p}\in\R^4\setminus\{0\}: p_0=\eta\,|p|\}$ endowed with the Lorentz invariant measure and  let $\mathcal{O}^0:=\mathcal{O}^{0,+}\cup\mathcal{O}^{0,-} =\{\mathfrak{p}\in \R^4\setminus\{0\}: |p_0|=|p|\}$ denote the light cone.\\

\hspace*{6mm}The $1$-dimensional representations $(\mathfrak{a},A)\mapsto \operatorname{e}^{\operatorname{i}(\eta a_0-a_3)} (A_{11})^{2 s}$ for  $s\in \mathbb{Z}/2$ of the inhomogeneous little group $I E(2)\sqsubset \tilde{\mathcal{P}}$ (recall $A\in E(2)$ if and only if $A\in SL(2,\C)$ with $A_{21}=0$, $A_{22}=\overline{A}_{11}$) induce the irreducible finite spinor dimensional massless representations of $\tilde{\mathcal{P}}$. They act on $L^2(\mathcal{O}^{0,\eta})$  by
\begin{equation}\label{MLIRPG}
\big(U^{0,s,\eta}(\mathfrak{a},A)\phi\big)(\mathfrak{p}):=   
\operatorname{e}^{\operatorname{i}\mathfrak{a}\cdot\mathfrak{p}}
k(\mathfrak{p},A)^{2 s}
\,\phi(A^{-1}\cdot \mathfrak{p})
\end{equation}
see e.g. \cite[Eq.\,(1.1.15)]{S70}. Here 
\begin{equation}\label{DEWRMLC}
k(\mathfrak{p},A):=H(\mathfrak{p},A)_{11}
= \frac{(|p|+\eta p_3)\overline{A_{22}}-\eta(p_1-\operatorname{i}p_2)\overline{A_{12}}}{|(|p|+\eta p_3)\overline{A_{22}}-\eta(p_1-\operatorname{i}p_2)\overline{A_{12}}|}
\end{equation}
(if the denominator $\ne 0$) with the Wigner rotation $H(\mathfrak{p},A):=H(\mathfrak{p})^{-1}A\,H(A^{-1}\cdot\mathfrak{p})$ with respect to the \textbf{helicity cross section}  $H(\mathfrak{p}):= B\big(\eta \,p \big)A_{\ln|p|}$,  cf. Eq.\,(\ref{CCSEF}) and recall $A_\rho=\operatorname{e}^{\frac{\rho}{2}\sigma_3}$.    (The canonical cross section is not defined for $\mathfrak{p}\in \mathcal{O}^{0,\eta}$). 
Since $H(\mathfrak{p})\cdot \big(\eta (e_0+e_3)\big)=\mathfrak{p}$,   one has $H(\mathfrak{p},A)\in E(2)$.\\

\hspace*{6mm}  
 As  expected, $\eta$  denotes the sign of the energy. Indeed,  the representation of $|H|^{-1}H$ by  $U^{0,s,\eta}$  equals $\eta I$. We are going to clarify the significance of $s$. By explicit computation one finds that for the representation $U^{0,s,\eta}$ the helicity operator $|P|^{-1}JP$ equals $\eta sI$.
 
\begin{Lem}\label{AUEMLR} $U^{0,s,-}$ is antiunitarily equivalent to    $U^{0,-s,+}$. More precisely, $U^{0,-s,-}=\mathcal{C}U^{0,s,+}
\mathcal{C}^{-1}$ for the antiunitary transformation $\mathcal{C}:L^2(\mathcal{O}^{0,+})\to L^2(\mathcal{O}^{0,-})$, $(\mathcal{C}\phi)(\mathfrak{p}):=\overline{\phi(-\mathfrak{p})}$. $\mathcal{C}$ is uniquely determined up to a constant phase.
\end{Lem} 
 
{\it Proof.} According to Eq.\,(\ref{DEWRMLC}), $k(-\mathfrak{p},A)=k(\mathfrak{p},A)$ holds for all $\mathfrak{p}\in \mathcal{O}^0$. Further recall that $U^{0,s,\eta}$ is irreducible. Then the assertion is easily verified.\qed\\
 
   We will denote  by $[0,s,\eta]$ any representation of  $\tilde{\mathcal{P}}$, which is unitarily equivalent to $U^{0,s,\eta}$. Hence in view of (\ref{AUEMLR}) one has the following consistent
 
 \begin{Int}\label{IRNE} $[0,s,+]$ represents a massless particle with \textbf{positive energy} and helicity $s\in \mathbb{Z}/2$, and 
$[0,-s,-]$ represents its antiparticle, which has \textbf{positive energy}, zero mass,  and helicity $s$, too.  
\end{Int}

For later use we  cast $U^{0,s,\eta }$ into a more convenient form. By the  isomorphism $X_\eta:L^2(\mathcal{O}^{0,\eta})\to L^2(\R^3)$, $(X_\eta \phi)(p):=(2|p|)^{-\frac{1}{2}}\phi(\eta |p|,p)$ one passes to the representation $X_\eta U^{0,s,\eta } X_\eta^{-1}$ on $L^2(\R^3)$. In the case of positive energy $\eta = +$ it has already the desired form
\begin{equation}\label{MSPERPG}
\big(W^{0,s,+}(t,b,A)\varphi\big)(p):=    (|q^+|/|p|)^\frac{1}{2}   
\operatorname{e}^{\operatorname{i}t|p|}   \operatorname{e}^{-\operatorname{i}bp}
k(\mathfrak{p}^+,A)^{2 s}
\,\varphi(q^+)
\end{equation}
In  the case of negative energy $\eta = -$ one carries out a further unitary transformation by $X_s:L^2(\R^3)\to L^2(\R^3)$, $(X_s\varphi)(p):=b(p)^{2s}\varphi(p)$. Then $X_s X_- U^{0,s,- } X_-^{-1}X_s^{-1}$ equals
\begin{equation}\label{MSNERPG}
\big(W^{0,s,-}(t,b,A)\varphi\big)(p):=    (|q^-|/|p|)^\frac{1}{2}   
\operatorname{e}^{-\operatorname{i}t|p|}   \operatorname{e}^{-\operatorname{i}bp}
\big(b(p)k(\mathfrak{p}^-,A)\overline{b(q^-)}\,\big)^{2 s}
\,\varphi(q^-)
\end{equation}
Recall  $b(p)=\frac{p_1+\operatorname{i}p_2}{|p_1+\operatorname{i}p_2 |}$ for $(p_1,p_2)\ne 0$, and 
$\mathfrak{p}^\eta= (\eta |p|,p)$ and $\mathfrak{q}^\eta= (\eta |q^\eta|,q^\eta)=A^{-1}\cdot \mathfrak{p}^\eta$. 

\begin{Lem}\label{SERLKG} Let $s\in\mathbb{Z}/2$. The representations 
\begin{equation}\label{SPERLKG}
\big(K^{s,+}(t,b,B)\varphi\big)(p):=      
\operatorname{e}^{\operatorname{i}t|p|}   \operatorname{e}^{-\operatorname{i}bp}
\kappa(p,B)^{2 s}
\,\varphi(B^{-1}\cdot p)
\end{equation}
and
\begin{equation}\label{SNERLKG}
\big(K^{s,-}(t,b,B)\varphi\big)(p):=      
\operatorname{e}^{-\operatorname{i}t|p|}   \operatorname{e}^{-\operatorname{i}bp}
\kappa(p,B)^{-2 s}
\,\varphi(B^{-1}\cdot p)
\end{equation}
on $L^2(\R^3)$ of the little kinematical group are extended by \emph{Eq.\,(\ref{MSPERPG})} and \emph{Eq.\,(\ref{MSNERPG})}, respectively, 
to representations of $\tilde{\mathcal{P}}$. Here $\kappa(p,B):=\big(B(p)^{-1} B\,B(B^{-1}\cdot p)\big)_{11}$ for $p\ne 0$.
\end{Lem}\\
{\it Proof.} Note that for $p\ne 0$  the Wigner rotation $B(p)^{-1} B\,B(B^{-1}\cdot p)\in SU(2)$ is diagonal as it leaves $(0,0,1)$ fixed. Since $A_{\ln|p|}$ is diagonal, too,  $H(\mathfrak{p}^+,B
) =B(p)^{-1} B\,B(B^{-1}\cdot p) =\operatorname{diag}\big(\kappa(p,B),\overline{\kappa(p,B)} \,\big)$ with $\kappa(p,B)=k(\mathfrak{p}^+,B)$. This already shows $W^{0,s,+}(t,b,B)=K^{s,+}(t,b,B)$.\\
\hspace*{6mm} For the negative energy case note first that $H(-\mathfrak{p})=H(\mathfrak{p})$ for all $\mathfrak{p}\in \mathcal{O}^0$. Hence $k(\mathfrak{p}^-,B)=\kappa(-p,B)$.  It remains to check
\begin{equation*} 
\kappa(-p,B)=\overline{b(p)}\,\overline{\kappa(p,B)}\,b(q),  \;q:=B^{-1}\cdot p\tag{1}
\end{equation*}
Indeed,  $\operatorname{diag}\big(\kappa(-p,B),\overline{\kappa(-p,B)}\,\big)= B(-p)^{-1} B\,B(-q) =C(p) B(p)^{-1} B\,B(q) C(q)^{-1}=C(p)   \operatorname{diag}\big(\kappa(p,B),\overline{\kappa(p,B)} \,\big)C(q)^{-1}=\operatorname{diag}\big(\overline{b(p)}\,\overline{\kappa(p,B)}\,b(q),b(p)\kappa(p,B)\,\overline{b(q)}\big) $ where 
\begin{equation*}
C(p):=B(-p)^{-1}B(p)=\left(\begin{array}{cc} 0&-\overline{b(p)} \\ b(p) & 0\end{array}\right),\;b(p)=\frac{p_1+ip_2}{|p_1+ip_2|}\tag{2}
\end{equation*}
\qed 

We will need the integral decomposition of the representations $K^{s,\eta}$ of the little kinematical group.

\begin{Lem}\label{DIRLKG}  There is the following set of mutually inequivalent irreducible representations of the little kinematical group: $U^{r,s,\eta}_{\R\otimes ISU(2)}$ on $L^2(S_r^2)$ with $S_r^2:=\{p\in \R^3:|p|=r\}$ for $r>0$, $s\in \mathbb{Z}/2$, and $\eta=\pm$, given by 
\begin{equation*}
\big(U^{r,s,\eta}_{\R\otimes ISU(2)}(t,b,B)\varphi\big)(p)=\operatorname{e}^{\operatorname{i}\eta t r}\operatorname{e}^{-\operatorname{i}bp} \kappa(p,B)^{2\eta s}\varphi(B^{-1}\cdot p)
\end{equation*}
Then the integral decomposition $K^{s,\eta}=\int_{\R^+}U^{r,s,\eta}_{\R\otimes ISU(2)} 4\pi r^2\operatorname{d}r$ holds.
\end{Lem}\\
{\it Proof.} The first part of the assertion follows from general representation theory for semi-direct products of locally compact groups. The second part is easily verified.\qed\\

Let us add a  result on covariance of the helicity cross section which we need in (\ref{PRMLC}).

\begin{Lem}\label{CRHCS}  For all $p\in\R^3\setminus\{0\}$ and $C=\operatorname{diag}(\gamma,\overline{\gamma})$ with $\gamma\in \C$, $|\gamma|=1$ one has
 $ B(C\cdot p)=CB(p)\,C^{-1}$. Also, $H(\mathfrak{p},C)=B(p)^{-1}CB(C^{-1}\cdot p)=C$ for all $\mathfrak{p}\in \mathcal{O}^0$, implying $k(\mathfrak{p},C)=\kappa(p,C)=\gamma$.
\end{Lem}

{\it Proof.} Explicit calculation yields 
$(C\cdot p)_1 =\frac{1}{2} (\gamma^2+\overline{\gamma}^2) p_1 +\frac{1}{2\operatorname{i}} (\gamma^2-\overline{\gamma}^2) p_2,\quad(C\cdot p)_2 =-\frac{1}{2\operatorname{i}} (\gamma^2-\overline{\gamma}^2) p_1 +\frac{1}{2} (\gamma^2+\overline{\gamma}^2) p_2,\quad(C\cdot p)_3=p_3$. Recall Eq.\,(\ref{CCSEF}). One finds $a_\pm(C\cdot p)=a_\pm(p)$ and $b(C\cdot p)=\overline{\gamma}^2b(p)$, whence the first part of assertion. --- Therefore $B(p)^{-1}CB(C^{-1}\cdot p)=C$ for all $p\ne 0$. Finally recall that $A_{\ln|p|}$ is diagonal and hence  commutes with $C$.  \qed\\

\subsection{Derivation of the massless SCT}\label{MLST} 
Regarding the notations see sec.\,\ref{IIMRPG}. Moreover let 
$$h^\chi(p):=\chi(\sigma_1p_1+\sigma_2p_2+\sigma_3p_3), \quad 
\pi^{\chi\eta}(p):=\frac{1}{2}\big(I_2+\frac{\eta}{|p|}h^\chi(p)\big)$$
 for $\chi\in\{+,-\}$. Particular attention requires the rotation $R_0(\mathfrak{p}, A)$ in 
 Eq.\,(\ref{RMLC}), which   replaces the Wigner rotation 
$R(\mathfrak{p},A)$ in Eq.\,(\ref{IMCS}) in the massive case.

\begin{Lem}\label{PRMLC} 
Let $\mathfrak{p}\in \mathcal{O}^0$ and $A\in SL(2,\C)$. Set $\mathfrak{q}:=A^{-1}\cdot \mathfrak{p}$. Choose $B,B'\in SU(2)$ and $\rho\in\R$ such that $A=B'A_\rho B$. Set
\begin{equation}\label{RMLC}
R_0(\mathfrak{p}, A):= B'B(B'^{-1}\cdot p)B(B\cdot q)^{-1} B
\end{equation}
Then $R_0$ is well-defined, i.e. the right hand side does not depend on the particular choice of $B',B,\rho$. Furthermore, \begin{itemize}
\item[$(\alpha)$] $R_0(\mathfrak{p}, A) \in SU(2)$
\item[$(\beta)$] $R_0(\mathfrak{p}, B)=B$
\item[$(\gamma)$] $R_0(\lambda\mathfrak{p}, A)=R_0(\mathfrak{p}, A)$ for all $\lambda>0$
\item[$(\delta)$] $R_0(\mathfrak{p}, A)=\lim_{m\to 0}R(\mathfrak{p}^{m}, A)$ with $\mathfrak{p}^{m}:=(\operatorname{sgn}(p_0) \sqrt{|p|^2+m^2},p)$, $m>0$
\item[$(\epsilon)$] $R_0(\mathfrak{p}, A)R_0(\mathfrak{q}, A')=R_0(\mathfrak{p}, AA')$ 
\end{itemize}
for all $\mathfrak{p}\in \mathcal{O}^0$, $A,A'\in SL(2,\C)$, and $B\in SU(2)$.
\end{Lem}\\
{\it Proof.} $(\alpha)$ is obvious. --- If $\rho=0$, i.e., $A=B'B\in SU(2)$, then $B(B\cdot q)^{-1} =B(B'^{-1}\cdot p)^{-1}$, whence  $R_0(\mathfrak{p}, A)=B'B=A$. So $(\beta)$ holds. \\
\hspace*{6mm} 
We examine the ambiguity of the  representation $A=B'A_\rho B$ for $A\in SL(2,\C)\setminus SU(2)$. Note that $A=(B'B)(B^{-1}A_\rho B)$ is the polar decomposition of $A$. This is unique. Therefore,  the eigenvalues $\operatorname{e}^{\rho/2}$ and $\operatorname{e}^{-\rho/2}$ of $B^{-1}A_\rho B$  determine   $\rho\ne 0$. Moreover, as the eigenvalues are not equal, $B$ is unique up to a diagonal matrix $C\in SU(2)$ multiplied from the left. But when $B$ is replaced with $CB$ then $B'$ has to be replaced by $B'C^{-1}$ in order to leave unchanged $B'B$. Conversely, $B'C^{-1} A_\rho CB$ still equals  $A$ as $A_\rho$ and $C$ commute. \\
\hspace*{6mm} Now we show that $R_0$ is well-defined. In view of $(\beta)$ assume at once $A\not\in SU(2)$. Consider the representation $A=B'C^{-1} A_\rho CB$. Then  the right hand side of 
Eq.\,(\ref{RMLC}) reads $B' C^{-1}B(C\cdot B'^{-1}\cdot p)B(C\cdot B\cdot q)^{-1} C B$, which actually does not depend on $C$ by (\ref{CRHCS}).\\
\hspace*{6mm} $(\gamma)$ holds as $B(\lambda p)=B(p)$ for $\lambda>0$, $p\in\R^3\setminus\{0\}$.\\
\hspace*{6mm} $(\delta)$ By definition $R_0(\mathfrak{p}, A_\rho)= B(p)B(q)^{-1}$ for $A=A_\rho$.  Recall $\alpha= \cosh\rho$, $\beta=\sinh \rho$ and  $\gamma=\cosh\frac{\rho}{2}$, $\delta=\sinh\frac{\rho}{2}$.  
Explicit computation using Eq.\,(\ref{CCSEF}) yields
\begin{equation}\label{RDBM0}
R_0(\mathfrak{p},A_\rho)=\big(|p|(\alpha|p|-\beta\eta p_3)\big)^{-1/2}\left( \begin{array}{cc} \gamma  |p| -\delta \eta p_3 & \delta \eta (p_1-\operatorname{i}p_2)     \\ -\delta \eta (p_1+\operatorname{i}p_2)     &\gamma |p| -\delta \eta p_3 \end{array} \right)
\end{equation}
which shows $(\delta)$ for $A=A_\rho$ by Eq.\,(\ref{WRDB}). As to the general case $A=B'A_\rho B$ note that $R(\mathfrak{p}^m, A)=B' R(B'^{-1}\cdot\mathfrak{p}^m, A_\rho)B$ by the analogous properties $(\beta)$ and $(\epsilon)$ for $R$, which are easy to verify for $R$ (in place of $R_0$).
Hence $R(\mathfrak{p}^m, A)\to B'R_0(B'^{-1}\cdot\mathfrak{p}^m, A_\rho)B= B'B(B'^{-1}\cdot p)B(B\cdot q)B=R_0(\mathfrak{p},A)$ by definition.\\
\hspace*{6mm} $(\epsilon)$ follows immediately from $(\delta)$, as $R$ satisfies the analogous property. \qed\\

It is interesting  that the limit  in (\ref{PRMLC})$(\delta)$ exists whereas   the limit of  $Q(\mathfrak{p}^m)$ for $m\to 0$ does not  if $p\ne 0$. Moreover one notes  $R_0(\mathfrak{p}, A)\cdot q =\frac{|q|}{|p|}p$  as an easy consequence of $|p|B(p)\cdot e_3=p$ for all $p\in\R^3\setminus\{0\}$.

\begin{The} \label{RCL}
For every  massless  WL with causal time evolution    there exists a representation of $\tilde{\mathcal{P}}$ extending the representation of the little kinematical group. \\
\hspace*{6mm} Let $\chi\in\{+,-\}$. For $(t,b,A) \in \tilde{\mathcal{P}}$ and $\varphi\in L^2(\R^3,\C^{2n})$  put 
\begin{equation*}
\big(W_n^{\chi \,mom}(t,b,A)\varphi\big)(p) :=\operatorname{e}^{-\operatorname{i}bp}
\sum_{\eta =\pm 1} \left( \operatorname{e}^{\operatorname{i}t \eta |p|} \pi^{\chi\eta}(p)s^\chi(A)^{*-1} \otimes D^{(\frac{n}{2}-\frac{1}{2})} \big(R_0(\mathfrak{p}^\eta,A) \big) \right)    \varphi(q^\eta)
\end{equation*}
with $s^+(A):=A$, $s^-(A):=A^{*-1}$. Then $W_n^\chi$ is a representation  of  $\tilde{\mathcal{P}}$, which extends $(V_n^\chi,U_n)$. 
\end{The}

{\it Proof.} Due to (\ref{CLC}) we may turn at once to the Weyl tensor-localization $(V_n^\chi,U_n,E_n)$. Obviously $W_n^{\chi \,mom}(B)=U_n^{mom}(B)$ for $B\in SU(2)$, and  $\operatorname{e}^{\operatorname{i} t h^\chi(p)}\pi^{\chi \eta}(p)=\operatorname{e}^{\operatorname{i}t\eta |p|}\pi^{\chi \eta}(p)$ holds. Hence  $W_n^\chi$ is an extension of $(V_n^\chi,U_n)$. Then, using the relation
\begin{equation}\label{VSPI}
\pi^{\chi \eta}(p) s^\chi(A)^{*-1}=(|q^\eta|/|p|) \,s^\chi(A) \pi^{\chi\eta}(q^\eta)
\end{equation}
which is mainly a consequence of the fundamental relation $A(\sum_{\mu=0}^3\mathfrak{p}_\mu\sigma_\mu)A^*=\sum_{\mu=0}^3(A\cdot\mathfrak{p})_\mu\sigma_\mu$ with $\sigma_0:=I_2$ for $A\in SL(2,\C)$, $\mathfrak{p}\in\R^4$,
and by means of  the property (\ref{PRMLC})$(\epsilon)$, one verifies the group multiplication law $W_n^{\chi \,mom}(\mathfrak{a}+A\cdot \mathfrak{a}', AA')=W_n^{\chi \,mom}(\mathfrak{a},A)W_n^{\chi \,mom}(\mathfrak{a}',A')$. Finally one checks unitarity of  $W_n^{\chi \,mom}(A)$.\qed

\begin{Def}\label{CHWS} $(W^{\textsc{w}\chi}, E^{\textsc{w}}):=(W_1^\chi,E_1)$, $\chi\in\{+,-\}$, is called the $\chi$-handed \textbf{Weyl system}.
\end{Def}

The representation $W_n^{\chi}$  of $\tilde{\mathcal{P}}$ in (\ref{RCL}) can be  constructed as follows giving an insight into its composition. By the unitary transformation $(Y_0\varphi)(p):= D(B(p)^{-1})\, \varphi(p)$ with $D:=D^{(\frac{1}{2})}   \otimes D^{(\frac{n}{2}-\frac{1}{2})}$ we go to the energy representation $Y_0(V_n^{\chi mom},U_n^{mom})Y_0^{-1}$ of  $(V_n^\chi ,U_n)$. It reads
\begin{equation}\label{MLERLKG}
\big(V_n^{\chi erg}(t)U_n^{erg}(b,B)\varphi\big)(p) =\big(\operatorname{e}^{\operatorname{i} t\chi |p|\sigma_3}\otimes I_n\big)\operatorname{e}^{-\operatorname{i} bp}D\left(\begin{array}{cc} \kappa(p,B) & 0\\  0 &\overline{\kappa(p,B)}\end{array} \right)\varphi(B^{-1}\cdot p)
\end{equation}
Hence $(V_n^{\chi erg}, U_n^{erg}) \simeq \big(\bigoplus_{i=0}^{n-1} K^{\chi(\frac{n}{2}-i),\chi}\big)   \oplus \big( \bigoplus_{i=1}^nK^{-\chi(\frac{n}{2}-i),-\chi}\big)$. Each $K^{s,\eta}$ is extended to $\tilde{\mathcal{P}}$ by $W^{0,s,\eta}$ according to (\ref{SERLKG}). Thus one gets an extension   to $\tilde{\mathcal{P}}$ of $(V_n^{\chi erg}, U_n^{erg})$ by
\begin{equation}\label{DWTL} 
W_n^{\chi erg}\simeq \big(\bigoplus_{i=0}^{n-1} W^{0,\chi(\frac{n}{2}-i),\chi}\big)   \oplus \big( \bigoplus_{i=1}^nW^{0,-\chi(\frac{n}{2}-i),-\chi}\big)
\end{equation}
 It satisfies 
$\big(W_n^{\chi erg}(A_\rho)\varphi\big)(p)=\big(\sqrt{|q^\chi|/|p|}\,\varphi_1(q^\chi), \sqrt{|q^{-\chi}|/|p|}\,\varphi_2(q^{-\chi})\big)$
as $k(\mathfrak{p}, A_\rho)=1$ for all $\mathfrak{p}\in \mathcal{O}^0$ by Eq.\,(\ref{DEWRMLC}) and $b(p)=b(q^-)$ for $A=A_\rho$. By some computations one goes back to the momentum representation obtaining $W_n^{\chi \,mom}(A_\rho)$.
 For general $A=B'A_\rho B$ with $B',B\in SU(2)$  one obtains $W_n^{\chi \,mom}(A)$ in (\ref{RCL})  from $W_n^{\chi \,mom}(A)= U_n^{mom}(B') W_n^{\chi \,mom}(A_\rho)U_n^{mom}(B)$. \\

\begin{Lem}\label{ERLKG} Let  $(V,U)$ be a representation of the little kinematical group with  $C=0$ and finite helicity spectrum. Then all representations $W$ of  
 $\tilde{\mathcal{P}}$ extending $(V,U)$ are unitarily equivalent.
\end{Lem}\\
{\it Proof.} This is an immediate consequence of the representation theory of  $\tilde{\mathcal{P}}$ because the mass-squared operator $C$,  the sign of the energy operator   $\operatorname{sgn}(H)$, and the helicity operator  $|P|^{-1}JP$ are already determined by $(V,U)$.\qed

\begin{The}\label{RCLX} Let $(V,U,E)$ be a massless  WL with causal time evolution. Let $X$ be the associated causal position operator (recall \emph{Eq.\,(\ref{ACPO})}). Then there is a unique representation $W_E$ of $\tilde{\mathcal{P}}$ which extends $(V,U)$ such that its booster is given by
\begin{equation}\label{NCPO}
N=\frac{1}{2}\{X,H\} + \frac{\operatorname{sgn}(H)}{|H|} P \times \left(J- X\times P -\frac{\operatorname{i}}{4} [X,H] \times  [X,H] \right)
\end{equation}
The representation $W_n^\chi$ from \emph{(\ref{RCL})} satisfies this relation. Any  extension $W$ of $(V,U)$ to $\tilde{\mathcal{P}}$ is unitarily equivalent to $W_E$.
\end{The}

{\it Proof.} Uniqueness of $W_E$ is obvious  as the representation of all generators of the Poincar\'e group is  set. Due to   (\ref{CLC})  it suffices to prove the existence of $W_E$ for $(V_n^\chi,U_n)$ of the Weyl tensor-localization. It is not hard to show that the booster of $W_n^\chi$ from (\ref{RCL}) satisfies the formula for $N$. In particular check $\frac{d}{dt}R_0(p^\eta,\operatorname{e}^{\frac{t}{2}\sigma_k})|_{t=0}=\frac{\operatorname{i} \eta}{|p|}\frac{1}{2}(p_l\sigma_m-p_m\sigma_l)$ with $(k,l,m)$ a cyclic permutation of $(1,2,3)$ and verify that
$I_2\otimes \frac{1}{2}\mathfrak{d}^{(\frac{n}{2}-\frac{1}{2})}(\sigma_1,\sigma_2,\sigma_3)$ 
is given by  $ J- X\times P -\frac{\operatorname{i}}{4} [X,H] \times  [X,H]$.
The last part of the assertion holds true by  (\ref{ERLKG}). \qed\\

The next result, which follows from  (\ref{RCLX}), shows how a general massless SCT is  built of the systems $(W_n^\chi,E_n)$. Recall (\ref{SCTER}).

\begin{The}\label{CMLST} Every massless SCT $(W,E)$ is unitarily related to $(W_E,E)$. The latter is unitarily equivalent to a finite orthogonal sum of the systems $(W_n^\chi,E_n)$, $\chi\in\{+,-\}$, $n\in\N$  with uniquely determined multiplicities.
\end{The}

As to (\ref{CMLST}) note that a massless SCT $(W,E)$  in general is not unitarily equivalent to $(W_E,E)$. The example (\ref{ISCTURRSCT}) shows that there are massless SCT $(W,E)$ which are irreducible although $(W_E,E)$ is not. In view of the result (\ref{FDWTL}) we study the following special case.

\begin{Lem}\label{ISTRMWI} Let 
$\chi\in\{+,-\}$ and  let  $(W,E)$ be a massless SCT unitarily related to a finite multiple of $(W^{\textsc{w}\chi}, E^{\textsc{w}})$.
Let $Q$ be an orthogonal projection commuting with $(W,E)$. Then there is a unitary transformation $S$ such that $W=SW_ES^{-1}$ and which commutes with $Q$.
\end{Lem}\\
{\it Proof.} By assumption there are unitary $S$ and $U$ such that  $W=SW_ES^{-1}$ and $(W_E,E)=U\,\big(I_\nu\otimes (W^\chi_1,E_1)\big)\,U^{-1}$ with multiplicity $\nu\in\N$ and  $I_\nu$ the identity on $\C^\nu$. Therefore it is no  restriction assuming   $W=S\,(I_\nu \otimes W^\chi_1) \,S^{-1}$ and $E=I_\nu\otimes E_1$. Recall that $S$ commutes with $W|_\mathcal{K}=(I_\nu\otimes W^\chi_1)|_\mathcal{K}$, where $\mathcal{K}$ denotes the little kinematical group $\R\otimes ISU(2)$.
\\
\hspace*{6mm} 
  As $Q$ commutes with  $(W,E)$, it commutes with its restriction to the Euclidean group $ISU(2)$,  which is $I_\nu\otimes (U_1,E_1)$. Now, $(U_1,E_1)$ is the system of imprimitivities induced by $D=D^{\frac{1}{2}}$ on $SU(2)$ and therefore  irreducible. Hence $Q=Q_\nu\otimes I$ with $Q_\nu$ an orthogonal projection matrix on $\C^\nu$ and $I$ the identity on $L^2(\R^3,\C^2)$.\\
\hspace*{6mm} 
Recall that $Q$ commutes with $W$. Hence $S^{-1}QS$ commutes with $I_\nu\otimes W^\chi_1$. By Eq.\,(\ref{DWTL}), $I_\nu\otimes W_1^{\chi erg} =(I_\nu\otimes W^{0,\chi/2,\chi})\oplus (I_\nu\otimes W^{0,\chi/2,-\chi})$. Recall that $W^{0,s,\eta}$ is $[0,s,\eta]$ and, hence, irreducible and mutually inequivalent. Consequently one has $S^{ erg -1}QS^{erg}= (Q_\nu^\chi\otimes I') \oplus  (Q_\nu^{-\chi}\otimes I')=(Q_\nu^\chi \oplus  Q_\nu^{-\chi})\otimes I'$ with orthogonal projection matrices $Q_\nu^{\chi}$, $Q_\nu^{-\chi}$ on $\C^\nu$ and $I'$ the identity on $L^2(\R^3,\C)$.\\
\hspace*{6mm} 
Similarly, as $S^{erg}$ commutes with $I_\nu\otimes W_1^{\chi erg} |_\mathcal{K}=(I_\nu\otimes K^{\chi/2,\chi}) \oplus (I_\nu\otimes K^{\chi/2,-\chi})$, one infers from (\ref{DIRLKG}) that $(S^{erg}\varphi)(p)=s(|p|)\varphi(p)$
with unitary matrices $s(r)=s^\chi(r)\oplus s^{-\chi}(r)$ on $\C^\nu \oplus \C^\nu$ for $r\ge 0$.\\
\hspace*{6mm} 
Therefore $s^\chi(r)^{-1}Q_\nu\, s^\chi(r)=Q^\chi_\nu$  for almost all $r\ge 0$, and also for $-\chi$ in place of $\chi$. Hence there is $r_0\ge 0$ such that $s(r)s(r_0)^{-1}$ commutes with $Q_\nu^\chi \oplus  Q_\nu^{-\chi}$  for almost all $r\ge 0$. 
Then $(S^{erg}_Q\varphi)(p):=s(|p|)s(r_0)^{-1}\varphi(p)$, $(S^{erg}_0\varphi)(p):=s(r_0)\varphi(p)$ define unitary transformations $S_Q$, $S_0$ such that $S=S_QS_0$. Moreover, $S_0$ commutes with $I_\nu\otimes W^\chi_1$, whence $W=S(I_\nu\otimes W^\chi_1)S^{-1}= S_Q(I_\nu\otimes W^\chi_1)S_Q^{-1}$. Finally, $S_Q$ commutes with $Q$, thus concluding the proof.\qed 

\begin{Exam}\label{ISCTURRSCT} for an \textbf{irreducible} massless SCT $(W,E)$ unitarily related to  $2\,(W^{\textsc{w}\chi}, E^{\textsc{w}})$.
 For its construction we refer to   the proof of (\ref{ISTRMWI}). 
Let $s^\pm(r):= I_2 \cos r  \pm \operatorname{i}\sigma_2 \sin r$ for $r\ge 0$, which depends continuously on $r$.   Choose $r_0= 0$.
 Then  $0,I_2$ are the only   projections invariant under $\{s^\pm(r)s^\pm(0)^{-1}:r\ge 0\}$. Hence $(W,E)$ is irreducible by (\ref{ISTRMWI}).\\
\hspace*{6mm} 
For  the momentum representation $S^{mom}=Y_h^{-1}S^{erg}Y_h$ of $S$ one gets $(S^{mom}\varphi)(p)=s(p)\varphi(p)$ with $s(p):=\exp(\operatorname{i}\sigma_2\otimes \sum_{k=1}^3\sigma_kp_k)$.
\end{Exam}

\subsection{Determination of the massless causal systems}\label{MLCS}  It does not suffice to examine   the $(W^\chi_n,E_n)$ for  $n\in\N$, $\chi\in\{+,-\}$, but all irreducible massless SCT have to be considered which as in the above example are unitarily related to finite orthogonal sums of the former.

\begin{The}\label{FDWTL} Let $(W,E)$ be an irreducible  massless SCT, which  localizes frame-independently. Then $(W,E)$ is unitarily related to a finite multiple of $(W^{\textsc{w}\chi},E^{\textsc{w}})$ for some $\chi\in\{+,-\}$. In particular,   $(W_n^\chi,E_n)$, $\chi\in\{+,-\}$,  $n>1$ do not localize frame-independently. 
\end{The}\\
{\it Proof.} Recall $A_\rho=\operatorname{e}^{\frac{\rho}{2}\sigma_3}$. 
By (\ref{CMLST}) and  (\ref{RCL}), up to unitary equivalence, the momentum representation of
$W(A_\rho)$  reads
\begin{equation*}
\big(W^{mom}(A_\rho)\varphi\big)(p) =s(p)^{-1}
\sum_{\eta =\pm 1} w^\eta(p) s(q^\eta)    \varphi(q^\eta) \tag{$1$}
\end{equation*}
for almost all $p\in\R^3$. Here, $w^\eta(p):=\bigoplus_\chi \pi^{\chi\eta}(p)s^\chi(A_\rho)^{*-1} \otimes D^\chi \big(R_0(\mathfrak{p}^\eta,A_\rho) \big)$ with some finite dimensional representation $D^\chi$ of $ISU(2)$, and, according to (\ref{SCTER}), there  is  some unitary operator $S$ commuting with $W|_{\R\times ISU(2)}$ so that $(S^{mom}\varphi)(p)=s(p)\varphi(p)$ with unitary matrices $s(p)$ acting on spinor space. 
\\ 
\hspace*{6mm}
Now let $\varphi$ be localized in a bounded region. By assumption so is $\varphi_\rho:=W^{mom}(A_\rho)\varphi$. Hence, by the Paley-Wiener Theorem,  these functions are the restriction of entire functions on $\C^3$, still denoted by $\varphi$ and $\varphi_\rho$. We need to exploit only their continuity, in particular at $p=0$.\\
\hspace*{6mm}
Since $S$ is rotation-invariant, one has $s(p)=D(B(p))s(|p|e_3)D(B(p))^{-1}$,  where $D:=(D^{(\frac{1}{2})}\otimes D^+) \oplus (D^{(\frac{1}{2})}\otimes D^-)$. Let $I^\chi:=D^\chi(I_2)$. As $S$ is also time translation-invariant, $s(|p|e_3)$  commutes with $(\operatorname{e}^{\operatorname{i}t|p|\sigma_3} \otimes I^+)\oplus  (\operatorname{e}^{-\operatorname{i}t|p|\sigma_3} \otimes I^-)$ for all $t$ (see the energy representation Eq.\,(\ref{MLERLKG})). Hence, equivalently, $s(re_3)$, $r>0$ commutes with $P^\eta$, where $P^+:=(\big(\tiny{\begin{array}{cc}1& 0\\ 0 & 0 \end{array}}\big)\otimes I^+)\oplus (\big(\tiny{\begin{array}{cc}0& 0\\ 0 & 1 \end{array}}\big)\otimes I^-)$ and $P^-:=I-P^+$.\\
\hspace*{6mm}
One has $D(B(p))^{-1} w^\eta(p)D(B(q^\eta))=\sqrt{|q^\eta|/|p|} \,P^\eta$ (see the energy representation of $W^{\chi }_n(A_\rho)$ after Eq.\,(\ref{DWTL})). Let $t^\eta(p):=s(|p|e_3)^{-1}s(|q^\eta|e_3)\sqrt{|q^\eta|/|p|} \,P^\eta$ for $p\ne 0$. Clearly $t^\eta$ commutes with $P^\pm$ and  $t^\eta(p)=t^\eta(p')$  if $|p|=|p'|$ and $|q^\eta|=|q'^\eta|$. By $(1)$,
\begin{equation*}
\varphi_\rho(p)=\sum_{\eta=\pm 1}D(B(p)) t^\eta(p) D(B(q^\eta))^{-1}   \varphi(q^\eta) \tag{$2$}
\end{equation*}
holds for almost all $p\in\R^3$. 

\hspace*{6mm}
(a) First  we will   redefine  $t^\eta$ on some null set such that it becomes  continuous on $\R^3\setminus Z$ with $Z:= \{0\}\times \{0\}\times  \R$.\\
\hspace*{6mm}
 Fix $p_*\in\R^3$. One has $\varphi(p)=(2\pi)^{-3/2}\int \operatorname{e}^{-\operatorname{i}px}\psi(x)\operatorname{d}^3 x$ for  $\psi:=\mathcal{F}^{-1}\varphi$  with compact support. Hence, if necessary
it suffices to alter $\psi$ a bit on its support in order to achieve  $\varphi(q^\eta_*)\ne 0$ for $\eta=\pm$. Choose   matrices $u_i$ acting on the spinor space such that the squared matrix 
$\phi(p):=(u_1\varphi(p),u_2\varphi(p),\dots,u_n\varphi(p))$,  where $n$ is the dimension of the spinor space, is invertible at $p=q_*^+$ and $p=q_*^-$. By continuity there is a neighborhood $N(p_*)$ of $p_*$ such that $\phi(q^\eta)$ is still invertible for all $p\in N(p_*)$, $\eta=\pm$.\\
\hspace*{6mm}
Clearly, the functions $u_i\psi(x)$ have compact support with $u_i\varphi(p)$ their Fourier transforms. Define the matrix $\phi_\rho(p)$  replacing $\varphi$ with $\varphi_\rho$ in   $\phi(p)$. Then $(2)$ holds for $\phi_\rho(p)$ and $\phi(q^\eta)$ in place of $\varphi_\rho(p)$ and $\varphi(q^\eta)$, respectively. Hence, as $t^\eta(p)$ commutes with $P^+$ and $P^-$, one finds
\begin{equation*}
t^\eta(p)=P^\eta D(B(p))^{-1}\,\phi_\rho(p)\phi(q^\eta)^{-1}D(B(q^\eta)) \tag{$3$}
\end{equation*}
for almost all $p\in N(p_*)$.\\
\hspace*{6mm}
Assume $p_*\not\in Z$. Then the right hand side of $(3)$ is continuous on $N(p_*)\setminus Z$ because of Eq.\,(\ref{CCSEF}) and since $q^\eta_1=p_1$, $q^\eta_2=p_2$. Hence, redefining $t^\eta$ by $(3)$, one gets $t^\eta$ continuous on $\R^3\setminus Z$ and $t^\eta(p)= s(|p|e_3)^{-1}s(|q^\eta|e_3)\sqrt{|q^\eta|/|p|} \,P^\eta$ for all $p\in \R^3\setminus L$ for some Lebesgue null set  $L\subset \R^3$.

\hspace*{6mm}
(b) Let $p,p'\in \R^3\setminus Z$ and $\eta=\pm$. Clearly,  $t^\eta(p)$ still commutes with $P^+$ and $P^-$.  We show that  $t^\eta(p)=t^\eta(p')$ still holds if $|p|=|p'|$ and $|q^\eta|=|q'^\eta|$ or,  equivalently,  if $|p|=|p'|$ and $p_3=p'_3$, since $q^\eta=(p_1,p_2,\alpha p_3-\beta \eta |p|)$ and hence $|q^\eta|=\alpha |p|-\beta \eta p_3$. \\
\hspace*{6mm}
Indeed, assume the contrary. Then there are $p_*,p_*'\in \R^3\setminus Z$ with $|p_*|=|p_*'|$, $p_{*3}=p'_{*3}$, and  $t^\eta(p_*)\ne t^\eta(p_*')$. By continuity $t^\eta(p)\ne t^\eta(p')$ holds true for all $p\in N$ and $p'\in N'$, where $N$ and $N'$ are some neighborhoods in $\R^3\setminus Z$ of $p_*$ and  $p'_*$, respectively. Using cylindrical coordinates $z,r,\varphi$ for $p\in\R^3$, there is an open disc $Q\subset \R\times ]0,\infty[$ around $(z_*, r_*)$ and open intervals $I$ and $I'$ around $\varphi_*$ and  $\varphi'_*$, respectively, such that $Q\times I\subset N$ and $Q\times I'\subset N'$. By Tonelli's theorem there is $\varphi\in I$ and a  Lebesgue null set $M\subset Q$ with $(Q\setminus M)\times \{\varphi\}\subset N\setminus L$. $\varphi'\in I'$ and $M'\subset Q$ are determined analogously. Hence there is  $q\in Q\setminus (M\cup M')$.
Then $p$ and $p'$ given by $(q,\varphi)$ and $(q,\varphi')$, respectively, yield the contradiction $t^\eta(p) = t^\eta(p')$.

\hspace*{6mm}
(c) Consider $p\in \R^3\setminus Z$. Then $\lambda p\in N(0)\setminus Z$ for  all small $\lambda>0$. So, by $(3)$, $t^\eta(\lambda p)=P^\eta D(B(\lambda p))^{-1}\phi_\rho(\lambda p)\phi(\lambda q^\eta)\,D(B(\lambda q^\eta))
\to T^\eta(p)$ for $\lambda \to 0$ with 
\begin{equation*}
T^\eta(p):=P^\eta D(B( p))^{-1}X\,D(B( q^\eta))\tag{$4$}
\end{equation*} 
and $X:=\phi_\rho(0)\phi(0)^{-1}$ independent of $p$. It follows from (b) that $T^\eta(p)$ commutes with $P^+$ and $P^-$, and that  $T^\eta(p')=T^\eta(p)$ if $|p'|=|p|$, $p'_3=p_3$. We will infer from (4) and   these properties  that $D^+$ and $D^-$ are multiples of $D^{(0)}$. 

\hspace*{6mm}
(d) Write $X$ as a $2\times 2$-block matrix according to the orthogonal sum  of $D$ with entries $X_{\chi \chi'}$, $\chi,\chi'=\pm$. Then 
$T^+(p)_{++}=\left(\big(\tiny{\begin{array}{cc}1& 0\\ 0 & 0 \end{array}}\big)B(p)^{-1}\otimes D^+(B(p))^{-1}\right)\,X_{++}\,\big(B(q^+)\otimes D^+(B(q^+))\big)$. According to the tensor structure of the factors write $X_{++}$ as a $2\times 2$-block matrix with entries $U_{\varsigma \varsigma'}$, $\varsigma,\varsigma'=\pm$. (The indices refer to the helicity values $s=\varsigma\frac{1}{2}$.) So,  $T^+(p)=T^+(p)P^+$ implies $T^+(p)_{++,+-}=D^+(B(p))^{-1} \big((a_+(p)U_{++}+\overline{b}a_-(p)U_{-+})(-\overline{b})a_-(q^+)+(a_+(p)U_{+-}+\overline{b}a_-(p)U_{--})a_+(q^+)\big)D(B(q^+))=0$ using Eq.\,(\ref{CCSEF}). Equivalently, $(|p|^2-p^2_3)C=(-p_1+\operatorname{i}p_2)(|p|-p_3)A+(p_1+\operatorname{i}p_2)(|p|+p_3)B$
with $A:=\operatorname{e}^{\rho/2}U_{-+}$, $B:=\operatorname{e}^{-\rho/2}U_{+-}$, and $C:=\operatorname{e}^{\rho/2}U_{++}-\operatorname{e}^{-\rho/2}U_{--}$. As $A,B,C$ are independent of $p$, and since the equation holds for all $p\in \R^3\setminus Z$, A=B=C=0 follows. Hence $U_{+-}=U_{-+}=0$ and  $U_{--}= \operatorname{e}^{\rho} U_{++}$. Let $U$ denote the constant matrix $\operatorname{e}^{\rho/2}U_{++}$. Then a short computation yields
\begin{equation*}
T^+(p)_{++}=\sqrt{|q^+|/|p|}\left(\begin{array}{cc}D^+(B( p))^{-1}UD^+(B( q^+)) & 0\\ 0 & 0 \end{array}\right) \tag{$5$}
\end{equation*} 
Similarly one computes $T^+(p)_{+-}=0$, $T^+(p)_{-+}=0$ and
\begin{equation*}
T^+(p)_{--}=\sqrt{|q^+|/|p|}\left(\begin{array}{cc} 0 & 0 \\ 0 & D^-(B( p))^{-1}VD^-(B( q^+))  \end{array}\right) \tag{$6$}
\end{equation*} 
for some constant matrix $V$. From the definition of $t^+(p)$ it follows that $U$ and $V$ are unitary. Analogous formulae hold for $T^-(p)$.

\hspace*{6mm}
(e) Next we exploit  that $T^\eta(p)=T^\eta(p')$ if  $p,p'\in \R^3\setminus Z$ with $|p|=|p'|$, $p_3=p'_3$. It implies by $(5)$ that the matrices $D^+\big(B(q'^+)B(q^+)^{-1}\big)$ and $D^+\big(B(p')B(p)^{-1}\big)$ are unitarily equivalent by means of $U$. Hence they have the same spectrum.\\
\hspace*{6mm}
Choose $p,p'$ such that $b(p) =b(q^+)=1$  (i.e., $p_2=0$) and $b(p')=b(q'^+)=\operatorname{e}^{\operatorname{i}\varphi}$ for $\varphi\in\R\setminus \{0\}$. Then the trace of $B(p')B(p)^{-1}$ is $2-2a_-(p)^2(1-\cos \varphi)$. Hence its eigenvalues are $\lambda(p), \overline{\lambda(p)}$ with $|\lambda(p)|=1$ and $\Re(\lambda(p))=1-a_-(p)^2(1-\cos \varphi)$. Therefore $B(p')B(p)^{-1}$ is unitarily equivalent to $\operatorname{diag}(\lambda(p),\overline{\lambda(p)})$ by some $B\in SU(2)$. Consequently, $D^+\big(B(p')B(p)^{-1}\big)$ 
is unitarily equivalent to $D^+\big(\operatorname{diag}(\lambda(p),\overline{\lambda(p)})\big)$ by $D^+(B)$. As $D^{(j)}\big(\operatorname{diag}(\lambda(p),\overline{\lambda(p)})\big)=\operatorname{diag}(\lambda(p)^{2j}, \lambda(p)^{2j-2},\dots,\lambda(p)^{-2j})$ for $j\in\N_0/2$, the spectrum $\Sigma(p)$ of $D^+\big(B(p')B(p)^{-1}\big)$ equals
$\{\lambda(p)^{2m}:m=-j_f, -j_f+1,\dots,j_f\}$ for some half-integer $j_f$ or equals $\{\lambda(p)^{2m}:m=-j_b, -j_b+1,\dots,j_b\}$ for some integer $j_b$ or equals the union of  these sets. Replacing $p$ by $q^+$ one obtains the spectrum $\Sigma(q^+)$ of $D^+\big(B(q'^+)B(q^+)^{-1}\big)$.

\hspace*{6mm}
(f) Now we claim that $D^+$ is a multiple of $D^{(0)}$. Check that $\lambda(p)\ne 1$, $\lambda(q^+)\ne 1$ and $\Re(\lambda(p))\ne \Re(\lambda(q^+))$ as $\varphi\ne 0$. Hence $\lambda(p)\ne \lambda(q^+)$ and $\lambda(p)\ne \overline{\lambda(q^+)}$. Choose $\varphi\in\R$ such that $\lambda(p)$ is not a root of unity. This is possible as $a_-(p)\ne 0$.\\  
\hspace*{6mm}
For the proof  assume the contrary. As shown above,  $\Sigma(p)$ and $\Sigma(q^+)$ coincide.  Hence, if $\lambda(p) \in \Sigma(p)$, then 
there are $a, b\in\mathbb{Z}\setminus\{0,-1\}$ such that $\lambda(p)^{2a+1}=\lambda(q^+)$ and $\lambda(q^+)^{2b+1}=\lambda(p)$, and the contradiction $\lambda(p)^{(2a+1)(2b+1)-1}=1$ follows. If 
$\lambda(p)^2 \in \Sigma(p)$, then there are $a, b\in\mathbb{Z}\setminus\{-1,1\}$ such that $\lambda(p)^{2a}=\lambda(q^+)^2$ and $\lambda(q^+)^{2b}=\lambda(p)^2$, and the contradiction $\lambda(p)^{2ab-2}=1$ follows. This proves the claim.\\
\hspace*{6mm}
Analogously one shows that $D^-$ is a multiple of $D^{(0)}$.

\hspace*{6mm} 
(g) By the result  achieved so far, $W=\nu^+W^+_1 \oplus \nu^-W^-_1$ with multiplicities $\nu^\chi\in\N_0$, $\chi=\pm$. It remains to show that $S$ decomposes accordingly. Let $\mathcal{K}$ denote the little kinematical group $\R\otimes ISU(2)$. As $S$ commutes with $W|_{\mathcal{K}}$, it suffices to show that $W^+_1|_\mathcal{K}$ does not contain an irreducible subrepresentation which is equivalent to a subrepresentation of  $W^-_1|_\mathcal{K}$. Hence we need the decomposition of $W^\chi_1|_\mathcal{K}$. By Eqs.\,(\ref{DWTL}), (\ref{MLERLKG}), $W^\chi_1|_\mathcal{K}\simeq K^{\chi/2,\chi} \oplus K^{\chi/2,-\chi} $ holds, whence the assertion by (\ref{DIRLKG}).\qed\\

This result (\ref{FDWTL}), due to  (\ref{FIDCWL}), definitely proves that the higher Weyl tensor-systems $(W^\chi_n,E_n)$, $\chi\in\{+,-\}$, $n>1$ are mathematical artefacts, interim results  in determining the massless causal systems.

In sec.\,\ref{FIWL} it is shown that the Weyl systems $(W^{\textsc{w}\chi},E^{\textsc{w}})$, $\chi\in\{+,-\}$ are causal.  The following example concerns an irreducible massless SCT unitarily related to $(W^{\textsc{w}\chi},E^{\textsc{w}})$, which localizes frame-independently without being causal.

\begin{Exa}\label{ESTFINC} Let $W:=SW^{\textsc{w}\chi}S^{-1}$ with $S:=W^{\textsc{w}\chi}(r)$ for $r\equiv (r,0,I_2)$ with $r\in\R\setminus\{0\}$, and $E:=E^\textsc{w}$. Then obviously $(W,E)$ is an irreducible  massless SCT unitarily related to 
$(W^{\textsc{w}\chi},E^\textsc{w})$. One has $W(h)E(\Gamma)W(h)^{-1}=W^{\textsc{w}\chi}(rhr^{-1})E^\textsc{w}(\Gamma)W^{\textsc{w}\chi}(rhr^{-1})^{-1}$ for all $h\in\tilde{\mathcal{P}}$ and $\Gamma\in\mathcal{S}$ , whence
\begin{equation*}  
W(h)E(\Gamma)W(h)^{-1}\le E(\Gamma_{rhr^{-1}})\tag{1}
\end{equation*}
 by causality of $(W^{\textsc{w}\chi},E^{\textsc{w}})$. Then\\
\hspace*{6mm}
(a)    {\it $(W,E)$ localizes frame-independently.} \\
\hspace*{6mm}
(b) {\it $(W,E)$ is not causal.}\\
{\it Proof.}  (a) is obvious by (1). ---  (b) In (1) specify $h=A_\rho$ for $\rho\ne 0$ and, for the moment, $\Gamma=\{0\}$ as subset of the spacelike hyperplane  $\{0\} \times\R^3 \equiv \R^3$. Then $A_\rho\cdot 0=0$, whence $\Gamma_h=\{0\}$. Furthermore, $rhr^{-1}=(r(1-\cosh \rho),0,0,-r\sinh \rho, A_\rho)$ and $rhr^{-1}\cdot \Gamma =\{(r(1-\cosh \rho),0,0,-r\sinh \rho)\}$, whence $x\in \Gamma_{rhr^{-1}}$ $\Leftrightarrow$ $r^2(\cosh \rho-1)^2-x_1^2-x_2^2-(x_3+r\sinh\rho)^2\ge 0$.
 For $x\in \Gamma_{rhr^{-1}}$,  therefore,  $x_3$ lies in the closed interval with endpoints $r(\operatorname{e}^{-\rho}-1)$ and $r(1-\operatorname{e}^\rho)$, which are either both positive or both negative, thus excluding $x_3=0$. Hence $\Gamma_h \cap  \Gamma_{rhr^{-1}}=\emptyset$. By continuity this still holds true if $\Gamma=\{0\}$ is replaced by $\Gamma=\{y\in\R^3:|y|<\epsilon\}$ for $\epsilon>0$ small enough. This implies $W(h)E(\Gamma)W(h)^{-1}\not\le E(\Gamma_h)$ confirming the claim.\qed 
  \end{Exa}

  \begin{Lem} Let the massless causal system $(W,E)$ be unitarily related to $\nu\,(W^{\textsc{w}\chi},E^{\textsc{w}})$ for some $\chi\in\{+,-\}$ with multiplicity $\nu\in\N$ by the unitary transformation $S$. Suppose  $SE(B)S^{-1}\le E(B')$, where $B$, $B'$ are two  balls with equal radius. Then $(W,E)$ is unitarily equivalent to $\nu\,(W^{\textsc{w}\chi},E^{\textsc{w}})$.
    \end{Lem}

{\it Proof.} Without restriction $W=S\,\nu W_1^{\chi mom}\, S^{-1}$ and $E=\nu\,E_1^{mom}=\mathcal{F}E^{can}\mathcal{F}^{-1}$. Recall that then $(S\varphi)(p)=s(p)\varphi(p)$, where $s:\R^3\to\C^{2\nu\times 2\nu}$ is measurable and $s(p)$ unitary. 

\hspace*{6mm}
(a) First we show that $s(p)=\operatorname{e}^{\operatorname{i}(a-a')p}s_0$, where $a$ and $a'$ are the midpoints of $B$ and $B'$, respectively, and $s_0$ is some constant. 
Indeed, let $R$ denote the radius of $B$ and $B'$, let $B_R$ be the open ball around the origin with radius $R$ and set $S':=W(a')^{-1}SW(a)$. Then $a+B_R=B$, $B_R=B'-a'$, whence $S'E(B_R)S'^{-1}=W(a')^{-1}SW(a)E(B_R)W(a)^{-1}S^{-1}W(a')=W(a')^{-1}SE(B)S^{-1}W(a')\le W(a')^{-1}E(B')W(a')=E(B_R)$. Moreover check by (\ref{RCL}) for $n=1$ that $(S'\varphi)(p)=\operatorname{e}^{-\operatorname{i}(a-a')p}s(p)\varphi(p)$.
 Now, following the proof of \cite[Theorem 10]{CL15} for $\delta=0$ one infers that $(S'\varphi)(p)=s_0\varphi(p)$ for some constant matrix $s_0$. This shows the assertion.

\hspace*{6mm}
(b) Now one shows that $S$ commutes with $\nu W_1^{\chi mom}\equiv W_1^{\chi mom}\otimes I_\nu$ thus concluding the proof.
 Indeed, as $S$ commutes with $\nu W_1^{\chi mom}|_{SU(2)}$ one has $s(B\cdot p)=(B\otimes I_\nu)s(p)(B^{-1}\otimes I_\nu)$, whence $\operatorname{e}^{\operatorname{i}bp}s_0=(B\otimes I_\nu)s_0(B^{-1}\otimes I_\nu)$ with $b:=\big((B^{-1}-I_2)\cdot (a-a')\big)p$, for almost all $p\in\R^3$ and all $B\in SU(2)$. This implies 
$(B^{-1}-I_2)\cdot (a-a')=0$ and hence $a=a'$. So $s_0$ commutes with $B\otimes I_\nu$ for all $B\in SU(2)$. Therefore there is $s_\nu\in\C^{\nu\times\nu}$ such that $s_0=I_2\otimes s_\nu$, whence the assertion.
\qed\\
 
  \hspace*{6mm}
It remains to find out those irreducible SCT unitarily related to multiples of   $(W^{\textsc{w}\chi},E^{\textsc{w}})$ for  some $\chi\in\{+,-\}$ that are causal.
 The foregoing considerations clearly suggest  that up to unitary equivalence the only ones 
 are the two Weyl systems  
 $(W^{\textsc{w}\chi},E^{\textsc{w}})$, $\chi\in\{+,-\}$  studied   in  sec.\,\ref{OWL}.

\section{Appendix}

\subsection{A total set in $L^p_\mu(\R^d,\C^m)$ of radially symmetric functions}\label{A:TSLPS}    The Lebesgue measure on  $\R^d$ is a Radon measure. The following result is valid for any Radon measure  on $\R^d$. As to the definition of a Radon measure on a topological Hausdorff space see e.g. \cite[2.1.1\,(9)]{CR08}, cf. also  \cite[1.5 Definition]{M95}. A subset $L$ of a topological vector space $X$ is said to be \textbf{total} if the linear hull $\langle L\rangle$ of $L$ is dense in $X$.

\begin{The}\label{TSLPS} Let $\mu$ be a Radon measure on $\R^d$. Let $m,d\in\N$ and $1\le p<\infty$. Let $e_1,\dots,e_m$ denote the elements of the standard basis of $\C^m$. Then $\{1_B e_i: B\subset \R^d \textnormal{ closed ball},\; i=1,\dots,m\}$ is total in $L^p_\mu(\R^d,\C^m)$.  
\end{The}\\
{\it Proof.} (a) {\it Let $L$ be total in the topological vector space $X$. Let $M\subset X$ with $L\subset
 \overline{\langle M\rangle}$. Then $M$ is total in $X$.} This is  obvious as $X= \overline{\langle L\rangle}\subset\overline{\langle M\rangle}$.\\
\hspace*{6mm}
(b) {\it Let also $L'$ be total in the topological vector space $X'$. Then $M:=(L\times\{0\}) \cup (\{0\}\times L')$ is total in $X\times X'$.} This is obvious as $ \langle M\rangle=\langle L\times L'\rangle =\langle L\rangle\times 
\langle L'\rangle$, which is dense in $X\times  X'$.\\
\hspace*{6mm}
Due to (b) the assertion is readily reduced to the claim that $M:=\{1_B : B\subset \R^d \textnormal{ closed ball}\}$ is total in $L^p:=L^p_\mu(\R^d,\C)$.\\
\hspace*{6mm}
  It follows from Lebesgue's ladder  that $\{1_A: A \textnormal{ measurable},\;\mu(A)<\infty\}$ is total in $L^p$. As $\mu$ is finite at compact sets, for $\varepsilon >0$ and $A$ measurable with $\mu(A)<\infty$ there is $R>0$ great enough such $\mu(A\setminus B_R)<\varepsilon$. Therefore even $\{1_A: A \textnormal{ measurable bounded}\}$ is total  in $L^p$. Furthermore, 
due to the outer regularity of $\mu$, for $\varepsilon >0$ and $A$ measurable bounded there is an open bounded $U\supset A$ with $\mu(U\setminus A)<\varepsilon$. Consequently,  $L:=\{1_U: U \textnormal{ open bounded}\}$ is total  in $L^p$.\\ 
\hspace*{6mm}
Thus due to (a) it remains to show that $L\subset \overline{\langle M\rangle}$.  So for $U$ open bounded consider the set $\mathcal{B}:=\{B: B\subset U, B\textnormal{ closed ball}\}$. Then according to Vitali's covering theorem \cite[2.8 Theorem]{M95} there  are countably many mutually disjoint $B_i\in\mathcal{B}$ with $\mu(U\setminus \bigcup_iB_i)=0$. Hence, for $\varepsilon>0$ there is $n\in\N$ with $\mu(U\setminus V) <\varepsilon^p$ for $V:=\bigcup_{i=1}^nB_i$. Consequently, $\norm{1_U-\sum_{i=1}^n1_{B_i}}_p=\norm{1_U-1_V}_p=\norm{1_{U\setminus V}}_p=\mu(U\setminus V)^{1/p}<\varepsilon$, accomplishing the proof.\qed\\

\subsection{Imprimitivity Theorem}\label{A:IT} Let $G$ be a locally compact group with countable base acting 
on  a locally compact space $X$. Let $H$ be a separable Hilbert space. Then a system of imprimitivity $(V,E)$ for $(G,X)$ on $H$ 
is a representation $V$ of $G$ on $H$, an action of $G$ on $X$,  and a PM $E$ on $X$ in $H$ such that $V(g)E(M)V(g)^{-1}=E(g M)$  for all $g\in G$ and Borel sets $M\subset X$.\\
\hspace*{6mm} 
Let $G$ act continuously and transitively on $X$ and let $x_0\in X$ be any point. 
Then every system of imprimitivity $(V,E)$ for $(G,X)$ on $H$ is unitarily equivalent to a canonical system $(U^{\textsc{g}},E^{can})$ on $L^2_m(X,H_{\textsc{u}})$. Here
\begin{equation*} 
E^{can}(M)\psi:=1_M\psi\tag{1} 
\end{equation*}
for Borel set $M\subset X$ is the canonical PM on $X$, and $U$ is a representation of the stabilizer subgroup $G_{x_0}$ in $x_0$ on the Hilbert space $H_{\textsc{u}}$ and $U^{\textsc{g}}$ the representation of $G$ induced by $U$, i.e.,
\begin{equation*}
(U^{\textsc{g}}(g)\psi)(x):=\sqrt{(\operatorname{d}m^g/ \operatorname{d}m)(x)} \,U\big(R(x,g)\big)\psi(g^{-1}x)\tag{2}
\end{equation*}
where $m$ is the up equivalence unique  non trivial regular Borel measure on $X$, which is quasi-invariant under $G$, $m^g(M):=m(g^{-1}M)$ for Borel set $M\subset X$, $\operatorname{d}m^g/ \operatorname{d}m$ the Radon-Nikodym derivative, $q:X\to G$ a Borel section with respect to the surjection $G\to X$, $g\mapsto gx_0$, and the Wigner rotation $R(x,g):=q(x)^{-1}g\,q(g^{-1}x)$.\\
\hspace*{6mm} 
Two canonical systems of imprimitivity for $(G,X)$ are equivalent, respectively irreducible, if and only if the corresponding representations of the stabilizer subgroup are equivalent, respectively irreducible.

\subsection{Systems of imprimitivity for $(\tilde{\mathcal{P}}_\chi,\chi)$}\label{A:SI} The aim is   to extend a causal WL to non-timelike not spacelike hyperplanes. As studied in sec.\,\ref{ECLNTLHP},  a rep of the lattice of the $\perp'$-complete regions (i.e., of the causal logic) would require such an extension.\\
\hspace*{6mm}
  By Poincar\'e covariance it suffices to consider the hyperplane $\chi=\{\mathfrak{x}: x_0=x_3\}$. Then by (\ref{IGNSLHP}) 
the invariance subgroup $\tilde{\mathcal{P}}_\chi\subset \tilde{\mathcal{P}}$ equals 
  $$IST(2)=\{(\mathfrak{a},A):\mathfrak{a}\in\chi, A\in SL(2,\C), A_{21}=0\}$$  
One gets the canonical systems of imprimitivity for $(\tilde{\mathcal{P}}_\chi,\chi)$ by (\ref{A:IT}). First note that the Lebesgue measure $\lambda_\chi$ on $\chi$ is quasi-invariant under $\tilde{\mathcal{P}}_\chi$. Indeed, let $j:\R^3\to\chi$, $j(x):=(x_3,x)$. Then $\lambda_\chi=j(\lambda)$ with $\lambda$ the Lebesgue measure on $\R^3$. Hence $\lambda_\chi^{(\mathfrak{a},A)}(M)=\lambda\big(-j^{-1}(A^{-1}\cdot\mathfrak{a})+j^{-1}(A^{-1}\cdot M)\big)=\det\big(j^{-1}\Lambda(A^{-1})j\big)\,\lambda_\chi(M)=|A_{11}|^{-2}\lambda_\chi(M)$, where by  (\ref{IGNSLHP}) the last equation holds since $j^{-1}\Lambda(\operatorname{e}^{-\rho\sigma_3/2})j=\operatorname{diag}(1,1,\operatorname{e}^{-\rho})$ and
\begin{displaymath}
j^{-1}\Lambda\big(\left(\begin{array}{cc} 1& w\\ 0 & 1\end{array}\right)\big)j=\left(\begin{array}{ccc} 1& 0& 0 \\ 0 &1 & 0 \\  u & -v & 1\end{array}\right)
\end{displaymath}
In particular, $\operatorname{d}\lambda_\chi^{(\mathfrak{a},A)}/\operatorname{d}\lambda_\chi=|A_{11}|^{-2}$ is constant. --- Next we choose $0\in\chi$ to be the fixed point. Then  $\tilde{\mathcal{P}}_\chi\to \chi$, $(\mathfrak{a},A)\mapsto (\mathfrak{a},A)\cdot 0=\mathfrak{a}$ is the natural surjection and obviously $q:\chi\to\tilde{\mathcal{P}}_\chi$, $q(\mathfrak{a}):=(\mathfrak{a}, I_2)$  is a Borel section of it. Moreover, $ST(2)$
 is the stabilizer subgroup in $0$, and the Wigner rotation is $R\big(\mathfrak{x},(\mathfrak{a},A)\big)=A$. Indeed, $q(\mathfrak{x})^{-1}(\mathfrak{a},A)\,q\big((\mathfrak{a},A)^{-1}\cdot \mathfrak{x}\big)=(-\mathfrak{x},I_2)(\mathfrak{a},A)\,q\big(-A^{-1}\cdot \mathfrak{a},A^{-1})\cdot \mathfrak{x}\big)=(-\mathfrak{x}+\mathfrak{a},A)\,q(-A^{-1}\cdot \mathfrak{a}+A^{-1}\cdot \mathfrak{x})=(-\mathfrak{x}+\mathfrak{a},A)(-A^{-1}\cdot \mathfrak{a}+A^{-1}\cdot \mathfrak{x},I_2)=\big(-\mathfrak{x}+\mathfrak{a}+A\cdot(-A^{-1}\cdot \mathfrak{a}+A^{-1}\cdot \mathfrak{x}), A\big)=(0,A)\equiv A$. \\
\hspace*{6mm} 
  Now
let $U$ be any representation of $ST(2)$ on the Hilbert space $H_{\textsc{u}}$. Then the representation of $\tilde{\mathcal{P}}_\chi$ induced by $U$  acting on $L^2(\chi,H_{\textsc{u}})$, here called $V^{can}$, reads
\begin{equation*}
 \big(V^{can}(\mathfrak{a},A)\psi\big)(\mathfrak{x})=|A_{11}|^{-1}U(A)\psi\big(A^{-1}\cdot(\mathfrak{x}-\mathfrak{a})\big)
 \end{equation*}
and $(V^{can},P^{can})$, where $P^{can}$ denotes the canonical PM, is a system of imprimitivity. \\ 

\hspace*{6mm}
\textbf{Shell representation.} Recall that $\mathcal{O}^{m,\eta}$ for $m>0$, $\eta=\pm$ denotes the mass shell $\{\mathfrak{p}\in\R^4: p_0=\eta \epsilon(p)\}$ and let  $\mathcal{O}^m:=\mathcal{O}^{m,+}\cup\mathcal{O}^{m,-}$ be the double shell $\{\mathfrak{p}\cdot \mathfrak{p}=m^2\}$ endowed with the Lorentz invariant measure. We shall use 
the  shell representation $(V^{shell},P^{shell})$, which is equivalent to $(V^{can},P^{can})$. The former acts on $L^2(\mathcal{O}^m,H_{\textsc{u}})$ by
\begin{itemize}
 \item[(a)] $\big(V^{shell}(\mathfrak{a},A)\phi\big)(\mathfrak{p})=\operatorname{e}^{\operatorname{i}\mathfrak{a}\cdot \mathfrak{p}}U(A)\phi(A^{-1}\cdot\mathfrak{p})$
 \end{itemize}
The  form of  $P^{shell}(M)$, $M\subset \chi$ is  complicated. For $M:=\{\mathfrak{x}\in\chi: 0\le \varsigma x_3\le \alpha\}$,  $\alpha>0$, $\varsigma=\pm 1$ and $\phi\in L^1 \cap L^2$ we get explicitly
\begin{itemize}
\item[(b)]  $\big(P^{shell}(M)\phi\big)(\mathfrak{p})=\textrm{\footnotesize{$\sqrt{|p_0-p_3|}$}}\sum_{\eta'=+,-}\int k(\mathfrak{p},\mathfrak{p}')\frac{\sqrt{|p'_0-p'_3|}}{2|p'_0|}\phi(\mathfrak{p}')\operatorname{d}p'_3$
\end{itemize}
where $p'_1:=p_1$, $p'_2:=p_2$, $p'_0:=\eta'\epsilon(p')$,  and $k(\mathfrak{p},\mathfrak{p}'):=\frac{\exp (\operatorname{i}\alpha\varsigma(p_0-p_3-p'_0+p'_3))-1}{\operatorname{i}\pi\varsigma(p_0-p_3-p'_0+p'_3)}$. A similar formula regarding $P^{shell}(M)$ is available if $M$ is a Borel set with finite Lebesgue measure.\\

\hspace*{6mm}
Let us sketch how to get the shell representation. Let $\mathfrak{e}:=(1,0,0,1)$, consider the bijection $l: \R^3\to \R^4/\R\mathfrak{e}$, $l(p):=(0,p)+\R\mathfrak{e}$ and endow $\R^4/\R\mathfrak{e}$ with the  measure $l(\lambda)$, where $\lambda$ is the Lebesgue measure on $\R^3$. Then $\psi\mapsto \varphi:= \mathcal{F}(\psi\circ j)\circ l^{-1}$ defines an  isomorphism $\iota: L^2(\chi, H_{\textsc{u}})\to
L^2(\R^4/\R\mathfrak{e}, H_{\textsc{u}})$. For integrable $\psi$ one has $\varphi(\mathfrak{p}+\R\mathfrak{e})=\big(\frac{1}{2\pi}\big)^{3/2}\int \operatorname{e}^{\operatorname{i}\mathfrak{y}\cdot\mathfrak{p}}\psi(\mathfrak{y})\operatorname{d}^3y$ with $\mathfrak{y}:=j(y)$.  Let $\hat{V}:=\iota V^{can}\iota^{-1}$.  It is easy to verify $\big(\hat{V}(\mathfrak{a},A)\varphi\big)(\mathfrak{p}+\R\mathfrak{e})=|A_{11}|\operatorname{e}^{\operatorname{i}\mathfrak{a}\cdot \mathfrak{p}}U(A)\varphi(A^{-1}\cdot\mathfrak{p}+\R\mathfrak{e})$.\\
\hspace*{6mm}
Put $\mathcal{K}^\eta:=\{\mathfrak{p}+\R\mathfrak{e}:\operatorname{sgn}(p_0-p_3)=\eta\}$ and $\mathcal{K}:=\mathcal{K}^+\cup\mathcal{K}^-$. Then $\mathcal{K}=( \R^4/\R\mathfrak{e})\setminus (\chi/\R\mathfrak{e})$, where $\chi/\R\mathfrak{e}$ is a $l(\lambda)$-null set.\\
\hspace*{6mm}
 Now we pass to the double shell by the bijection $k:\mathcal{O}^m\to \mathcal{K}$, $k(\mathfrak{p}):=\mathfrak{p}+\R\mathfrak{e}$. Indeed,  $k|_{\mathcal{O}^{m,\eta}}\to \mathcal{K}^\eta$ is a bijection since $\operatorname{sgn}(p_0)=\operatorname{sgn}(p_0-p_3)$ for every $\mathfrak{p}\in \mathcal{O}^m$ and since $\mathcal{O}^{m,\eta}\cap (\mathfrak{p}+\R\mathfrak{e})=
 \{\mathfrak{p}+\frac{m^2-\,\mathfrak{p}\cdot  \mathfrak{p}}{2(p_0-p_3)}\mathfrak{e}\}$ is a one-point set (and not empty) just for every $\mathfrak{p}\in\R^4$ with $\operatorname{sgn}(p_0-p_3)=\eta$. Moreover, $k^{-1}( \mathfrak{p}+\R\mathfrak{e})=\mathfrak{p}+\frac{m^2-\,\mathfrak{p}\cdot  \mathfrak{p}}{2(p_0-p_3)}\mathfrak{e}$ for every $\mathfrak{p}\not\in\chi$.\\
\hspace*{6mm}
 One claims that $\kappa: L^2(\R^4/\R\mathfrak{e}, H_{\textsc{u}})\to L^2(\mathcal{O}^m,H_{\textsc{u}})$, $\kappa(\varphi):=\sqrt{2|p_0-p_3|}\varphi\circ k$ is an isomorphism. Indeed, note first $\kappa^{-1}(\phi)=\frac{1}{\sqrt{2|p_0-p_3|}}\phi\circ k^{-1}$. Then, on the one hand $\norm{\varphi}^2=\int_\mathcal{K}\norm{\varphi(\cdot)}^2\operatorname{d}l(\lambda)=\int_{\R^3\setminus\{p_3=0\}}\norm{\varphi\circ l(\cdot)}^2\operatorname{d}\lambda$. On the other hand $\norm{\kappa(\varphi)}^2=\sum_\eta\int_{\R^3}2(\epsilon(p)-\eta p_3)$ $
 \norm{\varphi\big((\eta\epsilon(p),p)+\R\mathfrak{e}\big)}^2\frac{\operatorname{d}^3p}{2\epsilon(p)}=\sum_\eta\int_{\R^3}\frac{\epsilon(p)-\eta p_3}{\epsilon(p)}
 \norm{\varphi\circ l(T^\eta(p)}^2\operatorname{d}^3p$ with the coordinate transformation $T^\eta(p):=p-\eta\epsilon(p)e_3$. Note  $T^+(\R^3)=\R\times\R\times ]-\infty,0[$,
$T^-(\R^3)=\R\times\R\times ]0, \infty[$, and the Jacobian determinant equals $1-\eta\frac{p_3}{\epsilon(p)}$. Hence by the substitution rule one has  $\norm{\kappa(\varphi)}^2=\int_{\R\times\R\times]-\infty,0[}\norm{\varphi\circ l(p')}^2\operatorname{d}^3p'+ \int_{\R\times\R\times]0,\infty[}\norm{\varphi\circ l(p')}^2\operatorname{d}^3p'$, whence the claim.\\
\hspace*{6mm}
Finally, one easily verifies $V^{shell}=\kappa \hat{V} \kappa^{-1}$ as indicated in (a). Also the formula regarding $P^{shell}(M)=\kappa \iota \,P^{can}(M) (\kappa \iota)^{-1}$ in (b) follows by a similar computation.

\hspace*{6mm}
\textbf{Cone representation.} We will need also the cone representation $(V^{cone},P^{cone})$  
acting on $L^2(\mathcal{O}^0,H_{\textsc{u}})$ equivalent to $(V^{can},P^{can})$. Recall the light half-cone $\mathcal{O}^{0,\eta}$  for $\eta=\pm$ and  the light cone $\mathcal{O}^0$ endowed with the Lorentz invariant measure. The above formulae (a) and (b) for the shell representation hold literally for the equivalent cone representation simply putting $m=0$. In particular, $\epsilon(p)$ becomes $|p|$.\\
\hspace*{6mm} 
 One passes to the cone  replacing $k$ by  $k_0:\mathcal{O}^0\setminus \R\mathfrak{e}\to \mathcal{K}$, $k_0(\mathfrak{p}):=\mathfrak{p}+\R\mathfrak{e}$. Note that $k_0|_{\mathcal{O}^{0,\eta}}\to \mathcal{K}^\eta$ is a bijection since $\operatorname{sgn}(p_0)=\operatorname{sgn}(p_0-p_3)$ for every $\mathfrak{p}\in \mathcal{O}^0\setminus\R\mathfrak{e}$ and since $(\mathcal{O}^{0,\eta}\setminus \R\mathfrak{e})\cap (\mathfrak{p}+\R\mathfrak{e})=
 \{\mathfrak{p}-\frac{\mathfrak{p}\cdot  \mathfrak{p}}{2(p_0-p_3)}\mathfrak{e}\}$ is a one-point set (and not empty) just for every $\mathfrak{p}\in\R^4$ with $\operatorname{sgn}(p_0-p_3)=\eta$. Moreover, $k_0^{-1}( \mathfrak{p}+\R\mathfrak{e})=\mathfrak{p}-\frac{\mathfrak{p}\cdot  \mathfrak{p}}{2(p_0-p_3)}\mathfrak{e}$ for every $\mathfrak{p}\not\in\chi$, and note that $\R\mathfrak{e}$ in a null set of $\mathcal{O}^0$. Then like in the shell case one shows that $\kappa_0: L^2(\R^4/\R\mathfrak{e}, H_{\textsc{u}})\to L^2(\mathcal{O}^m,H_{\textsc{u}})$, $\kappa_0(\varphi):=\sqrt{2|p_0-p_3|}\varphi\circ k_0$ is an isomorphism with $\kappa_0^{-1}(\phi)=\frac{1}{\sqrt{2|p_0-p_3|}}\phi\circ k_0^{-1}$.\\
 \hspace*{6mm}
Finally, one  verifies $V^{cone}=\kappa_0 \hat{V} \kappa_0^{-1}$ and $P^{shell}(M)=\kappa_0 \iota \,P^{can}(M) (\kappa_0 \iota)^{-1}$ as claimed.

 \subsection{Energy and shell representation of the Dirac system and of $W^D|_{\tilde{\mathcal{P}}_\chi}$}\label{A:ESRDS} 
 The energy representation $(W^{\textsc{d}\,erg},E^{\textsc{d}\,erg})$ of the Dirac system is related to the momentum representation  Eq.\,(\ref{RDS}) by the unitary self-adjoint transformation $Y(=Y^{-1})$ on $L^2(\R^3,\C^4)$
given by 
\begin{displaymath} 
(Y\varphi)(p)=Y(p)\varphi(p), \;Y(p):=\left(\frac{m}{2\epsilon(p)}\right)^{1/2}
\left(\begin{array}{cc} Q(\mathfrak{p}^+)& Q(\mathfrak{p}^+)^{-1}\\ Q(\mathfrak{p}^+)^{-1} & -Q(\mathfrak{p}^+)\end{array}\right)
 \end{displaymath}
where  $\mathfrak{p}^+=(\epsilon(p),p)$ and  $Q(\mathfrak{p}^+)$ denotes the canonical cross section given in (\ref{PCCS}). The transformation $Y$ commutes with the representation of the Euclidean group due to the covariance $\operatorname{diag}(B,B)Y(p)\operatorname{diag}(B^*,B^*)=Y(B\cdot p)$, and diagonalizes the time translations since $h(p)=Y(p)\operatorname{diag}\big(\epsilon(p),-\epsilon(p)\big)Y(p)^{-1}$. For $(W^{\textsc{d}\,erg},E^{\textsc{d}\,erg}):=Y^{-1}(W^{\textsc{d}\,mom},E^{\textsc{d}\,mom})Y$ one has
\begin{equation*} W^{\textsc{d}\,erg}=\oplus_\eta W^{\textsc{d}\,\eta},\; \big(W^{\textsc{d}\,\eta}(t,b,A)\varphi_\eta\big)(p)=\sqrt{\epsilon(q^\eta)/\epsilon(p)}\,\operatorname{e}^{\operatorname{i}(\eta t \epsilon(p)-bp)}R(\mathfrak{p}^\eta,A)\varphi_\eta(q^\eta)
\end{equation*}
where $W^{\textsc{d}\,\eta}$ acts on $L^2(\R^3,\C^2)$, cf. Eq.\,(\ref{RIMS}) for $j=1/2$. The computations use the formulae regarding the canonical cross section in (\ref{PCCS}) and the identity 
\begin{displaymath} 
\pi^\eta(p)=\frac{1}{2}\big(I_2+\frac{\eta}{\epsilon(p)}h^{\textsc{d}}(p)\big)=\frac{m}{2\epsilon(p)}
\left(\begin{array}{cc} Q(\mathfrak{p}^+)^{2\eta}& \eta I_2\\ \eta I_2 & Q(\mathfrak{p}^+)^{-2\eta}\end{array}\right)
\end{displaymath}

\hspace*{6mm}
As to the shell representation consider the isomorphism $\xi:L^2(\mathcal{O}^m,\C^2)\to L^2(\R^3,\C^4)$ with  $(\xi\phi)(p):=\frac{1}{\sqrt{2\epsilon(p)}}\big(\phi(\mathfrak{p}^+),\phi(\mathfrak{p}^-)\big)$ and $(\xi^{-1}\varphi)(\mathfrak{p})=\sqrt{2\epsilon(p)}\varphi_\eta(p)$ with $\varphi=(\varphi_+,\varphi_-)$ and $\varphi_\eta\in L^2(\R^3,\C^2)$. Then $(W^{\textsc{d}\,shell},E^{\textsc{d}\,shell}):=\xi^{-1}(W^{\textsc{d}\,erg},E^{\textsc{d}\,erg})\xi$ acts on $L^2(\mathcal{O}^m,\C^2)$ and
\begin{equation*} \big(W^{\textsc{d}\,shell}(\mathfrak{a},A)\phi\big)(\mathfrak{p})=\operatorname{e}^{\operatorname{i}\mathfrak{a}\cdot \mathfrak{p}}R(\mathfrak{p},A)\phi(A^{-1}\cdot \mathfrak{p})
\end{equation*}

\hspace*{6mm}
A last unitary transformation is performed in order to cast $W^{\textsc{d}}|_{\tilde{\mathcal{P}}_\chi}$ into the form $V^{shell}$ in sec.\,\ref{A:SI}, namely $(W^{'\textsc{d}\,shell},E^{'\textsc{d}\,shell}):=Z^{-1}(W^{\textsc{d}\,shell},E^{\textsc{d}\,shell})Z$ with $Z:L^2(\mathcal{O}^m,\C^2)\to L^2(\mathcal{O}^m,\C^2)$
\begin{equation*} (Z\phi)(\mathfrak{p})=Z(\mathfrak{p})\phi(\mathfrak{p}), \; Z(\mathfrak{p}):=(m/|p_0-p_3|)^{1/2}\left(Q(\mathfrak{p})\Tiny{\big(\begin{array}{cc} 0 & 0\\ 0 & 1\end{array}\big)}+\eta \,Q(\mathfrak{p})^{-1}\Tiny{\big(\begin{array}{cc} 1 & 0\\ 0 & 0\end{array}\big)}\right)
\end{equation*}
Check $I_2=Z(\mathfrak{p})Z(\mathfrak{p})^*$ using $(|p_0-p_3|/m) \,I_2=Q(\mathfrak{p})^2\Tiny{\big(\begin{array}{cc} 0 & 0\\ 0 & 1\end{array}\big)}+\,\Tiny{\big(\begin{array}{cc} 1 & 0\\ 0 & 0\end{array}\big)}Q(\mathfrak{p})^{-2}$. Finally one finds
\begin{displaymath}(W^{'\textsc{d}\,shell}(\mathfrak{a},A)\phi)(\mathfrak{p})=\operatorname{e}^{\operatorname{i}\mathfrak{a}\cdot \mathfrak{p}}U(A)\phi(A^{-1}\cdot\mathfrak{p})
\end{displaymath}
for $(\mathfrak{a},A)\in\tilde{\mathcal{P}}_\chi$ and $U(A):=\operatorname{diag}\big(\frac{A_{11}}{|A_{11}|}, \frac{\overline{A}_{11}}{|A_{11}|}\big)$,  hence  $W^{'\textsc{d}\,shell}|_{\tilde{\mathcal{P}}_\chi}= V^{shell}$, see sec.\,\ref{A:SI} (a).

 \subsection{Energy and cone representation of the Weyl systems and of $W^{\textsc{w}\chi}|_{\tilde{\mathcal{P}}_\chi}$}\label{A:ESRWS} 
  The energy representation $(W^{\textsc{w}\chi\,erg},E^{\textsc{d}\,erg})$ of the Weyl system is 
defined as   
$$(W^{\textsc{w}\chi\,erg},E^{\textsc{w}\,erg}):=(Y^\chi)^{-1}(W^{\textsc{w}\chi\,mom},E^{\textsc{w}\,mom})Y^{\chi}$$
where the  momentum representation  is  given in Eq.\,(\ref{2WS}) and the unitary  transformation $Y^\chi$ on $L^2(\R^2,\C^2)$ reads
\begin{displaymath}
(Y^+\varphi)(p)=B(p)\Tiny{\big(\begin{array}{cc}1&0\\ 0& b(p)\end{array}\big)} \varphi(p),  \,(Y^-\varphi)(p)=B(p)\Tiny{\big(\begin{array}{cc}0&1\\ 1&0\end{array}\big)}   \Tiny{\big(\begin{array}{cc}1&0\\ 0& \overline{b(p)}\end{array}\big)}    \varphi(p)
\end{displaymath}
with $B(p)$ from Eq.\,(\ref{CCSEF}). Cf.  (\ref{RCL}) and (\ref{DWTL}). Note that $Y^\chi$ relatively to $Y_0^{-1}$ has an additional unitary transformation yielding $W^{\textsc{w}\chi\,erg}$ particularly clear.
Explicitly one finds for $\varphi=(\varphi_+,\varphi_-)\in L^2(\R^3,\C^2)$
\begin{eqnarray*}
\big(W^{\textsc{w}\chi\,erg}(t,b,A)\Tiny{\big(\begin{array}{c}\varphi_+\\\varphi_-\end{array}\big)}\big)(p) &=&\operatorname{e}^{-\operatorname{i}bp}\left(\begin{array}{c}(|q^+|/|p|)^{1/2}\;\operatorname{e}^{\operatorname{i}t|p|}\;k(\mathfrak{p}^+,A)^\chi\;\varphi_+(q^+)\\ (|q^-|/|p|)^{1/2}\operatorname{e}^{-\operatorname{i}t|p|}k(\mathfrak{p}^-,A)^\chi \;\varphi_-(q^-)\end{array}\right)
\end{eqnarray*}

For the computations use Eq.\,(\ref{CRCS}), the fact that $B(\eta p)^{-1}AB(\eta q)=A_{\ln|p|}H(\mathfrak{p},A)A_{\ln|q|}^{-1}\in E(2)$ by definition of $H(\mathfrak{p},A)$, and  $B(-p)=C(p)^{-1}B(p)$,\, $C(p)\operatorname{diag}(\alpha,\beta)=\operatorname{diag}(\beta,\alpha)C(p)$ for $C(p)=-C(p)^{-1}$ in
Eq.\,(\ref{SERLKG})(2). \\
\hspace*{6mm}
One passes to the cone representation by the isomorphism $\xi_0:L^2(\mathcal{O}^0)\to L^2(\R^3,\C^2)$, $(\xi_0\phi)(p):=\frac{1}{\sqrt{2|p|}}\big(\phi(\mathfrak{p^+}),\phi(\mathfrak{p^-})\big)$, $(\xi_0^{-1}\varphi)(\mathfrak{p})=\sqrt{2|p|}\,\varphi_\eta(p)$. For $(W^{\textsc{w}\chi\,cone},E^{\textsc{w}\,cone}):=\xi^{-1}_0(W^{\textsc{w}\chi\,erg},E^{\textsc{w}\,erg})\xi_0$ one finds 
$$\big(W^{\textsc{w}\chi\,cone}(\mathfrak{a},A)\phi\big)(\mathfrak{p})=\operatorname{e}^{\operatorname{i}\mathfrak{p}\cdot \mathfrak{a}}k(\mathfrak{p},A)^\chi\phi(A^{-1}\cdot\mathfrak{p})$$ 
(cf. Eq.\,(\ref{MLIRPG})).\\
\hspace*{6mm}
A last unitary transformation is performed in order to cast $W^{\textsc{w}\chi}|_{\tilde{\mathcal{P}}_\chi}$ into the form  $V^{cone}$ in sec.\,\ref{A:SI}, namely $(W^{'\textsc{w}\chi\,cone},E^{'\textsc{w}\,cone}):=(Z^\chi)^{-1}(W^{\textsc{w}\chi\,cone},E^{\textsc{w}\,cone})Z^\chi$ with $Z^\chi:L^2(\mathcal{O}^0)\to L^2(\mathcal{O}^0)$, $(Z^\chi\phi)(\mathfrak{p}):=\overline{b(\mathfrak{p})}^\chi\phi(\mathfrak{p})$ with $b(\mathfrak{p}):=b(p)$.  Then 
$\big(W^{'\textsc{w}\chi\,cone}(\mathfrak{a},A)\phi\big)(\mathfrak{p})=\operatorname{e}^{\operatorname{i}\mathfrak{p}\cdot \mathfrak{a}}k'(\mathfrak{p},A)^\chi\phi(A^{-1}\cdot\mathfrak{p})$ 
with $k'(\mathfrak{p},A):=b(\mathfrak{p})k(\mathfrak{p},A)\overline{b(A^{-1}\cdot \mathfrak{p})}$. Now, by the remarkable property  $k'(\mathfrak{p},A)=\frac{\overline{A}_{11}}{A_{11}}$ for all  $\mathfrak{p}\in\mathcal{O}^0$,  $A\in ST(2)$ shown below one has
$$W^{'\textsc{w}\chi\,cone}|_{\tilde{\mathcal{P}}_\chi}=V^{cone}$$
for $U(A):=\operatorname{diag}\big(\frac{\overline{A}_{11}}{|A_{11}|},\frac{A_{11}}{|A_{11}|}\big)^\chi$.\\
\hspace*{6mm}
It remains to show $k'(\mathfrak{p},A)=\overline{A}_{11}/A_{11}$. Let $A=CA_\rho A'$ with $C=\operatorname{diag}(\gamma,\overline{\gamma})$, $|\gamma|=1$ and $A'_{11}=1$ according to 
(\ref{IGNSLHP}). Then $k(\mathfrak{p},A)=k(\mathfrak{p},CA_\rho)k(\mathfrak{q},A')$ for $\mathfrak{q}=(CA_\rho)^{-1}\cdot \mathfrak{p}$. By Eq.\,(\ref{DEWRMLC}), $k(\mathfrak{p},CA_\rho)=\gamma$. Explicit computation using the formula for  $\Lambda(A')$ in (\ref{IGNSLHP})  shows $k(\mathfrak{q},A')=b(q')\overline{b(q)}$ for $\mathfrak{q}':=A^{'-1}\cdot\mathfrak{q}$. 
Now the assertion follows, since $\mathfrak{q}':=A^{-1}\cdot \mathfrak{p}$ and since $q=C^{-1}\cdot p$, whence $b(q)=\gamma^2b(p)$ by a short computation. \\

\subsection{Extension of a positive-operator-valued content to a POM}\label{A:EPOCPOM} The main result is (\ref{EPOCPOM}). By physical reasons we are interested in the corollary (\ref{EPOMPOVC}).

\begin{The}\label{EPOCPOM}
Let  $\mathcal{Q}$ be an algebra of subsets of a set $X$, $\hat{\mathcal{Q}}$ the $\sigma$-algebra generated by $\mathcal{Q}$, and $G$ a positive-operator-valued content on $\mathcal{Q}$. Let $m$ be a normalized content on $\mathcal{Q}$ controlling  $G$, i.e., 
$$m(M_i) \to 0\quad \Leftrightarrow\quad G(M_i)\to 0$$
 as strong limit for every sequence $(M_i)_i$ in $\mathcal{Q}$. Then there is a POM $\hat{G}$ on $\hat{\mathcal{Q}}$ extending $G$ if and only if $m$ is $\sigma$-additive. The extension $\hat{G}$ is unique. Finally, if $G$ is projection-valued then  $\hat{G}$ is a PM.
\end{The}\\
{\it Proof.}  Let $\hat{G}$ be a POM extending $G$. By $\sigma$-additivity of $\hat{G}$ one has $\lim_iG(M_i)=\lim_i\hat{G}(M_i)=0$
for $(M_i)$ in $\mathcal{Q}$ with $M_i\downarrow\emptyset$. Hence $\lim_i m(M_i)=0$. Therefore $m$ is $\sigma$-additive (see e.g.\,\cite[3.2]{B68}).\\
\hspace*{6mm}
Conversely, let $m$ be $\sigma$-additive.   There exists a unique extension $\hat{m}$ of $m$ on $\hat{\mathcal{Q}}$ satisfying 
$\hat{m}(\Delta)=\inf\big\{\sum_jm(K_j): K_j\in \mathcal{Q}, \Delta\subset \bigcup_jK_j\big\}$ $\forall\; \Delta\in \hat{\mathcal{Q}}$ (see e.g.\,\cite[5.2, 5.7]{B68}).
Put
\begin{equation*}
 \mathcal{C}(\Delta):=\big\{(M_{ij}): M_{ij}\in  \mathcal{Q},\;  M_{ij}\downarrow_i,\;M_{ij}\uparrow_j\cup_jM_{ij} \supset \Delta,\; \hat{m}(\Delta)=\hat{m}(\cap_i\cup_jM_{ij})\big\}\tag{1}
\end{equation*}  
It follows $\mathcal{C}(\Delta)\ne \emptyset$. Indeed, note first that there are $K_{ij}\in\mathcal{Q}$ with $\Delta\subset \cup_jK_{ij}$ and $\hat{m}(\Delta)=\inf_i\big(\sum_jm(K_{ij})\big)$. Then $\hat{m}(\Delta)\le\hat{m}(\cup_jK_{ij})\le\sum_jm(K_{ij})$, whence $\hat{m}(\Delta)=\inf_i\hat{m}(\cup_jK_{ij})$. Put $L_{ij}:=\cup_{l=1}^jK_{il}$ and finally let $M_{ij}:=\cap_{l=1}^iL_{lj}$. Now verify $\cup_jM_{ij}=\cap_{l=1}^i\cup_jL_{lj}=\cap_{l=1}^i\cup_jK_{lj}$. Then it is easy to show $(M_{ij})\in\mathcal{C}(\Delta)$. ---
From (1) it follows immediately
\begin{equation*}
\forall\, (M_{ij})\in\mathcal{C}(\Delta),\;  i,n\in\N\;\;\exists\; i_n,\,j_{n,i}\in\N: \;\;\;\hat{m}(\Delta \vartriangle M_{ij})\le\textnormal{\tiny{$\frac{1}{n}$}}\;\;\;\forall\;i\ge i_n,\,j\ge j_{n,i}\tag{2}
\end{equation*}
Put 
\begin{equation*}
\hat{G}(\Delta):=\lim_i\lim_jG(M_{ij}) \quad\forall \Delta\in \hat{\mathcal{Q}},\; (M_{ij})\in\mathcal{C}(\Delta)\tag{3}
\end{equation*}
According to \cite[4.28 (b)]{W76} the strong limits exist and yield bounded operators. Since $0\le G(M_{ij})\le I$ also $0\le \hat{G}(\Delta)\le I$ holds. If $G(M_{ij})$ are projections then $\hat{G}(\Delta)$ is a projection by \cite[4.32]{W76}.\\
\hspace*{6mm}
(a) First one shows that $\hat{G}(\Delta)$ in (3) does not depend on the choice of $(M_{ij})\in\mathcal{C}(\Delta)$.
Let $(M^l_{ij})\in\mathcal{C}(\Delta)$, $l=1,2$. Then $(M^0_{ij})\in\mathcal{C}(\Delta)$ for $M^0_{ij}:=M^1_{ij}\cap M^2_{ij}$. Let $\hat{G}^l(\Delta):=\lim_i\lim_jG(M^l_{ij})$, $l=0,1,2$. Obviously  $\hat{G}^0(\Delta)\le \hat{G}^1(\Delta)$. Assume $\hat{G}^0(\Delta)< \hat{G}^1(\Delta)$. Then there exists a vector $\varphi$ with 
$\langle \varphi, 
\hat{G}^1(\Delta)\varphi\rangle-\langle \varphi, \hat{G}^0(\Delta)\varphi\rangle=:2c>0$. From  (3) for $\hat{G}^l(\Delta)$, $l=0,1$ one infers that there exists $i_0\in\N$  and for every $i\in\N$ a $j_i\in\N$ such that $\langle\varphi,G(M^1_{ij}\setminus M^0_{ij})\varphi\rangle\ge c$ for all $i\ge i_0$ and $j\ge j_i$. This and (2) applied to $(M^l_{ij})$, $l=0,1$ imply that for every $n\in\N$ there are $i_n,j_n\in\N$ such that $L^l_n:=M^l_{i_nj_n}$ satisfies
$\langle\varphi,G(L^1_n\setminus L^0_n)\varphi\rangle\ge c$ and
$m(L^1_n\setminus L^0_n)=\hat{m}(\Delta\vartriangle L^1_n\vartriangle \Delta \vartriangle L^0_n)\footnote{ The symmetric difference $\vartriangle$ is a commutative and associative 
set-theoretic operation.}\le \hat{m}(\Delta\vartriangle L^1_n)+\hat{m}(\Delta\vartriangle L^0_n)\le\frac{2}{n}$ contradicting the assumption that $m$ controls $G$. Similarly one shows $\hat{G}^0(\Delta)= \hat{G}^2(\Delta)$.
\\
\hspace*{6mm}
(b) Obviously (a) implies $\hat{G}|_{\mathcal{Q}}=G$. Hence, in particular,   $\hat{G}(\emptyset)=0$,  $\hat{G}(X)=I$.
\\
\hspace*{6mm}
(c) Now additivity of $\hat{G}$ is shown. Let $\Delta^l\in\hat{\mathcal{Q}}$ and $(M^l_{ij})\in\mathcal{C}(\Delta^l)$ for $l=1,2$.
Put $M^l:=\cap_i\cup_jM^l_{ij}$.
Further put $L_{ij}:=M^1_{ij}\cup M^2_{ij}$ and  $K_{ij}:=M^1_{ij}\cap M^2_{ij}$. Keep in mind $M_{ij}\downarrow_i$,   $M_{ij}\uparrow_j$ and  equally for $(L_{ij})$ and $(K_{ij})$.
\\
\hspace*{6mm}
Note $\Delta^1\cup\Delta^2\subset \cap_i \cup_jL_{ij}
=M^1\cup M^2$. As to the less trivial inclusion $\subset$  of the equality let $x\not\in  
M^1\cup M^2$.
 Then for $l=1,2$ there exists $i_l$ with $x\not\in\cup_jM^l_{i_lj}$. Hence $x\not\in(\cup_jM^l_{ij})\cup(\cup_jM^l_{ij})=\cup_jL_{ij}$ for $i\ge i_1,i_2$, whence $x\not\in \cap_i \cup_jL_{ij}$.
\\
\hspace*{6mm}
Note $\Delta^1\cap\Delta^2\subset \cap_i \cup_jK_{ij}=M^1\cap M^2$. Indeed, clearly 
 $\Delta^1\cap\Delta^2\subset M^1\cap M^2$.   If $x\in M^1\cap M^2$, then, for every $i$, $x\in (\cup_j M^1_{ij})\cap (\cup_j M^2_{ij})$, whence $x\in M^1_{ij_1}\cap M^2_{ij_2}$ for some $j_1,j_2$ and hence $x\in K_{ij}$ for $j\ge j_1,j_2$. Therefore $x\in \cap_i \cup_jK_{ij}$. ---  Conversely, if 
$x\not\in M^1\cap M^2$, then  $x\not\in M^l$ for some $l$. Hence  $x\not\in \cap_i \cup_jK_{ij}$.
\\
\hspace*{6mm} 
Now, by these inclusions, $\hat{m}(\Delta^1)+\hat{m}(\Delta^2)=\hat{m}(\Delta^1\cup\Delta^2)+\hat{m}(\Delta^1\cap\Delta^2)\le 
\hat{m}(M^1\cup M^2)+\hat{m}(M^1\cap M^2)= \hat{m}(M^1)+\hat{m}( M^2)=\hat{m}(\Delta^1)+\hat{m}(\Delta^2)$. It follows  $\hat{m}(\Delta^1\cup\Delta^2)=\hat{m}(M^1\cup M^2)$ and $\hat{m}(\Delta^1\cap\Delta^2)=\hat{m}(M^1\cap M^2)$. This shows that $(L_{ij})\in\mathcal{C}(\Delta^1\cup\Delta^2)$ and $(K_{ij})\in\mathcal{C}(\Delta^1\cap\Delta^2)$.\\
\hspace*{6mm} 
By the last result  one may perform the limits in (3) for each summand of  $G(L_{ij})+G(K_{ij})=G(M^1_{ij})+G(M^2_{ij})$ thus obtaining the additivity of $\hat{G}$.  \\ 
\hspace*{6mm}
(d) The claim is $\lim_{l\to\infty}\hat{G}(\Delta^l)=0$ for  $(\Delta^l)$ in $\hat{\mathcal{Q}}$ with $\Delta^l\downarrow_l\emptyset$, which together with additivity is equivalent to $\sigma$-additivity of $\hat{G}$. Assume $\lim_{l\to\infty}\hat{G}(\Delta^l)>0$. Then there exists a vector $\varphi$ with 
$\langle \varphi, 
\hat{G}(\Delta^l)\varphi\rangle=:2c>0$ for all $l\in\N$. Let $(M^l_{ij})\in\mathcal{C}(\Delta^l)$. Then, according to (3), for  $l$ there exists $i_l$ and for every $i$ a $j_{l,i}$   such that $\langle \varphi, 
G(M^l_{ij})\varphi\rangle\ge c$ for all $i\ge i_l$ and $j\ge j_{l,i}$. 
This and (2) applied to $(M^l_{ij})$ imply that for every $n\in\N$ there are $l_n,i_n,j_n\in\N$ such that $\hat{m}(\Delta^{l_n})\le\frac{1}{n}$ and such that
$L_n:=M^{l_n}_{i_nj_n}$ satisfies
$\langle\varphi,G(L_n)\varphi\rangle\ge c$ and
$m(L_n)=\hat{m}(\Delta^{l_n}\vartriangle L_n\vartriangle \Delta^{l_n})\le \hat{m}(\Delta^{l_n}\vartriangle  L_n)+\hat{m}(\Delta^{l_n})\le\frac{2}{n}$. This  contradicts the assumption that $m$ controls $G$.  \\
\hspace*{6mm}
(e) It remains to show uniqueness of the extension $\hat{G}$. For every vector $\varphi$, 
$M\mapsto \langle \varphi, G(M)\varphi\rangle$ is a finite $\sigma$-additive content on $\mathcal{Q}$. It allows a unique extension to a measure on $\hat{\mathcal{Q}}$ 
(see e.g.\,\cite[5.7]{B68}). Therefore this extension coincides with $\Delta\mapsto \langle \varphi, \hat{G}(\Delta)\varphi\rangle$. This ends the proof.\qed\\

A normalized content  controlling $G$ is easily available, namely

\begin{Lem}\label{NCCPOVC} Let  the Hilbert space be separable and $\{\varphi_n:n\in\N\}$  a total set  of unit vectors.
Then  $m(M):=\sum_n2^{-n}\langle \varphi_n,G(M)\varphi_n\rangle$  defines a normalized content $m$ controlling $G$ in \emph{(\ref{EPOCPOM})}.
\end{Lem}\\
{\it Proof.}
 $m(M_i) \to 0\quad \Leftrightarrow\quad G(M_i)\varphi_n\to 0$ for every $n$ by dominated convergence, since generally $\langle \varphi,G(M)\varphi\rangle=\norm{\sqrt{G(M)}\varphi}^2$ and $\norm{G(M)\varphi}\le \norm{\sqrt{G(M)}\varphi}$ using $\norm{G(M)}\le 1$. Hence
$G(M_i)\varphi\to 0$ for a dense set of vectors $\varphi$, whence $G(M_i)\to 0$.\qed\\

Recall that the unions of finitely many disjoint boxes $\langle a, b[ :=\{x\in\R^d: a_k\le x_k<b_k, k=1,\dots,d\}$, $a,b\in\overline{\R}^d$   form an algebra $\mathcal{Q}^d$ in $\R^d$, which generates the Borel $\sigma$-algebra of $\R^d$ (cf. the proof of (\ref{NGENTLNSLHP})).

\begin{Lem}\label{SAC} Let $m$ be a finite content on $(\R^d,\mathcal{Q}^d)$, which is translation quasi-invariant, i.e., for every $b\in\R^d$ and every sequence $(M_i)_i$ in $\mathcal{Q}^d$
$$m(M_i)\to 0\quad \Leftrightarrow\quad m(b+M_i)\to 0$$
Moreover, for every box $B$ and every $\epsilon>0$ there is a bounded box $B'\subset B$ such that $m(B\setminus B')<\epsilon$. Then $m$ is $\sigma$-additive. 
\end{Lem}\\
{\it Proof.}
(a) Let $B:=\langle a,b[ \,\in \mathcal{Q}^d$ be a non-empty box  with $a,b\in\R^d$. Let $\epsilon >0$. The claim is that there is $b'\in\R^d$ with  $a_k<b'_k<b_k$, $k=1,\dots,d$ such that  for $B':=\langle a,b'[$
$$m(B)\le m(B')+\epsilon$$
 Indeed, let $c\in\R^d$ with $c_k< b_k$, $k=1,\dots,d$ and put $\Delta(c,b):=\langle a,b[\,\setminus \langle a,c[$.  Obviously, the result follows if
 $\inf_{c} m(\Delta(c,b))=0$.   Assume the contrary and a fortiori   $a=(-\infty,\dots,-\infty)$. Then by translation quasi-invariance $\inf_{c}m(d+\Delta(c,b))>0$ for every $d\in\R^d$. Choose $d:=tf$, $t\in\R$, $f:=(1,1,\dots,1)$. Then $d+\Delta(c,b)=\Delta(c+tf,b+tf)$. Since $]0,1[$ is uncountable, there are $\delta>0$ and a strictly monotonic sequence $(t_n)$ in $ ]0,1[$ such that $\inf_{c}m(\Delta(c+t_nf,b+t_nf))\ge \delta $ for all $n$. If $(t_n)$ is increasing then $\Delta\big( b+\frac{1}{2}(t_{n+1}+t_n)f,b+t_{n+1}f\big)$, $n\in\N$ are mutually disjoint and contained in $\langle a,b+f[$. Each set has content $\ge\delta$. Hence additivity of $m$  implies the contradiction $m(\langle a,b+f[\,)=\infty$. If $(t_n)$ is decreasing then consider $\big(\Delta( b+\frac{1}{2}(t_{n+1}+t_n)f, b+t_{n}f )\big)_n$. \\
\hspace*{6mm} 
(b) Due to (a) and by the assumption on $m$ it follows that generally for every non-empty box $B=\langle a,b\,[$ and every $\epsilon>0$ there are $a',b'\in\R^d$ with $a_k\le a'_k<b'_k<b_k$ $\forall$ $k$ and  $m(B)\le m(B')+\epsilon$. Hence the closure of $B'$ is compact and contained in $B$. Now, by a well-known compactness argument we show  for $m$  the continuity from above which is equivalent to $\sigma$-additivity.
 \\
 \hspace*{6mm}  
(c) Let $(M_n)_n$ be a decreasing sequence in $\mathcal{Q}^d$ such that $\delta:=\lim_nm(M_n)>0$. One has to show $\bigcap_n M_n\ne\emptyset$.  
Recall that the figure $M_n$ is the union of finitely many disjoint boxes $B$. Replace every $B$ by  some $B'$ as above forming a new figure $M'_n\subset M_n$ such that 
\begin{equation*}
m(M_n)-m(M_n')\le 2^{-n}\delta 
\end{equation*}
Let $L_n:=M_1'\cap\dots\cap M_n'$. Then $(L_n)_n$ is a decreasing sequence in $\mathcal{Q}^d$ with $\overline{L}_n\subset\overline{M}_n'\subset M_n$. In particular,  $\overline{L}_n$ is compact. It suffices to show 
$\overline{L}_n\ne\emptyset$ for every $n$, since then by the finite intersection property $\emptyset\ne \bigcap_n\overline{L}_n\subset \bigcap_nM_n$.\\
 \hspace*{6mm} 
 Assume $\overline{L}_n=\emptyset$ and hence $L_n=\emptyset$ for some $n$. Then $\delta\le m(M_n)=m(M_n\setminus L_n)=m\big(\bigcup_{k=1}^n(M_n\setminus M_k')\big)\le \sum_{k=1}^nm(M_n\setminus M_k')\le \sum_{k=1}^nm(M_k\setminus M_k')\le \sum_{k=1}^n2^{-k}\delta$, which is a contradiction. Hence the result follows.\qed\\

\begin{Cor}\label{EPOMPOVC} Let the Hilbert space be separable. Let $G$ be a positive-operator-valued content on  $(\R^d,\mathcal{Q}^d)$. Let $U(b)$ be unitary with $U(b)G(M)U(b)^{-1}=G(b+M)$ for all $M\in\mathcal{Q}^d$, $b\in\R^d$. Moreover, let for every box $B$, every vector $\varphi$, and every $\epsilon>0$ exist a bounded box $B'\subset B$ with $\langle \varphi, G(B\setminus B')\varphi\rangle\le \epsilon$.\\
\hspace*{6mm}
 Then there is a unique POM $T$ on the Borel sets of $\R^d$ with $G=T|_{\mathcal{Q}^d}$. If $G$ is projection-valued, then $T$ is a PM. If $S$ is a Borel-algebra homomorphism with $S(\mathcal{Q}^d)\subset \mathcal{Q}^d$ and $U(S)$ unitary with $U(S)G(M)U(S)^{-1}=G\big(S(M)\big)$ for all $M\in\mathcal{Q}^d$, then one has $U(S)T(\Delta)U(S)^{-1}=T\big(S(\Delta)\big)$ for all Borel sets $\Delta\subset \R^d$.
\end{Cor}\\
{\it Proof.} Let $m$ be the normalized content on $\mathcal{Q}^d$ controlling $G$ from (\ref{NCCPOVC}). Then $m$ is translation quasi-invariant. Indeed, let $(M_i)_i$ be a sequence  in $\mathcal{Q}^d$, $b\in\R^d$,  then $\lim_i m(M_i)=0\Rightarrow \lim_i G(M_i)=0\Rightarrow \lim_i G(b+M_i)=\lim_i U(b)G(M_i)U(b)^{-1}=0\Rightarrow \lim_i m(b+M_i)=0$.\\
 \hspace*{6mm} 
 We like to apply (\ref{SAC}). Let $\epsilon>0$. Let $n'\in\N$ with $\sum_{n> n'}2^{-n}<\frac{\epsilon}{2}$. Let $B$ be a box. By assumption there is a bounded box $B_1\subset B$ with 
 $\langle \varphi_1, G(B\setminus B_1)\varphi_1\rangle\le \frac{\epsilon}{2}$.Then, by monotony of $G$, there is a bounded box $B_1\subset B_2\subset B$ with $\langle \varphi_2, G(B\setminus B_2)\varphi_2\rangle\le \frac{\epsilon}{2}$, and so on. Hence there is a bounded box $B'\subset B$ such that $\langle \varphi_n, G(B\setminus B')\varphi_n\rangle\le \frac{\epsilon}{2}$ for all $n=1,\dots,n'$. It follows $m(B\setminus B')<\epsilon$.
Hence $m$ is $\sigma$-additive by (\ref{SAC}).\\
\hspace*{6mm} 
Thus, by (\ref{EPOCPOM}), there is the unique POM $T$ as asserted, which is an PM if $G$ is projection-valued.\\
\hspace*{6mm}
As to the last part of the assertion one has $m(M_i)\to 0\;\Leftrightarrow\; m(S(M_i))\to 0$ for all $(M_i)$ in ${\mathcal{Q}^d}$. Let $\Delta\subset \R^d$ be Borel, $(M_{ij})\in\mathcal{C}(\Delta)$, and $(L_n):=(M_{i_nj_n})$ such that $\hat{m}(\Delta \vartriangle L_n)\to 0$ (cf.\;the proof of (\ref{EPOCPOM}), in particular (1) and (2)). Therefore $\hat{m}\big(S(\Delta) \vartriangle S(L_n)\big)\to 0$, whence easily $\big(S(M_{ij})\big)\in\mathcal{C}\big(S(\Delta)\big)$. Therefore, by (3) of the proof of (\ref{EPOCPOM}), $T\big(S(\Delta)\big)=\lim_i\lim_jT\big(S(M_{ij})\big)=\lim_i\lim_jU(S)T(M_{ij})U(S)^{-1}=U(S)T(\Delta)U(S)^{-1}$.\qed\\

The following example  shows that $G$ in (\ref{EPOMPOVC}) need not be $\sigma$-additive without the assumption that $G$ is determinated by its values at bounded boxes. More precisely, there is a translation-covariant non-$\sigma$-additive projection-valued content $G$ on $(\R^d,\mathcal{Q}^d)$ in any Hilbert space $\ne\{0\}$. Hence, in particular, \cite[Theorem A5]{W62} is not generally valid. \\
\hspace*{6mm}   
   Let $R_n:=\{x:x_i\ge n,  i=1,\dots,d\}, n\in\N$. For $M\in\mathcal{Q}^d$ set $G(M):=I$ if $R_n\subset M$ for some $n$ and $G(M):=0$ otherwise. Now let $M=B_1\cup\dots\cup B_k$ with disjoint boxes $B_l$. The claim is that $R_{n_0}\subset M$ for some $n_0$ if and only if there is just one box $B_l$ containing  some $R_m$. A moment of reflection shows that this implies the additivity of $G$. --- As to the claim note first that at most one box may contain some $R_m$ since the boxes are disjoint. Now let $S:=\{(m,\dots,m):m\in\N, m\ge n_0\}$. As $S\subset M$ there is some $B_l=\langle a,b[$ such that $B_l\cap S$ is infinite. This implies for every $i$ that $m<b_i$ for infinitely many $m\in\N$. Hence $b_i=\infty$ for $i=1,\dots,d$. Thus $R_m\subset B_l$ for $m\ge\max_i a_i$ proving the claim. --- Now let $U(b)$ for $b\in\R^d$ be any unitary operator. Since obviously $G(b+M)=0 \Leftrightarrow G(M)=0$, translational covariance holds. --- Finally, $G$ is not $\sigma$-additive, as e.g.  the boxes $B_n:=\langle (n-1,\dots,n-1),(n,\dots,n)[$, $n\in\N$ are disjoint, their union is $B:=\langle (0,\dots,0),(\infty,\dots,\infty)[$, but $G(B_n)=0$, $G(B)=I$.\\

\section*{List of Notions and Symbols}

$\tilde{\mathcal{P}} =ISL(2,\C)$ $\hfill$  covering  group  of the Poincar\'e group, sec.\,\ref{NWPOWL}, sec.\,\ref{POCOPO}\\
$W$ $\hfill$ denotes a unitary representation of $\tilde{\mathcal{P}}$, sec.\,\ref{NWPOWL}\\
$W^{\textsc{d}}$  $\hfill$ Dirac representation, sec.\,\ref{DS}\\
 $W^{\textsc{w}\chi}$ $\hfill$ representation of the Weyl fermion, sec.\,\ref{OWL}\\

$ISU(2)$ $\hfill$covering group of Euclidean motions, sec.\,\ref{NWPOWL}\\
$U$ $\hfill$ denotes  a unitary representation of $ISU(2)$, sec.\,\ref{NWPOWL}\\
$V$,  $V(t)= \operatorname{e}^{itH}$  $\hfill$  denotes a unitary representation of the time translations, sec.\,\ref{PELS}, sec.\,\ref{CPOL}\\

 $\tilde{\mathcal{P}}_\chi=IST(2)$ $\hfill$ sec.\,\ref{ECLNTLHP}\\

PM  $\hfill$projection valued measure, sec.\,\ref{NWPOWL}\\
WL $\hfill$ Wightman localization, sec.\,\ref{NWPOWL}\\
unitarily related  WL  $\hfill$sec.\,\ref{NWPOWL}\\
NWL $\hfill$Newton-Wigner localization, sec.\,\ref{NWPOWL}, sec.\,\ref{FDNWL}\\
WL with causal time evolution  $\hfill$ sec.\,\ref{CPOL}\\
Poincar\'e covariant WL$\hfill$ sec.\,\ref{GCPCPOL}\\
SCT $\hfill$ system with causal time evolution,  sec.\,\ref{DCRS}\\
unitarily related  SCT $\hfill$ sec.\,\ref{DCRS}\\
causal WL$\hfill$sec.\,\ref{GCPCPOL}\\
causal system  $\hfill$ sec.\,\ref{SeCWL}\\

$E$ $\hfill$ denotes a projection valued measure, sec.\,\ref{NWPOWL}\\
 $E^{\textsc{nw}}$ $\hfill$ projection valued measure of Newton-Wigner localization, sec.\,\ref{NWPOWL}, sec.\,\ref{FDNWL}\\
$X^{\textsc{nw}}$ $\hfill$  Newton-Wigner position operator, sec.\,\ref{NWPOWL}, sec.\,\ref{PCPO}, sec.\,\ref{FDNWL}\\ 
 $E^{\textsc{d}}$   $\hfill$ projection valued measure of Dirac localization, sec.\,\ref{DS}\\
  $X^{\textsc{d}}$   $\hfill$ Dirac position operator, sec.\,\ref{DS}\\
 $E^{\textsc{w}}$ $\hfill$ projection valued measure of Weyl localization, sec.\,\ref{OWL}\\
 $X^{\textsc{w}}$ $\hfill$ Weyl position operator, sec.\,\ref{OWL}\\
 
 $(V,U,E)$ $\hfill$ denotes a WL with causal time evolution, sec.\,\ref{CPOL}\\
$(W,E)$  $\hfill$ denotes a causal system and more generally an SCT,  sec.\,\ref{SeCWL}, sec.\,\ref{DCRS}\\
$(W^{\textsc{d}},  E^{\textsc{d}})$  $\hfill$ Dirac system, sec.\,\ref{DS}\\ 
$(W^{\textsc{w}\chi},E^{\textsc{w}})$ $\hfill$ Weyl system, sec.\,\ref{OWL}\\

POM $\hfill$positive operator valued measure, sec.\,\ref{SePOL}\\
separated POM $\hfill$ sec.\,\ref{SPLSS}\\
POM admitting dilational covariance $\hfill$  sec.\,\ref{POMADC}\\
POL  $\hfill$positive operator valued localization, sec.\,\ref{SePOL}\\
POL with causal time evolution $\hfill$ sec.\,\ref{CPOL}\\
Poincar\'e covariant  POL  $\hfill$ sec.\,\ref{SPCPOL}\\
causal POL $\hfill$sec.\,\ref{GCPCPOL}\\

$T$ $\hfill$ denotes a positive operator valued measure, sec.\,\ref{SePOL}\\
$T^e$ $\hfill$ positive operator valued measure for the Dirac electron, sec.\,\ref{IPOL}\\
 $T^{\chi\eta}$ $\hfill$ positive operator valued measure for the Weyl fermion, sec.\,\ref{IPOL}\\

$(V,U,T)$ $\hfill$ denotes a POL with causal time evolution, sec.\,\ref{CPOL}\\

{\it pos} $\hfill$  superscript denoting the position representation, sec.\,\ref{DCRSPR}\\ 
{\it mom} $\hfill$ superscript denoting the momentum representation, sec.\,\ref{MRTE}\\
{\it can} $\hfill$  superscript denoting the  canonical representation, sec.\,\ref{DCRSPR}, sec.\,\ref{A:IT}\\

$\Delta$,   $\Gamma$   $\hfill$flat spatial region, sec.\,\ref{SePCWL}\\
$\sigma$, $\tau$ $\hfill$spacelike hyperplane, sec.\,\ref{SePCWL}, sec.\,\ref{PPCS}\\
 $\varepsilon$ $\hfill$ spacelike hyperplane $\{0\}\times \R^3$, sec.\,\ref{SPCPOL}\\
  $\mathfrak{S}$   $\hfill$ set of all  flat spatial regions, sec.\,\ref{SPCPOL}\\
 $\Delta_t$,  $\Gamma_\sigma$,   $\Delta_\tau$  $\hfill$    region of influence, sec.\,\ref{CPOL}, sec.\,\ref{GCPCPOL}, sec.\,\ref{PPCS}\\

$\mathcal{F}$ $\hfill$ Fourier transformation sec.\,\ref{FDNWL}\\
$A_\rho$,     $A_{\rho e}$     $\hfill$boost, sec.\,\ref{FDNWL}, sec.\,\ref{GCPCPOL}\\
$\mathfrak{R}$ $\hfill$reference frame,  sec.\,\ref{SePCWL}, sec.\,\ref{FDNWL}, sec.\,\ref{DLB}\\


  $\mathcal{M}$$\hfill$ lattice of  causally complete regions,  sec.\,\ref{HLCCR}\\
  $\mathcal{M}'$ $\hfill$ lattice generated by the non-timelike relation, sec.\,\ref{LGNTLR}, sec.\,\ref{LMMBOR}\\
   $\mathcal{M}^{'borel}$ $\hfill$ sublattice of $\mathcal{M}'$ of the Borel sets, sec.\,\ref{LMMBOR}\\
$(W,F')$,  $\hfill$ denotes a representation (rep) of  $\mathcal{M}^{'borel}$,  sec.\,\ref{SRLBSGNTLR}\\
 $(W',F')$ with $W'=d^{\tilde{\mathcal{P}}}$$\hfill$  rep of $\mathcal{M}^{'borel}$, sec.\,\ref{SRLBSGNTLR}\\
 $\mathfrak{S}_\infty$ $\hfill$  sec.\,\ref{ECLNTLHP}\\

\end{document}